\documentclass[12pt]{article}

\usepackage[utf8]{inputenc}
\usepackage{cancel}
\usepackage{amssymb}
\usepackage{soul}
\usepackage{tabularx}
\usepackage{xcolor}
\usepackage{booktabs}
\usepackage{subcaption} 
\usepackage{colortbl}
\usepackage{authblk}
\DeclareUnicodeCharacter{00A0}{ }
\renewcommand{\arraystretch}{1.3}
\newcolumntype{C}{>{\centering\arraybackslash}X}
\usepackage[T1]{fontenc}
\usepackage{epsfig}
\usepackage{latexsym}
\usepackage{graphicx}
\usepackage{amsmath}
\usepackage{amsfonts}   % if you want the fonts
\usepackage{amssymb}    % if you want extra symbols
\usepackage{float}
\usepackage{bm}
\usepackage{url}
\usepackage{hyperref} 
\usepackage[nodisplayskipstretch]{setspace}
\def\lsim{\raise0.3ex\hbox{$\;<$\kern-0.75em\raise-1.1ex\hbox{$\sim\;$}}}
\def\gsim{\raise0.3ex\hbox{$\;>$\kern-0.75em\raise-1.1ex\hbox{$\sim\;$}}}
\def    \beq            {\begin{equation}}
\def    \eeq            {\end{equation}}
\def    \bea           {\begin{eqnarray}}
\def    \eea           {\end{eqnarray}}

\def \mn{\mu\nu{\rm SSM}}

\def\g2{{\rm GeV}^2}
\def\tb{{\rm tan}\beta}

\def\sw2{sin^2 \theta_w}

\def\a^tau{\alpha_{\tau}}

\def\beq{\begin{equation}}
\def\eeq{\end{equation}}
\def\beqa{\begin{eqnarray}}
\def\eeqa{\end{eqnarray}}

\newcommand{\newc}{\newcommand}
\newc\BR{BR}
\newc{\akappa}{A_{\kappa} }
\newc\deltagmtwo{\delta (g-2)_{\mu}} 
\newc\deltaamu{\Delta a_{\mu}}

\def\anti{\overline}

\def\la{\lambda}
\def\bla{\bm \lambda}
\def\ka{\kappa}

\newc{\haa}{BR\(h_1\to a_1 a_1\)}
\newc{\abb}{BR\(a_1\to b\anti{b}\)}
\newc{\hbb}{BR\(h_1\to b\anti{b}\)}
\newc{\abund}{\Omega h^2}
\newc\bsgamma{b\rightarrow s \gamma }
\newc\bxsgamma{\overline{B}\rightarrow X_{s}\gamma}
\newc\brbsgamma{\BR(\overline{B}\rightarrow X_s\gamma)}

\allowdisplaybreaks

\usepackage{array, makecell}
\usepackage{boldline}
\usepackage[nosort]{cite}
\usepackage[outdir=./]{epstopdf}
\usepackage{cleveref}

\title{\bf{
Impact of Higgs physics on the parameter space of the $\mn$}}
\author[a,b]{Essodjolo Kpatcha\thanks{kpatcha.essodjolo@uam.es}}
\author[c,d]{Daniel~E.~L\'opez-Fogliani\thanks{daniel.lopez@df.uba.ar}}
\author[a,b]{Carlos~Mu\~noz\thanks{c.munoz@uam.es}} 
\author[e]{Roberto~Ruiz~de~Austri\thanks{rruiz@ific.uv.es}}

\affil[a]{Departamento de F\'{\i}sica Te\'{o}rica, Universidad Aut\'{o}noma de Madrid (UAM),
Campus de Cantoblanco, 28049 Madrid, Spain}
\affil[b]{Instituto de F\'{\i}sica Te\'{o}rica (IFT) UAM-CSIC, 
  Campus de Cantoblanco, 28049 Madrid, Spain}
  \affil[c]{Instituto de F\'isica de Buenos Aires UBA \& CONICET, Departamento de F\'isica,
 Facultad de Ciencia Exactas y Naturales, Universidad de Buenos Aires, 
1428 Buenos Aires, Argentina}
\affil[d]{
{Pontificia Universidad Cat\'olica Argentina, 
1107 Buenos Aires, Argentina}}
  \affil[e]{Instituto de F\'{\i}sica Corpuscular CSIC--UV, 
c/ Catedr\'atico Jos\'e Beltr\'an 2, 46980 Paterna,
 Valencia, Spain}

\date{}

\begin{document}
%\epstopdfsetup{outdir=./}

\maketitle

\begin{abstract}
Given the increasing number of experimental data, 
together with the precise measurement of the
properties of the Higgs boson at the LHC, the parameter space of
supersymmetric models starts to be constrained. We carry out a detailed analysis of this issue in the framework of the $\mu\nu$SSM. 
In this model, three families of right-handed neutrino superfields are present in order to solve the $\mu$ problem and simultaneously reproduce neutrino physics. 
The new couplings and sneutrino vacuum expectation values in the $\mu\nu$SSM induce new mixing of states, and, in particular,
the three right sneutrinos can be substantially mixed with the neutral Higgses.
After diagonalization, 
the masses of the corresponding three singlet-like eigenstates can be smaller or larger than the mass of the Higgs, or even degenerated with it.
We analyze whether these situations are still compatible with the experimental results.
To address it we scan the parameter space of the Higgs sector of the model. In particular, we sample the $\mu\nu$SSM using a powerful likelihood data-driven method, paying special attention to satisfy {the constraints
coming from Higgs sector measurements/limits (using {\tt HiggsBounds} and {\tt HiggsSignals}),
as well as a class of flavor observables
such as $B$ and $\mu$ decays, while muon $g-2$ is briefly discussed}. We find that large regions of the parameter space of the $\mu\nu$SSM are viable, containing an interesting phenomenology that could be probed at the LHC.
\end{abstract}


%%%%%%%%%%%%%%%%%%%%%%%%%%%%%%%%%%%%%%%%%%%%%%%%%%%%%%%%%%%%%%%%%%
\clearpage
\tableofcontents 
%\listoffigures
%\listoftables
%%%%%%%%%%%%%%%%%%%%%%%%%%%%%%%%%%%%%%%%%%%%%%%%%%%%%%%%%%%%%%%%%%

%\clearpage

\section{Introduction}
\label{introduction}

The measurements of the properties and signal rates of the discovered scalar boson at the LHC~\cite{Chatrchyan:2012xdj,Aad:2012tfa}, indicate that it is compatible with the expectations of the Standard Model (SM). Besides, no hints for new physics have been detected yet despite of numerous searches and tremendous efforts of the experimental collaborations. As a consequence, extensions of the SM such as low-energy Supersymmetry (SUSY) are being severely constrained, namely the parameter space of SUSY models 
is shrinking considerably. 
This renders the detailed analyses of Higgs properties, signal rates and couplings to SM particles as important as the search for new particles.

Concerning the latter, the search of SUSY particles 
has been focused mainly on signals with missing transverse energy inspired by $R$-parity conserving (RPC) models, such as the
Minimal Supersymmetric Standard 
Model (MSSM)~\cite{Nilles:1983ge,Haber:1984rc,Gunion:1984yn,Martin:1997ns}. There, significant bounds on sparticle masses have been obtained~\cite{Tanabashi:2018oca}, especially for strongly interacting sparticles whose masses must be above about 1 TeV.
%\cite{Aaboud:2017vwy, Sirunyan:2017kqq}.
Less stringent bounds of about 100 GeV have been obtained for weakly interacting sparticles, with the bino-like neutralino basically not constrained due to its small pair production cross section.
Qualitatively similar results have also been obtained in the analysis of simplified 
$R$-parity violating (RPV) scenarios with trilinear lepton- or baryon-number 
violating 
terms~\cite{Barbier:2004ez}, assuming a single channel available for the decay of the LSP into leptons.
However, this assumption is not possible in other RPV scenarios, such as 
the `$\mu$ from $\nu$' supersymmetric standard 
model ($\mn$)~\cite{LopezFogliani:2005yw},
where the several decay branching ratios (BRs) of the lightest supersymmetric particle (LSP) significantly decrease the signal.
This implies that the extrapolation of the usual bounds on sparticle masses to the $\mn$ is not applicable. 
For example, it was shown in Refs.~\cite{Lara:2018rwv,Kpatcha:2019gmq} that the LEP lower bound on masses of slepton LSPs of about 90 GeV obtained in trilinear RPV~\cite{Abreu:1999qz,Abreu:2000pi,Achard:2001ek,Heister:2002jc,Abbiendi:2003rn,Abdallah:2003xc} is not applicable in the $\mn$.
%The most recent analyses of signals at the LHC for LSP candidates in the
%$\mn$ have been dedicated 
%to the left sneutrino~\cite{Ghosh:2017yeh,Lara:2018rwv,Kpatcha:2019gmq}, and to 
For the bino LSP,\footnote{The phenomenology of a neutralino LSP was analyzed 
in the past in 
Refs.~\cite{Ghosh:2008yh,Bartl:2009an,Ghosh:2012pq,Ghosh:2014ida}.}
%In the latter case, 
only a small region of the parameter space of the $\mn$ 
was excluded~\cite{Lara:2018zvf} when the left sneutrino is the next-to-LSP (NLSP) and hence a suitable source of binos. In particular, 
the region of bino (sneutrino) masses $110-150$ ($110-160$) GeV.
%it was found a tri-lepton signal compatible with the local excess reported by ATLAS~\cite{Aaboud:2018sua}.
%If this excess were due to a statistical fluctuation,\footnote{{The recent emulated recursive jigsaw reconstruction~\cite{ATLAS-CONF-2019-020} confirmed the 3$\sigma$ excess with 36 fb$^{-1}$, but sees only a small 1.27$\sigma$ excess of data with respect to predictions with full 139 fb$^{-1}$.}} 

%\noindent
{Concerning Higgs physics, the $\mn$ expands the singlet superfield of the NMSSM~\cite{Ellwanger:2009dp,Maniatis:2009re} to three right-handed neutrino superfields~\cite{LopezFogliani:2005yw,Escudero:2008jg}.
In the NMSSM, various works, using different methods (see e.g.~\cite{Cao:2012fz,Gherghetta:2012gb,Cao:2012yn,King:2014xwa,Beskidt:2016egy,Zhou:2016nkd,Beskidt:2019mos}), have been dedicated to sample the parameter space in the light of a given set experimental data, and vast regions have been explored.} 
%However concerning the methods employed, so far to our knowledge a somewhat simplified approaches have been widely used. Concretely, on the one hand these methods are either based on random grid scans of subsets of parameter space imposing the experimental constraints based on the “pass-or-fail” methods. With these methods the viable benchmark points are simply requested to lie within a certain confidence level cuts of the corresponding experimental data set. On the other hand the approaches that relied on a chi-square methods remain however of limited use as they do not allow to perform a full scan over all relevant parameters.
%The point is that these approaches do not allow for a probabilistic interpretation of results or to fully explore properties of a given model of interest or to quantitatively and qualitatively investigate regions of the parameter space that are still compatible with data. As a results these methods can optimistically allow or exclude some corners of the parameter spaces.
%For example \cite{King:2014xwa} in the context of the NMSSM found that for $\tan\beta \lesssim 10$ and
%$0.05 \lesssim \lambda \lesssim 0.5 $ the SM Higgs boson cannot be the lightest scalar and we 
%wonder whether this is general or just a limitation of the method they employed.
%\noindent
In this work, we use a powerful likelihood data-driven method based on the algorithm called {\tt MultiNest}~\cite{Feroz:2007kg,Feroz:2008xx,Feroz:2013hea} for sampling the Higgs sector of the $\mn$.
Since three families of right-handed neutrino superfields are present in the model in order to solve the $\mu$ problem and simultaneously reproduce neutrino physics, 
the new couplings and sneutrino vacuum expectation values (VEVs) produce a substantial mixing among
the three right sneutrinos and the doublet-like Higgses.
Although a detailed analysis of this sector was performed in Ref.~\cite{Escudero:2008jg},
finding viable regions that avoid false minima and tachyons, as well as fulfill the Landau pole constraint, it was carried out  
prior the discovery of the SM-like Higgs boson, and therefore the issue of reproducing Higgs data was missing.
In Ref.~\cite{Ghosh:2014ida}, this issue was taking into account 
to perform an analytical estimate of all the new two-body decays for the SM-like Higgs in the presence of light scalars, pseudoscalars and neutralinos.  
More recently, in Refs.~\cite{Biekotter:2017xmf,Biekotter:2019gtq}, in addition to performing the complete one-loop renormalization of the neutral scalar sector of the $\mn$, interesting benchmark points (BPs) with singlet-like eigenstates lighter than the SM-like Higgs boson were studied.

Given the increasing data 
including the properties of the SM-like Higgs and the {exclusion limits 
on scalars from extended Higgs-sneutrino sectors} 
provided by the combined 7-, 8- and 13-TeV searches at the LHC, 
and also by other results such as flavor observables, 
it appears relevant to re-investigate the $\mn$ parameter space %\cite{Escudero:2008jg,Ghosh:2014ida} 
to simultaneously accommodate this new scalar and its properties, the exclusion limits and to explore the phenomenological consequences respecting various experimental results. 
To carry this out, 
%In this work, we use a powerful likelihood data-driven method based on the algorithm called {\tt Multinest}~\cite{Feroz:2008xx}, for sampling the Higgs sector of the 
%\R{`$\mu$ from $\nu$' supersymmetric standard model ($\mn$)~\cite{LopezFogliani:2005yw}
%which offers a framework for a simultaneous solution to the $\mu$-problem of the minimal supersymmetric standard model (MSSM) and explanation of neutrino masses and mixing angles by adding to the MSSM superfields content three generations of right-handed neutrino superfields, singlets under the SM gauge group.}
the likelihood data-driven method used in our analysis presents advantages over traditional ones 
such as those based on random grid scans or chi-square methods, 
since it is much more efficient in the computational effort required to explore a parameter space. Also, since it uses a Bayesian approach, it allows to take easily into account all relevant sources of uncertainties in the likelihood. In addition, given the accumulation of data from various experimental collaborations, this method provides a convenient approach to qualitatively explore beyond standard models compared to simplified methods.

%\noindent
The paper is organized as follows. In Sec.~\ref{themodel}, we will briefly review the $\mn$ and its relevant parameters for our analysis of the Higgs sector.
This sector will be studied in detail in Sec.~\ref{SMhiggs}, where the mixing among doublet-like Higgses and right and left sneutrinos will be explained.
We will pay special attention to accommodate the correct mass of the SM-like Higgs, depending on the values of the couplings $\lambda$ among right sneutrinos and doublet-like Higgses, and the masses of the singlet-like eigenestates. 
%, present its main features namely the superpotential and neutral scalar potential. 
%In Sec.~\ref{EW-stability}, we will study the existence and stability of the minimum. 
Subsequently, in Sec.~\ref{methodology} we will discuss the strategy that we will employ to
perform scans searching for points of the parameter space of our scenario compatible with current experimental data on Higgs physics, as well as flavor observables.
The results of these scans will be presented 
in Section~\ref{results-scans}, and applied to show 
that there are large viable regions of the parameter space of the $\mn$.
Our conclusions and prospects for future work are left for Section~\ref{conclusions}. 
Finally, useful formulae, figures, and BPs are given in the Appendices. In Appendix~\ref{Apendix:Sneutrino-masses}, the Higgs-right sneutrino mass submatrices are written. 
In Appendix~\ref{figuresscans}, results from the $\lambda-\kappa$ plane are shown for different values of the other parameters, using several figures for each scan performed.
In Appendix~\ref{BPs}, several BPs showing interesting characteristics of the model are given.

%\section{The $\mn$}
\section{The $\mn$}

% and neutrino physics}
\label{themodel}

The $\mn$~\cite{LopezFogliani:2005yw,Escudero:2008jg} is 
a natural extension of the MSSM where the $\mu$ problem is solved and, simultaneously, the neutrino data can be 
reproduced~\cite{LopezFogliani:2005yw,Escudero:2008jg,Ghosh:2008yh,Bartl:2009an,Fidalgo:2009dm,Ghosh:2010zi}. This is obtained through the presence of trilinear 
terms in the superpotential involving right-handed neutrino superfields $\hat\nu^c_i$, which relate the origin of the $\mu$-term to the origin of neutrino masses and mixing angles. 
The simplest superpotential of the $\mn$~\cite{LopezFogliani:2005yw,Escudero:2008jg,Ghosh:2017yeh} with three right-handed neutrinos is the following: 
%\bea
%W &=&
%\sum_{a,b}\sum_{i,j,k}\left\{
%\epsilon_{ab} \left(
%Y_{e_{ij}}
%\, \hat H_d^a\, \hat L^b_i \, \hat e_j^c +
%Y_{d_{ij}} 
%\, 
%\hat H_d^a\, \hat Q^{b}_{i} \, \hat d_{j}^{c} 
%+
%Y_{u_{ij}} 
%\, 
%\hat H_u^b\, \hat Q^{a}
%\, \hat u_{j}^{c}
%\right)\right.
%\nonumber\\
%&+& 
%\left.\epsilon_{ab} \left(
%Y_{{\nu}_{ij}} 
%\, \hat H_u^b\, \hat L^a_i \, \hat \nu^c_j
%-
%\lambda \, \hat \nu^c_i\, \hat H_u^b \hat H_d^a
%\right)
%+
%\frac{1}{3}
%\kappa_{ijk}
%\hat \nu^c_i\hat \nu^c_j\hat \nu^c_k\right\}\,,
%\label{superpotential}
%\eea
\bea
W =
&&
%&=&
\sum_{a,b}\sum_{i,j}
\epsilon_{ab} \left(
Y_{e_{ij}}
%Y^e_{ij} 
\, \hat H_d^a\, \hat L^b_i \, \hat e_j^c +
Y_{d_{ij}} 
%Y^d_{ij} 
\, 
%\delta_{\alpha\beta}\, 
\hat H_d^a\, \hat Q^{b}_{i} \, \hat d_{j}^{c} 
+
Y_{u_{ij}} 
%Y^u_{ij} 
\, 
%\delta_{\alpha\beta}\, 
\hat H_u^b\, \hat Q^{a}_i
%_{i\alpha} 
\, \hat u_{j}^{c}
\right)
\nonumber\\
% &+&
% \epsilon_{ab} Y^{\nu}_{ij} \, \hat H_u^b\, \hat L^a_i \, \hat \nu^c_j -
&+& 
\sum_{a,b}\sum_{i,j}\epsilon_{ab} \left(
Y_{{\nu}_{ij}} 
%Y^{\nu}_{i} 
\, \hat H_u^b\, \hat L^a_i \, \hat \nu^c_j
-
%\epsilon_{ab}
\lambda_i \, \hat \nu^c_i\, \hat H_u^b \hat H_d^a
\right)
+
\sum_{i,j,k}
\frac{1}{3}
\kappa_{ijk}
\hat \nu^c_i\hat \nu^c_j\hat \nu^c_k\,,
\label{superpotential}
\eea
%where the summation convention is implied on repeated indices, 
with 
$a,b=1,2$ $SU(2)_L$ indices with
$\epsilon_{ab}$ the totally antisymmetric tensor $\epsilon_{12}=1$,
and $i,j,k=1,2,3$ the usual family indices of the SM. 

The simultaneous presence of the last three terms in 
Eq.~\eqref{superpotential} makes it impossible to assign $R$-parity charges consistently to the 
right-handed neutrinos ($\nu_{iR}$), thus producing explicit RPV (harmless for proton decay). Note nevertheless, that in the limit of neutrino Yukawa couplings
$Y_{{\nu}_{ij}} 
\to 0$, $\hat \nu^c_i$ can be identified in the superpotential 
%of Eq.~(\ref{superpotential}) 
as 
pure singlet superfields without lepton number, similar to the singlet of the NMSSM,
%~\cite{Ellwanger:2009dp,Maniatis:2009re}, 
and therefore $R$ parity is restored.
%NMSSM
%where one extra singlet is added to the spectrum of the MSSM and $R_p$ is not broken.
Thus, $Y_{\nu}$ are the parameters which control the amount of RPV in the $\mn$, and as a consequence
%in the superpotential of Eq.~(\ref{superpotential}), 
this violation is small
since the size of $Y_{\nu}\lsim 10^{-6}$ is determined by the electroweak-scale 
seesaw of the $\mn$~\cite{LopezFogliani:2005yw, Escudero:2008jg}.

The tree-level neutral scalar potential $V=V_{\text{soft}} + V_F + V_D$, receives in addition to the usual $D$- and $F$-term contributions that can be found e.g. in Refs.~\cite{Escudero:2008jg,Ghosh:2017yeh}, the following contribution from the soft SUSY-breaking Lagrangian:
%The couplings of the superpotential together with the corresponding soft SUSY-breaking terms give rise to the tree-level neutral scalar potential $V=V_{\text{soft}} + V_F + V_D$, where the $F$- and $D$-terms can be found e.g. in Refs.~\cite{Escudero:2008jg,Ghosh:2017yeh} and
\bea
V_{\text{soft}}  =&&
\sum_{i,j,k}\left(
T_{{\nu}_{ij}} \, H_u^0\,  \widetilde \nu_{iL} \, \widetilde \nu_{jR}^* 
- T_{{\lambda}_{i} }\, \widetilde \nu_{iR}^*\, H_d^0  H_u^0
+ \frac{1}{3} T_{{\kappa}_{ijk}} \, \widetilde \nu_{iR}^* 
\widetilde \nu_{jR}^* 
\widetilde \nu_{kR}^*\
+
\text{h.c.} \right)
\nonumber\\
&+&
\sum_{i,j}
\left(
m_{\widetilde{L}_{ij}}^2\widetilde{\nu}_{iL}^* \widetilde\nu_{jL}
+
%\left(m_{\widetilde{L}_L}^2\right)_{ij} \widetilde{\nu}_{iL}^* \widetilde\nu_{jL}
%+
m_{\widetilde{\nu}_{ij}}^2 \widetilde{\nu}_{iR}^* \widetilde\nu_{jR}
\right)
%+
%\left(m_{\widetilde{\nu}_R}^2\right)_{ij} \widetilde{\nu}_{iR}^* \widetilde\nu_{jR} 
+
m_{H_d}^2 {H^0_d}^* H^0_d + m_{H_u}^2 {H^0_u}^* H^0_u.
\label{akappa}
\eea
If we follow the assumption based on the breaking of supergravity that all the trilinear parameters are proportional to their corresponding couplings in the superpotential~\cite{Brignole:1997dp}, we can write
\bea
%T^{e}_{ij} &=& A^{e}_{ij} Y^{e}_{ij}\ , \;\;
%T^{d}_{ij} = A^{d}_{ij} Y^{d}_{ij}\ , \;\;
%T^{u}_{ij} = A^{u}_{ij} Y^{u}_{ij}\ ,
%\label{tyukawa}
%\\
T_{{\nu}_{ij}} = A_{{\nu}_{ij}} Y_{{\nu}_{ij}}, \;\;
T_{{\lambda}_i}= A_{{\lambda}_i} \lambda_i, \;\;
T_{{\kappa}_{ijk}}= A_{{\kappa}_{ijk}} \kappa_{ikj},
\label{tmunu}
%\\
%T^{\lambda}_{ijk} &=& A^{\lambda}_{ijk} \lambda_{ijk}\ , \;\;
%T^{\lambda'}_{ijk}= A^{\lambda'}_{ijk} \lambda'_{ijk}\ ,
\label{trilinear}
\eea
and the parameters $A$ substitute the $T$ as the most representative. We will use both type of parameters in our discussions.
{It is worth noticing here that we do not use the summation convention on repeated indices throughout this work.}

%
%and the $F$- and $D$-terms can be found e.g. in Refs.~\cite{Escudero:2008jg,Ghosh:2017yeh}.
The soft terms of Eq.~(\ref{akappa})
%of the order of the TeV 
induce the electroweak symmetry breaking (EWSB) in the $\mn$.
The minimization equations, with the choice of CP conservation,\footnote{The $\mu\nu$SSM with spontaneous 
CP violation was studied in Ref.~\cite{Fidalgo:2009dm}.}
can also be found in Refs.~\cite{Escudero:2008jg,Ghosh:2017yeh}.
For
neutral Higgses ($H^0_{u,d}$) and right ($\widetilde \nu_{iR}$) and 
left ($\widetilde \nu_{iL}$) sneutrinos defined as
\bea
%H_d^0 &=& \frac{1}{\sqrt 2} (H_{dR}+ i H_{dI}) + v_d\ ,
H_d^0 
=
%&=&
\frac{1}{\sqrt 2} \left(H_{d}^\mathcal{R} + v_d + i\ H_{d}^\mathcal{I}\right),\quad
%\\
%\label{vevd}
%\nonumber
%\\
H^0_u 
=
%&=& 
\frac{1}{\sqrt 2} \left(H_{u}^\mathcal{R}  + v_u +i\ H_{u}^\mathcal{I}\right) ,  
\\
\label{vevu}
\nonumber 
\\
\widetilde{\nu}_{iR} 
=
%&=&
%     \frac{1}{\sqrt 2} (\widetilde{\nu}^c_{iR}+ i \widetilde{\nu}^c_{iI})+\nu_i^c,  
      \frac{1}{\sqrt 2} \left(\widetilde{\nu}^{\mathcal{R}}_{iR}+ v_{iR} + i\ \widetilde{\nu}^{\mathcal{I}}_{iR}\right),  
      \quad
%\\
%\label{vevnuc}
%\nonumber
%\\
  \widetilde{\nu}_{iL} 
  =
%  &=& 
  \frac{1}{\sqrt 2} \left(\widetilde{\nu}_{iL}^\mathcal{R} 
  + v_{iL} +i\ \widetilde{\nu}_{iL}^\mathcal{I}\right),
\label{vevnu}
\eea
the following vacuum expectation values (VEVs) are developed: 
\begin{eqnarray}
\langle H_{d}^0\rangle = \frac{v_{d}}{\sqrt 2},\quad 
\langle H_{u}^0\rangle = \frac{v_{u}}{\sqrt 2},\quad 
\langle \widetilde \nu_{iR}\rangle = \frac{v_{iR}}{\sqrt 2},\quad 
\langle \widetilde \nu_{iL}\rangle = \frac{v_{iL}}{\sqrt 2},
\end{eqnarray}
with $v_{iR}\sim$ TeV {whereas 
%$v_i\sim Y_{\nu} v_u\lsim 10^{-4}$ GeV
$v_{iL}\sim 10^{-4}$ GeV}. The latter result is because of 
the contributions proportional to 
$Y_{\nu}$ to the $v_{iL}$ minimization equations. They enter through $V_F$ and
$V_{\text{soft}}$ (assuming $T_\nu$ as in Eq.~(\ref{trilinear})), and 
are small due to the electroweak-scale seesaw mentioned before that
determines $Y_{\nu}\lsim 10^{-6}$.
%$Y_{\nu} \lsim 10^{-6}$ 
%which enters through $V_F$ and
%$V_{\text{soft}}$ (assuming $T_\nu$ as in Eq.~(\ref{trilinear})).
%with the size of $Y_{\nu}$ determined by the electroweak-scale 
%seesaw of the $\mn$~\cite{LopezFogliani:2005yw, Escudero:2008jg}.
Note in this respect that the last term in the superpotential~\eqref{superpotential} generates dynamically Majorana masses for the right-handed neutrinos~$\sim$ TeV:
\bea
%m_{{\mathcal M}_{ij}}
{\mathcal M}_{ij}
=\sum_k {2}\kappa_{ijk} \frac{v_{kR}}{\sqrt 2}.
\label{majorana}
\eea
%$m_{{\mathcal M}_{ij}}={2}\kappa_{ijk} \frac{v_{kR}}{\sqrt 2}\sim$ TeV.
On the other hand, the fifth term generates an effective
$\mu$-term~$\sim$ TeV:
%\mu=\la_i \frac{v_{iR}}{\sqrt 2}\sim$ TeV.
\begin{equation}
\mu=\sum_i \la_i \frac{v_{iR}}{\sqrt 2}.
\label{mu}    
\end{equation}

Given the structure of the scalar potential,
the free parameters in the neutral scalar sector of the $\mn$
at the low scale $M_{EWSB}= \sqrt{m_{\tilde t_l} m_{\tilde t_h}}$, 
%where $m_{\tilde t_l}$ and $m_{\tilde t_h}$ 
%correspond to the lightest and heaviest stop mass eigenvalues measured at $M_{EWSB}$,
are therefore:
$\lambda_i$, $\kappa_{ijk}$, $Y_{{\nu}_{ij}}$, $m_{H_{d}}^{2}$, $m_{H_{u}}^{2}$,  
$m_{\widetilde{\nu}_{ij}}^2$,
$m_{\widetilde{L}_{ij}}^2$, 
$T_{{\lambda}_i}$, $T_{{\kappa}_{ijk}}$ and $T_{{\nu}_{ij}}$.
Using diagonal sfermion mass matrices, in order to avoid the
strong upper bounds upon the intergenerational scalar mixing (see e.g. Ref.~\cite{Gabbiani:1996hi}), from the eight minimization conditions with respect to $v_d$, $v_u$,
$v_{iR}$ and $v_{iL}$
one can eliminate
the above
%low-energy 
soft masses 
%$m_{H_{d}}^{2}$, $m_{H_{u}}^{2}$, $m^2_{\widetilde{\nu}_{iR}}$ and
%$m^2_{\widetilde{L}_{iL}}$ 
in favor
of the VEVs.
In addition, using 
$\tan\beta\equiv {v_u}/{v_d}$ 
%$\tanb$ 
and the SM Higgs VEV, $v^2 = v_d^2 + v_u^2 + \sum_i v^2_{iL}={4 m_Z^2}/{(g^2 + g'^2)}\approx$ (246 GeV)$^2$
with the electroweak gauge couplings estimated at the $m_Z$ scale by
$e=g\sin\theta_W=g'\cos\theta_W$, 
one can determine the SUSY Higgs 
VEVs, $v_d$ and $v_u$. 
Since $v_{iL} \ll v_d, v_u$, one has 
$v_d\approx v/\sqrt{\tan^2\beta+1}$.
Besides, we can use diagonal neutrino Yukawa couplings,
%i.e. $Y_{{\nu}_{i}}\equiv Y_{{\nu}_{ii}}$ and vanishing otherwise.
%In fact,
since data on neutrino physics
%~\cite{Capozzi:2017ipn,deSalas:2017kay,deSalas:2018bym,Esteban:2018azc} 
can easily be reproduced at tree level in the $\mn$
%~\cite{LopezFogliani:2005yw,Escudero:2008jg,Ghosh:2008yh,Bartl:2009an,Fidalgo:2009dm,Ghosh:2010zi} 
with such structure,
%~\cite{Ghosh:2008yh,Fidalgo:2009dm,Gomez-Vargas:2016ocf,Kpatcha:2019gmq}, 
as we will discuss below.
Finally, assuming for simplicity that the 
off-diagonal elements of $\kappa_{ijk}$
and soft trilinear parameters $T$ vanish,
we are left with  
the following set of variables as independent parameters in the neutral scalar sector:
\bea
\lambda_i, \, \kappa_{i}, \, Y_{{\nu}_{i}}, \, \tan\beta, \, v_{iL}, \, v_{iR}, \, T_{{\lambda}_i}, \, T_{{\kappa}_{i}}, \, T_{{\nu}_i},
\label{softfree}
\eea
where $\kappa_i\equiv\kappa_{iii}$,
$Y_{{\nu}_i}\equiv Y_{{\nu}_{ii}}$,
$T_{{\nu}_i}\equiv T_{{\nu}_{ii}}$ and
$T_{{\kappa}_i}\equiv T_{{\kappa}_{iii}}$.
Note that now the Majorana mass matrix is diagonal, with the non-vanishing entries given by
\bea
{\mathcal M}_{i}
={2}\kappa_{i} \frac{v_{iR}}{\sqrt 2}.
\label{majorana2}
\eea
The rest of (soft) parameters of the model, namely the following gaugino masses,
%$M_{1,2,3}$, 
scalar masses, 
%$m^2_{\tilde Q_i,\tilde u^c_i,\tilde d^c_i,\tilde e^c_i}$,
and trilinear parameters:
%, $A_{u_{ij},d_{ij},e_{ij}}$, 
%
\bea
M_1, \, M_2,\, M_3, \, m_{\tilde Q_{iL}},\, 
m_{\tilde u_{iR}}, \, m_{\tilde d_{iR}}, \,
m_{\tilde e_{iR}}, \,
T_{u_{i}}, \, T_{d_{i}}, \, T_{e_{i}},
\label{freeparameterssoft}
\eea
are also taken as free parameters and specified at low scale.

A further sensible simplification that we will also use in the next sections when necessary,
is to assume universality of the parameters in Eq.~(\ref{softfree}) with the exception of those connected directly with neutrino physics such as $Y_{{\nu}_i}$ and $v_{iL}$, that must be non-universal to generate correct neutrino masses and mixing angles. Neither 
we will impose universality for $T_{{\nu}_i}$, since they are connected with sneutrino physics as we will discuss in the next section, and a hierarchy of masses in that sector can be phenomenologically interesting~\cite{Kpatcha:2019gmq}.
We are then left with the following set of low-energy free parameters:
\bea
\lambda, \, \kappa, \, Y_{{\nu}_{i}}, \, \tan\beta, \, v_{iL}, \, v_{R}, \, T_{\lambda}, \, T_{\kappa}, \, T_{\nu_i},
\label{freeparameters00}
\eea
where $\lambda\equiv \lambda_i$, $\kappa\equiv \kappa_i$, $v_R\equiv v_{iR}$,
$T_{\lambda}\equiv T_{{\lambda}_i}$ and 
$T_\kappa \equiv T_{{\kappa}_i}$.
%and $T_{{\nu}_i}=T_{{\nu}$. 
In this case, the three non-vanishing Majorana masses
are equal ${\mathcal M}_i={\mathcal M}$, with
\bea
{\mathcal M}
={2}\kappa \frac{v_{R}}{\sqrt 2},
\label{majorana3}
\eea
and the $\mu$-term is given by
\begin{equation}
\mu=3\la \frac{v_{R}}{\sqrt 2}.
\label{mu2}    
\end{equation}

%If we follow the usual assumption based on the breaking of supergravity that all the trilinear parameters are proportional to their corresponding couplings~\cite{Brignole:1997dp}, then we can write for example:
%\bea
%T_{{\nu}_{i}} = A_{{\nu}_{i}} Y_{{\nu}_{i}}, \;\;
%T_{{\lambda}_i}= A_{{\lambda}_i} \lambda_i, \;\;
%T_{{\kappa}_{i}}= A_{{\kappa}_{i}} \kappa_{i},
%\label{tmunu}
%\label{trilinear}
%\eea
%and the parameters $A$ substitute the $T$ as the most representative. We will use both type of parameters throughout this work.
%It is worth noticing here that we do not use the summation convention on repeated indices throughout this work.

%Let us finally remark that 
The new couplings and sneutrino VEVs in the $\mn$ induce new mixing of states.
%, and in particular there are eight neutral scalars and seven neutral pseudoscalars (Higgses-sneutrinos).
The associated mass matrices were studied in detail in
Refs.~\cite{Escudero:2008jg,Bartl:2009an,Ghosh:2017yeh}.
Summarizing, 
%in the case of one $\hat\nu^c$
%right-handed neutrino superfield, 
there are 
eight neutral scalars and pseudoscalars (Higgses-sneutrinos),
where after rotating away the pseudoscalar would be Goldstone boson we are left with seven pseudoscalar states. There are also
eight charged scalars (charged Higgses-sleptons),
five charged fermions (charged leptons-charginos), and
ten neutral fermions (neutrinos-neutralinos). 
%{In our analysis of the electroweak sector below, we are mainly interested in the scalars/pseudoscalars and neutral fermions.} 

Since reproducing neutrino data
%~\cite{Capozzi:2017ipn,deSalas:2017kay,deSalas:2018bym,Esteban:2018azc} 
is an important asset of the $\mn$, in the following we will briefly review this issue.
%a few words on the subject are worth it.
The neutral fermions have the flavor 
composition 
$(\nu_{iL},\widetilde B^0,\widetilde W^0,\widetilde H_{d}^0,\widetilde H_{u}^0,\nu_{iR})$. Thus,
with the low-energy bino and wino soft masses, $M_1$ and $M_2$, of the order of the TeV, and similar values for $\mu$ and $\mathcal{M}$ as discussed above, this generalized seesaw
%mixing left and right-handed neutrinos with neutralinos
produces three light neutral fermions dominated by the left-handed neutrino ($\nu_{iL}$) flavor composition. 
In fact, as mentioned before,
%Because of this structure, 
data on neutrino physics~\cite{Capozzi:2017ipn,deSalas:2017kay,deSalas:2018bym,Esteban:2018azc} can easily be reproduced at tree level~\cite{LopezFogliani:2005yw,Escudero:2008jg,Ghosh:2008yh,Bartl:2009an,Fidalgo:2009dm,Ghosh:2010zi}, even with diagonal Yukawa couplings~\cite{Ghosh:2008yh,Fidalgo:2009dm}, i.e.
$Y_{{\nu}_{ii}}=Y_{{\nu}_{i}}$ and vanishing otherwise as used already in 
Eq.~(\ref{softfree}).
A simplified formula 
for the effective mixing mass matrix of the 
light neutrinos is~\cite{Fidalgo:2009dm}:
\begin{eqnarray}
\label{Limit no mixing Higgsinos gauginos}
(m_{\nu})_{ij} 
%\simeq 
=&&
%\frac{Y_{{\nu}_{i}} Y_{{\nu}_{j}} (v_u/\sqrt 2)^2}
\frac{m_{{\mathcal{D}_i}} m_{{\mathcal{D}_j}} }
%{6\sqrt 2 \kappa v_{R}}
{3{\mathcal{M}}}
                   (1-3 \delta_{ij})
                   -\frac{(v_{iL}/\sqrt 2)(v_{jL}/\sqrt 2)}
                   {2M^{\text{eff}}}
%                   \frac{v_{iL}}{\sqrt 2} \frac{v_{jL}}{\sqrt 2}
\nonumber\\
&-&
\frac{
m_{{\mathcal{D}_i}}m_{{\mathcal{D}_j}}
}
{2M^{\text{eff}}}
\frac{1}{3\lambda\tan\beta}
\left(
%m_{{\mathcal{D}_i}}\frac{v_{jL}}{\sqrt 2}
%   +m_{{\mathcal{D}_j}}\frac{v_{iL}}{\sqrt 2}\right)
\frac{v_{iL}/\sqrt 2}{m_{{\mathcal{D}_i}}}
   +
   \frac{v_{jL}/\sqrt 2}{m_{{\mathcal{D}_j}}}
   + \frac{1}{3\lambda\tan\beta}
   \right),
\label{neutrinoph}
  \end{eqnarray}     
where we have defined the Dirac mass for neutrinos as
\bea
m_{{\mathcal{D}_i}}\equiv Y_{{\nu}_{i}} 
\frac{v_u}{\sqrt 2},
\label{diracmass}
\eea 
and 
\begin{eqnarray}
\label{effectivegauginomass}
 M^{\text{eff}}\equiv M -
 \frac{\left(v/\sqrt 2\right)^2
 %(v/\sqrt 2)^2
 }
 {2\mu
 \left({\mathcal{M}} \frac{v_R}{\sqrt 2}+ 2\lambda 
 \left(\frac{v}{\sqrt 2}\right)^2 
 \frac{\tan\beta}{1+\tan^2\beta}\right)
        %\ 3 \lambda v_R
        }
        \left[2{\mathcal{M}} 
        \frac{v_R}{\sqrt 2} 
        %\left(\frac{v}{\sqrt 2}\right)^2 
        \frac{\tan\beta}{1+\tan^2\beta}
        +\lambda \left(\frac{{v}}{\sqrt 2}\right)^2\right],
        \label{first}
%\nonumber\\
\end{eqnarray} 
with
\begin{eqnarray}
\label{effectivegauginomass2}
\frac{1}{M} = \frac{g'^2}{M_1} + \frac{g^2}{M_2}.
%\nonumber\\
\label{gauginom}
\end{eqnarray} 
Here
%$M_{1,2}$ are the bino and wino soft masses, 
%and
%$v^2 = v_d^2 + v_u^2 + \sum_i v^2_{i}={4 m_Z^2}/{(g^2 + g'^2)}\approx$ (246 GeV)$^2$.
%${M}= \frac{M_1 M_2}{g'^2 M_2 + g^2 M_1}$. 
%and 
we have assumed universal $\lambda_i = \lambda$, $v_{iR}= v_{R}$, and
$\kappa_{i}=\kappa$ as in Eq.~(\ref{freeparameters00}).
The first term of Eq.~(\ref{Limit no mixing Higgsinos gauginos}) is generated
through the mixing 
of $\nu_{iL}$ with 
$\nu_{iR}$-Higgsinos, and the other two
%The rest of them 
also include the mixing with gauginos.
{These are the so-called $\nu_{R}$-Higgsino seesaw and gaugino seesaw, respectively~\cite{Fidalgo:2009dm}.}

We are then left in general with the following subset of variables of
Eqs.~(\ref{softfree}) and~(\ref{freeparameterssoft})
as independent parameters in the neutrino sector:
\bea
\lambda_i,\, \kappa_i,\, Y_{\nu_i}, \tan\beta, \, v_{iL}, \, v_{iR}, \, M,
\label{freeparametersn}
\eea
{In the numerical analyses of the next sections, it will be enough for our purposes to consider
the sign convention where all these parameters are positive.}

Under several assumptions, the formula for $(m_{\nu})_{ij}$ can be further simplified.
Notice first that the third term is inversely proportional to $\tan\beta$,
%in brackets in Eq.~(\ref{Limit no mixing Higgsinos gauginos}) 
%are proportional to $v_d$, 
and therefore negligible in the limit of large or even moderate $\tan\beta$ provided that 
$\lambda$ is not too small. Besides,
the first piece inside the brackets in the second term of Eq.~(\ref{first}) %for $M^{\text{eff}}$
%in brackets in Eq.~(\ref{effectivegauginomass}) 
is also negligible in this limit, and for typical values of the parameters involved in the seesaw also the second piece, thus
$M^{\text{eff}}\sim M$.
Under these assumptions, the second
term for $(m_{\nu})_{ij}$ 
%in Eq.~(\ref{Limit no mixing Higgsinos gauginos}) 
is generated only through the mixing of left-handed neutrinos with gauginos. Therefore, we arrive to a very simple formula where only the first two terms survive with $M^{\text{eff}}= M$ {in Eq.~(\ref{neutrinoph}):
\begin{eqnarray}
\label{Limit no mixing Higgsinos gauginos}
(m_{\nu})_{ij} 
%\simeq 
=
%\frac{Y_{{\nu}_{i}} Y_{{\nu}_{j}} (v_u/\sqrt 2)^2}
\frac{m_{{\mathcal{D}_i}} m_{{\mathcal{D}_j}} }
%{6\sqrt 2 \kappa v_{R}}
{3{\mathcal{M}}}
                   (1-3 \delta_{ij})
                   -\frac{(v_{iL}/\sqrt 2)(v_{jL}/\sqrt 2)}
                   {2M}.
%                   \frac{v_{iL}}{\sqrt 2} \frac{v_{jL}}{\sqrt 2}
%\nonumber\\
%&-&
%\frac{
%m_{{\mathcal{D}_i}}m_{{\mathcal{D}_j}}
%}
%{2M^{\text{eff}}}
%\frac{1}{3\lambda\tan\beta}
%\left(
%\frac{v_{iL}/\sqrt 2}{m_{{\mathcal{D}_i}}}
 %  +
 %  \frac{v_{jL}/\sqrt 2}{m_{{\mathcal{D}_j}}}
  % + \frac{1}{3\lambda\tan\beta}
  % \right),
\label{neutrinoph2}
  \end{eqnarray}     
%Under these assumptions, 
%the second
%term for $(m_{\nu})_{ij}$ 
%is generated only through the mixing of left-handed neutrinos with gauginos. Therefore, we arrive to a very simple formula where only the first two terms survive with $M^{\text{eff}}= M$ in Eq.~(\ref{neutrinoph}). It 
This expression} can be used to understand easily the
seesaw mechanism in the $\mn$ in a qualitative way.
%As we can understand from these equations, neutrino physics in the $\mn$ is
%closely related to the parameters and VEVs of the model, 
%since the values chosen for them must reproduce current data on neutrino masses
%and mixing angles.
From this discussion,
it is clear that $Y_{\nu_i}$, $v_{iL}$ and $M$ are crucial parameters to determine the neutrino physics.

Let us finally point out that the accommodation of the SM-like Higgs boson discovered at the LHC is mandatory for any SUSY model. 
In the next section, we will review this subject in some detail in the context of the $\mn$. In order to carry it out, we will study the Higgs sector of the model and its enhanced particle content.
Although the  parameter space of the model given by Eqs.~(\ref{softfree}) and~(\ref{freeparameterssoft}) is large, we will see in the next sections that some of these parameters are not relevant for the study of the Higgs sector, as also occurs for the neutrino sector, and therefore its analysis is simplified.

%%%%%%%%%%%%%%%%%%%%%%%%%%%%%%%%%%%%%%%%%%%%%%%%%%%%%%%%%
%\section{The SM-like Higgs in the \texorpdfstring{$\mn$}{Lg}}
\section{
%The SM-like Higgs in the $\mn$
The 
%neutral 
Higgs sector of the $\mn$}
\label{SMhiggs}
%%%%%%%%%%%%%%%%%%%%%%%%%%%%%%%%%%%%%%%%%%%%%%%%%%%%%%%%%

%As already mentioned, 
In the $\mn$, doublet-like Higgses are mixed with left and right
sneutrinos, giving rise to $8\times 8$ (`Higgs') mass matrices for scalar and pseudoscalar states.
However, the $5\times 5$ Higgs-right sneutrino submatrix is almost decoupled from the $3\times 3$ left sneutrino submatrix due to the small values of $Y_{{\nu}_{ij}}$ and $v_{iL}$ in the off-diagonal entries~\cite{Escudero:2008jg,Ghosh:2008yh}.
Thus, to accommodate the SM-like Higgs in the $\mn$, we can focus on the analysis of the Higgs-right sneutrino mass submatrix.
The tree-leel entries of the scalar and pseudoscalar mass submatrices are shown in Appendix~\ref{Apendix:Sneutrino-masses}
using the parameters of Eq.~(\ref{softfree}).
Upong diagonalization of the scalar submatrix, one obtains the
SM-like Higgs, the heavy doublet-like neutral Higgs, and three singlet-like states.
Similarly, upon diagonalization of the pseudoscalar submatrix,
and after rotating away the pseudoscalar would be Goldstone boson, we are left with the doublet-like neutral pseudoscalar, and three singlet-like pseudoscalar states. 

In what follows, we will concentrate on the study of the properties of the SM-like Higgs, given the amount of experimental data, and on the
ones of the new states with respect to the MSSM, i.e. the right sneutrino-like states.
Although not relevant for accommodating the SM-like Higgs, for completeness we will also review the left sneutrino states and the charged Higgs sector.

%In the following, we will see that the left sneutrinos are basically decouple from the rest of states, and that therefore they are not relevant for the crucial issue of accommodating the SM-like Higgs. 

%Then we will discuss how to disentangle the latter from the singlet-like states upon
%diagonalization.
%Although similar to the left sneutrinos, the charged Higgs sector of the model is not relevant for accommodating the SM-like Higgs, we will briefly review it at the end for completeness. 

\subsection{The SM-like Higgs}
\label{smlike}

%First of all, it is crucial to notice that in the $8\times 8$ mass matrices, the $5\times 5$ Higgs-right sneutrino submatrix is almost decoupled from
%the $3\times 3$ left sneutrino submatrix due to the very small values of $Y_{{\nu}_{ij}}$ and $v_{iL}$ in the off-diagonal entries \cite{Escudero:2008jg,Ghosh:2008yh}.
As explained before, we can focus on the analysis of the Higgs-right sneutrino mass submatrix of Appendix~\ref{SubAppendix-scalarmasses}
to accommodate the SM-like Higgs in the $\mn$. 
%The scalar and pseudoscalar mass submatrices are shown in 
%Appendix~\ref{Apendix:Sneutrino-masses}
%using the parameters of Eq.~(\ref{softfree}).
Through the mixing with the right sneutrinos, which 
appears through $\la_i$, the tree-level mass of the lightest doublet-like Higgs 
receives an extra contribution with respect to the MSSM.
We want to 
emphasize that this analysis has a notable similarity with that of the NMSSM 
%Higgs sector
(although in the NMSSM one has only one singlet), however 
RPV and an enhanced particle content offer a novel and unconventional
phenomenology for 
the Higgs-right sneutrino sector of the $\mn$~\cite{Escudero:2008jg,Ghosh:2008yh,Bartl:2009an,
Bandyopadhyay:2010cu,Fidalgo:2011ky,Ghosh:2012pq,Ghosh:2014rha,Ghosh:2014ida,Biekotter:2017xmf,Biekotter:2019gtq}.
%
%~\cite{Escudero:2008jg,Ghosh:2008yh,Bartl:2009an,
%Bandyopadhyay:2010cu,Fidalgo:2011ky,Ghosh:2012pq,Ghosh:2014rha,Ghosh:2014ida,Ghosh:2017yeh,Lara:2018rwv,Lara:2018zvf,Biekotter:2017xmf,Biekotter:2019gtq,Kpatcha:2019gmq}.
%After rotating away the pseudoscalar would be Goldstone boson, we are left with seven pseudoscalar states. 
%$m^{tree}_{h}$
%From the above discussion, it is obvious that the $5\times 5$ Higgs-right sneutrino submatrix is the relevant one for analyzing the SM-like Higgs in the $\mn$.
%Since reproducing correctly its mass is a crucial task for this work,  
%we will discuss now in some detail how to get it.
Taking into account all the contributions, the mass of the SM-like Higgs in the $\mn$ can be schematically written as~\cite{Escudero:2008jg,Ghosh:2014ida}:
%
%Through the mixing with the right sneutrinos, which 
%appears through $\la_i$, the tree-level mass of the lightest doublet-like %Higgs $m^{tree}_{h}$
%receives an extra contribution
%in the $\mu\nu$SSM in such a way that the SM-like Higgs mass can be written as
%%%%%%%%%%%%%%%%%%%%%%%%%%%%%%%%%%%%%%%%%%%%%%%%%%%%%%%%%
\bea
m^{2}_{h} = 
m^2_{0h} 
%\cos^2 2\beta + \left(\frac{v}{\sqrt 2}\right)^2 {\bm \lambda}^2 \sin^2 2\beta
%\left( m^2_Z  \cos^2 2\beta + \frac{3}{2} {\lambda}^2 v^2 \sin^2 2\beta \right)  
+  \Delta_{\text{mixing}} + \Delta_{\text{loop}},
\label{boundHiggs}
\eea
%%%%%%%%%%%%%%%%%%%%%%%%%%%%%%%%%%%%%%%%%%%%%%%%%%%%%%%%%%%%%%%%%%%%%
% with ${\bm \la}\equiv \sqrt{\la^2_1 + \la^2_2 + \la^2_3}=\sqrt{3}\la$.
where  
\bea
m^{2}_{0h} =
m^2_Z \cos^2 2\beta + 
%\left(\frac{v}{\sqrt 2}\right)^2 
({v}/{\sqrt 2})^2\
{\bm \lambda}^2 \sin^2 2\beta\;
%\left( m^2_Z  \cos^2 2\beta + \frac{3}{2} {\lambda}^2 v^2 \sin^2 2\beta \right)  
%+  \Delta_{\text{mixing}} + \Delta_{\text{loop}},
\label{boundHiggs2}
\eea
corresponds to neglect the mixing of the SM-like Higgs with the other states in the mass squared matrix, 
$\Delta_{\text{mixing}}$ encodes those mixing effects
lowering (raising) the mass if it mixes with heavier (lighter) states, and $\Delta_{\text{loop}}$ refers to the radiative corrections. 
%\R{In Eq.~(\ref{boundHiggs2}), $v^2 = v_d^2 + v_u^2 + \sum_i v^2_{iL}={4 m_Z^2}/{(g^2 + g'^2)}\approx$ (246 GeV)$^2$
%with the electroweak gauge couplings estimated at the $m_Z$ scale by
%$e=g\sin\theta_W=g'\cos\theta_W$, and $\tan\beta \equiv v_u/v_d$.
%Since $v_{iL} \ll v_d, v_u$ we have
%$v_d\approx v/\sqrt{\tan^2\beta+1}$.} 
Note that $m^{2}_{0h}$ contains 
two terms, where the first is characteristic of the MSSM and the second of the $\mn$ with 
\begin{equation}
%{\bm \la}\equiv \sqrt{\la^2_1 + \la^2_2 + \la^2_3}=\sqrt{3}\la,
{\bm \la}\equiv \left({\sum_i \la^2_i}\right)^{1/2}=\sqrt{3}\ \la,
\label{lambda123}
\end{equation}
where the last equality is obtained if one assumes universality of the parameters
$\lambda_i=\lambda$.
 
We can write $m^{2}_{0h}$ in a more
elucidate form for our discussion below as
\bea
m^{2}_{0h} = {m^2_Z}
\left\{
\left(\frac{1 - {\rm tan}^2\beta}{1 + {\rm tan}^2\beta}\right)^2  +  
%8(g^2+g'^2)^{-1}\
%\frac{4}{m^2_Z}
\left(\frac{v/\sqrt 2}{m_Z}\right)^2\
{\bm \lambda}^2 
\left(\frac{{\rm 2\ tan\beta}}{1 + {\rm tan}^2\beta}\right)^2
\right\},
\label{boundHiggs1}
\eea
where the factor
%8(g^2+g'^2)^{-1}\simeq 14.5$, 
%$\left(\frac{v/\sqrt 2}{m_Z}\right)^2\approx 3.63$,
$({v/\sqrt 2 m_Z})^2\approx 3.63$,
and 
we see straightforwardly that the second term 
%contribution characteristic of the NMSSM the extra piece of contribution 
grows with small tan$\beta$ and large 
${\bm \la}$.
%For convenience, we have explicitly introduced in Eq.~(\ref{boundHiggs}) the term $\Delta_{\text{mixing}}$ encoding the mixing effects on the SM-like Higgs mass. In addition, the term $\Delta_{\text{loop}}$ refers to the radiative corrections. 
%In the following we denote by $m^{tree}_{h}$ that the terms in the brackets represent the tree level mass without the mixing of the SM-like Higgs boson.
In the case of the MSSM this term 
%in Eq.~(\ref{boundHiggs1}) 
is absent, hence the maximum possible tree-level mass is about $m_Z$ for
$\tb \gg 1$ and, consequently, a contribution 
from loops is essential to reach the target of a SM-like Higgs in the mass region around 125 GeV. 
This contribution is basically determined by the soft parameters
$T_{u_3}, m_{\widetilde u_{3R}}$ and $m_{\widetilde Q_{3L}}$.
%\bea
%T_{u_3}, \, m_{\widetilde u_{3R}}, \,
%m_{\widetilde Q_{3L}},
%\label{freeparameterstop}
%\eea
%which therefore must be added to the list of (tree-level) 
% \R{parameters of Eq.~(\ref{freeparameters})} crucial for the analysis.
On the contrary, 
%as has already been mentioned in
%Ref.~\cite{Escudero:2008jg}, 
in the $\mu\nu$SSM one can reach this mass solely 
with the tree-level contribution for large 
values of ${\bm \la}$~\cite{Escudero:2008jg}.
Following the work of Ref.~\cite{Ghosh:2014ida}, we choose for this analysis three regions in ${\bm \la}$ values.
%following the work of Ref.~\cite{Ghosh:2014ida}:
%as we will discuss in some detail below.
% figure~\ref{fig:lamHigg}, 
In particular, for convenience of the discussion of Sec.~\ref{results-scans},
where the last equality of Eq.~(\ref{lambda123}) is used, our regions are:

%For the choice of the scans, although inspired by the ranges of $\bla$ ($\equiv \sqrt 3 \lambda$) discussed in Section~\ref{SMhiggs}, we will choose slightly different values for convenience of the discussion.

\vspace*{0.25cm}
\noindent
{\it (a) Small to moderate}
%${\bm \la}$, $\textit{i.e.}$ 
($0.01 \leq {\bm \la}/\sqrt 3 < 0.2$)

\noindent 
In this range, 
the maximum value of $m_{0h}$ using Eq.~(\ref{boundHiggs1}) with
$\bla/\sqrt 3=0.2$ goes as $\approx 78.9$~GeV for $\tb=2$, which is $\approx 18$ GeV
more compared to a similar situation in the MSSM. 
It is thus essential to have additional contributions to raise $m^{2}_{0h}$
up to around 125 GeV.
A possible source of extra tree-level mass can arise
when the right sneutrinos are lighter compared to the lightest
doublet-like Higgs. 
In this situation, the 
later  
feels a push away effect from
the former
states characterized by $\Delta_{\text{mixing}}>0$,   
pushing $m_h$ a bit further towards 125 GeV.
Unfortunately, for most of this range
of $\bla$ values, away from the upper end, the push-up effect is normally small 
owing to the small singlet-doublet mixing which is driven by $\lambda_i$
%$\bla$ 
(see Eq.~(\ref{mixingevenR2}) of Appendix) \cite{Escudero:2008jg,Fidalgo:2011ky}.
The additional contribution to accommodate the 125 GeV 
doublet-like Higgs is coming then from loop effects.
The situation is practically similar to that of the MSSM, where
large masses for the third-generation squarks and/or a large trilinear $A$-term
%soft-SUSY breaking 
are essential \cite{Hall:2011aa,Heinemeyer:2011aa,Draper:2011aa}. 
A small trilinear $A_{u_3}$-term is possible only by decoupling the scalars to at least 
5 TeV \cite{Draper:2011aa}.
%\R{CHECK THESE NUMBERS WITH THE OUTPUT OF OUR CODE}. 
A light third generation squark, especially a stop, 
is natural in the 
so-called maximal mixing scenario \cite{Haber:1996fp},
where
{$|X_{u_3}/m_{\widetilde Q_{3L}}| \approx \sqrt{6}$ with $X_{u_3}\equiv A_{u_3}
- \mu/\tan\beta$.}
%All in all, we conclude that, in addition to 
%the \R{parameters of Eq.~(\ref{freeparameters})} crucial for the tree-level analysis, the loop effects introduce other relevant parameters:
%\bea
%T_{u_3}, \, m_{\widetilde u_{3R}}, \,
%m_{\widetilde Q_{3L}},
%\label{freeparameterstop}
%\eea
%Of course, this is also valid for the other ranges of $\bla$ to be discussed below.

These issues indicate that the novel signatures
from SUSY particles (e.g., from a light stop or sbottom)
are less generic in this region of $\bla$. 
{Nevertheless, novel differences are feasible for Higgs decay phenomenology,
especially in the presence of singlet-like lighter states~\cite{Bartl:2009an,Bandyopadhyay:2010cu,Fidalgo:2011ky,Ghosh:2012pq,Ghosh:2014rha,Ghosh:2014ida,Biekotter:2017xmf,Biekotter:2019gtq}.}

\vspace*{0.25cm}
\noindent
{\it (b) Moderate to large} ($0.2 \leq \bla /\sqrt 3 < 0.5$)

\noindent
For this range of $\bla$
values, $m_{0h}$ can go beyond $m_Z$, especially
for $\tb\lsim 5$ and $\bla/\sqrt 3 \gsim 0.29$. 
%This is also clear from figure~\ref{fig:lamHigg}. 
% In fact, depending on $\bla$,
% in this region the maximum of $\mht$ can remain close to the 125 GeV
% target. 
For example with $\bla/\sqrt 3\approx 0.4$, $\tb=2~(5)$ gives
$m_{0h}\sim 112~(96)$ using Eq.~(\ref{boundHiggs1}). 
This 
is $\approx 100\%~(14\%)$ enhancement compared to the MSSM scenario
with the same $\tb$. 
% Thus, one needs $\sim 12\%~(30\%)$ contribution
% from other sources to reach the 125 GeV milestone. 
%However, for $\bla=0.2$ and $\tb=2~(5)$, $m_{0h}$ is estimated as
%$\sim 61~(85)$~GeV.
% and one needs rather large, $\sim 100\%~(47\%)$
%contribution over the tree-level mass to achieve 125 GeV.
%One 
%should note that $\bla=0.2$ and $0.7$ are translated as
%$\la\sim 0.12$ and $\sim 0.4$.
%, assigning a universality for $\la_i$.
%With an intermediate value, say $\bla=0.5$, $\tb=2~(5)$
%gives $m_{0h}$ as $\sim 88~(90)$~GeV.
% and thus, $\sim 40\%$ extra contribution over the $\mht$ is needed. 
% It is interesting
% to see from the last calculation and also from figure \ref{fig:lamHigg}
% that the $\bla=0.5$ line is almost overlapping to the $M_Z$ line
% and consequently magnitude of $\mht$ or the amount of extra
% contribution to reach 125 GeV remains practically
% the same for all $\tb$ values. Being quantitative, as
% $\tb$ changes from 2 to 10, the requirement of an extra
% contribution over the $\mht$ to reach the goal of 125 GeV
% changes by an amount of $\sim 4\%$. 
Note that this value $\bla\approx 0.7$ ($\lambda\approx$ 0.4) is the maximum
possible value of ${\bm \la}$ maintaining its perturbative
nature up to the scale of a grand unified theory (GUT), 
$M_{\text{GUT}}\sim 10^{16}$ GeV.
As discussed in Ref.~\cite{Escudero:2008jg}, using the
renormalization group equations (RGEs) for $\lambda$ and $\kappa$
between $M_{\text{GUT}}$ and the low scale $\sim$ 1 TeV,
neglecting the contributions from the top and gauge couplings, one can arrive straightforwardly to the simple formula:
$2.35\ \kappa^2 + 1.54\ \bla^2\lsim 1$.
This gives the bound $\bla\lsim 0.8$, similar but slightly larger than the one of 0.7 mentioned before. Nevertheless, one should expect a final bound slightly stronger when all contributions to the RGEs are taken into account.
The numerical analysis indicates that a better approximate formula is
\bea
2.77\ \kappa^2 + 2\ \bla^2\lsim 1,
\label{perturbativity1}
\eea
which produces the bounds $\bla\lsim 0.7$ and $\kappa\lsim 0.6$.
%, since this condition imposes at low scale approximately
%\bea
%2.35\ \kappa^2 + 1.54\ \bla^2\lsim 1.
%\label{perturbativity11}
%\eea
%Notice that this translates in the bound for each coupling 
%$\lambda_i=\lambda\lsim 0.7/\sqrt 3\approx 0.4$~\cite{Escudero:2008jg}, assigning a universality for $\lambda_i$ as in the last equality of Eq.~(\ref{lambda123}).
%It is straightforward to arrive to this formula following the arguments and renormalization group equations (RGEs) for $\lambda$ and $\kappa$ of Ref.~\cite{Escudero:2008jg}.
%For $\kappa = 0$ ($\lambda = 0$), Eq.~(\ref{perturbativity1}) gives the bound 
%$\bla\lsim 0.8$ ($\kappa\lsim 0.65$). Nevertheless,
%this equation is obtained neglecting the contributions from the top and gauge couplings to the RGEs, and therefore one should expect final bounds slightly stronger. The numerical analysis of Ref.~\cite{Escudero:2008jg} indicates 
%\bea
%2.77\ \kappa^2 + 2\ \bla^2\lsim 1,
%\label{perturbativity1}
%\eea
%producing the bounds $\bla\lsim 0.7$ and $\kappa\lsim 0.6$.

%Perturbativity up to the GUT scale imposes the following rough constraint on $\kappa$:
%$\kappa^2 + \bla^2\leq (0.7)^2$.
For this region of $\bla$ the singlet-doublet mixing is no
longer negligible as we will see in the next subsection, particularly as 
$\bla/\sqrt 3\to 0.4$. Thus,
a state lighter than 125 GeV with the leading singlet composition
appears difficult without a certain degree of tuning of the other parameters,
e.g. $\ka_i$, $v_{iR}$, $T_{\ka_i}$, $T_{\lambda_{i}}$, etc. 
In this situation,
the extra contribution to the tree-level value of $m_h$ is favourable through a push-up
action from the singlet states compared to small to 
moderate $\bla$ scenario. 
%However, a sizable doublet
%impurity can make it hard for these states to escape
%the collider constraints. The situation is ameliorated
%with $\bla$ as small as possible in the range studied.
%say around $0.2$ or $0.3$.
Once again a contribution from the loops is needed to reach
the 125 GeV target. However, depending on the values
of $\bla$ and $\tb$ the requirement sometime is much softer
compared to small to moderate $\bla$ scenario. 
Thus the
necessity of very heavy third-generation squarks and/or
large trilinear soft-SUSY breaking term may not be so
essential for this region \cite{Hall:2011aa}. 
%For example, 
%for the scenario studied in Ref. \cite{Ghosh:2012pq},
%where $\tb=3.7$ and $\la_i=0.11$ (i.e., $\bla \approx 0.2$), one needs 
%$A_{u_3}=2.4$ TeV and stop masses about $1$ TeV.
%
%Moving towards $\bla\sim 0.7$, on the contrary, room for 
%the third-generation squarks lighter than 1 TeV
%is possible. 
%
%For example, with $\tb=2$ and $\bla=0.7$, stop masses and 
%A-terms of about $300$ GeV are sufficient to raise the 
%Higgs mass to 125 GeV \cite{Hall:2011aa}. \R{CHECK THIS NUMBERS WITH THE OUTPUT OF OUR CODE}.
It is also worth noticing that the naturalness is therefore improved 
with respect to the MSSM or smaller values of $\bla$.

When $\bla/\sqrt 3\to 0.5$, $m_{0h}$
as evaluated from Eq.~(\ref{boundHiggs1}) can be larger than 125 GeV.
For example, for the upper bound of this range
$\bla/\sqrt 3= 0.5$, $\tb= 2$ gives $m_{0h}\sim 132$ GeV.
In this case we have to relax the idea of perturbativity up to the GUT scale, as we will discuss below.

\vspace*{0.25cm}
\noindent
{\it (c) Large ${\bm \la}$} ($0.5 \leq \bla /\sqrt 3 < 1.2$)

\noindent 
%One can  relaxes the idea of perturbativity up to the GUT scale.
Assuming e.g. a scale
of new physics around $10^{11}$ GeV, and following similar analytical arguments as above using the RGEs,
%Eq.~(\ref{perturbativity11}),
the perturbative limit gives approximately $1.48\ \kappa^2 + 0.96\ \bla^2\lsim 1$
producing the bounds $\bla\lsim 1$ and $\kappa\lsim 0.82$.
%\bea
%1.48\ \kappa^2 + 0.96\ \bla^2\lsim 1, 
%\label{perturbativity0}
%\eea
%producing the bounds 
%$\bla\lsim 1$ and $\kappa\lsim 0.82$.
For $\bla$, taking into account the contributions from the top and gauge couplings to the RGEs as above, one can
find numerically~\cite{Escudero:2008jg} the final bound $\bla\lsim 0.88$, i.e. 
$\bla/\sqrt 3 \lsim 0.5$.

%gives $\bla/\sqrt 3 \sim 0.58$.
%(i.e., $\la\sim 0.58$). 
Pushing the scale of new physics further below to 10~TeV,
the approximate analytical analysis
gives the perturbative limit 
\bea
0.25\ \kappa^2 + 0.14\ \bla^2\lsim 1, 
\label{perturbativity2}
\eea
producing now the bounds
$\bla\lsim 2.6$  (i.e. $\bla/\sqrt 3\lsim 1.5$) and $\kappa\lsim 2$.
%this limit gives $\bla/\sqrt 3\sim 1.2$.
%(i.e., $\la\sim 1.2$) \cite{Escudero:2008jg}. 
Given that the full numerical analysis produces typically stronger bounds,
we will use 
$\bla/\sqrt 3\lsim 1.2$ in our scan of Sec.~\ref{results-scans}.
A similar scenario 
in the context of the NMSSM has been popularised as $\la$-SUSY 
\cite{Barbieri:2006bg}. The constraint in this case~\cite{Farina:2013fsa} is slightly different than ours
%from the one in Eq.~(\ref{perturbativity2}), 
because of the presence of only one singlet. 
%$0.26 \kappa^2 + 0.17 \bla^2\lsim 1$ \cite{Farina:2013fsa}, 
%\bea
%0.26\ \kappa^2 + 0.17\ \bla^2\lsim 1, 
%\label{perturbativity2}
%\eea
%implying the upper bound $\kappa\lsim 2$.

In this region of $\bla$
values, $m_{0h}$
as evaluated from Eq.~(\ref{boundHiggs1})
can remain well above 125 GeV even up to $\tb\sim 8$ for
$\bla/\sqrt 3\sim 1.2$. 
%For $\bla/\sqrt =0.58$, a similar analysis gives $\tb\sim 2$ 
%as the upper limit. 
For $\bla/\sqrt 3 =0.58$, 
$m_{0h}$
%maximum of $\mht$ 
for $\tb=2,\,5$ and $10$
is estimated as $\sim 150$~GeV, $108$~GeV and $\sim 96$~GeV,
respectively. With $\bla/\sqrt 3=1.2$ these numbers increase further,
for example, $\sim 113$ GeV when $\tb=10$. The requirement of
an extra contribution to reach the target of 125 GeV is thus rather
small and even negative in this corner of the parameter space unless $\tb$
goes beyond 10 or 15 depending on the values of $\bla$.
A singlet-like state lighter than 125 
GeV is difficult in this corner of
the parameter space due to the large singlet-doublet mixing.
In fact even if one manages to get a scalar lighter
than 125 GeV with parameter tuning, a push-up action 
can produce a sizable effect to push the mass of the lightest
doublet-like state beyond 125 GeV, especially for 
$\tb \lsim 10$ taking $\bla/\sqrt 3=1.2$. Moreover, a huge doublet
component makes these light states hardly experimentally acceptable.
In this region of the parameter space a heavy singlet-like 
sector is more favourable which can push $m_h$ down towards
125 GeV, due to $\Delta_{\text{mixing}}<0$. 
In addition, 
for such a large $\bla$ value, new loop effects
from the right sneutrinos proportional to $\bla^2$
can also give a sizeable negative contribution
%generate an additional enhancement 
~\cite{Degrassi:2009yq,Gherghetta:2012gb}.
A set of very heavy singlet-like states, even with
non-negligible doublet composition is 
also experimentally less constrained.

It is needless to mention that the amount of the loop correction
is much smaller in this region compared to the two previous
scenarios. 
%For example, with $\tb=10$, one needs a loop
%effect $\sim 11\%$ and $30\%$ with $\bla/\sqrt 3=1.2$ and $0.58$,
%respectively. One should compare this with the 
%maximum value of $\bla$ keeping perturbative nature up to
%the GUT scale, i.e. $0.7$, where one needs $\sim 35\%$ contribution 
%over the tree-level mass for $\tb=10$. 
Following the above discussion {\it (b)} for large values of $\bla$, this 
region of the parameter space also favours third-generation 
squarks lighter than $1$ TeV, which can be produced
with enhanced cross sections and can lead to novel signatures
of this model with RPV at the LHC, 
%Note that the light third generation of squarks
%is still allowed by the LHC results, see e.g. Refs.~\cite{ATLAS-susy13,CMS-susy13}.
%
%This feature can produce new signatures at colliders
%with RPV for this region of $\bla$ values, 
even when the singlet-like states remain heavier, as stated earlier.

\vspace{0.25cm}

In Sec.~\ref{results-scans}, we will analyze these $\bla$ regions
using three scans, and we will check how much room is left for new physics 
%in the $\mn$ 
in the light of the current precise measurements of the SM-like Higgs properties.
Let us now study the right sneutrino-like sector which, as pointed out before, is crucial to determine the properties of the SM-like Higgs in the $\mn$.

\subsection{The right sneutrino-like states}
\label{rsls}

From the scalar and pseudoscalar mass submatrices in Appendix~\ref{Apendix:Sneutrino-masses},
it is clear 
%from these mass matrices
that $\kappa_i$ and $T_{\kappa_i}$ 
are crucial parameters to determine the masses of the singlet-like states, originating from the self-interactions.
The remaining parameters $\lambda_i$ and $T_{\lambda_i}$ ($A_{\lambda_i}$ assuming the supergravity relation $T_{\lambda_i}= \lambda_i A_{\lambda_i}$ of Eq.~(\ref{trilinear})) not only appear in the said interactions, but also control the mixing between the singlet and the doublet states and hence, contribute in determining the mass scale.
Note that the contributions of the parameters $T_{\nu_i}$ are negligible assuming $T_{\nu_i}=Y_{\nu_i} A_{\nu_i}$,
%The contributions proportional to 
%$T_{\nu_i}$ are multiplied by $v_u$ and are therefore typically small
%compared to the others
%as long as
%$v_{iR} \gg v_u$, even assuming e.g. $T_{\nu_i}\sim T_{\kappa_i}$.
%Actually, given the small values of the neutrino Yukawas, $T_{\nu_i}\ll T_{\kappa_i}$  under the assumption of Eq.~(\ref{trilinear}).
given the small values of neutrino Yukawas.
%the trilinear parameters are proportional to their corresponding Yukawa couplings, 
%i.e.
%$T_{\nu}=A_{\nu}Y_{\nu}$, as occurs in the breaking of supergravity.
We conclude, taking also into account the discussion below Eq.~(\ref{boundHiggs1}), that the relevant independent low-energy parameters
in the Higgs-right sneutrino sector are the following subset of parameters of
Eqs.~(\ref{softfree}) and~(\ref{freeparameterssoft}):
%\bea
%\lambda_i, \, \kappa_i, \tan\beta, \, v_{iR}, \, T_{\kappa_i},
%\, T_{\lambda_i}, T_{u_3}, \, m_{\widetilde u_{3R}}, \,
%m_{\widetilde Q_{3L}},
%\label{freeparameters}
%\eea
\bea
\lambda_i, \, \kappa_i, \tan\beta, \, v_{iR}, \, T_{\kappa_i},
\, T_{\lambda_i},
T_{u_3}, \, m_{\widetilde u_{3R}}, \,
m_{\widetilde Q_{3L}}.
\label{freeparameters}
\eea
%where we have included the last three third-generation soft terms given their importance through loop corrections.

%, and neglected the contribution of
%$T_{\nu_i}$ in the formulas. The latter is a sensible assumption when
%e.g. the trilinear parameters are proportional to their corresponding Yukawa couplings, i.e. $T_{\nu}=A_{\nu}Y_{\nu}$, as occurs in the breaking of supergravity.

In the limit of vanishingly small $\lambda_i$ (considering simultaneously very large
$v_{iR}$ 
%such that $\mu
%=3\lambda v_{R}/\sqrt 2 
in the case that we require 
the lighter chargino mass bound of RPC SUSY $\mu\gtrsim 100$ GeV),
%and $v_{iR} \gg v_d, v_u$, 
not only the off-diagonal 
entries of the right sneutrino submatrices~(\ref{evenR}) and~(\ref{evenI})
are negligible, but also the off-diagonal entries~(\ref{mixingevenR2}),~(\ref{mixingevenR}) and~(\ref{mixingevenI2}),~(\ref{mixingevenI}) 
of the Higgs-right sneutrino
matrices. 
%but also the off-diagonal entries of the right sneutrino submatrices~(\ref{evenR}) and~(\ref{evenI}).
%and therefore
%the mass squared eigenvalues for the right sneutrinos correspond to the diagonal entries. 
As a consequence of the latter,
the singlet states are decoupled from the doublets.
It is thus apparent, that $\lambda_i$
are undoubtedly the most relevant parameters for the analysis
of these states. Another aspect of the parameters $\lambda_i$, namely to yield additional contributions to the tree-level SM-like Higgs mass has already been discussed in the 
previous subsection.
Thus, one can write the right sneutrino masses as
\bea
m_{\widetilde{\nu}^{\mathcal{R}}_{iR} 
%\widetilde{\nu}^{\mathcal{R}}_{iR}
}^2
&=&
 \left(\frac{T_{{\kappa}_i}}{\kappa_i} 
 +2{\mathcal M}_{i}\right)\frac{{\mathcal M}_{i}}{2},
% =\left(
% A_{{\kappa}_i} 
% + 2{\mathcal M}_{i} \right) \frac{{\mathcal M}_{i}}{2},
\label{evenRR}
\\ 
\nonumber
\\
m_{\widetilde{\nu}^{\mathcal{I}}_{iR} 
}^2
&=&
-\frac{3}{2} \frac{T_{{\kappa}_i}}{\kappa_i} {\mathcal M}_{i},
%=
%-\frac{3}{2} A_{{\kappa}_i} {\mathcal M}_{i},
\label{evenII}
\eea
%where 
%the second equalities are obtained assuming the supergravity relation
%$T_{\kappa_i}=\kappa_i A_{\kappa_i}$.
where in the case of supergravity we can use the relation 
$T_{\kappa_i}/\kappa_i= A_{\kappa_i}$.
In addition, also in this limit ${\mathcal M}_{i}$ coincide approximately with
the masses of the right-handed neutrinos, since they are decoupled from the other entries of the neutralino mass matrix:
\bea
m_{\nu_{iR}} = {\mathcal M}_{i}.
\label{neutrinoR}
\eea
{
With the sign convention adopted in Sec.~\ref{themodel}, ${\mathcal M}_{i}>0$, and
from Eq.~(\ref{evenII}) we deduce that
%where the VEVs $v_{iR}$ are positive, 
negative values for $T_{\kappa_i}$ (or $A_{\kappa_i}$) 
%working with positive $\kappa_i$) 
are necessary in order to avoid tachyonic pseudoscalars.}
%Using now the formula for the Majorana masses of 
%Eq.~(\ref{majorana}) with our assumption of vanishing off-diagonal entries 
%%for $\kappa_{ijk}$
%\bea
%%m_{{\mathcal M}_{ij}}
%{\mathcal M}_{i}
%={2}\kappa_{i} \frac{v_{iR}}{\sqrt 2},
%\label{majorana2}
%\eea
%which in this limit coincide in a good approximation with
%the masses of the right-handed neutrinos, 
Using also that equation,
%Eq.~(\ref{evenII}), 
we can write~(\ref{evenRR}) as
\bea
m_{\widetilde{\nu}^{\mathcal{R}}_{iR}}^2
 &=&
 {\mathcal M}^2_{i} - \frac{1}{3} 
 m_{\widetilde{\nu}^{\mathcal{I}}_{iR} }^2.
\label{evenRR2}
\eea
%
%we can write the above expressions as
%\bea
%m_{\widetilde{\nu}^{\mathcal{R}}_{iR}}^2
% &=&
% {\mathcal M}^2_{i} - \frac{1}{3} 
% m_{\widetilde{\nu}^{\mathcal{I}}_{iR} }^2,
% \\
%\label{evenRR2}
%\nonumber
%\\
%m_{\widetilde{\nu}^{\mathcal{I}}_{iR}}^2
%&=&
%-\frac{3}{2} {\mathcal M}_{i} A_{{\kappa}_i}.
%\label{evenII2}
%\eea
Thus, the simultaneus presence of non-tachyonic scalars and pseudoscalars implies 
that~\cite{Ghosh:2014ida} 
\bea
m_{\widetilde{\nu}^{\mathcal{I}}_{iR}} &<& \sqrt 3 {\mathcal M}_{i},
\label{nonta1}
\\
%\nonumber
%\\
m_{\widetilde{\nu}^{\mathcal{R}}_{iR}} &<& {\mathcal M}_{i}.
\label{nonta2}
\eea
%$m_{\widetilde{\nu}^{\mathcal{R}}_{iR}} < {\mathcal M}_{i}$, and non tachyonic scalars implies that $m_{\widetilde{\nu}^{\mathcal{I}}_{iR}} < \sqrt 3 {\mathcal M}_{i}$
Hence, light scalars/pseudoscalar states are assured when light neutralinos (i.e. basically  the product $\kappa_i v_{iR}$) are present.
From Eq.~(\ref{evenRR}) we also see that the absence of tachyons implies the condition
\bea
\frac{-T_{{\kappa}_i}}{\kappa_i} 
%-A_{\kappa_i} 
< 2 {\mathcal M}_{i},
\label{nontach}
\eea
and therefore the value of 
%$A_\kappa$ 
$T_{{\kappa}_i}/\kappa_i$
and the product 
$\kappa_i v_{iR}$ have to be chosen appropriately to fulfill it.
%avoid tachyonic scalar right sneutrinos,
%i.e. $4 \kappa_i^2 v_{iR}/\sqrt 2 > - T_{\kappa_i}$.
%\bea
%\kappa_i v_{iR} >
%\frac{\sqrt 2}{4} |A_{\kappa_i}|,
%\label{tachyons}
%\eea
%where based on the supergravity framework we have already used $T_{\kappa_i}=A_{\kappa_i}\kappa_i$ in this equation for the discussion of Section~\ref{results-scans}.
%Singlet states lighter as well as heavier than the SM-like Higgs can be obtained.
%depending on the choice of the relevant parameters.
%On the one hand, 
Also from that equation 
%From Eq.~(\ref{evenRR}),
%and~(\ref{evenII}), 
we see that singlet scalars lighter than the SM-like Higgs
can be obtained when
%light right sneutrinos appear
%for small values of $-T_{\kappa_i}$ and/or $v_{iR}$, or tuning these parameters.
\bea
2 {\mathcal M}_{i}  - \frac{
2m^2_{\text{Higgs}}
%2m_h^2
}{{\mathcal M}_{i}}< 
%-A_{\kappa_i}. 
\frac{-T_{{\kappa}_i}}{\kappa_i}.
\label{sneutrinohiggs}
\eea
Thus, for a given value of ${\mathcal M}_{i}$, only a narrow range of values of
%$A_{\kappa_i}$ 
${-T_{{\kappa}_i}}/{\kappa_i}$ 
is able to fulfill simultaneously both conditions (\ref{nontach}) and (\ref{sneutrinohiggs}). We will come back to this issue in Sec.~\ref{results-scans}.

On the other hand, even in a region of small to moderate
$\lambda_i$,
to obtain approximate analytical formulas for tree-level scalar and pseudoscalar masses turn out to be rather complicated due to the
index structure of the parameters involved.
%$\kappa_i$.
%${\bm \la}$ ($0.01 \lsim {\bm \la} \leq 0.1$),
%It is thus apparent, that $\lambda_i$
%is undoubtedly the most relevant parameter for the analysis
%of these states. 
As discussed in detail in Ref.~\cite{Ghosh:2014ida},
the expressions for their masses can be simplified in the limit of 
complete degeneracy in all relevant parameters as in Eq.~(\ref{freeparameters00}), i.e.
when %Eq.~(\ref{freeparameters})
%is rewritten as
%\bea
%{\la}, \, \kappa, \tan\beta, \, v_{R}, \, T_{\kappa},
%\, T_{\lambda},
%\label{freeparametersu}
%\eea
%where we have defined $\lambda_i=\lambda$, $\kappa_i=\kappa$,
%$v_{iR}=v_R$, $T_{\kappa_i}=T_\kappa$ and $T_{\lambda_i}=T_\lambda$.
%\bea
%\lambda_i=\lambda, \,\, \kappa_i=\kappa, \,\, v_{iR}=v_R, \,\, T_{\kappa_i}=T_\kappa,
%\,\, T_{\lambda_i}=T_\lambda,
%\label{freeparametersu}
%\eea
$\lambda_i=\lambda$ ($\lambda = \bla / \sqrt 3$ as defined in Eq.~(\ref{lambda123})), $\kappa_i=\kappa, v_{iR}=v_R, T_{\kappa_i}=T_\kappa, T_{\lambda_i}=T_\lambda$.
%, i.e. as in Eq.~(\ref{freeparameters00}).
In this case, the $3\times 3$ scalar and pseudoscalar mass submatrices in Eqs~(\ref{evenR}) and~(\ref{evenI}) have the form
\begin{equation}
\begin{pmatrix}
a&b&b\\
b&a&b\\
b&b&a\\
\end{pmatrix},
\end{equation}
with the three eigenvalues given by $a-b$, $a-b$ and $a+2b$.
Then, it was shown that for both, scalars and pseudoscalars, the two mass eigenstates corresponding to the first 
two eigenvalues $a-b$
%of the right sneutrino mass matrix 
get decoupled and remain as pure singlet-like states without any doublet contamination. Using the values of $a$ and $b$ from 
Appendix~\ref{Apendix:Sneutrino-masses}, and neglecting 
$T_\nu$ under the supergravity assumption of being proportional to the small $Y_\nu$,
%~(\ref{evenR}) and~(\ref{evenI}) it is straightforward to 
one obtains the following degenerate masses:
\bea
m^2_{\widetilde{{\nu}}^{\mathcal{R}}_{1,2R}} &=& 
 \left(\frac{T_{{\kappa}}}{\kappa} 
 +2{\mathcal M}\right)\frac{{\mathcal M}}{2}
+  
%\bla^2 \frac{1}{\mu}\frac{T_\lambda }{\lambda} \,
%\frac{\tan\beta}{1+\tan^2\beta}\left(\frac{v}{\sqrt 2}\right)^2   
%-\bla^2
%\left(\frac{v}{\sqrt 2}\right)^2,
%\bla^2 
3\lambda^2
\left(\frac{v}{\sqrt 2}\right)^2 \left(\frac{1}{\mu}\frac{T_\lambda }{\lambda} \,
\frac{\tan\beta}{1+\tan^2\beta}-1\right),
\label{sps-approx-R}
\\
m^2_{\widetilde{{\nu}}^{\mathcal{I}}_{1,2R}} &=& 
-\frac{3}{2} \frac{T_{{\kappa}}}{\kappa} {\mathcal M}
+  
3\lambda^2
%\bla^2  
\left(\frac{v}{\sqrt 2}\right)^2
\left[
\left(\frac{1}{\mu}\frac{T_\lambda}{\lambda} 
+ \frac{4}{ 3}\frac{\kappa}{\lambda} \right) \frac{\tan\beta}{1+\tan^2\beta}
-1\right],
%\bla^2  \left(\frac{1}{\mu}\frac{T_\lambda}{\lambda} 
%+ \frac{4}{\sqrt 3}\frac{\kappa}{\bla} \right) \frac{\tan\beta}{1+\tan^2\beta}\left(\frac{v}{\sqrt 2}\right)^2 
%-\bla^2
%\left(\frac{v}{\sqrt 2}\right)^2,
\label{sps-approx-I}
\eea 
where now $\mu$ is defined in Eq.~(\ref{mu2}),
%$\mu=3\lambda v_R/\sqrt 2$, 
and the Majorana mass is given in
Eq.~(\ref{majorana3})
%\bea
%{\mathcal M}
%={2}\kappa \frac{v_{R}}{\sqrt 2},
%\label{majorana3}
%\eea
corresponding approximately also to two pure right-handed neutrino states decouple from the rest of the neutralinos:
\bea
m_{\nu_{1,2R}} = {\mathcal M}.
\label{neutrinoR2}
\eea
The degeneracy of these states
%the two scalar and pseudoscalar states (and the two right-handed neutrinos) 
can be broken by introducing mild splittings in 
$\kappa_i$ values~\cite{Fidalgo:2011ky,Ghosh:2014ida},
as it is obvious e.g. from Eqs.~(\ref{evenRR}), (\ref{evenII}) and~(\ref{neutrinoR}). 
%in the limit of small $\lambda_i$.
%On the other hand, 
One thing to highlight is that the first term of Eq.~(\ref{sps-approx-R}) is like the one of Eq.~(\ref{evenRR}) in the case of universality, and therefore a condition similar to~(\ref{nontach}) is welcome to avoid tachyons in the scalar spectrum.
%$T_{\kappa} + 4 \kappa^2v_{R}/\sqrt 2 > 0$.
Nevertheless, depending on the values chosen for the input parameters, the second term in~(\ref{sps-approx-R}) can be positive, relaxing this condition. The latter is especially true for large values of 
%$\bla$.
$\lambda$.

The mass eigenstate corresponding to the third
eigenvalue, namely the one which goes as $a+2b$, however mixes with the doublet-like states, and eventually its mass appears with a complicated form. In the case of the pseudoscalar this is given by 
\bea
m^2_{\widetilde{{\nu}}^{\mathcal{I}}_{3R}} = 
- \frac{3}{2}\frac{T_{\kappa}}{\kappa}\mathcal{M}
+  9\lambda
%\bla
\kappa\ 
\left(\frac{v}{\sqrt 2}\right)^2
\frac{T_{\lambda}/{\lambda}}
{\frac{T_\lambda}{\lambda} + \frac{{\mathcal M}}{2}}\ \frac{\tan\beta}{1+\tan^2\beta}.
\label{sps-approx-I3}
\eea 
In the case of the scalar, $m^2_{\widetilde{{\nu}}^{\mathcal{R}}_{3R}}$
appears with a much complicated form that can be found in
Ref.~\cite{Ghosh:2014ida}.
Similarly, the third right-handed neutrino-like state mixes with the other MSSM-like
neutralinos, and its mass is given by
\bea
m_{{\nu}_{3R}} = 
{\mathcal M}
%{2}\kappa \frac{v_{R}}{\sqrt 2}
-\frac{\lambda^2}{2\mu} \left(\frac{v}{\sqrt2}\right)^2
\left[\frac{1+\tan^2\beta}{\tan\beta} - \frac{4 M \mu}{\left({v}/{\sqrt2}\right)^{2}}
\right]
\left[ \frac{1+\tan^2\beta}{\tan\beta} \frac{M \mu}
{ \left({v}/{\sqrt2}\right)^{2}}-1
\right]^{-1},
\label{neutrinoR333}
\eea 
where $M$ is defined in Eq.~(\ref{gauginom}).

Depending on the input parameters chosen, the two degenerate states can be heavier or lighter than the third state. For example, for the pseudoscalars we see that the second term in Eq.~(\ref{sps-approx-I3})
is always positive whereas
the second term in Eq.~(\ref{sps-approx-I})
can be positive (larger or smaller than the previous one) or even negative.

It is also worthy to note that for further small $\lambda$ values (i.e. $\lsim 0.01$) or in the limit of a vanishingly small $\lambda$,
these formulas take simpler forms, and the three states are degenerate:
\bea
m_{\widetilde{\nu}^{\mathcal{R}}_{iR} 
%\widetilde{\nu}^{\mathcal{R}}_{iR}
}^2
&=&
 \left(\frac{T_{{\kappa}}}{\kappa} 
 +2{\mathcal M}\right)\frac{\mathcal M}{2},
% =\left(
% A_{{\kappa}_i} 
% + 2{\mathcal M}_{i} \right) \frac{{\mathcal M}_{i}}{2},
\label{evenRR2}
\\ 
\nonumber
\\
m_{\widetilde{\nu}^{\mathcal{I}}_{iR} 
}^2
&=&
-\frac{3}{2} \frac{T_{{\kappa}}}{\kappa} {\mathcal M},
%=
%-\frac{3}{2} A_{{\kappa}_i} {\mathcal M}_{i},
\label{evenII2}
\\ 
\nonumber
\\
m_{\nu_{iR}} 
&=&
{\mathcal M}.
\label{neutrinoR2}
\eea
As expected, they
coincide with Eqs.~(\ref{evenRR}),~(\ref{evenII}),
and~(\ref{neutrinoR}),
respectively, when written for universal parameters. 
%The same happens with the three right-handed neutrinos whose masses
%\bea
%{m}_{\nu_{iR}}
%={\mathcal{M}}
%\label{neutrinosR33}
%\eea
%coincide with Eq.~(\ref{neutrinoR}) when written again for universal %parameters.

Other simple formulas can be obtained in the limit
of $\tan\beta\rightarrow \infty$, where
\bea
m^2_{\widetilde{{\nu}}^{\mathcal{R}}_{1,2R}} &=& 
\left(\frac{T_{{\kappa}}}{\kappa} 
 +2{\mathcal M}\right)\frac{{\mathcal M}}{2}
%\left(
% T_{{\kappa}} 
% +4\kappa^2 \frac{v_{R}}{\sqrt 2} \right)
% \frac{v_{R}}{\sqrt 2}
%- (\sqrt 3 \lambda)^2 
-3\lambda^2
\left(\frac{v}{\sqrt 2}\right)^2,
\label{sps-approx-Rtan}
%\\
%m^2_{\widetilde{{\nu}}^{\mathcal{R}}_{3R}} &=& 
%\left(
% T_{{\kappa}} 
% +\frac{v_{R}}{\sqrt 2}\ 4\kappa^2 \right)
% \frac{v_{R}}{\sqrt 2}
%-\lambda^2 \frac{24\mu^2}{g^2+g'^2},
%\label{thirss}
\\
m^2_{\widetilde{{\nu}}^{\mathcal{I}}_{1,2R}} &=& 
- \frac{3}{2}\frac{T_{\kappa}}{\kappa}\mathcal{M}
%- 3T_{\kappa} \frac{v_R}{\sqrt 2}
%- (\sqrt 3 \lambda)^2 
-3\lambda^2
\left(\frac{v}{\sqrt 2}\right)^2,
\label{sps-approx-Itan}
\eea 
and 
\bea
m^2_{\widetilde{{\nu}}^{\mathcal{R}}_{3R}} &=& 
\left(\frac{T_{{\kappa}}}{\kappa} 
 +2{\mathcal M}\right)\frac{{\mathcal M}}{2}
%\left(
% T_{{\kappa}} 
% +4\kappa^2 \frac{v_{R}}{\sqrt 2} \right)
% \frac{v_{R}}{\sqrt 2}
%- (\sqrt 3 \lambda)^2 
-3\lambda^2
\left(\frac{v}{\sqrt 2}\right)^2 \left(\frac{2\mu}{m_Z}\right)^2,
%-\lambda^2 \frac{24\mu^2}{g^2+g'^2},
\label{thirss}
\\
m^2_{\widetilde{{\nu}}^{\mathcal{I}}_{3R}} &=& 
- \frac{3}{2}\frac{T_{\kappa}}{\kappa}\mathcal{M},
%- 3T_{\kappa} \frac{v_R}{\sqrt 2}.
%- 3 \lambda^2 \left(\frac{v}{\sqrt 2}\right)^2.
\label{thirsi}
\eea 
whereas for the righ-handed neutrinos one obtains
\bea
m_{\nu_{1,2R}} &=& \mathcal{M},
\label{thirssww}
\\
m_{{\nu}_{3R}} &=& 
{\mathcal M}
-\frac{
\lambda^2
%(\sqrt 3\lambda)^2
}
{2\mu^2M} \left(\frac{v}{\sqrt2}\right)^4.
\label{thirsiww}
\eea 
It is evident from this result that unless $\lambda$ is small to moderate,
%(i.e. $0.01 \lsim \bla/\sqrt 3 \lsim 0.2$), 
it is in general hard
to accommodate a complete non-tachyonic light spectrum (i.e. $\lsim m_{\text{Higgs}}/2$)
for both the scalars and pseudoscalars in the limit of large 
$\tan\beta$ without a parameter tuning.
In addition, this limit is severely constrained from diverse experimental results. This is because the BRs for some low-energy processes (e.g. $B^0_s\rightarrow \mu^+\mu^-$), depending on the other relevant parameters are sensitive to the high powers of $\tan\beta$ and thus, can produce large BRs for these processes in an experimentally unnacceptable way.
%\R{If present, they can significantly
%modify the SM-like Higgs decays, although still be compatible with Higgs data as discussed in Ref.~\cite{Ghosh:2014ida}.
%This happens when these light singlet-like states are almost pure and therefore do not affect Higgs decays.}
The other limit, i.e. small $\tan\beta$, on the contrary, is useful
from the viewpoint of raising the mass of the lightest doublet-like scalar towards 125 GeV, specially for moderate to
large $\lambda$ values .
%(${\bla} \gsim 0.2$).
However, as shown in the discussion of Eqs.~(\ref{sps-approx-R}),~(\ref{sps-approx-I}) 
and~(\ref{sps-approx-I3}), not all the mass formulas for the light states are simple structured in this region and a numerical analysis is convenient.

%{On the other hand, if some scalars are degenerate or quasi-degenerate, or with masses within
%a given experimental mass resolution with respect to the SM-like Higgs,
%the superposition of their signal rates with those of the SM-like Higgs has to be taken into account, as we will discuss in 
%Section~\ref{results-scans}.}

%To review this issue, let us work in the general case with the
%${\bm \la}$ parameter defined as in Eq.~(\ref{lambda123}),
%and following Ref.~\cite{Ghosh:2014ida} we
%choose the following three relevant
%ranges of values \R{(probably the discussion of the three cases below can be shortened)}:

\subsection{The left sneutrinos}

The behaviour of the left sneutrinos is very different from the one of the right sneutrinos, since the former are tightly associated to neutrino physics.

As discussed before, the $3\times 3$ scalar and pseudoscalar left sneutrino submatrices are decoupled from the $5\times 5$ Higgs-right sneutrino submatrices.
%As mentioned above, they are decouple from the $5\times 5$ Higgs-right sneutrino submatrices.
Besides, their off-diagonal entries
are negligible compared to the diagonal ones, since they
are suppressed by terms proportional 
to $Y^2_{\nu_{ij}}$ and $v^2_{iL}$. As a consequence, the mass
squared eigenvalues correspond to the diagonal entries, and in this approximation both states also have degenerate masses.
%the off-diagonal terms of the mass matrix mixing the left sneutrinos with Higgses and right sneutrinos are suppressed by $Y_{\nu}$ and $v_{iL}$,
%implying that 
%the $3\times 3$ left sneutrino submatrix is almost decouple from the rest producing almost pure states.
%The same happens for the pseudoscalar left sneutrino states $\widetilde{\nu}^{\mathcal{I}}_{iL}$, which have in addition degenerate masses with the scalars 
%{$m_{\widetilde{\nu}^{\mathcal{R}}_{iL}}
%\approx
% m_{\widetilde{\nu}^{\mathcal{I}}_{iL}}
%\equiv 
%m_{\widetilde{\nu}_{iL}}$. 
{Using the minimization equations for $v_{iL}$, we can write their tree-level 
values as~\cite{Escudero:2008jg,Ghosh:2008yh,Ghosh:2017yeh,Biekotter:2019gtq}
%
%\bea
%m_{\widetilde{\nu}^{\mathcal{R}}_{iL}}^2
%=
%m_{\widetilde{\nu}^{\mathcal{I}}_{iL}}^2
%=
%\frac{Y_{{\nu}_i}v_u}{v_{iL}} \frac{v_{iR}}{\sqrt 2}
%\left(\frac{-T_{{\nu}_i}}{Y_{{\nu}_i}}-\kappa_i\frac{v_{iR}}{\sqrt 2}
%+ \frac{\mu}{\tan\beta}+\lambda_i \frac{v_u}{v_{iR}}\frac{v_d}{\sqrt 2}\right).
%\label{evenLLL2}
%\eea
\bea
m_{\widetilde{\nu}^{\mathcal{R}}_{iL}}^2
=
m_{\widetilde{\nu}^{\mathcal{I}}_{iL}}^2
=
\frac{
m_{{\mathcal{D}_i}}}{v_{iL}/\sqrt 2}
%Y_{{\nu}_i}v_u}{v_{iL}}
%\frac{{\mathcal M}_{i}}{2\kappa_i}
\frac{v_{iR}}{\sqrt 2}
\left[\frac{-T_{{\nu}_i}}{Y_{{\nu}_i}}-
\frac{{\mathcal M}_{i}}{2}
%\kappa_i\frac{v_{iR}}{\sqrt 2}
+ \frac{\mu}{\tan\beta}+\lambda_i \frac{\left(v/\sqrt 2\right)^2}{v_{iR}/\sqrt 2}
%\left(\frac{v}{\sqrt 2}\right)^2
\frac{\tan\beta}{1+\tan^2\beta}\right].
\label{evenLLL2}
\eea
Therefore, 
%left sneutrino masses introduce 
in addition to the parameters of 
Eq.~(\ref{freeparametersn}) relevant for neutrino physics, 
the 
\bea
T_{\nu_i} 
\label{tnu}
\eea
are relevant parameters for the study of left sneutrino masses.
%
%where $T_{{\nu}_i}\equiv T_{{\nu}_{ii}}$ are the trilinear parameters in the soft potential, 
%$T_{{\nu}_{ii}} H_u^0 \widetilde \nu_{iL} \widetilde \nu_{iR}^*$ + h.c.,
%\R{where, in order to obtain this formula, we are assuming diagonal neutrino Yukawa couplings,
%i.e. $Y_{{\nu}_{i}}\equiv Y_{{\nu}_{ii}}$ and vanishing otherwise.}
%In fact,
%data on neutrino physics~\cite{Capozzi:2017ipn,deSalas:2017kay,deSalas:2018bym,Esteban:2018azc} can easily be reproduced at tree level in the $\mn$~\cite{LopezFogliani:2005yw,Escudero:2008jg,Ghosh:2008yh,Bartl:2009an,Fidalgo:2009dm,Ghosh:2010zi} with diagonal Yukawa couplings~\cite{Ghosh:2008yh,Fidalgo:2009dm,Kpatcha:2019gmq}.
The fourth term in Eq.~(\ref{evenLLL2}) 
can usually be neglected as long as $v_{iR} \gg v$ and/or $\lambda_i$ 
is small, and in the limit of
 moderate/large $\tan\beta$ one can also neglect the third term.
%as long as $v_{iR} \gg v_d, v_u$ and/or $\lambda_i$ is small.
%Therefore, left sneutrino masses introduce in addition to the parameters of Eq.~(\ref{freeparameters}), the 
%
%\bea
%T_{{\nu}_i},
%\label{tia}
%\eea
%\R{as other relevant parameters for our analysis.}
%In the limit of moderate/large $\tan\beta$ one can neglect also the third term, and 
Under these approximations, the condition for non-tachyonic left sneutrinos can be written as an upper bound on the Majorana masses
\bea
\frac{{\mathcal M}_{i}}{2}
 \lsim \frac{-T_{\nu_i}}{Y_{\nu_i}}. 
\label{sneulta}
\eea
{Given our sign convention of positive Majorana mass,
we will use negative values for $T_{{\nu}_i}$
in the numerical analyses of Sec.~\ref{results-scans}.
%~\ref{methodology} and~\ref{results-scans}.
}
%in order to avoid tachyonic left sneutrinos, 
%given our sign convention of positive Majorana mass.}
%where $\kappa_i$ are positive.
%In the case of universality of the parameters, the chargino lower bound 
%$\mu=3\lambda v_R/\sqrt 2 \gsim 100$ GeV together with the above upper bound on the Majorana mass give rise to the following constraint on the value of the right sneutrino VEVs:
%\bea
%\frac{100\ \text{GeV}}{3\lambda}  \lsim \frac{v_R}{\sqrt 2}
% < \frac{-T_{\nu_i}/Y_{\nu_i}}{\kappa},
%\label{vrbound}
%\eea
%which will be useful for our discussion of results in Section~\ref{results-scans}
%when studying the $\kappa-\lambda$ plane.
%Note that the chargino bound $\mu\gsim 100$ GeV, can in principle impose a significant limit on $\kappa$. 
%Using the supergravity relation $T_{\nu}=Y_{\nu} A_{\nu}$, this equation can be written as
%$\kappa <  -A_{\nu}3\lambda/\mu$.

Going back to Eq.~(\ref{evenLLL2}), we see clearly why 
left sneutrinos are special in the $\mn$ with respect to other SUSY models. 
Given 
%This is because
%, as 
%discussed in Eq.~(\ref{evenLLL2}), 
that their masses are determined by the minimization equations with respect to
$v_i$, they depend not only on left sneutrino VEVs but also on neutrino Yukawas, {\it unlike right sneutrinos}, and as a consequence neutrino physics is very relevant for them.
%In particular, 
%assuming the simplest situation that all the $A_{{\nu}_i}$ are naturally of the order of the TeV, neutrino physics determines sneutrino masses through the prefactor $m_{{\mathcal{D}_i}}/{v_{iL}}$.

%this hierarchy of Yukawas and VEVs determines from
%Eq.~(\ref{evenLLL22}) that 
%$m_{\widetilde{\nu}_{1}}$ is the smallest of all sneutrino masses.
Considering the normal ordering for the neutrino mass spectrum, which is nowadays favored by the analyses of neutrino data \cite{Capozzi:2017ipn,deSalas:2017kay,deSalas:2018bym,Esteban:2018azc}, 
and taking advantage of the 
dominance of the gaugino seesaw for some of the three neutrino families,
representative solutions for neutrino/sneutrino physics using diagonal neutrino Yukawas were summarized in Ref.~\cite{Kpatcha:2019gmq}.
Different hierarchies among the generations 
of left sneutrinos are possible, 
using different hierarchies among $Y_{\nu_i}$ (and also $v_{iL}$).
%since their masses depend on crucial parameters for neutrino physics such as 
%$Y_{\nu_i}$ and $v_{iL}$.

%Let us remark that
There is enough freedom in the parameter space of the $\mn$
in order to get heavy as well as light left sneutrinos from Eq.~(\ref{evenLLL2}), and the latter scenario 
with the left sneutrino as the 
lightest supersymmetry particle (LSP) was considered in Refs.~\cite{Ghosh:2017yeh,Lara:2018rwv,Kpatcha:2019gmq}. 
Due to the doublet nature of the left sneutrino, masses smaller than half of the mass of the SM-like Higgs were found to
be forbidden~\cite{Kpatcha:2019gmq} 
to avoid dominant decay of the latter into sneutrino pairs, leading to an inconsistency
with Higgs data.
{Let us finally remark that in those works  
negative values for $T_{u_3}$ were used,
in order to avoid too light left sneutrinos due to loop corrections.
Although we are not specially interested in light sneutrinos in this work,
we will maintain the same sign convention in what follows. To use positive values
for $T_{u_3}$ would have not modify our results, since their effect on the SM-like Higgs mass is similar.}

\subsection{The charged scalars}

The charged scalars have
a $8\times 8$ (`charged Higgs') mass matrix.
Similar to the 
neutral Higgs mass matrices where some sectors are decoupled, the $2\times 2$ 
charged Higgs submatrix is decoupled from the $6\times 6$ slepton 
submatrix.
Thus, as in the MSSM, the mass of the charged Higgs 
%$H^{\pm}$ 
is similar to the one of the doublet-like neutral pseudoscalar,
%$A$, 
specially when the latter is not very mixed with the right sneutrinos. In this case, both masses are also similar to the
one of the heavy doublet-like neutral Higgs.
%$H$.

%(see Eqs.~(\ref{pruebass11})--(\ref{olvidado})). 
Concerning the $6\times 6$ submatrix, the right sleptons are decoupled from the left ones,
%and ~(\ref{pruebass6}), 
since the mixing terms are suppressed by the electron-type Yukawa couplings
or $v_{iL}$. 
%(see Eq.~(\ref{pruebass5
Then, the masses of right and left sleptons are basically determined by their corresponding soft terms, 
$m_{\widetilde{e}_{iR}}^2$ and
$m_{\widetilde{L}_{i}}^2$, respectively.
Although the left sleptons are in the same $SU(2)$ doublet as the
left sneutrinos, they are a little heavier than the latter mainly due to the
mass splitting produced by the D-term contribution, $-m_W^2 \cos 2\beta$.

\section{Strategy for the scanning}
\label{methodology}
In this section we describe the methodology that we employed to search for points of our parameter space that are compatible with the latest experimental data on Higgs physics 
%as well as cosmological data on neutrinos masses. 
In addition, we demanded the compatibility with some flavor observables. To this end, we performed scans on the parameter space of the model, with the input parameters optimally chosen.

\subsection{Sampling the $\mn$}
%%%%%%%%%%%%%%%%%%%%%%%%%%%%%%%%%%%%%%%%%%%%%%%%

For the sampling of the $\mn$, we used a likelihood data-driven method employing the
{\tt MultiNest}~\cite{Feroz:2007kg,Feroz:2008xx,Feroz:2013hea} algorithm as optimizer. The goal is to find
regions of the parameter space of the $\mn$ that are compatible with a given experimental data. 
For it we have constructed the joint likelihood function: 
\begin{eqnarray}
 \mathcal{L}_{\text{tot}} =\mathcal{L}_{\text{Higgs}} 
% \times \mathcal{L}_{\text{neutrino}} 
 \times  \mathcal{L}_{\text{B physics}}
 \times \mathcal{L}_{\mu\text{ decay}} \times \mathcal{L}_{m_{\widetilde \chi^\pm}},
 \label{joint-likelihood}
\end{eqnarray}
where $\mathcal{L}_{\text{Higgs}}$ represents Higgs observables, %$\mathcal{L}_{\text{neutrino}}$ the upper bound on the sum of the masses of active neutrinos,  
$\mathcal{L}_{\text{B physics}}$ B-physics constraints, $\mathcal{L}_{\mu\text{ decay}}$ $\mu$ decay constraints, and $\mathcal{L}_{m_{\widetilde \chi^\pm}}$ constraints on the chargino mass.

To compute the spectrum and the observables we used {a suitably modified version of
{\tt SARAH} v4.5.9 code
\cite{Staub:2013tta} to generate a 
{\tt SPheno} v3.3.6~\cite{Porod:2003um, Porod:2011nf} version} for the model. 
We condition that each point is required not to have tachyonic eigenstates. 
For the points that pass this constraint, we compute the likelihood associated to each experimental data set and for each sample all the likelihoods are collected in the joint likelihood $\mathcal{L}_{\text{tot}}$ (see Eq.~(\ref{joint-likelihood}) above).
%%
%The likelihood functions we use have been discussed in Ref.~\cite{2019arXiv190702092K}.

%%%%%%%%%%%%%%%%%%%%%%%%%%%%%%%%%%%%%%%%%%%
\subsection{Likelihoods} \label{constraints}

%%%%%%%%%%%%%%%%%%%%%%%%%%%%%%%%%%%%%%%%%%%%%
\vspace{0.25cm}
The likelihood functions used have already been discussed in Ref.~\cite{Kpatcha:2019gmq}. 
%Subsequently we present the observables
%corresponding to each likelihood in Eq.(\ref{joint-likelihood}).
Summarizing, we used three types of likelihood functions in our analysis. 
For observables for which a measurement is available we use a 
Gaussian likelihood function defined as follows:
%the limit are provided as best fit and uncertainties,
%the likelihood function is a Gaussian,
\begin{eqnarray}
  \mathcal{L}(x) = \exp\left[-\frac{(x-x_0)^2}{2\sigma_T^2} \right],
  \label{eq:likelihood-Gaussian}
\end{eqnarray}
where $x_0 $ is the experimental best fit set on the parameter $x$, $\sigma_T^2= \sigma^2 +\tau^2$
with $\sigma$ and $\tau$ being respectively the experimental and theoretical uncertainties on
the observable $x$.

On the other hand, for any observable for which the constraint is set as lower or upper
limit, an example is the chargino lower mass bound, the likelihood function is defined as
\begin{eqnarray}
  \mathcal{L}(x) &=& \frac{\sigma}{\sigma_T}  [ 1-K(D(x)) ] 
  \exp\left[-\frac{(x-x_0)^2p}{2\sigma_T^2} \right] 
  + \frac{1}{ \tau} K((x-x_0)p), 
    \label{eq:likelihood-lower+upper}
\end{eqnarray}
where
%& \text{ where } & 
 \begin{eqnarray}
D(x) = \frac{\sigma}{\tau} \left( \frac{(x_0 - x)p}{ \sigma_T} \right), \; K(a) = \frac{1}{2} \text{erfc}\bigg(\frac{a}{\sqrt{2}}\bigg).
  \label{eq:likelihood-lower+upper2}
\end{eqnarray}
The variable $p$ takes $+1$ when $x_0$ represents the lower limit
and $-1$ in the case of upper limit, while \text{erfc} is the complementary error function.

The last class of likelihood function we used is a step function in such a way 
that the likelihood is one/zero if the constraint is satisfied/non-satisfied. 

%that is a 
%fixed $\chi^2$ value depending on whether the BP satisfies the limit or not. Note in this respect that the $\chi^2$ values are chosen to minimize the total $\chi^2_{\text{tot}}$ for viable points.

It is important to mention that in this work, unless explicitly mentioned, the theoretical uncertainties $\tau$ are unknown and therefore are taken to be zero.
Subsequently, we present each constraint used in this work together with the corresponding type of likelihood function.

\vspace{0.25cm}

\noindent
{\bf Higgs observables}

\noindent
Before the discovery of the SM-like Higgs boson, the negative searches of Higgs signals at the Tevatron, LEP and LHC, were transformed into exclusions limits that must
be used to constrain any model. Its discovery at the LHC added crucial constraints that
must be taking into account in those exclusion limits. 
We have considered all these constraints in the analysis of the $\mn$, where the Higgs sector is extended with respect to the MSSM as discussed in Sec.~\ref{themodel}.
For constraining the predictions in that sector of the model, we interfaced 
{\tt HiggsBounds} {{v}}5.3.2~\cite{Bechtle:2008jh,Bechtle:2011sb,Bechtle:2013gu,Bechtle:2013wla,Bechtle:2015pma} 
%{\tt HiggsBounds} {{v}}4.3.1~\cite{Bechtle:2008jh,Bechtle:2013wla}  
with {\tt MultiNest}.
First, several theoretical predictions in the Higgs sector (using a $\pm 3$ GeV theoretical uncertainty on the SM-like Higgs boson) are provided to determine which process has the highest exclusion power, according to the
list of expected limits {from LEP, Tevatron and LHC}. Once the process with the highest statistical
sensitivity is identified, the predicted production {cross sections of scalars and pseudoscalars are computed in the `effective coupling approximation' of {\tt HiggsBounds}
with the inputs given by {\tt SPheno}. The cross sections are then multiplied by the BRs 
%computed using {\tt SPheno} 
and compared with the limits set by
these experiments}. Finally, whether the corresponding point of the
parameter under consideration is allowed or not at 95\% confidence level is indicated.

In constructing the likelihood from HiggsBounds constraints, the likelihood function is taken to be a step function. Namely, it is set to one for points for which Higgs physics is realized, and zero otherwise. 
%This choice allows to minimize the $\chi^2$ for points that are allowed while penalizing the ones that are excluded.
%As already mentioned HiggsBounds algorithm does not offer the possibility of verifying whether
%a given Higgs scalar of a given model is in agreement with the signal that has been observed at
%CMS and ATLAS. 
In order to address whether a given Higgs scalar of the $\mn$ 
is in agreement with the signals 
observed by ATLAS and CMS, we interfaced 
{\tt HiggsSignals} 
{{v}}2.2.3~\cite{Bechtle:2013xfa,Stal:2013hwa,Bechtle:2014ewa} 
%{\tt HiggsSignals} 
%{{v}}1.4.0~\cite{Bechtle:2015pma,Bechtle:2013xfa} 
with {\tt MultiNest}.
%to test the model prediction against the measured mass and signal strength discovered by ATLAS and
%CMS collaborations. 
A $\chi^2$ measure is used to quantitatively determine the compatibility of the $\mn$ prediction with the measured signal strengths and mass. 
The experimental data used are those of the LHC with some complements from Tevatron. The details of the likelihood evaluation can be found in Refs.~\cite{Bechtle:2015pma,Bechtle:2013xfa}. 
\vspace{0.25cm}

\noindent
{\bf B decays}

\noindent
$b \to s \gamma$ is a flavour changing neutral current (FCNC) process, and hence it is forbidden
at tree level in the SM. However, its occurs at leading order through loop diagrams.
Thus, the effects of new physics (in the loops) on the rate of this 
process can
be constrained by precision measurements. 
In the combined likelihood, we used the average value of $(3.55 \pm 0.24) \times 10^{-4}$ provided in Ref.~\cite{Amhis:2012bh}. Note that the likelihood function is also a Gaussian
(see Eq.~(\ref{eq:likelihood-Gaussian})).
%%%%%%%%%%%%%%%%%%%%%%%%%%%%%%%%%%%%%%%%%%%%%%
Similarly to the previous process, $B_s \to \mu^+\mu^-$ and  $B_d \to \mu^+\mu^-$
%$b \to s \gamma$ process, its decays into a pair of muons 
are also forbidden at tree level in the SM but occur radiatively.
In the likelihood for these observables Eq.~(\ref{eq:likelihood-Gaussian}), we used the combined results of LHCb and CMS~\cite{CMS-PAS-BPH-13-007}\footnote{{While we were doing the scan this reference and the previous one for $b \to s \gamma$  had the latest data, nevertheless we have checked that our results for these three processes
%, as well as for $b \to s \gamma$, 
are $3\sigma$ compatible with the most recent data of 
Ref.~\cite{Amhis:2019ckw}.}}, %\cite{CMSandLHCbCollaborations:2013pla}, 
$ \text{BR} (B_s \to \mu^+ \mu^-) = (2.9 \pm 0.7) \times 10^{-9}$ and
$ \text{BR} (B_d \to \mu^+ \mu^-) = (3.6 \pm 1.6) \times 10^{-10}$.
Concerning the theoretical uncertainties for each of these observables, {we use the guesstimate $\tau= 10 \%$ of the corresponding best fit value}.
We denote by $\mathcal{L}_{\text{B physics}}$
the likelihood from $b \to s \gamma$, $B_s \to \mu^+\mu^-$ and $B_d \to \mu^+\mu^-$.

%%%%%%%%%%%%%%%%%%%%%%%%%%%%%%%%%%%%%%%%%%%%%
\vspace{0.25cm}

\noindent
{\bf $\mu $ decays}

\noindent
We also included in the joint likelihood the constraint from 
BR$(\mu \to e\gamma) < 5.7\times 10^{-13}$~\cite{Adam:2013mnn}\footnote{{We have checked that our results are in agreement with the most recent data of Ref.~\cite{TheMEG:2016wtm}.} }
and BR$(\mu \to eee) < 1.0 \times 10^{-12}$~\cite{Bellgardt:1987du}.
For each of these observables we defined the likelihood as a step function. As explained before, if a point is in agreement with the data the likelihood
$\mathcal{L}_{\mu\text{ decay}}$ is set to 1, and otherwise to 0.

{Let us point out here that we did not try to 
explain the interesting but not conclusive 3.5$\sigma$ discrepancy between the measurement of the
anomalous magnetic moment of the muon and the SM prediction, 
$\Delta a_{\mu}= a_{\mu}^{\text{exp}}-a_{\mu}^{\text{SM}}=(26.8 \pm 6.3\pm 4.3) \times 10^{-10}$~\cite{Tanabashi:2018oca}.}
%Since we decouple the rest of the SUSY spectrum with respect to the tau left sneutrino mass, we do not expect a large SUSY contribution over the SM value.
%$a^{\text{exp}}_\mu =11659209.1 \pm 5.4 \pm 3.3 \times 10^{-10}$ \cite{Tanabashi:2018oca}. 
%We checked for the points fulfilling all constrains discussed in Section~\ref{results-scans},
%that the extra contribution $a_{\mu}^{\text{SUSY}}$ is within the SM uncertainty.}
{Nevertheless, we will check
the level of compatibility with this value of
the points fulfilling all constraints discussed in Sec.~\ref{results-scans}}.
%have (have not) 
%the extra contribution $a_{\mu}^{\text{SUSY}}$} within  $\pm 1 \,\sigma$ of SM $a^{\text{SM}}_{\mu}$ uncertainty.}
%%%%%%%%%%%%%%%%%%%%%%%%%%%%%%%%%%%%%%%%%%
\vspace{0.25cm}

\noindent
{\bf Chargino mass bound}

\noindent
In RPC SUSY, the lower bound on the lightest chargino mass 
%of about $94$ GeV
depends on the spectrum of the model~\cite{Tanabashi:2018oca,Sirunyan:2018ubx}.
Although in the $\mn$ there is RPV and therefore this constraint does not apply, 
to compute $\mathcal{L}_{m_{\widetilde \chi^\pm}}$ we have chosen a conservative
limit of $m_{\widetilde \chi^\pm_1} > 92$ GeV. {For the theoretical uncertainty we use the guesstimate $\tau= 5 \%$ 
of the chargino mass}.

%%%%%%%%%%%%%%%%%%%%%%%%%%%%%%%%%%%%%%%%%%%%%%%%%%%%%%%%%%%%%%%%%%%%%%%%%%%%%%%%%%%%%%%%%%%%%%%%%%%
\subsection{Input parameters}
\label{scan} 

In order to efficiently scan for Higgs physics in the $\mn$, 
it is important to identify first
the 
%most relevant 
parameters to be used, and optimize their number and their ranges of values.
In Sect.~\ref{SMhiggs}, we found that the relevant parameters 
are those in
Eq.~(\ref{freeparameters}). 
However, to perform scans over 19 parameters we would have to run
{\tt MultiNest} a extremely long time making the task very computer resources demanding.
The analysis can be nevertheless much simplified assuming universality of the parameters as we did in the discussion below Eq.~(\ref{sneutrinohiggs}), without significantly modifying the conclusions. In addition, we will also assume in the scans for the sake of simplicity
$m_{\tilde Q_{3L}} = m_{\tilde u_{3R}}$.
Thus, we will perform scans over the 8 parameters 
%\bea
%\lambda_i=\lambda,\,\, 
%\kappa_i=\kappa,\,\, 
%\tan\beta,\,\, 
%v_{iR}=v_R,\,\, 
%T_{\kappa_i}=T_\kappa,\,\, 
%T_{\lambda_i}=T_\lambda,\,\, 
%T_{u_{3}},\,\,
%m_{\tilde Q_{3L}}=m_{\tilde u_{3}},
%\label{finalp}
%\eea
%
\bea
\lambda,\,\, 
\kappa,\,\, 
\tan\beta,\,\, 
v_R,\,\, 
T_\kappa,\,\, 
T_\lambda,\,\, 
T_{u_{3}},\,\,
m_{\tilde Q_{3L}}=m_{\tilde u_{3}},
\label{finalp}
\eea
as shown in Table~\ref{Scans-priors-parameters}, {using the sign conventions discussed in Sec.~\ref{themodel} and~\ref{SMhiggs}.} We will use log priors (in logarithmic scale) for all of the parameters, except for
$\tan\beta$ which is taken to be a flat prior (in linear scale).
Let us point out, nevertheless, that we do not assume exact universality of $\kappa_i$,
to avoid an artificial degeneracy in the masses of the two scalars/pseudoscalars (and two neutralinos) which appear in the spectrum without doublet contamination (see the discussion in Subsec.~\ref{rsls}).
%\R{CHECK: because on the one hand
%only one linear combination of the right-handed neutrinos mix in an efficient way with the MSSM neutralinos and on the other hand two of the singlets CP even and CP odd are degenerate in masses without doublet contamination.} 
Thus we take 
%$\kappa_3 = 1.04 \kappa_1$, $\kappa_2 = 1.02 \kappa_1$, and scan over $\kappa_1\equiv\kappa$.
\bea
\kappa_3 = 1.04\kappa_1,\,\, 
\kappa_2 = 1.02 \kappa_1,\,\, 
\kappa_1=\kappa, 
\label{kappanu}
\eea
and scan over $\kappa$.

\begin{table}[t!]
\begin{center}
\renewcommand{\arraystretch}{1.4}
\begin{tabular}{|l|l|l|}
\clineB{1-3}{4}
\multicolumn{1}{|l|}{ \bf Scan 1 ($S_1$)}&\multicolumn{1}{|l|}{\bf Scan 2 ($S_2$)}
&\multicolumn{1}{l|}{ \bf Scan 3 ($S_3$)} \\ 
\clineB{1-3}{4}
 $0.01 \le \lambda < 0.2$ & $0.2 \le \lambda < 0.5$  &
 $0.5 \le \lambda < 1.2$ \\ \hline
 \multicolumn{3}{|c|}{$0.01\le \kappa \le 2 $ } \\ 
 \multicolumn{3}{|c|}{$1 \le \tan\beta \le 40$  } \\
 \multicolumn{3}{|c|}{$100 \le {v_R}/{\sqrt 2} \le 7000$ }\\  
 \multicolumn{3}{|c|}{$0 < T_{\lambda} \le 500$ }   \\ 
 \multicolumn{3}{|c|}{$0 < -T_{\kappa} \le 500$ }   
 \\ 
 \multicolumn{3}{|c|}{$0 < - T_{u_{3}} \le  5000$ }  
 \\  
 \multicolumn{3}{|c|}{$200 \le m_{\widetilde Q_{3L}} = m_{\widetilde u_{3R}} \le 2000$  } 
 \\ 
\clineB{1-3}{4}
\end{tabular}
\renewcommand{\arraystretch}{1}
\end{center}
\caption{
Range of low-energy values of the input parameters in Eq.~(\ref{finalp}) that are varied in the three scans, where $\tan\beta$ is a flat prior whereas the others are
%$Y_{\nu_{i}}$, $v_i$, $T_{\nu_{3}}$ and $M_2$ are 
$\log$ priors.
%Universality of the parameters in
%Eq.~(\ref{freeparameters})
%is assumed, with 
%$\lambda_i=\lambda, \kappa_i=\kappa, v_{iR}=v_R, T_{\kappa_i}=T_\kappa$, and 
%$T_{\lambda_i}=T_\lambda$. 
The
VEVs $v_{R}$, and the 
soft parameters $T_{\lambda}$, $T_{\kappa}$, 
$T_{u_{3}}$, $ m_{\tilde Q_{3L}}=m_{\tilde u_{3R}}$ are given
in GeV.}
  \label{Scans-priors-parameters}
\end{table} 

%\vspace{3cm}

%   
\begin{table}[t!]
\begin{center}
\renewcommand{\arraystretch}{1.4}
\begin{tabular}{|c|c|c|}
\clineB{1-3}{4}
\multicolumn{1}{|c|}{\bf\quad
{\bf Scan 1 ($S_1$)}\quad}&\multicolumn{1}{c|}
{\bf\quad{\bf Scan 2 ($S_2$)}\quad}
&\multicolumn{1}{c|}{\bf\quad{\bf Scan 3 ($S_3$)}\quad}\\
\clineB{1-3}{4} 
\multicolumn{3}{|c|}{$m_{\widetilde Q_{1,2L}}=m_{\widetilde u_{1.2R}}
=m_{\widetilde d_{1.2,3R}}
=m_{\widetilde e_{1,2,3R}}=$ 1000} \\
\multicolumn{3}{|c|}{$T_{u_{1,2}}=T_{d_{1,2}}=T_{e_{1,2}}=0$,\; $T_{e_{3}}=$ 40,\; $T_{d_{3}}=$ 100} \\
\multicolumn{3}{|c|}{$-T_{\nu_{1,2}}=10^{-3}$,\; $-T_{\nu_{3}}=3\times 10^{-4}$}\\ 
\multicolumn{3}{|c|}{$M_1 = \frac{M_2}{2} = \frac{ M_3}{3}$ = 900} \\ 
\multicolumn{3}{|c|}{$Y_{\nu_{1}}=
2\times 10^{-7}$,\; $Y_{\nu_{2}}=4\times 10^{-7}$,\, $Y_{\nu_{3}}=0.5\times 10^{-7}$}\\
\multicolumn{3}{|c|}{$v_{1L} =1.5\times 10^{-4}$,\; $v_{2L}=4\times 10^{-4}$,\; 
$v_{3L}=5.5\times 10^{-4}$}
\\
\clineB{1-3}{4}
\end{tabular}
\renewcommand{\arraystretch}{1}
\end{center}
    \caption{
    Low-energy values of the input parameters that are fixed in the three scans.
The VEVs $v_{iL}$ and the soft parameters $T_{u,d,e}$, $m_{{\widetilde Q},{\widetilde u},{\widetilde d},{\widetilde e}}$, $M_{1,2,3}$
%trilinear parameters $T$, soft scalar masses $m$ and soft gaugino masses $M$
are given in GeV.}
  \label{Scans-fixed-parameters}
\end{table} 

%%%%%%%%%%%%%%%%%%%%%%%%%%%%%%%%%%%%%%%%%%%%%%%%%%%%%%%%%

For the choice of the scans, 
we will choose the ranges of $\bla$ ($\equiv \sqrt 3 \lambda$) discussed in Sec.~\ref{SMhiggs} for convenience of the discussion.
In particular, the sample denoted by $S_1$ corresponds to small/moderate values with
 $0.01 \le \lambda < 0.2$, $S_2$ to moderate/large values with $0.2 \le \lambda < 0.5$, and finally $S_3$ to large values $0.5 \le \lambda < 1.2$. 
%The inputs are summarized in Tables~\ref{Scans-priors-parameters} and \ref{Scans-fixed-parameters}.
For each scan, 
%and for each of the other parameter, 
the same ranges for the other parameters are considered.
In particular, the upper bound of $\kappa$ has been motivated in the discussion of Subsec.~\ref{smlike} by relaxing the idea of perturbativity up to the GUT scale, pushing the scale of new physics further below to 10 TeV (see Eq.~(\ref{perturbativity2})).
%\R{CHECK: because on the one hand
%only one linear combination of the right-handed neutrinos mix in an efficient way with the MSSM neutralinos and on the other hand two of the singlets CP even and CP odd are degenerate in masses without doublet contamination.} 
Concerning the range of $v_{R}$, the lower and upper bounds allow to have reasonable Majorana masses for right-handed neutrinos,
${\mathcal M_{i}}=2k_iv_{R}/\sqrt 2$ 
(see Eq.~(\ref{majorana2})), even when $\kappa_i$ are very large or very small, respectively.
%in $150 - 10000$ GeV. The upper and lower bounds allow to avoid
%too small Majorana masses for right handed neutrinos for $\kappa \sim 0.01$ and $\sim 2$. For instance $\kappa \approx 0.01$ and $v_R \approx 10000$ GeV ($\kappa \approx 2$ and $v_R \approx 150$ GeV) gives $\left(m_{\mathcal M}\right) \approx 141$ GeV (424 GeV).
The ranges of $T_{\lambda}$ and $T_{\kappa}$ are also natural following the supergravity framework of Eq.~(\ref{trilinear}).
%$T_{\lambda}=A_{\lambda}\lambda$, $T_{\kappa}=A_{\kappa}\kappa$. 
The lower bound on $m_{\widetilde Q_{3L}}$ of 200 GeV is chosen to avoid too light stops/sbottoms, and the upper bound of 2 TeV is enough not to introduce too large soft masses and therefore too heavy squarks. With this range of $m_{\widetilde Q_{3L}}$, we take the upper bound of $-T_{u_{3}}$ at 5 TeV to be able to reproduce in the small $\lambda$ limit the usual {maximal mixing scenario} 
%in stop sector 
when $m_{\widetilde Q_{3L}} \sim 2$ TeV. 
%\R{Recall that the maximal mixing in the stop sector is given by $|X_t /m_{\widetilde Q_{3L}}| \approx \sqrt{6}$, where $X_t \equiv T_{u_{3}}/Y_{u_{3}} - \mu/\tan\beta$.}

{Let us remark that the MSSM limits on squark masses cannot be applied to the $\mn$. For example, if the stop is the LSP it can decay only via RPV channels into top plus neutrino and bottom plus lepton, and these decays can be prompt or displaced depending on the region of the parameter space of the model. Thus, dedicated analyses are necessary for recasting ATLAS and CMS results to the many possible cases of the model, and we leave them for a forthcoming publication~\cite{kpatcha:2019xxxx}. In this sense, in the present work we choose to be conservative enough using the above lower bound of 200 GeV. We consider that with this value we cover all the potentially interesting range of the model. It is also worth noticing that the regions with small stop masses correspond to (light-red and light-blue) points not fulfilling the perturbative condition up to GUT scale, as shown in Figs.~\ref{S1Tu3MQ3},~\ref{S2Tu3MQ3} and~\ref{S3Tu3MQ3} below (this is also evident in Figs.~\ref{S1-2D-Lambda-Kappa-MQ3},~\ref{S2-2D-Lambda-Kappa-MQ3} and~\ref{S3-2D-Lambda-Kappa-MQ3}).}

%\noindent
%Therefore the parameters we considered in the present study are
%$ \lambda$, $\kappa$, $T_{\lambda}$, $T_{\kappa}$, $v_R$, $\tan\beta$, 
%$T_{u_{3}}$ and $ M_{\tilde Q_{3}} = M_{\tilde U_{3}} $ and 
%Also, we chose the sign convention which guarantees the minimum of the potential always exists as discussed in Section \ref{EW-stability}.

%\noindent
%For the ranges of $\lambda$ and $\kappa$ we do not require the absence of Landau pole  below the GUT scale which would restrict $\sqrt 3 \lambda$ and $\kappa$ to be less about 0.7. We chose the them in a way that to maintain the perturbativity at 10 TeV by allowing for larger values of $\sqrt{3}\lambda$ and $\kappa$ about 2.

The rest of the parameters of the model, which are less relevant
for the analysis, are fixed as shown in Table~\ref{Scans-fixed-parameters}.
For squarks, and right sleptons we choose a value of 1000 GeV.
Note that the rest of soft masses for Higgses, right sneutrinos and left sleptons, are fixed by the minimization conditions, as discussed in Section~\ref{themodel}.
The relations among gaugino masses $M_{1,2,3}$ are inspired by GUTs, and in particular we choose gluinos masses of 2.7 TeV.
As for the other trilinear parameters, 
the values of $T_{d_3}$ and $T_{e_3}$ have been chosen taking into account the supergravity relations and the corresponding Yukawa couplings.
Finally, 
the parameters $Y_{{\nu}_i}$, $v_{iL}$, and $T_{\nu_i}$ are mainly determined by neutrino and sneutrino physics (see Eqs.~(\ref{neutrinoph}) and~(\ref{evenLLL2})). 

%In addition, for the left sneutrinos whose tree level approximate mass is given in Eq.(\ref{tree-vL-approx}) the parameters $Y_{{\nu}_i}$ and $v_i$ are mainly determined by neutrino physics. We fix them but in addition to  $\lambda$ and $\tan\beta$, we vary $\kappa$ and $v_R$ and $\tan\beta$ in order to control the pre-factor, the second and the third terms of Eq.(\ref{tree-vL-approx}). In this setup $T_{{\nu}_i}$ can be fixed.
%We comment that, typically the left sneutrino masses can controlled by $T_{{\nu}_i}$, $Y_{{\nu}_i}$ and $v_i$ \cite{2019arXiv190702092K}, but in this work the most
%relevant parameters are $\lambda$, $\kappa$, $v_R$ and $\tan\beta$.

Since reproducing neutrino data
%~\cite{Capozzi:2017ipn,deSalas:2017kay,deSalas:2018bym,Esteban:2018azc} 
is an important asset of the $\mn$, a few words on the subject are worth it.
As explained in Sec.~\ref{themodel}, how the model reproduces the correct neutrino masses and mixing angles has been intensively addressed in the literature~\cite{Ghosh:2008yh,Fidalgo:2009dm,Ghosh:2010zi,Gomez-Vargas:2016ocf,Kpatcha:2019gmq}. Although the parameters in Eq.~(\ref{freeparametersn}),
$\lambda_i$, $\kappa_i$, $v_{iR}$, $\tan\beta$, $Y_{\nu_i}$, $v_{iL}$ and $M$,
are important for neutrino physics, the most crucial of them are $Y_{\nu_i}$, $v_{iL}$ and $M$,
%There, it was shown that the most relevant parameters that control
%neutrino physics are 
% $\lambda_i$, $\kappa_i$, $v_{iR}$, $\tan\beta$, 
%$Y_{\nu_i}$, $v_{iL}$, $M_1, M_2$.
%However, 
and they are essentially decoupled from the parameters in Eq.~(\ref{freeparameters}) controlling Higgs physics.
%, as we can deduced from the discussion in Section~\ref{rsls} (see 
Thus, for a suitable choice of $\lambda_i$, $\kappa_i$, $v_{iR}$ and $\tan\beta$ reproducing Higgs physics, there is still enough freedom to reproduce in addition neutrino data by playing with $Y_{\nu_i}$, $v_{iL}$ and $M$,
%the other parameters, 
as shown in Ref.~\cite{Kpatcha:2019gmq}.
As a consequence, we will not scan over the parameters 
$Y_{\nu_i}$, $v_{iL}$, $M_1$, $M_2$ in order to relax our already demanding computing task, and since it is not going to affect our results.
For our purposes, it will be sufficient 
to choose these parameters mimicking the type of solutions
of neutrino physics with normal ordering found in Ref.~\cite{Kpatcha:2019gmq},
imposing only the cosmological upper bound on the sum of the masses of the light active neutrinos given by  $\sum m_{\nu_i} < 0.12$ \cite{Aghanim:2018eyx}.

The same comment applies to the parameters $T_{\nu_i}$ in 
Eq.~(\ref{tnu}), which are only relevant to determine the left sneutrino masses, and therefore we fix them to mimic also the left sneutrino physics of Ref.~\cite{Kpatcha:2019gmq}.
In that work, 
%Ref.~\cite{Kpatcha:2019gmq} 
it was
easy for $M>0$ to find solutions with the gaugino seesaw as the dominant one for the third family. In this case, $v_{3L}$ determines the corresponding neutrino mass and $Y_{\nu_3}$ can be small.
On the other hand, the normal ordering for neutrinos determines that the first family dominates the lightest mass eigenstate implying that $Y_{\nu_{1}}< Y_{\nu_{2}}$ and $v_1 < v_2,v_3$, {with both $\nu_{R}$-Higgsino and gaugino seesaws contributing significantly to the masses of the first and second family}. Taking also into account that the composition of these two families in the second mass eigenstate is similar, we expect $v_2 \sim v_3$. 
%Now for this solution we will have $m_{\widetilde{\nu}_{3}}$ as the smallest of all the sneutrino masses.
Concerning left sneutrino physics, a light tau left sneutrino was required in 
Ref.~\cite{Kpatcha:2019gmq} 
%that work 
implying $-T_{\nu_{3}} < -T_{\nu_{2}}=-T_{\nu_{1}}$.
This pattern of hierarchies for $Y_{\nu_{i}}$, $v_{iL}$, and $T_{\nu_{i}}$
is used in Table~\ref{Scans-fixed-parameters}.

%In this we discuss the relevant parameters for scanning for Higgs physics in the $\mn$.
%We optimize the number of parameters to be used and this choice is motivated by the fact that scanning over large number of parameters is time and computational expensive.

% \clearpage
%%%%%%%%%%%%%%%%%%%%%%%%%%%%%%%%%%%%%%%%%%%%%%%%%%%%%%%%%%%%%%%%%%%%%%%%%%%%%%%%%%%%%
\section{Results }\label{results-scans}
By using the methods described in the previous section, we 
evaluate now the constraints on the parameter space of the $\mn$.
%using the 7-, 8- and 13-TeV LHC data. 

To find regions consistent with experimental observations we have performed about 160 million of spectrum evaluations in total, and the total amount of computer required for this was approximately 1110 CPU years.

\begin{figure}[t!]
 \centering
 \includegraphics[width=\linewidth, height= 0.28\textheight]{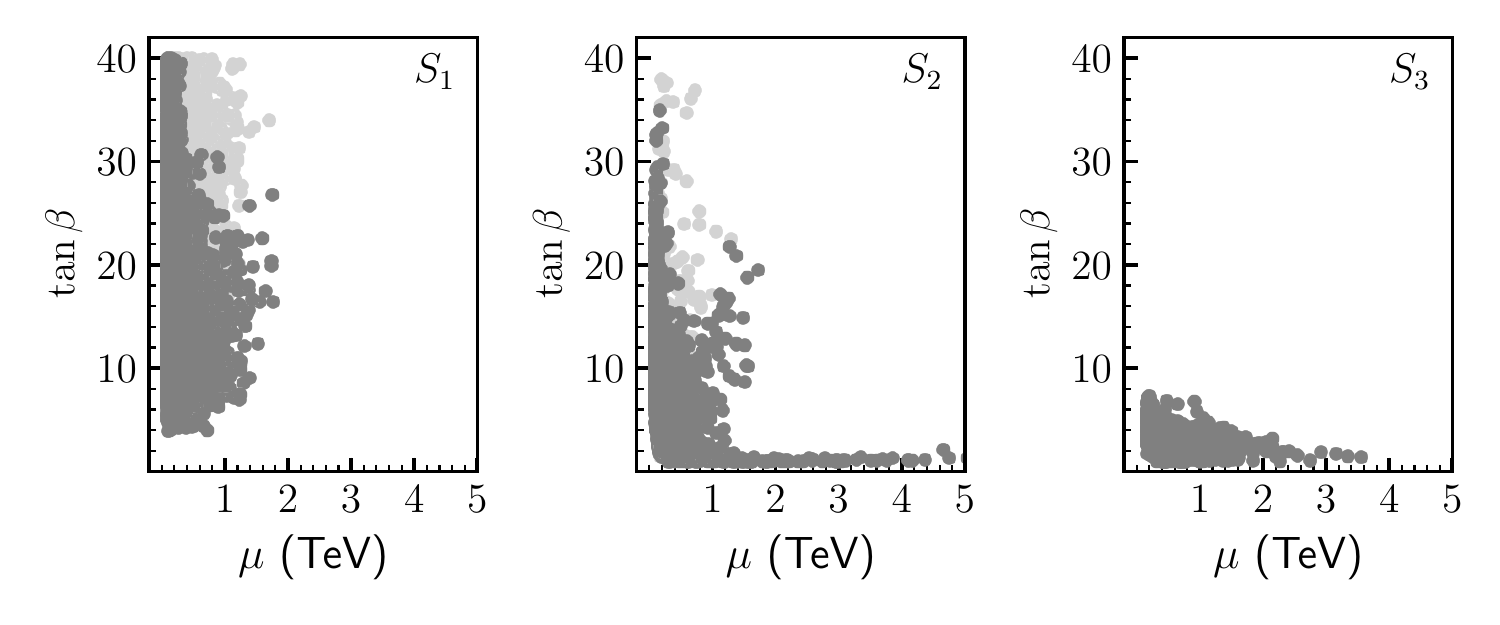}
  \caption{ Constraints from $b \to s \gamma$ in the $\tan\beta-\mu$ plane,
  for scans $S_{1,2,3}$. The grey (light-grey) color corresponds to points of the parameter space fulfilling Higgs physics that are (are not) compatible with the BR($b \to s \gamma$).
  }
  \label{BSG-Scans-1DFigs}
\end{figure}

{Differently from other studies, in this work the likelihood is used to drive {\tt MultiNest}. Then, the selection of the viable points is not based on the value of the likelihood but rather on the series of cuts that are applied on the samples, as will be discussed subsequently.}

To carry out the analysis, we first demand Higgs physics to be fulfilled. As already mentioned in Section~\ref{methodology}, we use {\tt HiggsBounds} and {\tt HiggsSignals} to take into account the constraints from 
7-, 8- and 13-TeV LHC data,
%the LHC data, 
%the 7, 8 and latest 13 TeV LHC data 
as well as those from LEP and Tevatron.
In particular, we require that the p-value derived by {\tt HiggsSignals} be larger than 5\%.
It is worth noticing here that, with the help of 
{\tt Vevacious}~\cite{Camargo-Molina:2013qva}, we have also checked that the EWSB vacua corresponding to the previous allowed points are viable.
Then, we select points that lie within $3\sigma$ from $b \to s \gamma$, $B_s \to \mu^+\mu^-$, and $B_d \to \mu^+\mu^-$. In the third step, the points that pass these cuts are required to also satisfy the upper limits of $\mu \to e \gamma$ and $\mu \to eee$, and the lower bound on the chargino mass inspired in RPC SUSY following 
Sec.~\ref{constraints}.
{After all these cuts are applied, about 91\% of the points survive.}
At last we require all the points that passed the above set of cuts to satisfy 
the cosmological upper bound on the sum of the masses of the light active neutrinos.

%In the next subsection,
%we discuss the corners of the parameter space that are affected by the constraints imposed, and in Subsection~\ref{viable-parameter-space} we present the viable regions of the parameter space.

%{\it Constraints on the parameter space}

%\label{Effects-constraints}

As 
we will explain below, after imposing the relevant constraints from Higgs physics, 
%Imposing all the cuts discussed above, 
%we will see that 
only $b \to s \gamma$ and (less importantly) the bound on neutrino masses put further constraints on the parameter space of the $\mn$.
%and we wish to discuss the regions of the parameter space that have been affected. 
As already mentioned in Sec.~\ref{constraints}, in our computation
we have not tried to explain the discrepancy between the measurement of the muon anomalous magnetic moment and the SM prediction $\Delta a_{\mu}$. Nevertheless, for completeness, we will discuss the level of compatibility of the SUSY contributions with this value, and possible improvements in this direction.

\vspace{0.25cm}

\noindent
\textbf{b$\to$ s$\gamma$}

\noindent 
The BR($b\to s\gamma$) puts some constraints on the parameters space of the $\mn$, as shown {in Fig.}~\ref{BSG-Scans-1DFigs}\footnote{{All plots in this work have been made using {\tt Matplotlib}~\cite{Hunter:2007ouj}}.}.
There we show the constraints from $b\rightarrow s\gamma$ for all points of the parameter space fulfilling Higgs physics. For instance, in our setup 
this BR can be too small in certain regions of the parameter space. 
%Note nevertheless that there are light-grey points (forbidden) on top of grey points (allowed), thus due to the several free parameters of the model we cannot say that concrete values of $\tan\beta$ and $\mu$ are always forbidden. We must check this constraint point by point.
Forbidden points occur for small to moderate values of $\lambda$, such as in $S_1$ and $S_2$, when $\tan\beta$ can be large while $m_{\widetilde Q_{3L}}$ can be small.
%
%\noindent
As is well known, the most important contributions
to the BR($b\to s\gamma$) come from chargino/stop and charged Higgs/top mediated processes~\cite{Ciuchini:1998xy}.
On the one hand, the charged Higgs contribution always tends to
increase the SM value while that of the charginos depends on the sign
of $M_2$, $T_{u_{3}}$ and $\mu$, 
where in our case $\mu=3\lambda v_R/\sqrt 2$.
Since we are working with $M_2, \mu > 0$ and $T_{u_{3}}<0$, 
the contribution 
from charginos in the loops acts destructively. Also for light
sparticles (here charginos, charged Higgs and stops) and/or 
large $\tan\beta$ the effects can be large.
This is actually what happens in our cases. For small/moderate $\lambda$,
large $\tan\beta$ favors increasing this effect. In the regime  of destructive contribution involving light stops (when $m_{\widetilde Q_{3L}}$ becomes small or in the maximal mixing scenario) and light Higgsinos (winos are moderately heavy since we fix $M_2$ to 1800 GeV), this effect is large and suppresses the BR($b\to s\gamma$). Note that for $S_3$ this does not occur. The reason is that large values of $\tan\beta$ are not needed, as we will see in detail in the next subsection, and in addition moderate values come together with relatively large values of $m_{\widetilde Q_{3L}}$. 

\begin{figure}[t!]
 \centering
 \includegraphics[width=\linewidth, height= 0.3\textheight]{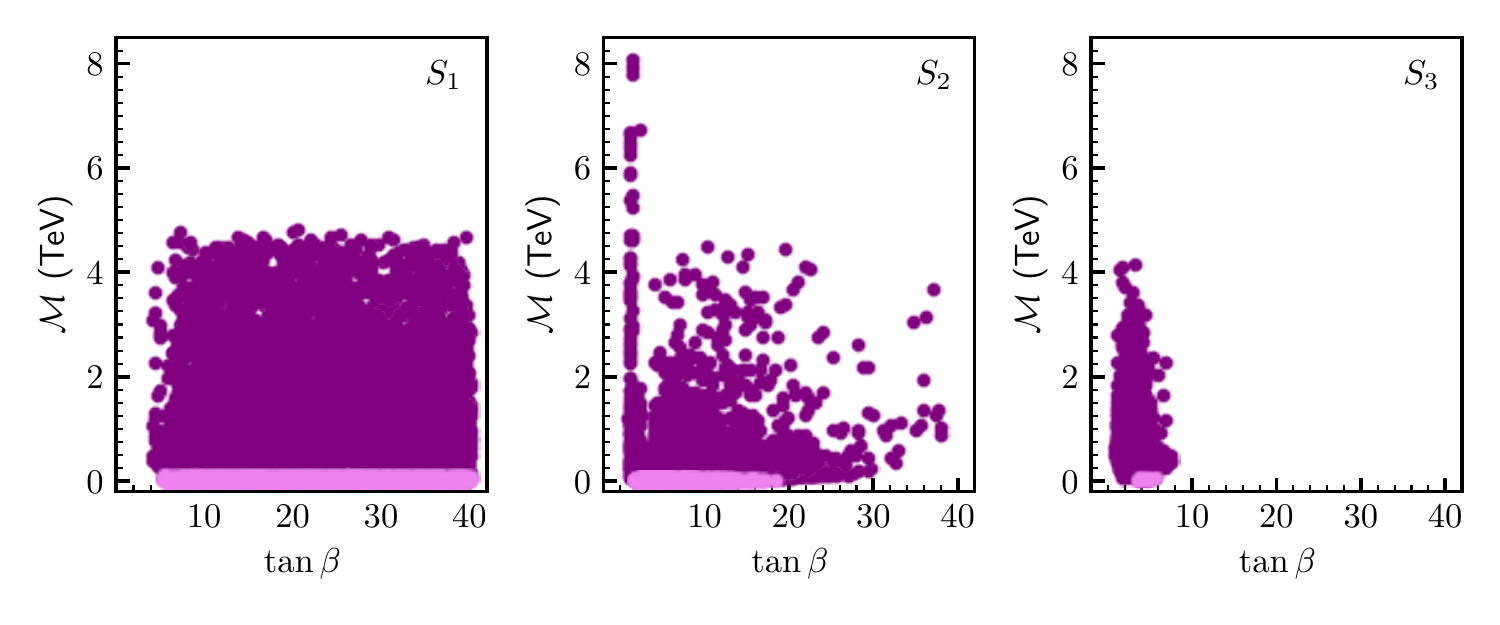}
  \caption{Constraints from $\sum m_{\nu_i} < 0.12$ eV in the $\mathcal{M}-\tan\beta$ plane, for scans $S_{1,2,3}$. The purple (light-purple) color corresponds to points of the parameter space fulfilling Higgs physics that are (are not) compatible with the cosmological upper bound on the sum of the masses of the light active neutrinos.}
  \label{nu-masses-KapvR-vu}
\end{figure}

\vspace{0.25cm}
%%%%%%%%%%%%%%%%%%%%%%%%%%%%%%%%%%%
\noindent 
\textbf{Sum of neutrino masses}

\noindent 
In Fig.~\ref{nu-masses-KapvR-vu}, we show the constraints on the parameter space fulfilling Higgs physics imposed by the requirement $\sum m_{\nu_i} < 0.12$ eV
in the $\mathcal{M}-\tan\beta$ plane, 
with $\mathcal{M}=2\kappa\ v_R/\sqrt 2$.
%in the $\tan\beta$ vs. $\kappa v_R$ plane.
We find that the sum of the masses of the three light neutrinos can exceed this upper bound when the Majoranna masses are small. This can be qualitatively explained using 
Eq.~(\ref{neutrinoph}) with the approximations discussed below 
Eq.~(\ref{freeparametersn}). Then, the gaugino seesaw contributions to
neutrino masses given by the second term in Eq.(\ref{neutrinoph}), with
$M^{\text{eff}}=M$, is fixed in our scans. In particular, 
using the values of Table~\ref{Scans-fixed-parameters} for 
$v_{iL}$ and $M = 2640.45$ GeV from the values of $M_{1,2}$,
we can compute these contributions to the diagonal entries of the mass
matrix $(m_{\nu})_{ii}$, which turn out to be in absolute value 0.002, 0.015, and 0.0286 eV
for $i=1,2,3$, respectively.
This indicates that for sizable $\nu_{R}$-Higgsino seesaw, i.e. the first term in Eq.~(\ref{neutrinoph}), the mass of the heaviest neutrino can easily be made too large. This occurs when $\mathcal{M}$ is small.
%making the first terms in Eq.(\ref{Limit no mixing Higgsinos gauginos2}) large.
For example, for $\tan\beta = 10$ and $\mathcal{M}=30$ GeV
%$\kappa \approx 0.05$ and $v_R \approx 500$ GeV,
the $\nu_{R}$-Higgsino seesaw contribution to the diagonal entries is in absolute value around
0.027, 0.108, and 0.0017 eV, respectively, and added to the gaugino seesaw
at least one neutrino mass would be larger than 0.12 eV.
Actually, in our scenarios the effect of $\tan\beta$ is not very relevant, and the size of $\mathcal{M}$ is the most important one. In particular, as shown in Fig.~\ref{nu-masses-KapvR-vu}, in scans $S_1$, $S_2$, and $S_3$,
for $\mathcal{M}$ below 123, 52, and 51 GeV, respectively, we find points excluded by the cosmological upper bound on neutrino masses.

\begin{figure}[t!]
 \centering
\includegraphics[width=\linewidth, height= 0.28\textheight]{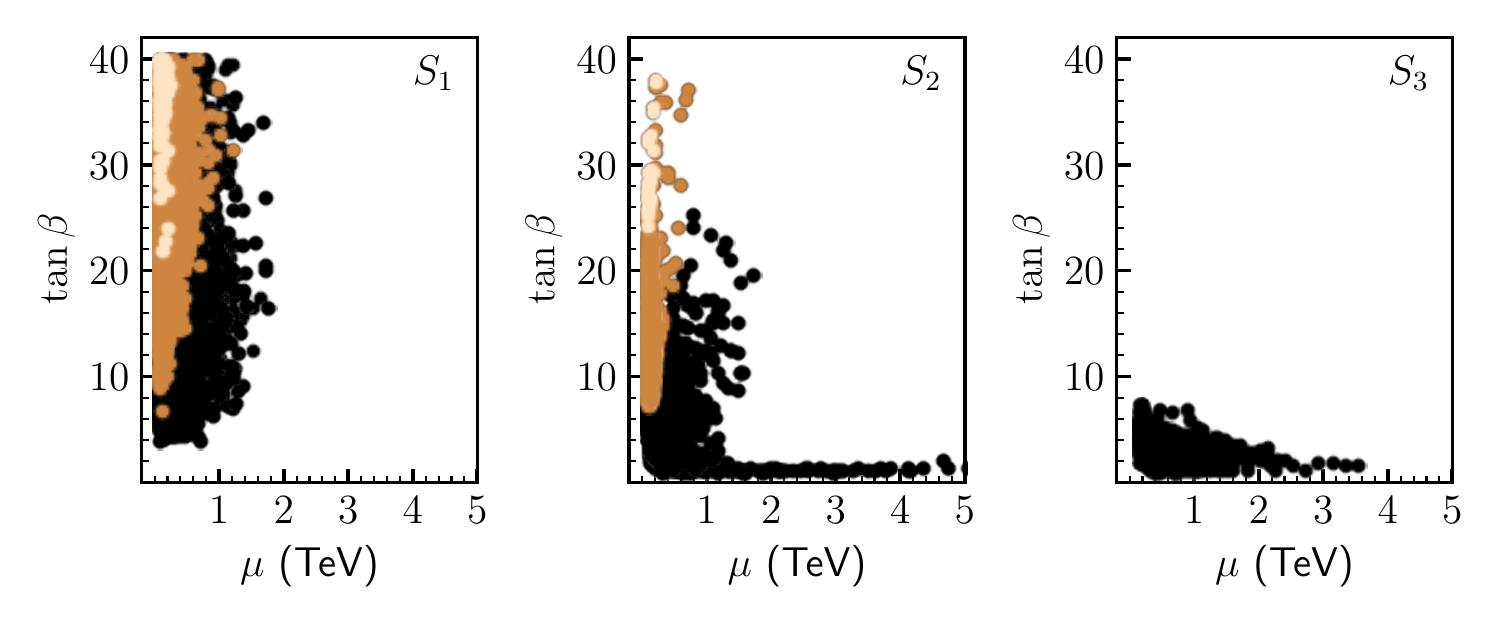}
  \caption{Analysis of $a^{\text{SUSY}}_\mu$ in the $\tan\beta-\mu$ plane 
  for scans $S_{1,2,3}$. The light-brown color 
  corresponds to points of the parameter space fulfilling Higgs physics compatible at
  $2 \sigma$ with $\Delta a_{\mu}$.
  Brown color corresponds to compatibility between $2$ and
  $3 \sigma$, whereas points with black color are compatible within 
  $3$ and
  $3.5 \sigma$.
%  are up to $2 \sigma$ deviations from experimental value, light-brown for points between $2 \sigma$ and $3 \sigma$ deviations from experimental value. Black points are in better agreement than the SM (close to $3.5 \sigma$ deviations) but beyond the $3\sigma$ deviations.
  % corresponds to
%  points of the parameter space 
  % have (have not) $a^{\text{SUSY}}_{\mu}$ within $\pm 1 \,\sigma$ of SM $a^{\text{SM}}_{\mu}$ uncertainty,
 }
  %The green points are those that are compatible at $\pm 1 \sigma$ with $\Delta a_{\mu}= a_{\mu}^{\text{exp}}-a_{\mu}^{\text{SM}}=26.8 \pm 6.3\pm 4.3 \times 10^{-10} $ \cite{Tanabashi:2018oca}}
  \label{Constraints-amu-mueff-vs-tanB}
\end{figure}

%% aexpμ= 11659209.1(5.4)(3.3)×10−10      = 11659209.1(6.3)×10−10  
%% aSMμ=  11659182.3(0.1)(3.4)(2.6)×10−10 = 11659182.3(4.281)×10−10

\vspace{0.25cm}

%%%%%%%%%%%%%%%%%%%%%%%%%%%%%%%%%%
\noindent
\textbf{Muon anomalous magnetic moment}

\noindent
The difference between the experimental measurement and the SM prediction
 $\Delta a_{\mu}= a_{\mu}^{\text{exp}}-a_{\mu}^{\text{SM}} = (26.8 \pm 6.3\pm 4.3) \times 10^{-10}$~\cite{Tanabashi:2018oca}, where the errors are from experiment and theory prediction (with all errors combined in quadrature), respectively, represents an interesting but not conclusive discrepancy of 3.5 times the combined 1$\sigma$ error.
SUSY contributions $a^{\text{SUSY}}_\mu$ 
%to the anomalous magnetic moment of the muon, 
can be large
in the presence of light muon sneutrino and charginos or light neutralino and smuons.
We found in our scans $S_1$, $S_2$ and $S_3$ that $a^{\text{SUSY}}_\mu$ is smaller than $16.96\times 10^{-10}$, $16.83\times 10^{-10}$, and $3.7 \times 10^{-10}$, respectively.
Thus, although none of the points of the parameter space is compatible at $1 \sigma$ with $\Delta a_{\mu}$
%= a_{\mu}^{\text{exp}}-a_{\mu}^{\text{SM}} = (26.8 \pm 6.3\pm 4.3) \times 10^{-10}$, %in some regions $a^{\text{SUSY}}_{\mu}$ is larger than the uncertainty of the SM value.
in some regions $a^{\text{SUSY}}_\mu$ is compatible at
%$a_{\mu}$ 
%is within the 
$2 \sigma$.
%deviation from the experimental mean value.
%
Note that we are neglecting the uncertainties in the SUSY computation. %and adding in quadrature the uncertainties of
%$\Delta a_{\mu}$
%($a^{\text{SM}}_{\mu} = (11659182.3 \pm 0.1 \pm 3.4 \pm 2.6) \times 10^{-10}$ \cite{Tanabashi:2018oca} and the experimental ones)
%to compute the total uncertainty.
The result is shown in Fig.~\ref{Constraints-amu-mueff-vs-tanB}.
% In Fig.~\ref{Constraints-amu-mueff-vs-tanB}, we show in brown (light-brown) color the points of the parameter space fulfilling Higgs physics where $a^{\text{SUSY}}_{\mu}$ is (is not) within $\pm 1 \,\sigma$ of the quadrature sum of the uncertainties of $a^{\text{SM}}_{\mu} = (11659182.3 \pm 0.1 \pm 3.4 \pm 2.6) \times 10^{-10}$ \cite{Tanabashi:2018oca}.
The largest contributions to $a^{\text{SUSY}}_{\mu}$ are found for small $\mu$ and large $\tan\beta$. In our scenarios, since bino- and wino-like (neutralino or chargino) eigenstates are heavy (in our scans $M_2=2M_1=1800$ GeV) the contributions involving them are suppressed. Besides, although the Higgsino-like eigenstates can be light when $\mu$ is relatively small, their contributions can be diluted by the small Yukawa coupling of the muon. Nevertheless, when $\tan\beta$ is very large this effect can be more important.
%\R{ Indeed it is possible to explain the 3.5$\sigma$ deviation between the measurement and the SM prediction of the anomalous magnetic moment of the muon, in some region of the parameter space. The green points in Fig.\ref{Constraints-amu-mueff-vs-tanB} are compatible at $\pm 1 \sigma$ with $\Delta a_{\mu}= a_{\mu}^{\text{exp}}-a_{\mu}^{\text{SM}}=26.8 \pm 6.3\pm 4.3 \times 10^{-10}$.  }
A way of explaining the discrepancy 
$\Delta a_{\mu}$ with $a^{\text{SUSY}}_\mu$
%between $a^{\text{exp}}_\mu$ and $a_\mu$ 
is to try to lower the muon left sneutrino mass, which in these scans is generically large given the input parameters chosen for neutrino physics. Changing the latter we could obtain smaller masses, and we leave the analysis of this interesting possibility for a forthcoming publication~\cite{kpatcha:2019xx}.

%\vspace{0.5cm}
%%\subsubsection*{Constraint from invisible decay widths of Z and W}

\subsection{ Viable regions of the parameter space}  \label{viable-parameter-space} 

Once  $b\to s\gamma$, and mainly Higgs physics, have determined 
the parameter space that is viable in the $\mn$, we will discuss it in detail.
In order to carry it out we will follow the division in the three different scans presented in Subsection~\ref{scan}.
%about the choice of the three relevant scans.
%let us concentrate first on the analysis of the results for $S_1$.

%%%%%%%%%%%%%%%%%%%%%%%%%%%%%%%%%%%%%%%%%%%%%%%%%%%%%%%%%%%%%%%%%%%%%%%%

%\vspace*{0.25cm}
%\noindent
%{\it Scan 1}
%($0.01 \leq \lambda < 0.2$)}

\begin{figure}[t!]
 \centering
 \includegraphics[width=0.8\linewidth, height= 0.35\textheight]{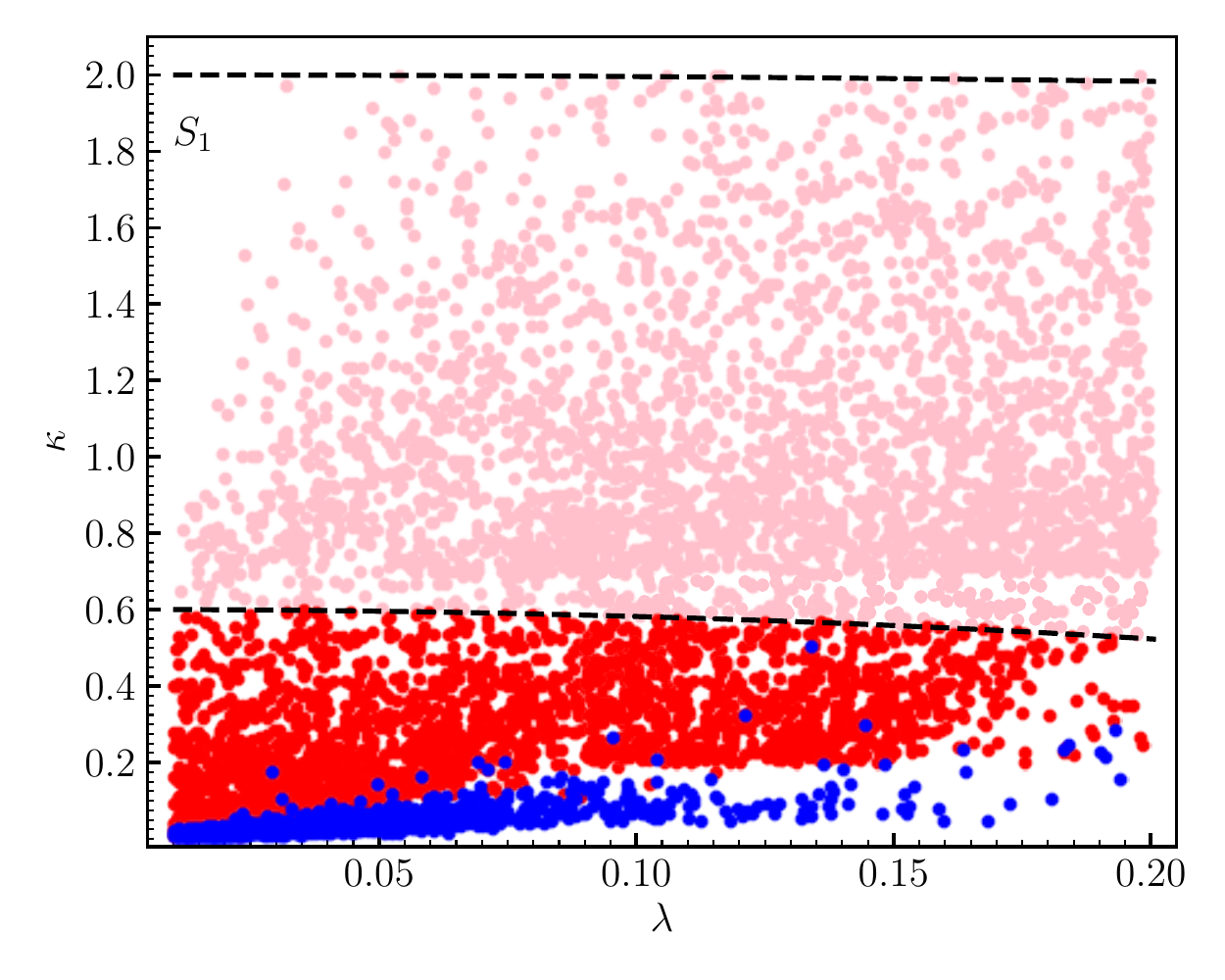}
  \caption{
 Viable points of the parameter space for $S_1$ in the $\kappa-\lambda$ plane.
% Universality of the parameters is assumed with 
%$\lambda_i=\lambda, \kappa_i=\kappa, 
%v_{iR}=v_R, T_{\kappa_i}=T_\kappa$, and 
%$T_{\lambda_i}=T_\lambda$. 
The red and light-red (blue) colours represent cases where the SM-like Higgs is (is not) the lightest scalar.
All red and blue points below the lower black dashed line fulfill the perturbativity condition up to GUT scale of Eq.~(\ref{perturbativity1}). 
Light-red points below the upper black dashed line fulfill the perturbativity condition up to 10 TeV of Eq.~(\ref{perturbativity2}).
%fulfilling perturbativity up to the GUT scale as in Eq.~(\ref{perturbativity1}). 
%For the light-red points the perturbativity condition is relaxed up to 10 TeV, as in Eq.~(\ref{perturbativity2}).
%\R{dibujar otra dashed line como en 3D plots?}
%The black dashed line separating the perturbative region from the non-perturbative one fulfills the equality in Eq.~(\ref{perturbativity2}).
%All points below the lower (upper) black dashed line fulfill the perturbativity condition in Eq.~(\ref{perturbativity1}) (Eq.~(\ref{perturbativity2})).
%  The same as in Fig.~\ref{S1LamtanB} but for $\kappa$ versus $\lambda$, where $\kappa=\kappa_i$ and $\lambda=\lambda_i$ 
}
\label{S1Lambda-kappa}
\end{figure}

\subsubsection{ {\it Scan 1} ($0.01 \leq \lambda < 0.2$)}
%\subsection{Scan $S_1$}
\label{scan01}

%\noindent 
%{\it (i) Scan $S_1$}

Let us concentrate first on the analysis of the results for Scan 1 ($S_1$).
We show in Fig.~\ref{S1Lambda-kappa} the viable points of the parameter space in the 
$\kappa - \lambda$ plane.
The red points represent cases where the SM-like Higgs boson 
is the lightest scalar. All of them fulfill perturbativity up to the GUT scale, and therefore $\kappa \lsim 0.6$. For the light-red points the SM-like Higgs boson is still the lightest scalar, but we have relaxed the perturbativity condition up to 10 TeV and therefore $0.6 \lsim \kappa \lsim 2$.
%The two black dashed lines correspond to the perturbativity conditions in Eqs.~(\ref{perturbativity1}) and~(\ref{perturbativity2}).
%Points above those lines do not fulfill the corresponding condition.
On the contrary, the blue points represent cases where the SM-like Higgs boson is not the lightest scalar. 
This figure can be considered as the summary of results for this scan, which we now discuss in detail.
%Let us now discuss them in detail.

\begin{figure}[t!]
 \centering
 \includegraphics[width=0.8\linewidth, height= 0.35\textheight]{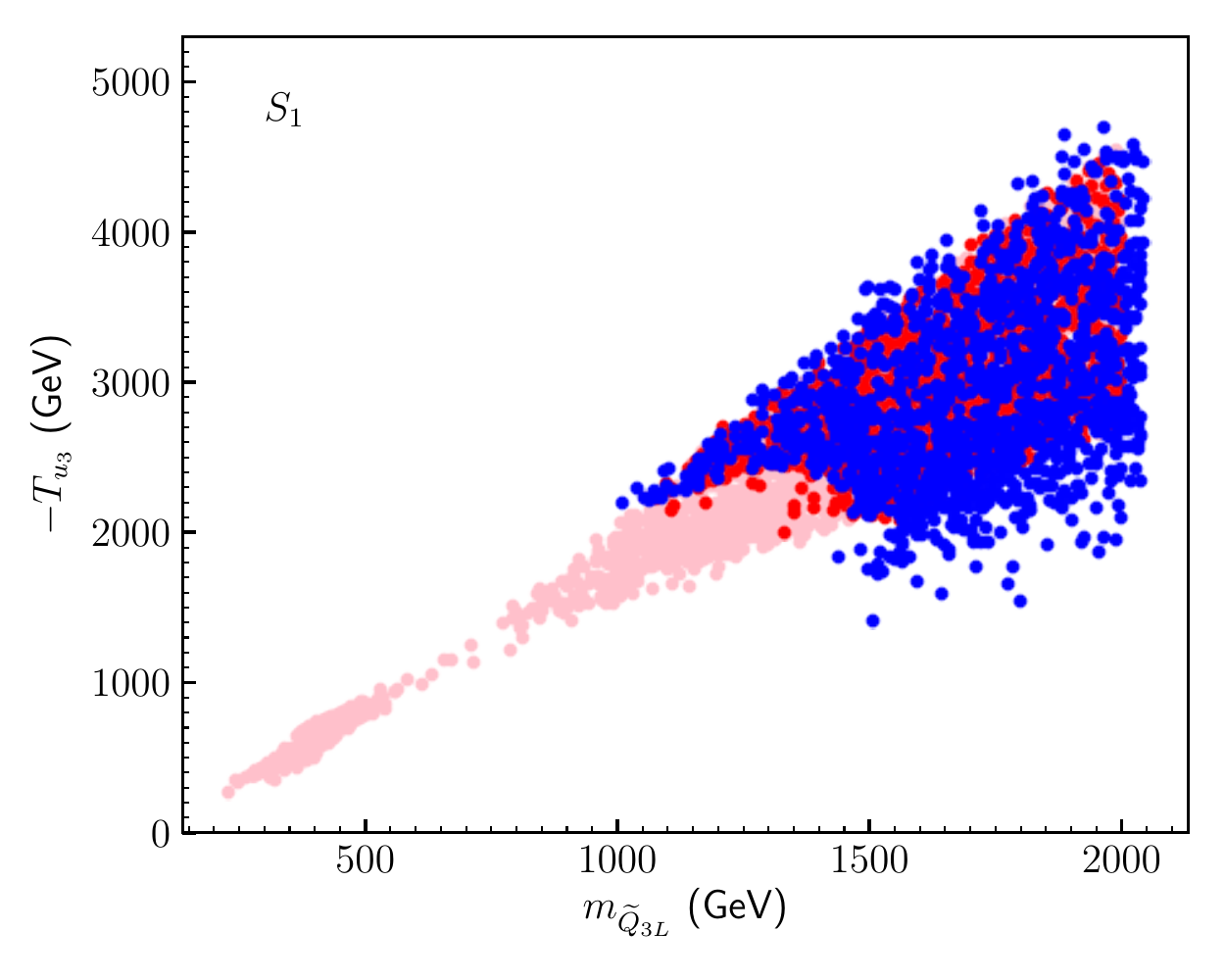}
  \caption{ Viable points of the parameter space for $S_1$ in the
  $-T_{u_3}$ versus $m_{\widetilde Q_{3L} }$ plane.
% Universality of the parameters is assumed with 
%$\lambda_i=\lambda, \kappa_i=\kappa, 
%v_{iR}=v_R, T_{\kappa_i}=T_\kappa$, and 
%$T_{\lambda_i}=T_\lambda$. 
The color code is the same as in Fig.~\ref{S1Lambda-kappa}.
%, but for $-T_{u_3}$ versus $m_{\widetilde Q_{3L} }$.
  %The color map is same as in Fig.\ref{Ak-kapvR}
  }
  \label{S1Tu3MQ3}
\end{figure}

As shown in the figure, we find viable solutions in almost the entire $\kappa - \lambda$ plane analyzed in $S_1$. 
The only small (white) region that becomes forbidden corresponds to 
very small values of $\lambda$ and very large (non-perturbative up to the GUT scale) values of $\kappa$.
This can be understood taking into account that we are asking to all the points to fulfill the chargino mass lower bound of RPC SUSY,
%of about 100 GeV, 
which corresponds to condition $\mu=3\lambda v_{R}/\sqrt 2 \gtrsim 100$ GeV.
Thus for a small $\lambda$, a large $v_R$ is needed
(see also Fig.~\ref{S1-2D-Lambda-Kappa-vR} in Appendix~\ref{figuresscans}\footnote{{We do not perform a statistical interpretation of the results. Thus, in all the plots shown in Appendix~\ref{figuresscans} the points are plotted on top of each other using {linear interpolation} {griddata} to make the filled contours.}}).
%For example, $\lambda=$ 0.01 implies 
%$v_{R}/\sqrt 2 \gtrsim$ 3300 GeV.
%$v_{R} >$ 5000 (1000) GeV.
%However, this gives rise to a large mixing between singlet- and SM-like states, as we can see in Eq.~(\ref{mixingevenR2}) where
%${\mathcal M}
%={2}\kappa {v_{R}/\sqrt 2}$ becomes too large when $\kappa$ is large.
However, this gives rise to a large value of 
%in Eq.~(\ref{mixingevenR2}) where
${\mathcal M}
={2}\kappa {v_{R}/\sqrt 2}$ and, as a consequence, the
condition in Eq.~(\ref{sneulta}) to avoid tachyonic left sneutrinos cannot be fulfilled for any value of $\kappa$. In particular, 
combining both conditions we can write 
$\frac{100\ \text{GeV}}{3\lambda}  \lsim \frac{v_R}{\sqrt 2}
 \lsim \frac{-T_{\nu_i}/Y_{\nu_i}}{\kappa}$, which cannot always be fulfilled.
%\bea
%\frac{100\ \text{GeV}}{3\lambda}  \lsim \frac{v_R}{\sqrt 2}
 %< \frac{-T_{\nu_i}/Y_{\nu_i}}{\kappa}.
%\label{vrbound}
%\eea
This is the case for the muon left sneutrino whose ratio 
$-T_{\nu_2}/{Y_{\nu_2}}=2500$ GeV
is the smallest of the three families, as can be deduced from Table~\ref{Scans-fixed-parameters}.
For example, $\lambda=$ 0.01 implies 
$v_{R}/\sqrt 2 \gtrsim$ 3300 GeV, and then
%For the above example with $\lambda=$ 0.01, 
it is straighforward to see that $\kappa\lsim 0.75$ to avoid tachyons.
%From Eq.~(\ref{sneulta2}), for an approximately fixed value of $\mu$ we can also deduce that there is a straight line separating the allowed region (to the right of the figure) from the forbidden tachyonic one (white).
Let us point out nevertheless, that this forbidden tachyonic region in Fig.~\ref{S1Lambda-kappa} turns out to be an artifact of our simplified assumption about the neutrino (sneutrino) pattern in order to relax the demanding computing task, as discussed in Subsection~\ref{scan}. Simply breaking the degeneracy between
$T_{\nu_1}$ and $T_{\nu_2}$, taking a larger value for $T_{\nu_2}$, we would recover this region as viable.
{It is apparent that there is a less point-dense area with $\kappa \lesssim 0.6$ and $\lambda \gtrsim 0.15$ (the same occurs for the area with $\lambda \gtrsim 0.45$ in Fig.~\ref{S2Lambda-kappa} to be discussed below).
This is just an artifact of the sampling, and it would have been filled out with more computing-time resources.}

\begin{figure}[t!]
 \centering
 \includegraphics[width=0.8\linewidth, height= 0.35\textheight]{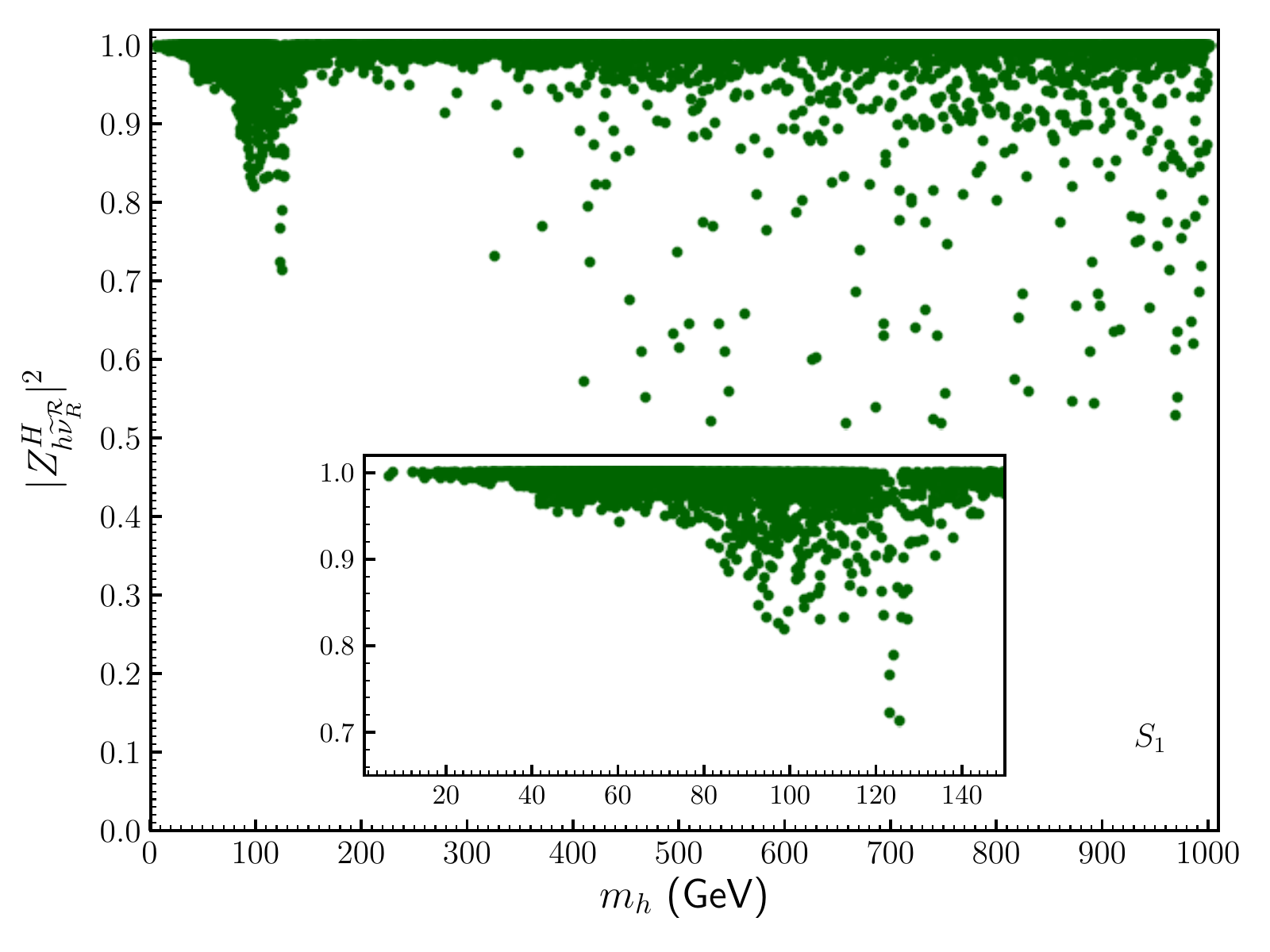}
  \caption{The singlet component $\sum_i|Z^H_{h
  \widetilde{{\nu}}^{\mathcal{R}}_{iR}}|^2$ of the singlet-like scalars $h$
versus their masses, for $S_1$.
We only show here viable points with scalar masses smaller than 1000 GeV.
In the lower part we zoom in the low-mass region.
}
\label{composition1}
\end{figure}

Going back to the values of $v_R$ in Fig.~\ref{S1-2D-Lambda-Kappa-vR},
%of Appendix~\ref{figuresscans}, 
it is worth noticing that
for large $\lambda$ and/or large $\kappa$ they are bounded,
$v_R/\sqrt 2 \lsim 2000$ GeV.
The reason is that for those points, to increase the value of $v_R$ would increase the mixing term
$m_{H^{\mathcal{R}}_{u} H^{\mathcal{R}}_{d}}^2$ in Eq.~(\ref{mixinghuhd})
of Appendix~\ref{Apendix:Sneutrino-masses}, decreasing 
therefore the SM-like Higgs mass, and eventually leading to the appearance of a negative eigenvalue. Note in this sense that the diagonal term
$m_{H^{\mathcal{R}}_{u} H^{\mathcal{R}}_{u}}^2$
($m_{H^{\mathcal{R}}_{d} H^{\mathcal{R}}_{d}}^2$) in Eq.~(\ref{mixinghu}) (Eq.~(\ref{mixinghd})) is small (large) for the large values of $\tan\beta$ present in this scan, as we will discuss below (see Fig.~\ref{S1-2D-Lambda-Kappa-tanB}).
The mixing terms with right sneutrinos
also increase with the value of $v_R$, as can be seen in 
Eqs~(\ref{mixingevenR2}) and~(\ref{mixingevenR}),
but much less than the above between Higgses, since the former go like $v_R$ whereas the latter as $v_R^2$.
As we can see in those equations, 
%of Appendix~\ref{Apendix:Sneutrino-masses}, 
the value of $T_\lambda$ is also important to determine the mixing among states.
In
Fig.~\ref{S1-2D-Lambda-Kappa-Tlambda}, we see that in most of the regions $T_\lambda$ has an upper bound of around $200$ GeV, and only for the lower right region with large $\lambda$, but small (perturbative up to the GUT scale) $\kappa$, it can reach up to 500 GeV. In the region to the left of the latter, although the values of 
$\kappa$ are also small, $v_R$ is large as discussed above, and smaller 
$T_\lambda$ is favoured.
%of Appendix~\ref{scan1}
On the other hand,
assuming the supergravity relation
$A_\lambda=T_\lambda/ \lambda$, one can check that 
in most of the regions $A_\lambda$ has the upper bound of around 2 TeV, 
as shown in Fig.~\ref{S1-2D-Lambda-Kappa-Alambda}
%although in a few small regions it can even reach 28 TeV. 

%----
%In the case of universality of the parameters, the chargino lower bound 
%$\mu=3\lambda v_R/\sqrt 2 \gsim 100$ GeV together with the above upper bound on the Majorana mass give rise to the following constraint on the value of the right sneutrino VEVs:
%which will be useful for our discussion of results in Section~\ref{results-scans}
%when studying the $\kappa-\lambda$ plane.
%----

Concerning the values of $\tan\beta$, we find
in $S_1$ that
%for the small/moderate values of $\lambda$ of scan $S_1$, we find 
%the lower bound 
$\tan\beta > 4$.
Such a lower bound is expected in order to maximize the tree-level SM-like Higgs mass for small/moderate values of $\lambda$, as discussed in Subsect.~\ref{smlike}}. We can see in Fig.~\ref{S1-2D-Lambda-Kappa-tanB} 
%of Appendix~\ref{scan1} 
that large values of $\tan\beta$ are welcome for this task, similarly to the MSSM.  
Given the small singlet-doublet mixing, 
significant loop contributions are the main
source to increase the tree-level mass of the SM-like Higgs.
The values of the masses of the third-generation squarks and 
trilinear soft term necessary to generate the large loop corrections are shown in Fig.~\ref{S1Tu3MQ3}. 
The white region in the upper left side is excluded by the mass of the SM-like Higgs or by the existence of tachyons when $-T_{u_3}$ is much larger than $m_{\widetilde Q_{3L}}$.
For the allowed regions,
%we show points in $-T_{u_3}$ vs. $M_{\widetilde Q_3}$. 
we can see first that for $\kappa$ perturbative up to 10 TeV (light-red points)
the values of
$-T_{u_3} \lesssim 2000$ GeV and $m_{\widetilde Q_{3L}}\lesssim 1000$ GeV are highly correlated.
Given the large value of $\kappa$, the push-down effect in these light-red points makes necessary the maximal mixing scenario to cancel it, bringing the mass of the SM-like Higgs to the correct value.
%This occurs for non-perturbative values of $\kappa$.
For the red points, where $\kappa$ is smaller, the push-down effect is not so large and the maximal mixing scenario can be relaxed. We can see that the lower right side of Fig.~\ref{S1Tu3MQ3} becomes populated.
The same argument applies to the blue points (note that most of them are on top of red points), but now for the push-up effect which is also small.
In Figs.~\ref{S1-2D-Lambda-Kappa-MQ3} and~\ref{S1-2D-Lambda-Kappa-Tu3}
of Appendix~\ref{figuresscans}, 
we show in the $\kappa-\lambda$ plane the values of 
$m_{\widetilde Q_{3L}}$ and $-T_{u_3}$, respectively.
As discussed, smaller values of these parameters are needed in the
perturbative region up to 10 TeV.

\begin{figure}[t!]
 \centering
 \includegraphics[width=0.8\linewidth, height= 0.35\textheight]{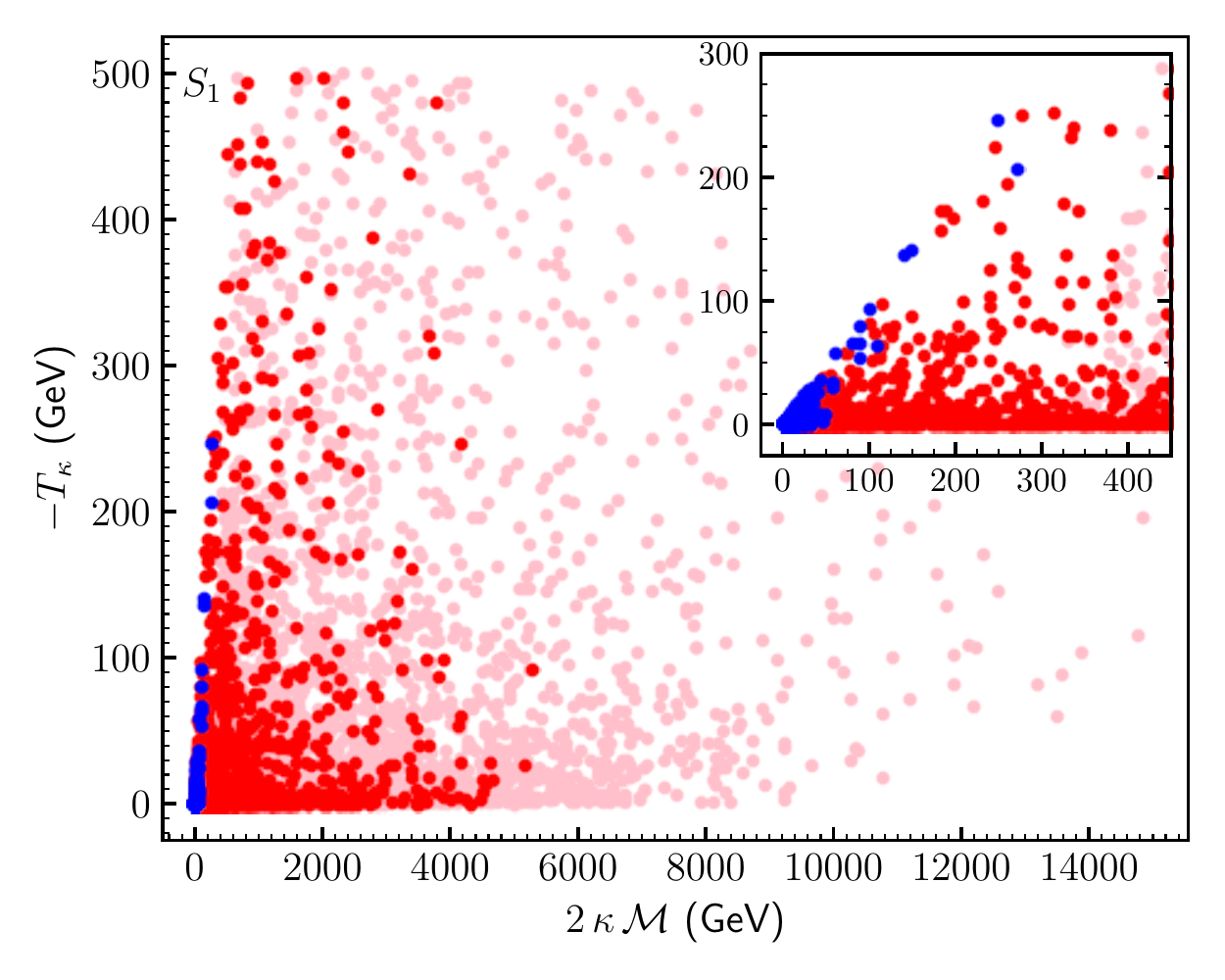}
  \caption{
  Viable points of the parameter space for $S_1$ in the
  $-T_{\kappa}$ versus $2\kappa {\mathcal{M}}$ plane.
% Universality of the parameters is assumed with 
%$\lambda_i=\lambda, \kappa_i=\kappa, 
%v_{iR}=v_R, T_{\kappa_i}=T_\kappa$, and 
%$T_{\lambda_i}=T_\lambda$. 
The color code is the same as in Fig.~\ref{S1Lambda-kappa}.
In the upper right we zoom in the region with blue points.
%  The same as in Fig.~\ref{S1Lambda-kappa}, but for 
%  $-A_\kappa$ versus the Majorana mass of right-handed neutrinos 
%  ${\mathcal{M}} = 2\kappa \frac{v_R}{\sqrt 2}$.
}
\label{S1Ak-kapvR}
\end{figure}

{
In Table~\ref{S1-R1} of Appendix~\ref{BPs}, we show the BP {\bf S1-R1}
%\footnote{\R{The SLHA-like output of {\tt SPheno} for all the benchmarks presented in the Appendix~\ref{BPs} can be found on the website of the $\mu\nu$SSM Working Group at \url{http://dark.ft.uam.es/mununiverse/index.php/preprints}.}} 
corresponding to the red region of Fig.~\ref{S1Lambda-kappa}, where the SM-like Higgs $h_1$ is the lightest scalar, and 
$h_{4,5,6}$ are the singlet-like states with masses larger than 900 GeV. Note nevertheless that the singlet-like pseudoscalars can be lighter than the SM-like Higgs,
%the SM-like Higgs, 
as shown in particular in 
this BP where they have masses around 40 GeV. 
As we can check from the fifth box of the table, the right sneutrinos are not very mixed among themselves because $\lambda$ is small and therefore the off-diagonal terms in Eq.~(\ref{evenR}) are negligible.
However, the singlet-like scalar $h_6$, with a mass similar to $h_7$ which has a dominant composition of $H^{\mathcal{R}}_d$, 
has a significant composition of the latter 
%$H^{\mathcal{R}}_d$ 
(22.5\%), whereas $h_4$ and $h_5$ are very pure singlets with dominant compositions
$\widetilde{{\nu}}^{\mathcal{R}}_{eR}$ and 
$\widetilde{{\nu}}^{\mathcal{R}}_{\mu R}$, respectively.}
This BP corresponds to one of those shown in the right-hand side of Fig.~\ref{composition1} with a large doublet composition. In that figure, the singlet component of the singlet-like states is shown versus their masses, and we can see that most (but not all) of them are almost pure singlets. The fact that 
only one of the three singlet-like states of {\bf S1-R1} has a large doublet composition is because of our assumption of almost degenerate $\kappa$'s implying that there are always two almost pure singlets.
In Table~\ref{S1-R2}, we show a different BP corresponding to the red region of Fig.~\ref{S1Lambda-kappa}, {\bf S1-R2}, where the three singlet-like scalars are very pure singlets.

It is worth remarking here that the masses of the left sleptons are determined by the parameters related to neutrino/left sneutrino physics, as discussed in Subsection~\ref{scan}, and therefore can be modified choosing different values for $Y_{\nu_i}$, $v_{iL}$, $M_{1,2}$ and $Y_{\nu_i}$ in Table~\ref{Scans-fixed-parameters}. This is true in general for all scans and applies therefore to all BPs studied in this work.

\begin{figure}[t!]
 \centering
 \includegraphics[width=0.8\linewidth, height= 0.35\textheight]{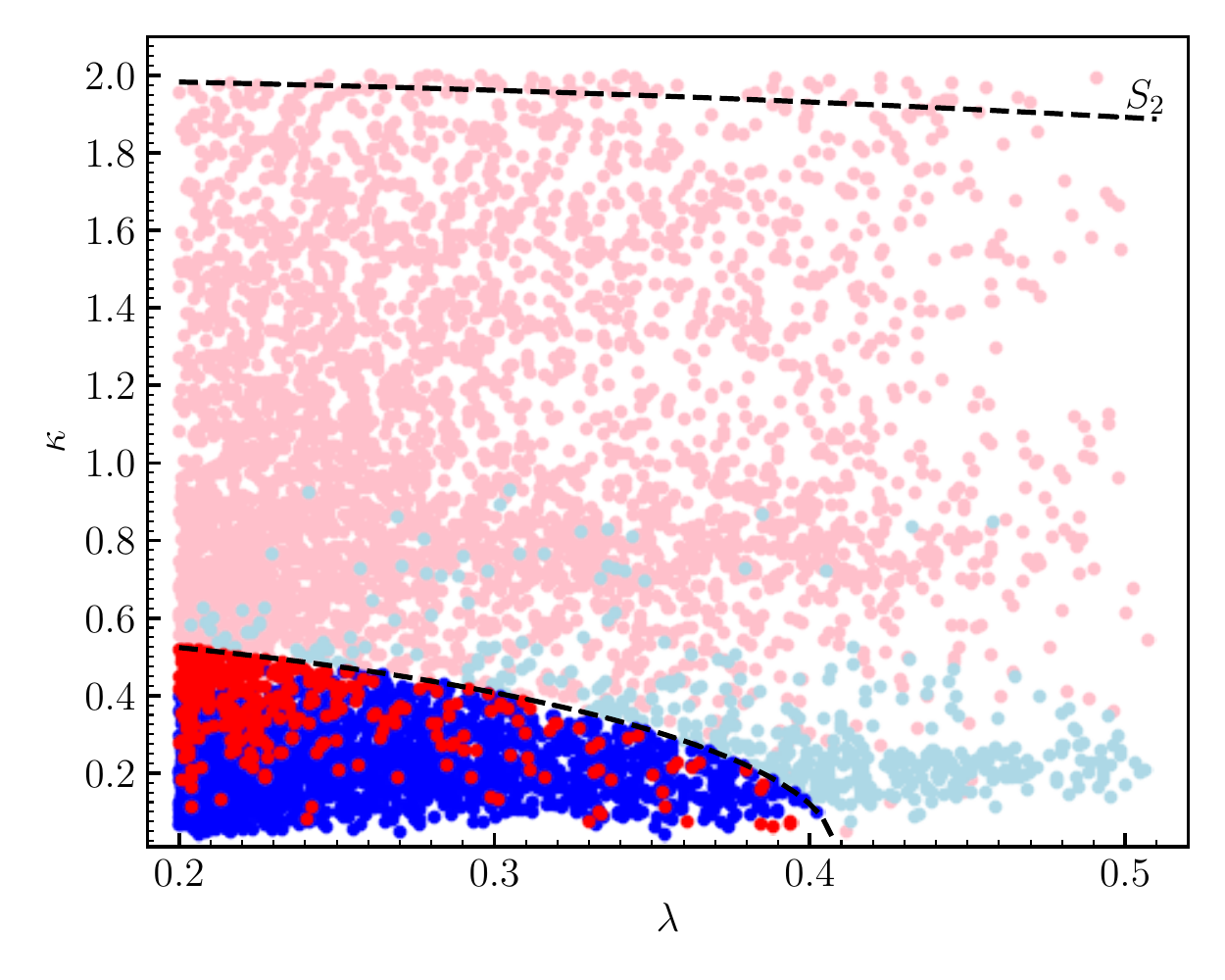}
  \caption{
%  The same as in Fig.~\ref{S1Lambda-kappa}, but for scan $S_2$.
  Viable points of the parameter space for $S_2$ in the $\kappa-\lambda$ plane.
% Universality of the parameters is assumed with 
%$\lambda_i=\lambda, \kappa_i=\kappa, 
%v_{iR}=v_R, T_{\kappa_i}=T_\kappa$, and 
%$T_{\lambda_i}=T_\lambda$. 
The red and light-red (blue and light-blue) colours represent cases where the SM-like Higgs is (is not) the lightest scalar.
All red and blue points below the lower black dashed line fulfill the perturbativity condition up to GUT scale of Eq.~(\ref{perturbativity1}). 
Light-red and light-blue points below the upper black dashed line fulfill the perturbativity condition up to 10 TeV of Eq.~(\ref{perturbativity2}).
%  Viable points of the parameter space for scan $S_2$ in the $\kappa-\lambda$ plane, where $\kappa=\kappa_i$ and $\lambda=\lambda_i$.
  %, for scan $S_1$ (left), $S_2$ (middle) and $S_3$ (right).
%The red (blue) colour represents cases where the SM-like Higgs is (is not) the lightest scalar. 
%All points below the lower (upper) black dashed line fulfill the perturbativity condition in Eq.~(\ref{perturbativity1}) (Eq.~(\ref{perturbativity2})).
%  The same as in Fig.~\ref{S2LamtanB} but for $\kappa$ versus $\lambda$, where $\kappa=\kappa_i$ and $\lambda=\lambda_i$ 
}
\label{S2Lambda-kappa}
\end{figure}

Let us now discuss in more detail the (narrow) region of
Fig.~\ref{S1Lambda-kappa} 
with blue points.
For small values of $\lambda$, the first term of Eq.~(\ref{sps-approx-R}) is a good approximation for right sneutrino masses.
Clearly, unless one makes a tuning between the two pieces in that term, $T_\kappa/\kappa$ and $2{\mathcal{M}}=4\kappa v_R/\sqrt 2$, one needs 
these two quantities to be small in order to obtain right sneutrinos lighter than the SM-like Higgs.
Now, since $v_R$ is typically large in this scan compared to the SM-like Higgs mass, small values of $\kappa$ are necessary for this task. 
This is what we observe in the blue region of Fig.~\ref{S1Lambda-kappa}, where
$\kappa\lsim 0.2$.
There we also see that for larger values of $\lambda$, larger values of $\kappa$ are allowed, because the values of $v_R$ decrease with $\lambda$ as shown in Fig.~\ref{S1-2D-Lambda-Kappa-vR}.
The correlation between the above two pieces for the blue points is also obvious from Eqs.~(\ref{nontach}) and~(\ref{sneutrinohiggs}).
We show explicitly this effect in Fig.~\ref{S1Ak-kapvR}, where
basically the line
$-T_{\kappa} = 2\kappa {\mathcal{M}}$
separates the tachyonic (white) region from the non-tachyonic one
with blue and 
red points, i.e. $-T_{\kappa}< 2\kappa {\mathcal{M}}$. 
Blue points have to be close to the line since 
they have to fulfill in addition the approximate condition~(\ref{sneutrinohiggs}).
In Fig.~\ref{S1-2D-Lambda-Kappa-Tkappa} of the Appendix, we show the different values of
$-T_\kappa$ in the $\kappa-\lambda$ plane. As we can see, for the small values of $\kappa$ corresponding to the blue region of 
Fig.~\ref{S1Lambda-kappa}, the values of $-T_\kappa$ are typically small.
For larger values of $\kappa$ corresponding to the regions with red and light-red points of 
Fig.~\ref{S1Lambda-kappa}, i.e. with masses of the singlet-like states larger than the SM-like Higgs mass, the tachyonic region can be avoided even with large values of $-T_\kappa$ (up to the upper bound of 500 GeV imposed in the scan), as shown in the figure. We show for completeness
in Fig.~\ref{S1-2D-Lambda-Kappa-Akappa} the different values of the supergravity parameter
$A_\kappa = T_\kappa/\kappa$ in the
$\kappa-\lambda$ plane. Due to this relation, values of $-A_\kappa$ as large as around 2.9 TeV can be obtained in regions with small $\kappa$.
Larger values of $-A_\kappa$ are not possible because the condition 
$-A_{\kappa}< 2 {\mathcal{M}}$ cannot be fulfilled since $v_R$ is bounded, and therefore 
tachyons would appear.

\begin{figure}[t!]
 \centering
 \includegraphics[width=0.8\linewidth, height=0.35\textheight]{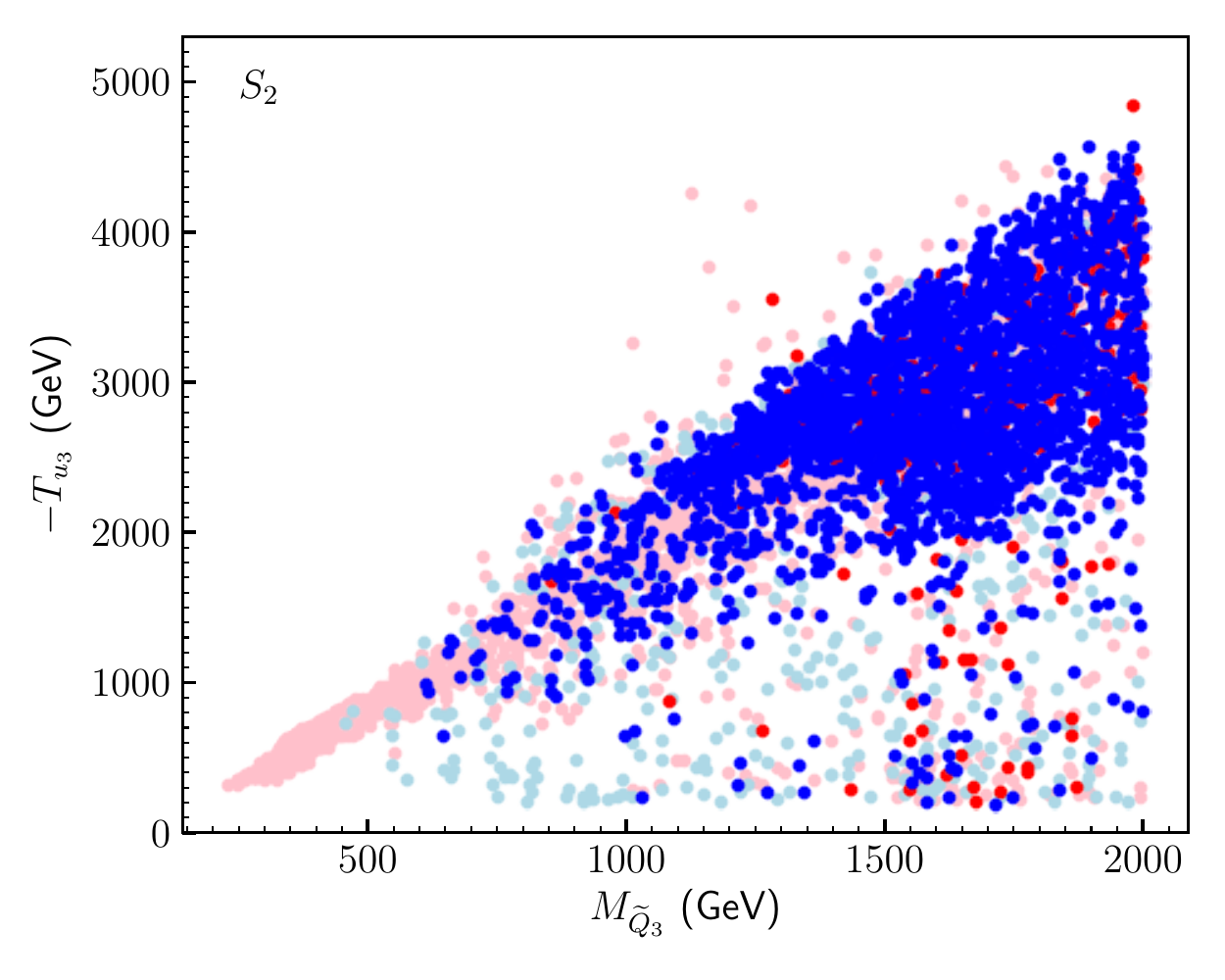}
  \caption{
    Viable points of the parameter space for $S_2$ in the
  $-T_{u_3}$ versus $m_{\widetilde Q_{3L} }$ plane.
% Universality of the parameters is assumed with 
%$\lambda_i=\lambda, \kappa_i=\kappa, 
%v_{iR}=v_R, T_{\kappa_i}=T_\kappa$, and 
%$T_{\lambda_i}=T_\lambda$. 
The color code is the same as in Fig.~\ref{S2Lambda-kappa}.
  %The same as in Fig.~\ref{S1Tu3MQ3}, but for scan $S_2$
  %$-T_{u_3}$ versus $m_{\widetilde Q_{3L} }$. 
  }
  \label{S2Tu3MQ3}
\end{figure}

Finally, it is worth noticing that
about {40\% of the blue points} correspond to cases where the singlet-like scalars have masses $\lsim m_{\text{Higgs}}/2$. As can be seen in Fig.~\ref{composition1}, most of these states are almost pure singlets and therefore do not affect the Higgs decays, surviving as viable points.
We show the BP {\bf S1-B1} with these characteristics in 
Table~\ref{S1-B1}. There we see that the three singlet-like states $h_{1,2,3}$
with masses around 50 GeV are lighter than the SM-like Higgs $h_4$.
Besides, these light states are significantly mixed among themselves because of the moderate value of $\lambda= 0.1$, which makes the off-diagonal terms in Eq.~(\ref{evenR}) significant.
It is also worth noticing that for {\bf S1-B1} the Majorana mass is small,
$\mathcal{M}=52.92$ GeV, giving rise to two almost degenerate light right-handed neutrinos of masses 49.75 and 53.28 GeV, and one heavier of mass 64.49 GeV. As a consequence, although the singlet composition of $h_4$ is small, this mixing is already sufficient to produce a significant decay of $h_4$ to two neutralinos (with dominant right-handed neutrino composition) with BR=0.15.

{The presence of light scalars ($h_i$),
pseudoscalars ($A_i$) and neutralinos ($\widetilde \chi^0_i$) such that $m_{h_i}+m_{A_j} < M_Z$, $m_{\widetilde \chi^0_i}+m_{\widetilde \chi^0_j} < M_Z$ or $m_{\widetilde \chi^0_i}+ m_{\widetilde \chi^\pm_j} < M_W$
(here $\widetilde \chi^\pm_j =e,\, \mu, \, \tau$)
opens up new on-shell decay modes for the $Z$ and $W$ bosons. The possible signs of new physics from these new decay modes in the $\mn$ have been studied in Ref.~\cite{Ghosh:2014rha}.
%but it is beyond the scope of this analysis.}

On the other hand,
when the masses of the singlet-like states are close to 125 GeV, 
%We can also see in that figure that 
it is possible to find solutions with a large doublet composition. Actually, as mentioned before, for each point of the parameter space only one of the three states has this property, given our assumption of almost degenerate $\kappa$'s implying that there are always two almost pure singlets.
For these solutions, if the SM-like Higgs and the singlet-like state with significant doublet composition have masses within the mass resolution of the experiment, they will have their signal rates superimposed,
and both will contribute to the resonance observed at 125 GeV~\cite{Bechtle:2013xfa}.
In this scan $S_1$, 
{about 0.4\% of the phenomenologically viable points found have singlet-like states with masses close to 125 GeV.
%, and around ??? of them have a significant doublet composition as to have their signals superimposed.}
We show in Table~\ref{S1-3-scalars} 
%of Appendix~\ref{BPs}
the
%phenomenologically viable 
BP {\bf S1-2h1} with these properties.
There we see that the right sneutrino $h_4$ has a large composition of $H^{\mathcal{R}}_u$ (27.58\%), whereas $h_1$ and $h_2$ are very pure singlets with dominant compositions
$\widetilde{{\nu}}^{\mathcal{R}}_{eR}$ and $\widetilde{{\nu}}^{\mathcal{R}}_{\mu R}$, respectively, and not contributing therefore to the superposition of Higgs-like states.
As already discussed for the BP {\bf S1-R1},
the singlet-like states are very little mixed among themselves given that $\lambda$ is small.
%and therefore the off-diagonal terms in Eq.~(\ref{evenR}) are negligible.

%%%%%%%%%%%%%%%%%%%%%%%%%%%%%%%%%%%%%%%%%%%%%%%%%%%%%%%%%%%%%%%%%%%%%%%%%%%%%%%%

%For a better understanding of the region with blue points, we show $- A_\kappa$ versus ${\mathcal{M}}$ in Fig.~\ref{S1Ak-kapvR}.
%For small values of $\lambda$ the condition of Eq.~(\ref{nontach}) constraining the value of $A_\kappa$, in order to avoid tachyonic singlet-like scalars, is a good approximation as discussed below
%Eq.~(\ref{neutrinoR2}).
%This can be seen in the figure,
%where the line
%$-A_{\kappa} = 2 {\mathcal{M}}$
%separates the tachyonic (white) region from the non-tachyonic one
%with blue and 
%red points, i.e. $-A_{\kappa}< 2 {\mathcal{M}}$. 
%Blue points have to be close to the line since 
%they have to fulfill approximately the condition
%$2{\mathcal M} -\frac{2m_{\text{Higgs}}^2}{\mathcal M}< -A_{\kappa}$
%for singlet-like scalars being lighter than the SM-like Higgs.
%Thus, to obtain very light singlet-like scalars one needs 
%small $-A_\kappa$ together with small ${\mathcal{M}}$,
%or a tuning among these quantities.
%Related to this, since $v_R$ is typically large in this scan as will be discussed below, small values of $\kappa$ are necessary in this region. Therefore, $A_\kappa = T_\kappa/\kappa$ can be large, as we can see in
%Fig.~\ref{S1Ak-kapvR}
%where it reaches up to 3000 GeV.

%\vspace*{0.25cm}
%\noindent
%{\it Scan 2}
%($0.2 \leq \lambda < 0.5$)

\begin{figure}[t!]
 \centering
 \includegraphics[width=0.8\linewidth, height= 0.35\textheight]{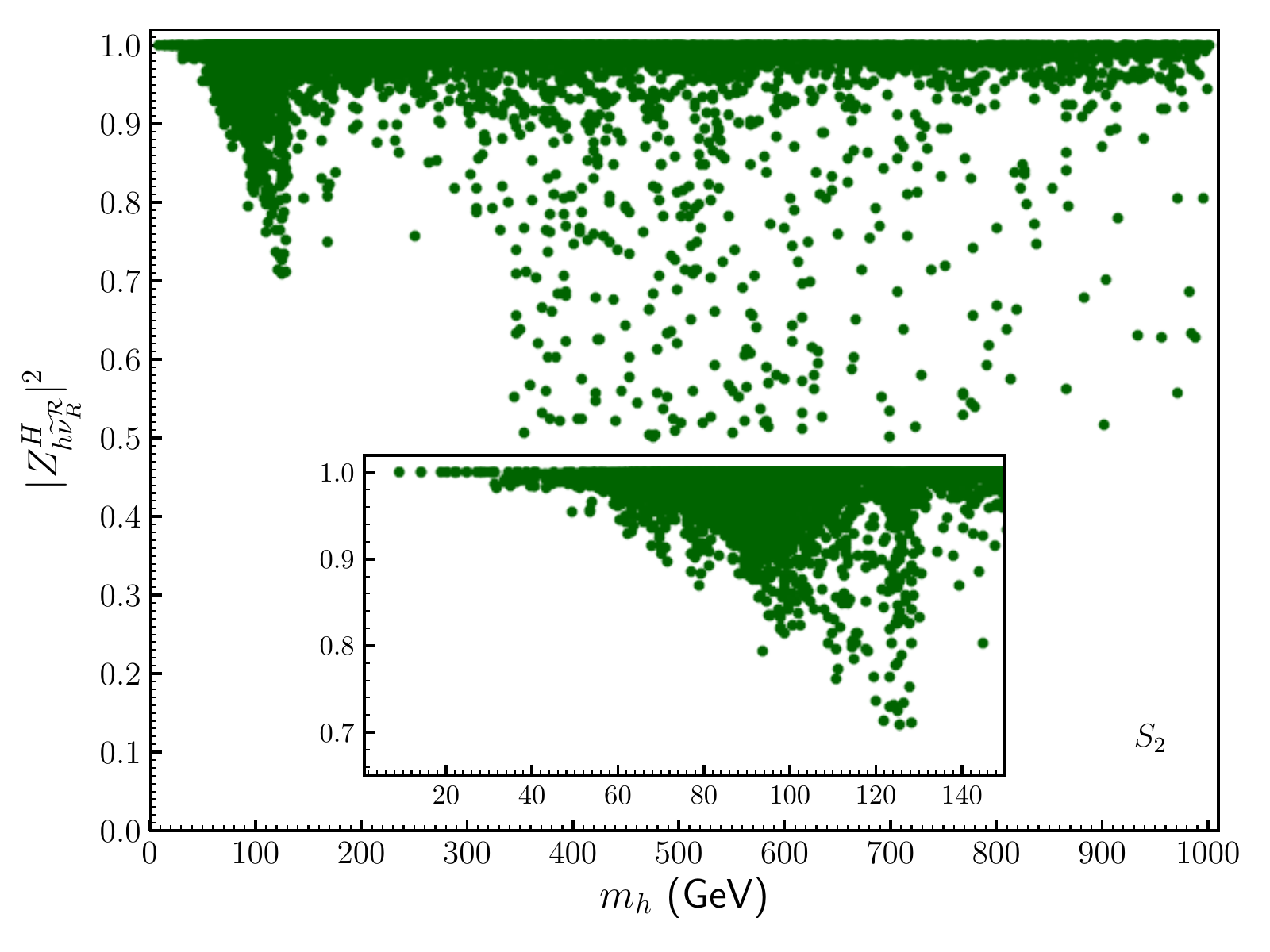}
  \caption{
  The singlet component $\sum_i|Z^H_{h
  \widetilde{{\nu}}^{\mathcal{R}}_{iR}}|^2$ of the singlet-like scalars $h$
versus their masses, for $S_2$.
We only show here viable points with scalar masses smaller than 1000 GeV.
In the lower part we zoom in the low-mass region.
%  The singlet component $|Z^H_{h
%  \widetilde{{\nu}}^{\mathcal{R}}_{R}}|$ of the singlet-like scalars $h$
%versus their masses, for $S_2$.
}
\label{composition2}
\end{figure}

\subsubsection{ {\it Scan 2} ($0.2 \leq \lambda < 0.5$)}
%\subsection{Scan $S_1$}
\label{scan02}

%\noindent 
%{\it (ii) Scan $S_2$}

Fig.~\ref{S2Lambda-kappa} summarizes our results for Scan 2 ($S_2$).
In this case, we find viable solutions in the entire $\kappa - \lambda$ plane,
since now $\lambda \geq 0.2$ and therefore the chargino mass lower bound can be fulfilled with low values of $v_R$, being safe from 
tachyonic left sneutrinos.
In fact, we see in Fig.~\ref{S2-2D-Lambda-Kappa-vR} that in most of the regions
$v_R/\sqrt 2 \lsim 1000$ GeV. 
This bound, in order to avoid a too large mixing term $m_{H^{\mathcal{R}}_{u} H^{\mathcal{R}}_{d}}^2$, is smaller than for $S_1$ because now we are working with moderate/large values of $\lambda$.
{Concerning the value of $T_\lambda$, in 
Fig.~\ref{S2-2D-Lambda-Kappa-Tlambda} we see that in regions with large $\kappa$,
and given the values of the other relevant parameters,
%{this value} %of $T_\lambda$ 
%\R{has an upper bound of around $200$ GeV.}
small values are preferred (for example, for $\kappa>1$ about 87\% of points have $T_\lambda < 200$ GeV), whereas for lower values of $\kappa$ the mixing term is smaller and larger values of $T_\lambda$ are favoured} (up to the upper bound of 500 GeV imposed in the scan).
Assuming the supergravity relation $A_\lambda= T_\lambda/\lambda$, we see in 
Fig.~\ref{S2-2D-Lambda-Kappa-Alambda} that
given the moderate/large values of $\lambda$ for this scan, the upper bound for $A_\lambda$ is typically smaller than for $S_1$ in all regions, with a maximum value of around 2.5 TeV. 

Concerning $\tan\beta$, since $\lambda$ is larger than in $S_1$ we find that smaller values are favoured to maximize the tree-level SM-like Higgs mass, as shown in Fig.~\ref{S2-2D-Lambda-Kappa-tanB} of Appendix. In addition, as a consequence of the moderate/large $\lambda$, the singlet-doublet mixing is larger and therefore the push-up effect for the blue points helps to increase the tree-level mass.
All these effects together produce that the
loop contributions to increase the tree-level mass of the SM-like Higgs can be relaxed.
One can observed this comparing Fig.~\ref{S2Tu3MQ3} and Figs.~\ref{S2-2D-Lambda-Kappa-MQ3} and~\ref{S2-2D-Lambda-Kappa-Tu3}
with the corresponding ones of $S_1$.

\begin{figure}[t!]
 \centering
 \includegraphics[width=0.8\linewidth, height= 0.35\textheight]{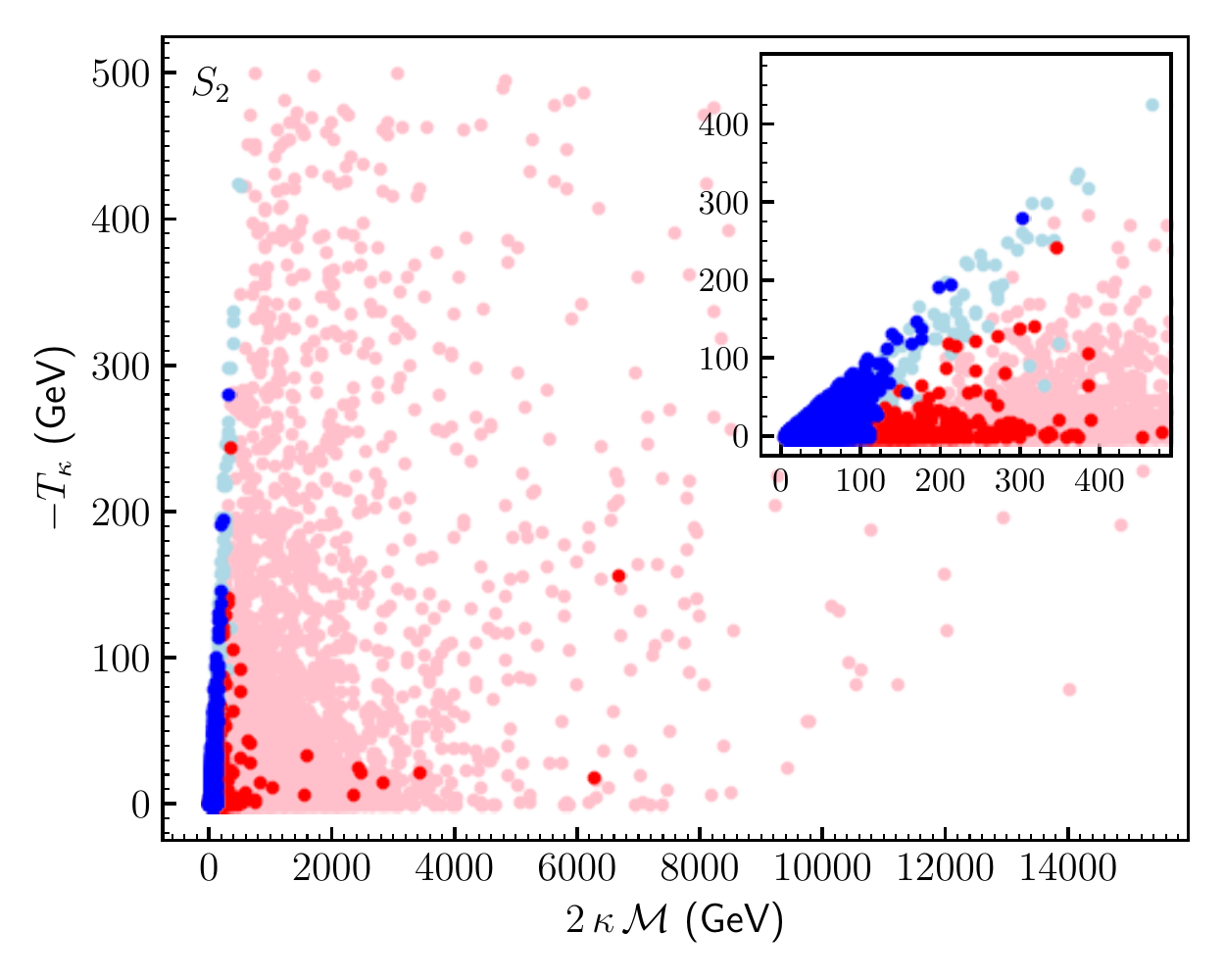}
  \caption{
   Viable points of the parameter space for $S_2$ in the
  $-T_{\kappa}$ versus $2\kappa {\mathcal{M}}$ plane.
% Universality of the parameters is assumed with 
%$\lambda_i=\lambda, \kappa_i=\kappa, 
%v_{iR}=v_R, T_{\kappa_i}=T_\kappa$, and 
%$T_{\lambda_i}=T_\lambda$. 
The color code is the same as in Fig.~\ref{S2Lambda-kappa}.
In the upper right we zoom in the region with blue points.
%  The same as in Fig.~\ref{S1Ak-kapvR}, but for scan $S_2$.
  %$-A_\kappa$ versus $\kappa v_R$, where
%$A_\kappa = A_{\kappa_i}$, $\kappa=\kappa_i$ and $v_{R}=v_{iR}$. 
}
\label{S2Ak-kapvR}
\end{figure}

Related to the above discussion, is the fact that in the perturbative region 
up to the GUT scale is more easy to find blue than red points. The push-down effect of the latter makes for them more difficult to reach the correct mass of the SM-like Higgs. Note also that blue points with all values of $\kappa$ are present, since $v_R$ is now smaller than for $S_1$.

In Tables~\ref{NonSK-benchmarks} and~\ref{NonSK-benchmarks2} of Appendix~\ref{BPs}, we show the two BPs {\bf S2-R1} and {\bf S2-R2}, respectively, corresponding to the red region of Fig.~\ref{S2Lambda-kappa}.
They have different singlet-like scalar masses, around 230 and 600 GeV, mainly due to the different values of $v_R$.
For {\bf S2-R1}, as we can see in the fifth box,
%in Table~\ref{NonSK-benchmarks}, 
the singlet-like states 
%$h_{2,3,4}$ 
are significantly mixed among themselves because of the moderate/large value of $\lambda$, and the eigenstate $h_5$ is the one having a significant composition of $H^{\mathcal{R}}_u$ (10.29\%).
The SM-like Higgs with a composition of $H^{\mathcal{R}}_d$ of 19.36\% is phenomenologically viable because $\tan\beta$ is as small as 2.31.
The same occurs for {\bf S2-R2},
%in Table~\ref{NonSK-benchmarks2}, 
where now the
SM-like Higgs composition of $H^{\mathcal{R}}_d$ is larger, 46.66\%, but
$\tan\beta$ is smaller, 1.08. For this BP the mixing among
right sneutrinos is larger, but no eigenstate has a  significant composition of $H^{\mathcal{R}}_u$ given their larger masses.
In Fig.~\ref{composition2}, we show the singlet component of the singlet-like scalars. As for $S_1$, for large masses we can find scalars with a very large composition of $H^{\mathcal{R}}_d$.

The correlation discussed for $S_1$ in Fig.~\ref{S1Ak-kapvR} is relaxed in this new scan, again because of the larger values of $\lambda$, as discussed below Eq.~(\ref{neutrinoR2}). We show this in 
Fig.~\ref{S2Ak-kapvR}.
In Figs.~\ref{S2-2D-Lambda-Kappa-Tkappa} and~\ref{S2-2D-Lambda-Kappa-Akappa}
of the Appendix, we can see the different values of
$-T_\kappa$ and $A_\kappa = T_\kappa/\kappa$, respectively, in the $\kappa-\lambda$ plane. 
In the perturbative region up to the GUT scale, except for areas with $\kappa$ close to its upper bound, $T_\kappa$ is typically small to avoid tachyonic right sneutrinos because $v_R$ is small. As a consequence, in the case of supergravity $A_\kappa$ is also typically small in this region.
In this scan, 
about 11\% of the blue points correspond to cases where the singlet-like scalars have masses $\lsim m_{\text{Higgs}}/2$.
In Table~\ref{S2-B1}, we show the BP {\bf S2-B1} corresponding to the blue region of Fig.~\ref{S2Lambda-kappa}, with 
singlet-like scalar masses $> m_{\text{Higgs}}/2$. Apart for that,
its characteristics are similar to the BP {\bf S1-B1} of scan $S_1$.

\begin{figure}[t!]
 \centering
 \includegraphics[width=0.8\linewidth, height= 0.35\textheight]{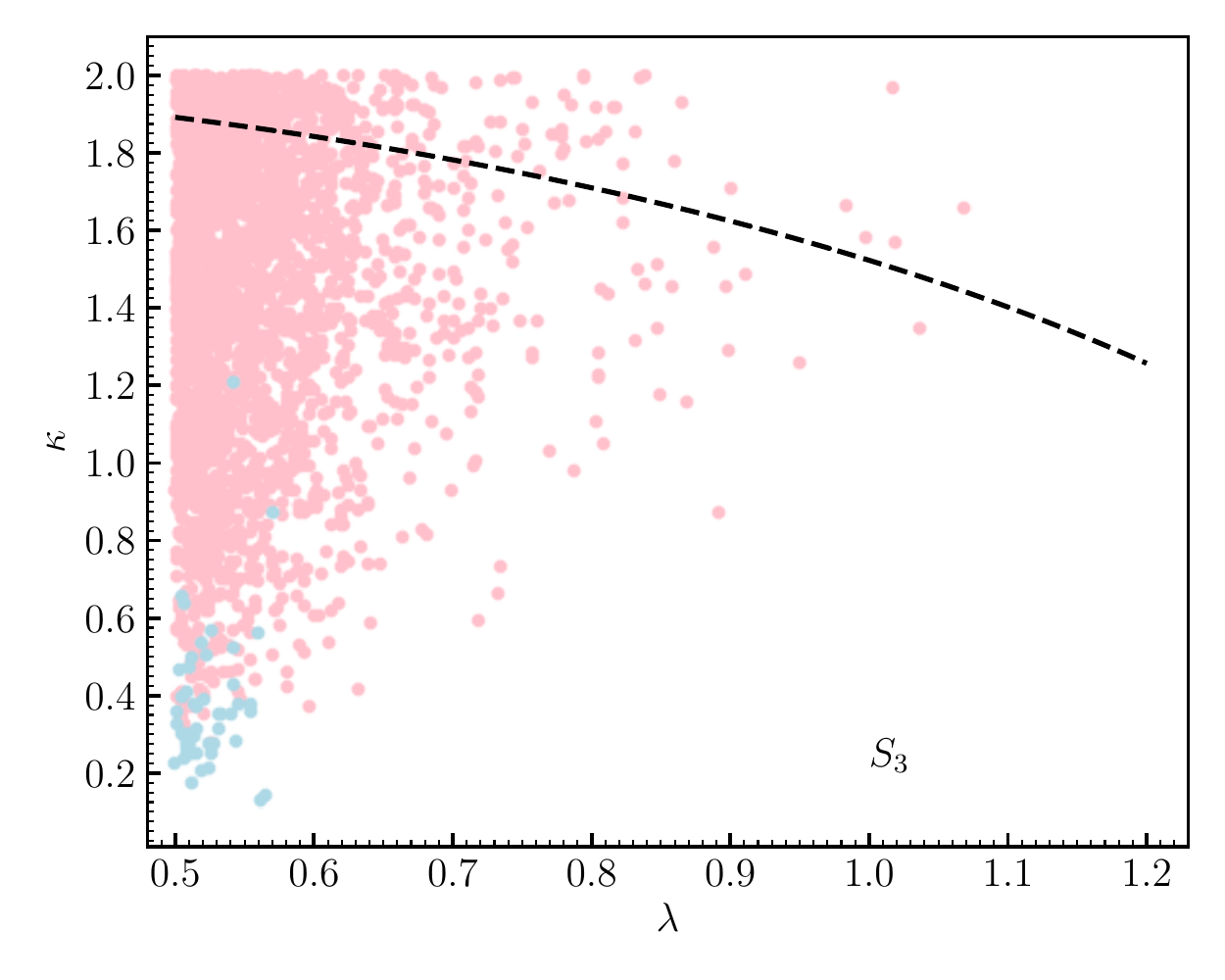}
  \caption{
    Viable points of the parameter space for $S_3$ in the $\kappa-\lambda$ plane.
% Universality of the parameters is assumed with 
%$\lambda_i=\lambda, \kappa_i=\kappa, 
%v_{iR}=v_R, T_{\kappa_i}=T_\kappa$, and 
%$T_{\lambda_i}=T_\lambda$. 
The light-red (light-blue) colour represents cases where the SM-like Higgs is (is not) the lightest scalar.
All light-red and light-blue points below the black dashed line fulfill the perturbativity condition up to 10 TeV of Eq.~(\ref{perturbativity2}).
%  Viable points of the parameter space for scan $S_3$ in the $\kappa-\lambda$ plane, where $\kappa=\kappa_i$ and $\lambda=\lambda_i$.
  %, for scan $S_1$ (left), $S_2$ (middle) and $S_3$ (right).
%The red (blue) colour represents cases where the SM-like Higgs is (is not) the lightest scalar. 
%All points below the black dashed line fulfill the perturbativity condition in 
%Eq.~(\ref{perturbativity2}).
}
\label{S3Lambda-kappa}
\end{figure}

Concerning solutions with singlet-like states with masses close to 125 GeV, %producing a superposition of signals with the SM-like Higgs,
{about 5\% of the phenomenologically viable points} found in this scan are of this type. However, not all of them have a significant doublet composition as to have their signals superimposed with that of the SM-like Higgs.
We show in Table~\ref{S2-3-scalars} the BP {\bf S2-2h1} as an
%of Appendix~\ref{BPs}
example of this situation.
%phenomenologically viable 
%BP with these properties.
As we can see, 
the right sneutrino $h_4$ has the largest doublet composition of 
$H^{\mathcal{R}}_{u}$ (3.77\%) and $H^{\mathcal{R}}_{d}$ (2.65\%), 
but insufficient as to contribute significantly to the Higgs signals.
%whereas $h_2$ and $h_3$ are very pure singlets as expected.
Unlike the BP {\bf S1-2h1} of $S_1$, now the three sneutrinos are very mixed because of the larger value of $\lambda$.
Similar to {\bf S1-B1}, for this BP also the Majorana mass is small,
$\mathcal{M}=55.8$ GeV, giving rise to two almost degenerate ligth right-handed neutrinos of masses 55.9 and 57.2 GeV, and one heavier of mass 76.7 GeV. As a consequence, the decay channel right sneutrino to two neutralinos (with dominant right-handed neutrino composition) opens, giving the most important contribution to the BRs. 
%with dominant compositions
%$\widetilde{{\nu}}^{\mathcal{R}}_{eR}}$ and $\widetilde{{\nu}}^{\mathcal{R}}_{\mu R}}$, respectively, and not contributing therefore to the superposition of Higgs-like states.
In Table~\ref{S2-ss1}, we show the BP {\bf S2-2h2} with the singlet-like state $h_3$ having the largest doublet composition of 
$H^{\mathcal{R}}_{u}$ (8.79\%) and contributing significantly to the superposition of signals
with the SM-like Higgs $h_4$,
unlike the previous case {\bf S2-2h1}.

%\vspace*{0.25cm}
%\noindent
%{\it Scan 3}
%($0.5 \leq \lambda < 1.2$)

\subsubsection{ {\it Scan 3} ($0.5 \leq \lambda < 1.2$)}
%\subsection{Scan $S_1$}
\label{scan03}

%\vspace*{0.25cm}

%\noindent 
%{\it (iii) Scan $S_3$}

The results for Scan 3 ($S_3$) are summarized in 
Fig.~\ref{S3Lambda-kappa}.
In this case with so large values of $\lambda$, the white region in the lower right is forbidden because of the too large mixing term $m_{H^{\mathcal{R}}_{u} H^{\mathcal{R}}_{d}}^2$ producing tachyons. 
To avoid that situation, in most of the allowed regions the right sneutrino VEVs take small values,  
$v_R/\sqrt 2 \lsim 500$ GeV, as 
shown in Fig.~\ref{S3-2D-Lambda-Kappa-vR}.
These small values imply that $\kappa\gsim 0.2$ to avoid tachyonic right sneutrinos.
We show in Fig.~\ref{S3-2D-Lambda-Kappa-Tlambda} the value of the corresponding $T_\lambda$, whereas in Fig.~\ref{S3-2D-Lambda-Kappa-Alambda}
the supergravity parameter $A_\lambda$ is shown. 

\begin{figure}[t!]
 \centering
 \includegraphics[width=0.8\linewidth, height= 0.35\textheight]{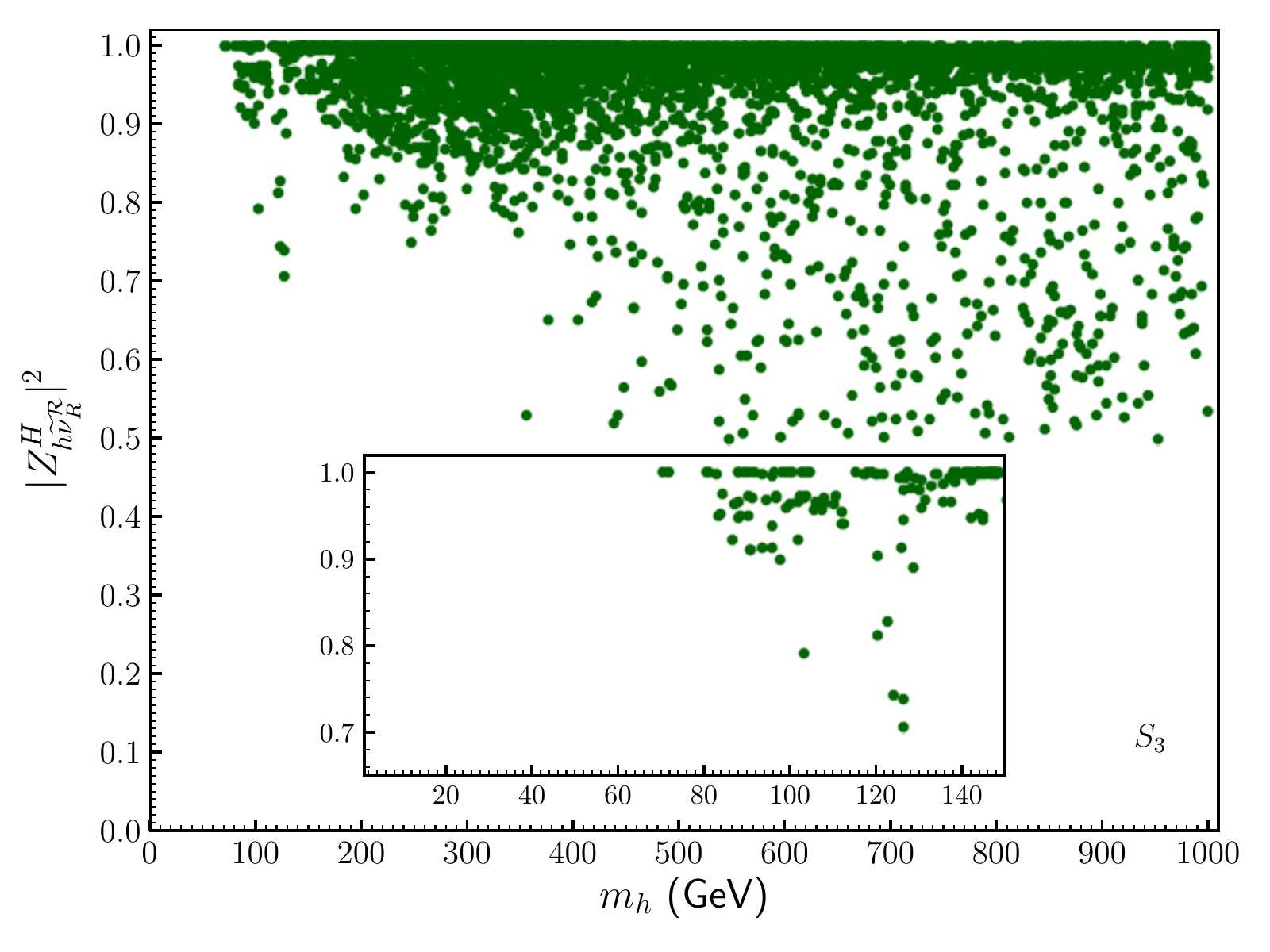}
  \caption{
  The singlet component $\sum_i|Z^H_{h
  \widetilde{{\nu}}^{\mathcal{R}}_{iR}}|^2$ of the singlet-like scalars $h$
versus their masses, for $S_3$.
We only show here viable points with scalar masses smaller than 1000 GeV.
In the lower part we zoom in the low-mass region.
 % The singlet component $|Z^H_{h
 % \widetilde{{\nu}}^{\mathcal{R}}_{R}}|$ of the singlet-like scalars $h$
%versus their masses, for $S_3$.
}
\label{composition3}
\end{figure}

In Fig.~\ref{S3-2D-Lambda-Kappa-tanB} we show $\tan\beta$ which, given the large value of $\lambda$, can take smaller values than in $S_2$.
This region of the parameter space also favours light third generation squarks, as shown in Fig.~\ref{S3Tu3MQ3} (see also Figs.~\ref{S3-2D-Lambda-Kappa-MQ3} and~\ref{S3-2D-Lambda-Kappa-Tu3}).

As discussed in Section~\ref{smlike}, the push-down effect (together with negative loop corrections) of a heavy singlet-like sector is more favourable to reproduce the SM-like Higgs mass.
In Table~\ref{S3-R1} of Appendix~\ref{BPs}, we show the BP
{\bf S3-R1}
corresponding to the light-red region of Fig.~\ref{S3Lambda-kappa}.
In this case, $h_2$ is the SM-like Higgs, and the right sneutrinos are very mixed as expected from the large value of $\lambda$.
In Fig.~\ref{composition3}, we show the singlet component of the singlet-like scalars. As for the other scans, scalars with large masses and with a very large doublet composition can also be present 
%\R{(check that it is $H^{\mathcal{R}}_d$).}

Although more difficult than in previous scans, we are also able to find in $S_3$ solutions with light singlet-like scalars (light-blue region). 
However, as we can see in Fig.~\ref{composition3},
no solutions with masses $\lsim m_{\text{Higgs}}/2$ are present.
{In Table~\ref{S3-1sB1}, we show the BP {\bf S3-B1} corresponding to the blue region of Fig.~\ref{S3Lambda-kappa}.}
In Fig.~\ref{S3Ak-kapvR}, we show the
correlation that is necessary to find these points, and
in  Figs.~\ref{S3-2D-Lambda-Kappa-Tkappa} and~\ref{S3-2D-Lambda-Kappa-Akappa} we show the different values of
$-T_\kappa$ and $A_\kappa = T_\kappa/\kappa$, respectively, en the $\kappa-\lambda$ plane. 
The upper bound for $A_\kappa$ is around 500 GeV in this case, but this is an artifact of our scan. If we had allowed in $S_3$ values of $T_\kappa$ up to 1 TeV, then the upper bound for $A_\kappa$ would have been around 1 TeV.
%\R{In $S_3$, 
%about half (?) of the light-blue points correspond to cases where the singlet-like scalars have masses $\lsim m_{\text{Higgs}}/2$.}

In this scan, solutions with singlet-like states with masses close to 125 GeV are more rare. %producing a superposition of signals with the SM-like Higgs,
{Only about 0.2\% of the phenomenologically viable points} found are of this type. 
%However, not all of them have a significant doublet composition as to produce a superposition of signals.
We show in Table~\ref{S3-3-scalars} the BP {\bf S3-2h1} as an
example.
%of this situation.
As we can see, 
the right sneutrino $h_1$ has the largest doublet composition of 
$H^{\mathcal{R}}_{u}$ (1.42\%) and $H^{\mathcal{R}}_{d}$ (4.78\%), but its mass if far away from 125 GeV.
%but insufficient as to contribute significantly to the Higgs signal.
%whereas $h_2$ and $h_3$ are very pure singlets as expected.
%Unlike the BP S1-ss1 of $S_1$, now the three sneutrinos are very mixed because of the larger value of $\lambda$.
For this BP also the Majorana mass is small as for {\bf S2-2h1},
$\mathcal{M}=50.3$ GeV, and there are three neutralinos dominantly right-handed neutrinos with masses of that order, 50.7, 51.8 and 64.8 GeV. As a consequence, the decay channel right sneutrino to two right-handed neutrinos opens for $h_3$ and $h_4$ (also for the SM-like Higgs $h_5$), but is not possible for $h_1$. The latter can decay to right-handed neutrino plus light neutrino, but with a very small BR. 
%with dominant compositions
%$\widetilde{{\nu}}^{\mathcal{R}}_{eR}}$ and $\widetilde{{\nu}}^{\mathcal{R}}_{\mu R}}$, respectively, and not contributing therefore to the superposition of Higgs-like states.
In Table~\ref{S3-ss1}, we give another BP of this kind, {\bf S3-2h2}, where now one of the singlet-like states, $h_1$, contributes to the superposition of signals with the SM-like Higgs $h_2$.

\begin{figure}[t!]
 \centering
 \includegraphics[width=0.8\linewidth, height= 0.35\textheight]{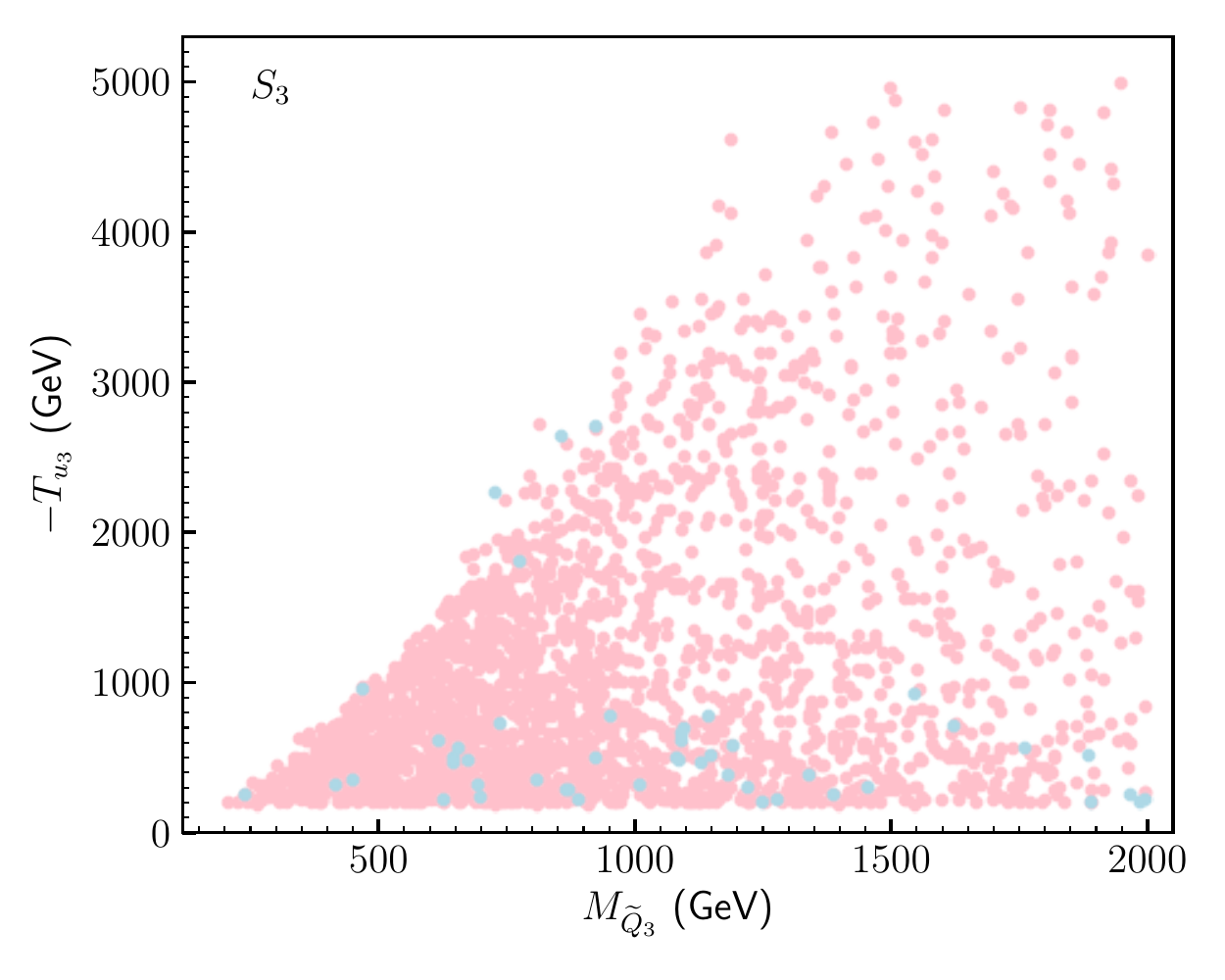}
  \caption{
     Viable points of the parameter space for $S_3$ in the
  $-T_{u_3}$ versus $m_{\widetilde Q_{3L}}$ plane.
% Universality of the parameters is assumed with 
%$\lambda_i=\lambda, \kappa_i=\kappa, 
%v_{iR}=v_R, T_{\kappa_i}=T_\kappa$, and 
%$T_{\lambda_i}=T_\lambda$. 
The color code is the same as in Fig.~\ref{S3Lambda-kappa}.
}
%  The same as in Fig.~\ref{S3LamtanB}, but for $-T_{u_3}$ versus $m_{\widetilde Q_{3L} }$. }
  \label{S3Tu3MQ3}
\end{figure}

%\begin{figure}[t!]
% \centering
% \includegraphics[width=0.8\linewidth, height= 0.35\textheight]{Figures/large-lambda/S3-Lambda-vR.pdf}
%  \caption{ The same as in Fig.~\ref{S3LamtanB} but for $v_R$ versus $\lambda$, where $v_R=v_{iR}$ and $\lambda=\lambda_i$ }
%\label{S3Lambda-vR}
%\end{figure}

%\begin{figure}[t!]
% \centering
% \includegraphics[width=0.8\linewidth, height= 0.35\textheight]{Figures/large-lambda/S3-Kappa-vR.pdf}
%  \caption{ The same as in Fig.~\ref{S3LamtanB} but for $v_R$ versus $\kappa$, where $v_R=v_{iR}$ and $\kappa=\kappa_i$}
%\label{S3Kappa-vR}
%\end{figure}

%\begin{figure}[t!]
% \centering
% \includegraphics[width=0.8\linewidth, height= 0.35\textheight]{Figures/large-lambda/S3-Lambda-TLambda.pdf}
%  \caption{ The same as in Fig.~\ref{S3LamtanB} but for $T_\lambda$ versus $\lambda$,
%  where $T_\lambda=T_{\lambda_i}$ and $\lambda=\lambda_i$}
%\label{S3Lambda-TLambda}
%\end{figure}

\begin{figure}[t!]
 \centering
 \includegraphics[width=0.8\linewidth, height= 0.35\textheight]{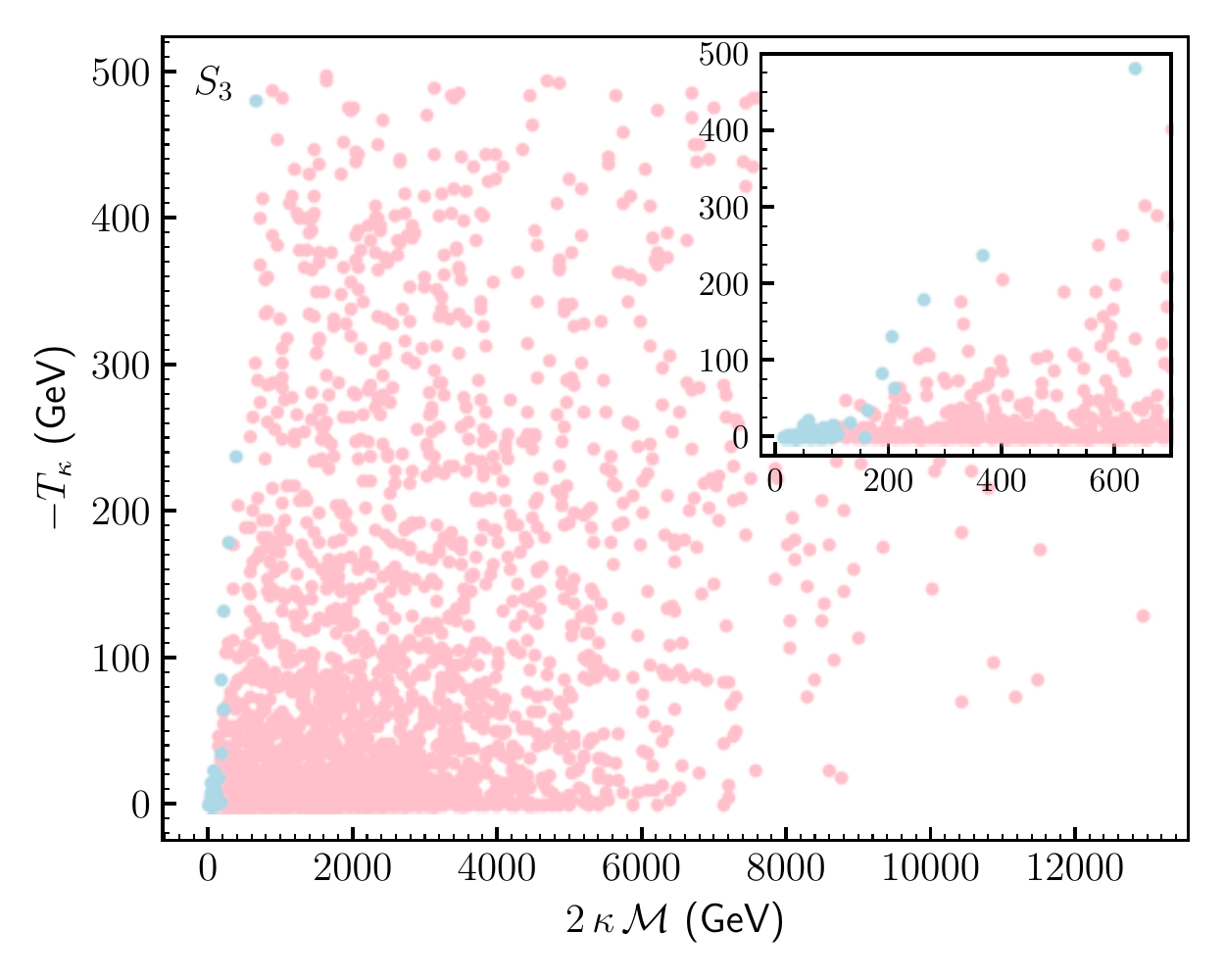}
  \caption{
    Viable points of the parameter space for $S_3$ in the
  $-T_{\kappa}$ versus $2\kappa {\mathcal{M}}$ plane.
% Universality of the parameters is assumed with 
%$\lambda_i=\lambda, \kappa_i=\kappa, 
%v_{iR}=v_R, T_{\kappa_i}=T_\kappa$, and 
%$T_{\lambda_i}=T_\lambda$. 
The color code is the same as in Fig.~\ref{S3Lambda-kappa}.
In the upper right we zoom in the region with light-blue points.
%  The same as in Fig.~\ref{S3LamtanB} but for
%  $-A_\kappa$ versus $\kappa v_R$, where
%$A_\kappa = A_{\kappa_i}$, $\kappa=\kappa_i$ and $v_{R}=v_{iR}$. 
}
\label{S3Ak-kapvR}
\end{figure}

\noindent

\clearpage

\section{Conclusions}
\label{conclusions} 

We performed a dedicated analysis of the parameter space of the $\mn$, in the light of the increasing data about the properties of the SM-like Higgs boson.
For sampling the Higgs sector, 
we used a powerful likelihood data-driven method based on the algorithm {\tt MultiNest}.
The states of the Higgs sector crucial for our analysis are the two Higgs doublets and the three right sneutrinos, which are mixed among themselves. After determining the relevant parameters related to this sector (see Eq.~(\ref{finalp})), 
we performed scans 
%on this parameter space of the model 
to search for points compatible with the latest experimental data on Higgs physics. 
For constraining the predictions of our extended Higgs sector,
we interfaced {\tt HiggsBounds} with {\tt MultiNest}, and to address whether a given Higgs scalar of the $\mn$ is in agreement with the signals observed by ATLAS and CMS we also interfaced {\tt HiggsSignals} with {\tt MultiNest}.
In addition, we demanded the compatibility of the points with observables such as $B$ and $\mu$ decays, and discussed the values of muon $g-2$ in different regions of the parameter space.

In this framework, we performed the three scans described in 
Table~\ref{Scans-priors-parameters}, which are determined by the range of  $\lambda$ couplings in the superpotential mixing Higgses and right sneutrinos,
$\sum_i \lambda_i \, \hat \nu^c_i\, \hat H_u \hat H_d$ (with $\lambda_i=\lambda$).
In particular, we considered the three ranges $\lambda\in$ $[0.01, 0.2)$, $[0.2, 0.5)$, and
$[0.5, 1.2)$.
Perturbativity up to the GUT scale for $\lambda$ is not imposed, and that is why we allow values of $\lambda$ larger than 0.4. Neither we imposed perturbativity up to the GUT scale for $\kappa$ couplings in the superpotential among right sneutrinos, 
$\sum_{i,j,k}
\kappa_{ijk}
\hat \nu^c_i\hat \nu^c_j\hat \nu^c_k$ (with $\kappa_{iii}=\kappa$ and vanishing otherwise),
considering therefore 
the range $\kappa\in$ $[0.01, 2]$.
The results are summarized in 
Figs.~\ref{S1Lambda-kappa},~\ref{S2Lambda-kappa}, and~\ref{S3Lambda-kappa} for the three scans.
Clearly, we find viable solutions in almost the entire 
$\kappa-\lambda$ plane with the exception of the scan $S_3$ in
Fig.~\ref{S3Lambda-kappa}, which is more constrained.
This is due to the large values of 
$\lambda\in$ $[0.5, 1.2)$ that can give rise to tachyons originated in the mixing between the two Higgs doublets. 

We have obtained therefore that the parameter space of the $\mn$ contains many viable solutions, including also many different phenomenological possibilities. For example, there are solutions where the SM-like Higgs is the lightest scalar (red and light-red points in the figures), but also solutions where right sneutrino-like states are lighter (blue and light-blue points). In the latter case, it is even possible to have these (singlet-like) scalars with masses $\lsim m_{\text{Higgs}}/2$.
In addition, 
we also find solutions where several scalars are degenerated with masses close to 125 GeV, and can have their signals rates superimposed contributing to the resonance observed at 125 GeV.

Given these results, it is then important to study in detail the collider phenomenology of the solutions found. In particular, the impact of the new states, not only the right but also the left sneutrinos, and the neutralinos containing right-handed neutrinos. Novel signals associated to them might help to probe the $\mn$ at the LHC. These analyses will be carried out in a fortcoming publication~\cite{kpatcha:2019xxx}.

\begin{acknowledgments}

We would like to thank Jes\'us Moreno for his collaboration during the early stages of this work, specially concerning the computing tasks carried out at CESGA. Also we would like to thank Florian Staub for his help with the implementation of the $\mn$ in SARAH.
The work of EK and CM was supported in part by the Spanish Agencia Estatal de Investigaci\'on 
through the grants FPA2015-65929-P (MINECO/FEDER, UE), PGC2018-095161-B-I00 and IFT Centro de Excelencia Severo Ochoa SEV-2016-0597.
The work of EK was funded by Fundación La Caixa under `La Caixa-Severo 
Ochoa' international predoctoral grant.
The work of DL was supported by the Argentinian CONICET, and also acknowledges the support of the Spanish grant FPA2015-65929-P (MINECO/FEDER, UE). 
RR acknowledges partial funding/support from the Elusives ITN (Marie Sklodowska-Curie grant agreement No 674896), the  
``SOM Sabor y origen de la Materia'' (FPA 2017-85985-P) and the Spanish MINECO Centro de Excelencia Severo Ochoa del IFIC 
program under grant SEV-2014-0398. EK, CM, DL and RR also acknowledge the support of the Spanish Red Consolider MultiDark FPA2017-90566-REDC.

\end{acknowledgments}

\clearpage

%%%%%%%%%%%%%%%%%%%%%%%%%%%%%%%%%%%%%%%%%%%%%%%%%%%%%%%%%%%%%%%%%%%%%%%%%%%
% \iffalse %%%%%%%%%%%%%%%%%%%%%%%
%%%%%%%%%%%%%\renewcommand\thefigure{\Alph{section}.\arabic{subsection}.\arabic{figure}}%%%%%%%%%%%%%%%%%%%%%%%%%%%%%%%%%%%%%%%%%%%%%%%%%%%%%%%%%%%%%%
\appendix
\numberwithin{equation}{section}
\numberwithin{figure}{section}
\numberwithin{table}{section}

\numberwithin{equation}{subsection}
\numberwithin{figure}{subsection}
\numberwithin{table}{subsection}

\section{Higgs-right sneutrino mass submatrices }
\label{Apendix:Sneutrino-masses}

%%%%%%%%%%%%%%%%%%%%%%%%%%%%%%%%%%%%%%%%%%%%%%%%%%%%%%%%%%%%%%%%%
%\footnotesize 
Using the parameters of Eq.~(\ref{softfree}), 
the minimization equations for $H_{d,u}$ and $v_{iR}$, and 
neglecting terms suppressed by the small $Y_{\nu_{ij}}$ and $v_{iL}$, the tree-level entries 
of the $5\times 5$ Higgs-right 
sneutrino submatrices~\cite{Escudero:2008jg,Ghosh:2008yh,Bartl:2009an,Ghosh:2017yeh,Biekotter:2019gtq}
%, and the mixing with the doublet fields 
can be approximated as follows:

%\noindent
%The scalar mass matrices generated in the $\mu\nu$SSM were computed 
%in Appendix A.1 of Ref.~\cite{Escudero:2008jg} with the assumption of CP conservation for simplicity.
%In this Appendix, we write those equations and replace the values of the soft masses obtained through the minimization conditions in Eqs.~(\ref{tadpoles1})-(\ref{tadpoles4}). To carry it out we assume that the slepton soft mass matrices are diagonal in flavor space. 

%\vspace{0.25cm}
%\noindent
%%%%%%%%%%%%%%%%%%%%%%%%%%%%%%%%%%%%%%%%%%%%%%%%%%%%%%%%%%%%%%%%%

% \resizebox{1.0\hsize}{!}{ $ 

%%%%%%%%%%%%%%%%%%%%%%%%%%%%%%%%%%%%%%%%%%%%%%%%%%%%%%%%%%%%%%%%
\subsection{Scalars
%Mass Matrix for CP-even neutral scalars
%CP-even neutral scalars
}
\label{SubAppendix-scalarmasses}
%%%%%%%%%%%%%%%%%%%%%%%%%%%%%%%%%%%%%%%%%%%%%%%%%%%%%%%%%%%%%%%%%

{\footnotesize
\bea
m_{H^{\mathcal{R}}_{d} H^{\mathcal{R}}_{d}}^2
%\approx  
%m_{\widetilde{\nu}^{\mathcal{I}}_{iL} \widetilde{\nu}^{\mathcal{I}}_{iL}}^2
&=&
 \tan\beta \sum_i \frac{v_{iR}}{\sqrt 2}\left(T_{{\lambda}_i} + \lambda_i
 \frac{\mathcal{M}_i}{2}
 %\kappa_i \frac{v_{iR}}{\sqrt 2}
 \right)
+
\left(\frac{v}{\sqrt 2}\right)^2\frac{1}{{1+\tan^2\beta}}\ \frac{1}{2}(g^2+g'^2),
\label{mixinghd}
\\ 
\nonumber
\\
m_{H^{\mathcal{R}}_{u} H^{\mathcal{R}}_{u}}^2
%m_{\widetilde{\nu}^{\mathcal{I}}_{iL} \widetilde{\nu}^{\mathcal{I}}_{iL}}^2
&=&
 \frac{1}{\tan\beta} \sum_i \frac{v_{iR}}{\sqrt 2}\left(T_{{\lambda}_i} + \lambda_i
  \frac{\mathcal{M}_i}{2}
% \kappa_i\frac{v_{iR}}{\sqrt 2}
\right)
+
\left(\frac{v}{\sqrt 2}\right)^2\frac{\tan^2\beta}{{1+\tan^2\beta}}\ \frac{1}{2}(g^2+g'^2),
\label{mixinghu}
\\ 
\nonumber
\\
m_{H^{\mathcal{R}}_{d} H^{\mathcal{R}}_{u}}^2
%\approx  
%m_{\widetilde{\nu}^{\mathcal{I}}_{iL} \widetilde{\nu}^{\mathcal{I}}_{iL}}^2
&=&
%& \approx & 
-\sum_i \frac{v_{iR}}{\sqrt 2}\left(T_{{\lambda}_i} + \lambda_i
 \frac{\mathcal{M}_i}{2}
%\kappa_i \frac{v_{iR}}{\sqrt 2}
\right)
+
\left(\frac{v}{\sqrt 2}\right)^2\frac{\tan\beta}{1+\tan^2\beta}
\left[-\frac{1}{2}(g^2+g'^2)  +2
\sum_i \lambda_i^2\right],
\label{mixinghuhd}
\\ 
\nonumber
\\
m_{\widetilde{\nu}^{\mathcal{R}}_{iR} H^{\mathcal{R}}_{u}}^2
&=&
-\frac{v}{\sqrt 2}\frac{1}{\sqrt{1+\tan^2\beta}}
\left[
 T_{{\lambda}_i}\ + \lambda_i\left({\mathcal M}_{i}
 -2\mu\tan\beta\right)
 \right],
 \label{mixingevenR2}
% \frac{v_{u,d}}{\sqrt 2}
%\frac{\tan\beta}{\sqrt{1+\tan^2\beta}}
%- T_{{\lambda}_i}\ \frac{v_{u,d}}{\sqrt 2}- \lambda_i{\mathcal M}_{i}
% +2\lambda_i\mu\ \frac{v_{d,u}}{\sqrt 2},
\\ 
\nonumber
\\
m_{\widetilde{\nu}^{\mathcal{R}}_{iR} H^{\mathcal{R}}_{d}}^2
%\approx  
%m_{\widetilde{\nu}^{\mathcal{I}}_{iL} \widetilde{\nu}^{\mathcal{I}}_{iL}}^2
&=&
%& \approx & 
-\frac{v}{\sqrt 2}\frac{\tan\beta}{\sqrt{1+\tan^2\beta}}
\left[
 T_{{\lambda}_i}\ + \lambda_i\left({\mathcal M}_{i}
 -\frac{2\mu}{\tan\beta}\right)
 \right],
\label{mixingevenR}
\\ 
\nonumber
\\
m_{\widetilde{\nu}^{\mathcal{R}}_{iR} \widetilde{\nu}^{\mathcal{R}}_{jR}}^2
&=&
\delta_{ij}\left\{
 \left(\frac{T_{{\kappa}_i}}{\kappa_i} 
 +2{\mathcal M}_{i}\right)\frac{{\mathcal M}_{i}}{2}
%\right.
%\nonumber\\ 
%  &&
%  \left.
+\frac{\lambda_i\mu}{v_{iR}/\sqrt 2}\left(\frac{v}{\sqrt 2}\right)^2\left(
\frac{1}{\mu}\frac{
T_{{\lambda}_i}}{\lambda_i}\frac{\tan\beta}{1+\tan^2\beta}-1
\right)\right.
\nonumber\\
&& 
\left.
-T_{{\nu}_i}
\frac{v}{\sqrt 2} 
\frac{\tan\beta}{\sqrt{1+\tan^2\beta}}
\right\}
%\nonumber\\ 
% &&
+
\lambda_i \lambda_j\left(\frac{v}{\sqrt 2}\right)^2,
\label{evenR}
%\\ 
%\nonumber
%\\
%m_{\widetilde{\nu}^{\mathcal{R}}_{iR} \widetilde{\nu}^{\mathcal{R}}_{jR}}^2
%&=&
%\lambda_i \lambda_j\left(\frac{v}{\sqrt 2}\right)^2
%+\delta_{ij}\left\{
% \left(\frac{T_{{\kappa}_i}}{\kappa_i} 
% +2{\mathcal M}_{i}\right)\frac{{\mathcal M}_{i}}{2}
%\right.
%\nonumber\\ 
%  &&
%  \left.
%-\frac{1}{v_{iR}/\sqrt 2}\left(\frac{v}{\sqrt 2}\right)^2\left[\lambda_i\mu
%-T_{{\lambda}_i}
%\frac{\tan\beta}{1+\tan^2\beta}\right]
%-T_{{\nu}_i}
%\frac{v}{\sqrt 2} 
%\frac{\tan\beta}{\sqrt{1+\tan^2\beta}}
%\right\},
%\label{evenR}
%\\ 
%\nonumber
%\\
%m_{\widetilde{\nu}^{\mathcal{R}}_{iR} H^{\mathcal{R}}_{d}}^2
%&=&
%-\frac{v}{\sqrt 2}\frac{\tan\beta}{\sqrt{1+\tan^2\beta}}
%\left[
% T_{{\lambda}_i}\ + \lambda_i\left({\mathcal M}_{i}
% -\frac{2\mu}{\tan\beta}\right)
% \right],
%\label{mixingevenR}
%\\ 
%\nonumber
%\\
%m_{\widetilde{\nu}^{\mathcal{R}}_{iR} H^{\mathcal{R}}_{u}}^2
%&=&
%-\frac{v}{\sqrt 2}\frac{1}{\sqrt{1+\tan^2\beta}}
%\left[
% T_{{\lambda}_i}\ + \lambda_i\left({\mathcal M}_{i}
% -2\mu\tan\beta\right)
% \right].
%\label{mixingevenR2}
\eea
}
{
%\footnotesize 
where $\mu=\sum_i \lambda_i \frac{v_{iR}}{\sqrt 2}$ %corresponds to the $\mu$-term of Eq.~(\ref{mu}) 
and ${\mathcal M}_{i} = 2\kappa_i\frac{v_{iR}}{\sqrt 2}$. 
%to the Majorana masses of Eq.~(\ref{majorana2}).
}

%\vspace{0.25cm}
%\noindent
%%%%%%%%%%%%%%%%%%%%%%%%%%%%%%%%%%%%%%%%%%%%%%%%%%%%%%%%%%%%%%%%%

\subsection{Pseudoscalars 
%Higgs-right sneutrino submatrix
%Mass Matrix for CP-odd neutral scalars 
}
\label{SubAppendix:NeutrealScalar}
%%%%%%%%%%%%%%%%%%%%%%%%%%%%%%%%%%%%%%%%%%%%%%%%%%%%%%%%%%%%%%%%%

{\footnotesize
\bea
m_{H^{\mathcal{I}}_{d} H^{\mathcal{I}}_{d}}^2
%\approx  
%m_{\widetilde{\nu}^{\mathcal{I}}_{iL} \widetilde{\nu}^{\mathcal{I}}_{iL}}^2
&=&
 \tan\beta \sum_i \frac{v_{iR}}{\sqrt 2}\left(T_{{\lambda}_i} + \lambda_i
 \frac{\mathcal{M}_i}{2}
 %\kappa_i \frac{v_{iR}}{\sqrt 2}
 \right)
%+
%\left(\frac{v}{\sqrt 2}\right)^2\frac{1}{{1+\tan^2\beta}}\ \frac{1}{2}(g^2+g'^2),
\label{mixingIhd}
\\ 
\nonumber
\\
m_{H^{\mathcal{I}}_{u} H^{\mathcal{I}}_{u}}^2
%m_{\widetilde{\nu}^{\mathcal{I}}_{iL} \widetilde{\nu}^{\mathcal{I}}_{iL}}^2
&=&
 \frac{1}{\tan\beta} \sum_i \frac{v_{iR}}{\sqrt 2}\left(T_{{\lambda}_i} + \lambda_i
  \frac{\mathcal{M}_i}{2}
% \kappa_i\frac{v_{iR}}{\sqrt 2}
\right)
%+
%\left(\frac{v}{\sqrt 2}\right)^2\frac{\tan^2\beta}{{1+\tan^2\beta}}\ %\frac{1}{2}(g^2+g'^2),
\label{mixingIhu}
\\ 
\nonumber
\\
m_{H^{\mathcal{I}}_{d} H^{\mathcal{I}}_{u}}^2
%\approx  
%m_{\widetilde{\nu}^{\mathcal{I}}_{iL} \widetilde{\nu}^{\mathcal{I}}_{iL}}^2
&=&
%& \approx & 
\sum_i \frac{v_{iR}}{\sqrt 2}\left(T_{{\lambda}_i} + \lambda_i
 \frac{\mathcal{M}_i}{2}
%\kappa_i \frac{v_{iR}}{\sqrt 2}
\right)
%+
%\left(\frac{v}{\sqrt 2}\right)^2\frac{\tan\beta}{1+\tan^2\beta}
%\left[-\frac{1}{2}(g^2+g'^2)  +2
%\sum_i \lambda_i^2\right],
\label{mixingIhuhd}
\\
\nonumber
\\
m_{\widetilde{\nu}^{\mathcal{I}}_{iR} H^{\mathcal{I}}_{u}}^2
&=&
 \frac{v}{\sqrt 2}\frac{1}{\sqrt{1+\tan^2\beta}}
\left(
T_{{\lambda}_i}\ - \lambda_i{\mathcal M}_{i}
 \right),
\label{mixingevenI2}
\\ 
\nonumber
\\ 
m_{\widetilde{\nu}^{\mathcal{I}}_{iR} H^{\mathcal{I}}_{d}}^2
&=&
 \frac{v}{\sqrt 2}\frac{\tan\beta}{\sqrt{1+\tan^2\beta}}
\left(
T_{{\lambda}_i}\ - \lambda_i{\mathcal M}_{i}
 \right),
\label{mixingevenI}
\\ 
\nonumber
\\ 
m_{\widetilde{\nu}^{\mathcal{I}}_{iR} \widetilde{\nu}^{\mathcal{I}}_{jR}}^2
&=&
\delta_{ij}\left\{
-\frac{3}{2} \frac{T_{{\kappa}_i}}{\kappa_i} {\mathcal M}_{i}
%+4\lambda_i\kappa_i
%\left(\frac{v}{\sqrt 2}\right)^2
%\frac{\tan\beta}{1+\tan^2\beta}
%\right.
%\nonumber\\ 
%  &&
%  \left.
  +\frac{\lambda_i\mu}{v_{iR}/\sqrt 2}\left(\frac{v}{\sqrt 2}\right)^2
  \left[\frac{1}{\mu}\left(\frac{T_{{\lambda}_i}}{\lambda_i} + 2{\mathcal M}_{i} \right)
\frac{\tan\beta}{1+\tan^2\beta}-1\right]
\right.
\nonumber\\ 
  &&
  \left.
-T_{{\nu}_i}
\frac{v}{\sqrt 2}
\frac{\tan\beta}{\sqrt{1+\tan^2\beta}}
\right\}
%\nonumber\\ 
%  &&
+\lambda_i \lambda_j\left(\frac{v}{\sqrt 2}\right)^2.
\label{evenI}
%\\ 
%\nonumber
%\\ 
%m_{\widetilde{\nu}^{\mathcal{I}}_{iR} H^{\mathcal{I}}_{d}}^2
%&=&
% \frac{v}{\sqrt 2}\frac{\tan\beta}{\sqrt{1+\tan^2\beta}}
%\left(
%T_{{\lambda}_i}\ - \lambda_i{\mathcal M}_{i}
% \right),
%\label{mixingevenI}
%\\
%\nonumber\\
%m_{\widetilde{\nu}^{\mathcal{I}}_{iR} H^{\mathcal{I}}_{u}}^2
%&=&
% \frac{v}{\sqrt 2}\frac{1}{\sqrt{1+\tan^2\beta}}
%\left(
%T_{{\lambda}_i}\ - \lambda_i{\mathcal M}_{i}
% \right).
%\label{mixingevenI2}
\eea
}
%{\footnotesize In these equations, $\mu=\sum_i \lambda_i \frac{v_{iR}}{\sqrt 2}$ corresponds to 
%the $\mu$-term of Eq.~(\ref{mu}) and 
%${\mathcal M}_{i} = 2\kappa_i\frac{v_{iR}}{\sqrt 2}$ to the Majorana masses of Eq.~(\ref{majorana2}).}

%$}  %% end resizebox

%\clearpage
%\newpage

\vspace{-1.8cm}
\section{Results from the $\lambda-\kappa$ plane}
\label{figuresscans}

{
%\footnotesize
Here we show several figures for each scan, where the viable points of the parameter space can be seen in the $\kappa-\lambda$ plane for different values of the other parameters.}
%: $v_R$, $T_\lambda$, $A_\lambda$, $\tan\beta$, $m_{\widetilde Q_{3L}}$, $T_{u_3}$, $T_\kappa$, and $A_\kappa$.}

\subsection{ {\it Scan 1} ($0.01 \leq \lambda < 0.2$)}
%\subsection{Scan $S_1$}
\label{scan1}

\begin{figure}[H]
 \centering
 \includegraphics[width=0.8\linewidth, height= 0.35\textheight]{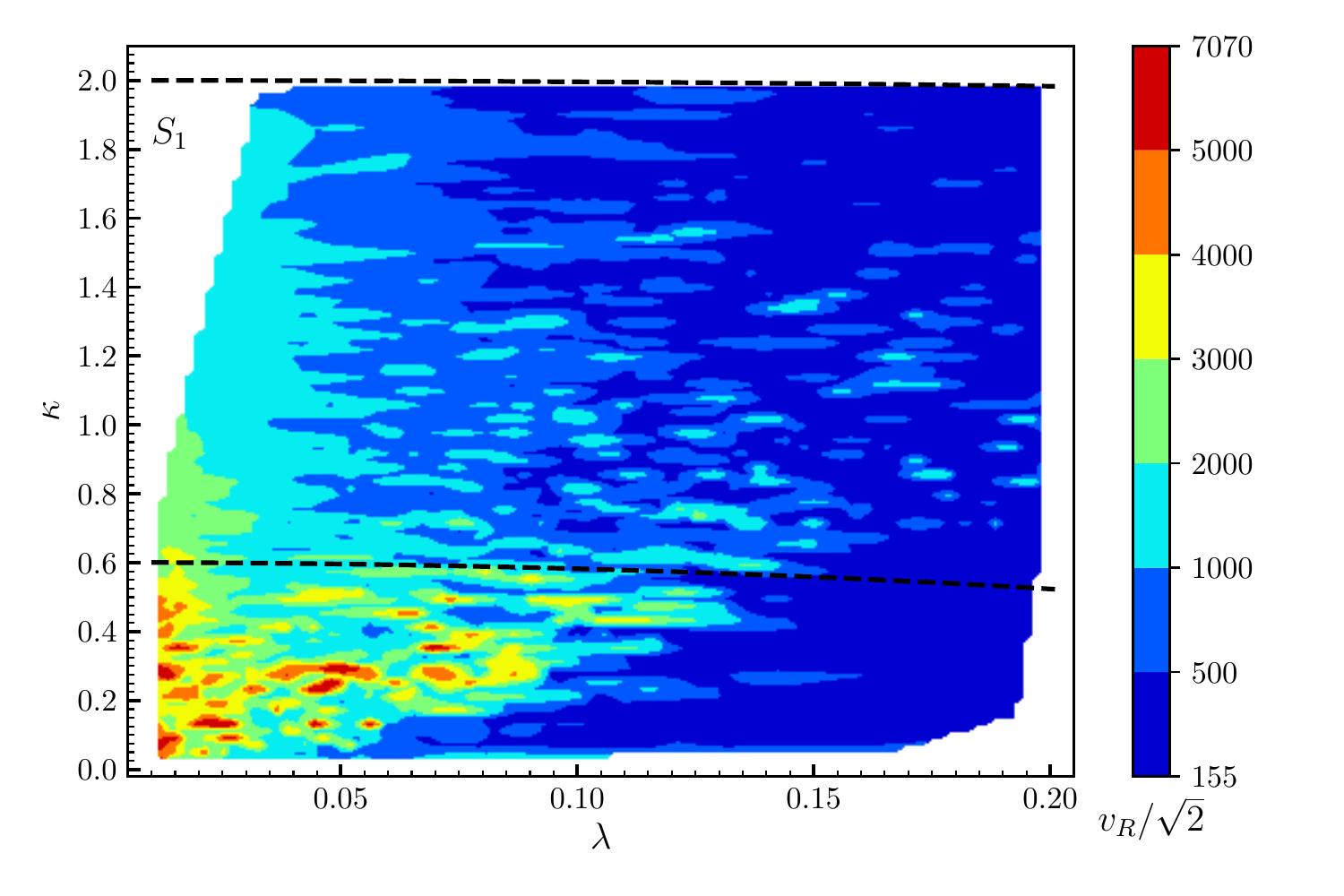}
  \caption{\footnotesize{
   Viable points of the parameter space for $S_1$ in the $\kappa-\lambda$ plane.
%    Universality of the parameters is assumed with 
%$\lambda_i=\lambda, \kappa_i=\kappa, 
%v_{iR}=v_R, T_{\kappa_i}=T_\kappa$, and 
%$T_{\lambda_i}=T_\lambda$. 
The points below 
the lower black dashed line fulfill the condition of 
Eq.~(\ref{perturbativity1}), where perturbativity is assumed up to the GUT scale. All points below the upper dashed line fulfill the condition of
Eq.~(\ref{perturbativity2}), where perturbativity is relaxed up to 10 TeV.
%The points between the lower and upper dashed lines fulfill
%separating the
%perturbative region (up to the GUT scale) from the non-perturbative one.
The colours indicate different values of the right sneutrino VEVs $v_R/\sqrt 2$. 
% assuming universality of the parameters as in Eq.~(\ref{freeparametersu}).
%The red (blue) colour represents cases where the SM-like Higgs is (is not) the lightest scalar, fulfilling perturbativity up to the GUT scale as in Eq.~(\ref{perturbativity1}). 
%For the light-red points the perturbativity condition is relaxed up to 10 TeV as in Eq.~(\ref{perturbativity2}).
%All points below the lower (upper) black dashed line fulfill the perturbativity condition in Eq.~(\ref{perturbativity1}) (Eq.~(\ref{perturbativity2})).
%  The same as in Fig.~\ref{S1LamtanB} but for $\kappa$ versus $\lambda$, where $\kappa=\kappa_i$ and $\lambda=\lambda_i$ 
 } }
\label{S1-2D-Lambda-Kappa-vR}
 \includegraphics[width=0.8\linewidth, height= 0.35\textheight]{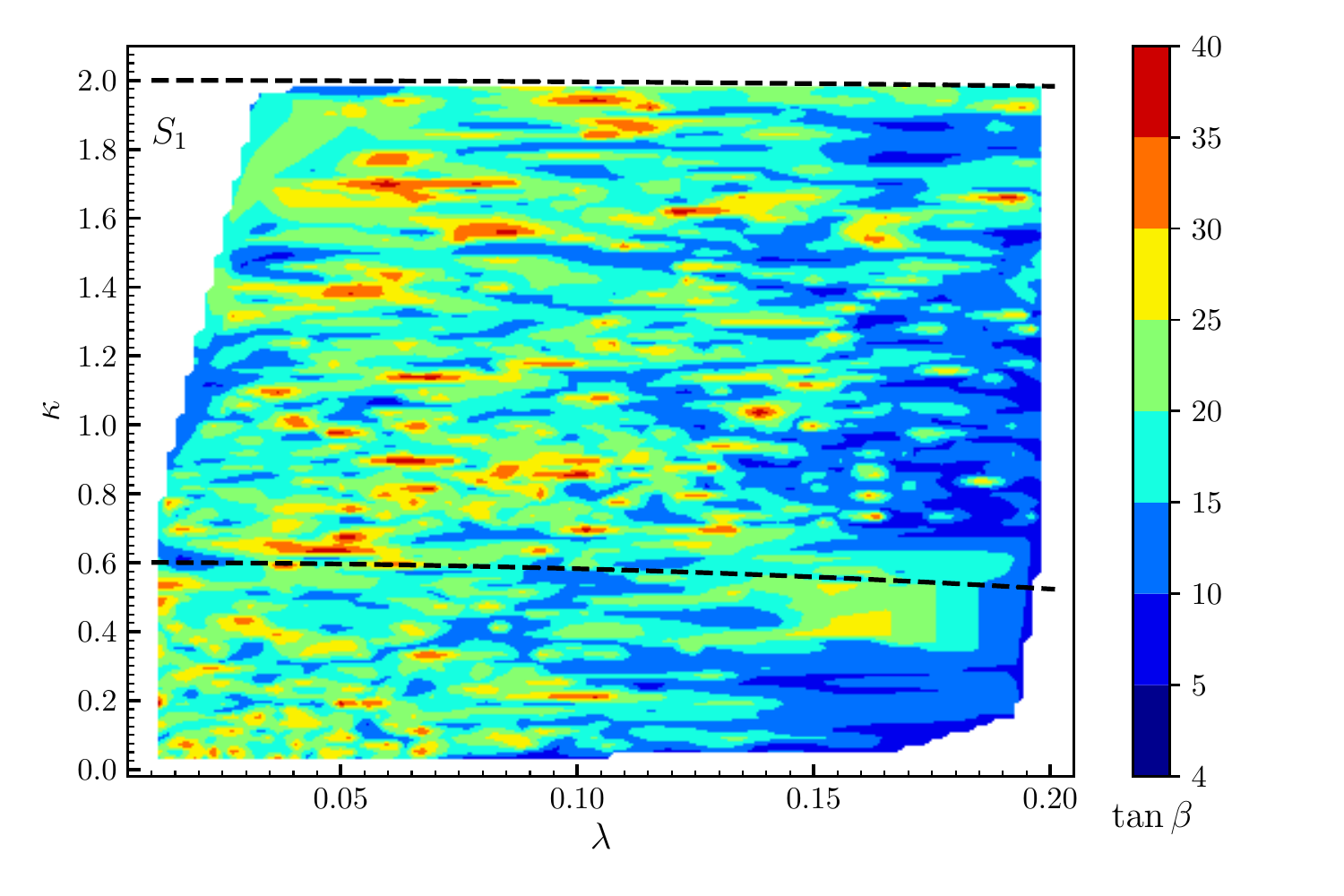}
\caption{\footnotesize{The same as in Fig.~\ref{S1-2D-Lambda-Kappa-vR}, but the colours
  indicate different values of $\tan\beta$.}
  }
\label{S1-2D-Lambda-Kappa-tanB}
\end{figure}

%\iffalse
%\begin{figure}[b!]
% \centering
% \includegraphics[width=0.8\linewidth, height= 0.32\textheight]{Figures/small-lambda/3D/S1-2D-Lambda-Kappa-tanB.pdf}
%\caption{\footnotesize{The same as in Fig.~\ref{S1-2D-Lambda-Kappa-vR}, but the colours
%  indicate different values of $\tan\beta$.}
%  }
%\label{S1-2D-Lambda-Kappa-tanB}
%\end{figure}
%\fi

\begin{figure}[t!]
 \centering
 \includegraphics[width=0.8\linewidth, height= 0.35\textheight]{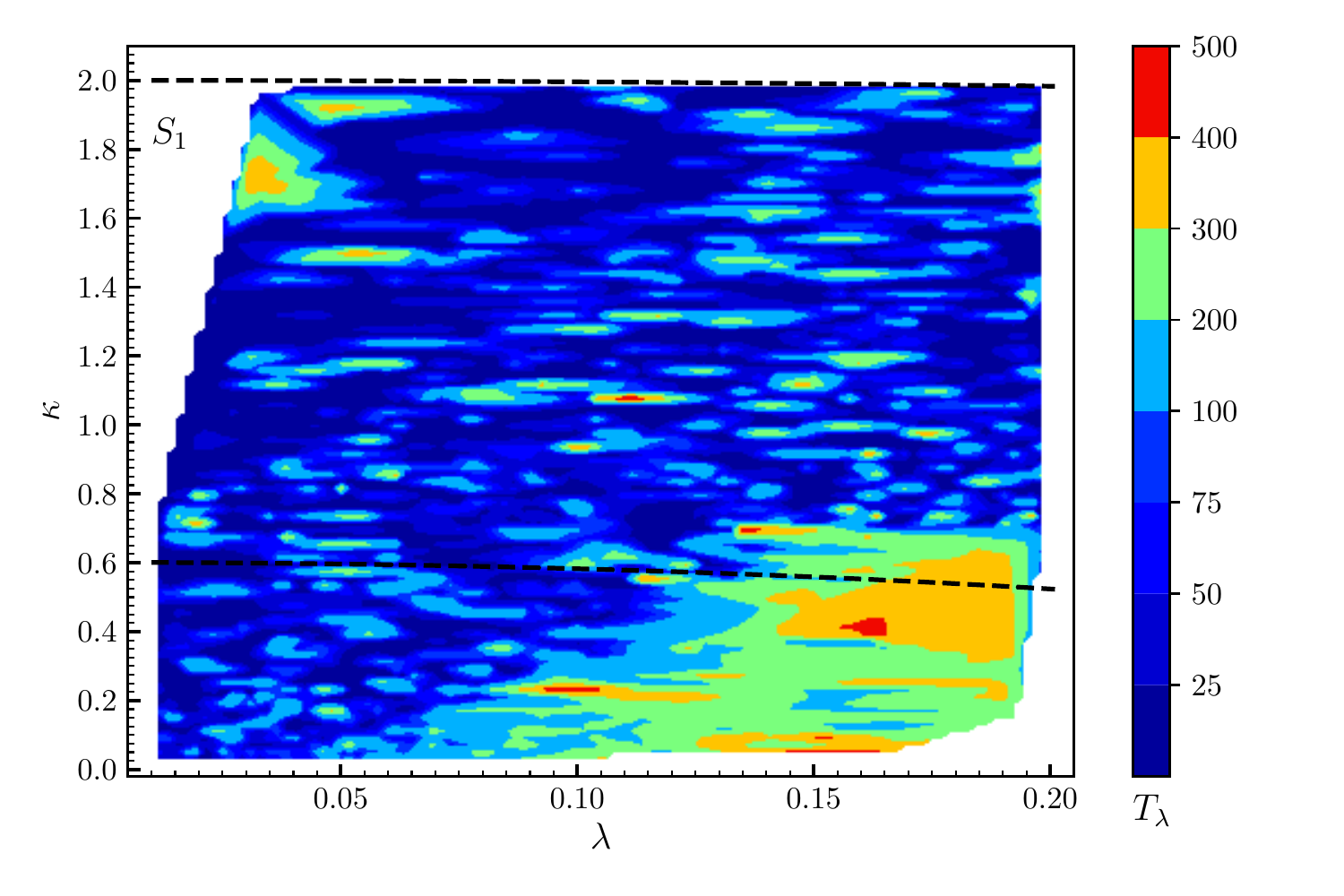}
  \caption{\footnotesize{The same as in Fig.~\ref{S1-2D-Lambda-Kappa-vR}, but the colours
  indicate different low-energy values of the trilinear soft terms $T_\lambda$.}  }
\label{S1-2D-Lambda-Kappa-Tlambda}
 \includegraphics[width=0.8\linewidth, height= 0.35\textheight]{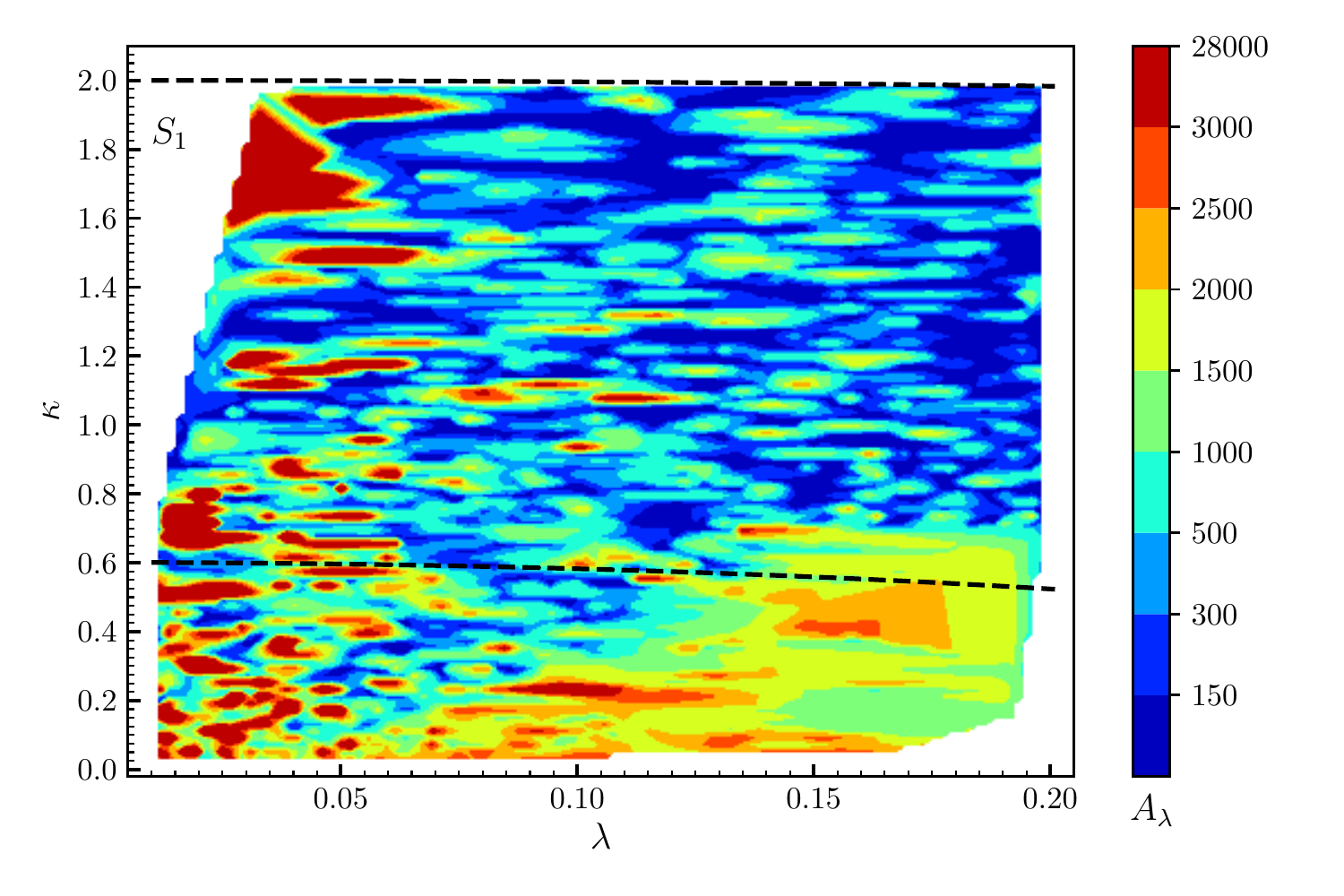}
  \caption{\footnotesize{The same as in Fig.~\ref{S1-2D-Lambda-Kappa-vR}, but the colours
  indicate different low-energy values of the trilinear soft terms $A_\lambda$, assuming the supergravity relation $A_\lambda = T_\lambda/\lambda$.  }}
\label{S1-2D-Lambda-Kappa-Alambda}
\end{figure}

%\begin{figure}[t!]
% \centering
% \includegraphics[width=0.8\linewidth, height= 0.35\textheight]{Figures/small-lambda/3D/S1-2D-Lambda-Kappa-Alambda.pdf}
%  \caption{\footnotesize{The same as in Fig.~\ref{S1-2D-Lambda-Kappa-vR}, but the colours indicate different low-energy values of the trilinear soft terms $A_\lambda$, assuming the supergravity relation $A_\lambda = T_\lambda/\lambda$.  }}
%\label{S1-2D-Lambda-Kappa-Alambda}
%\end{figure}

%\clearpage

\begin{figure}[t!]
 \centering
 \includegraphics[width=0.8\linewidth, height= 0.35\textheight]{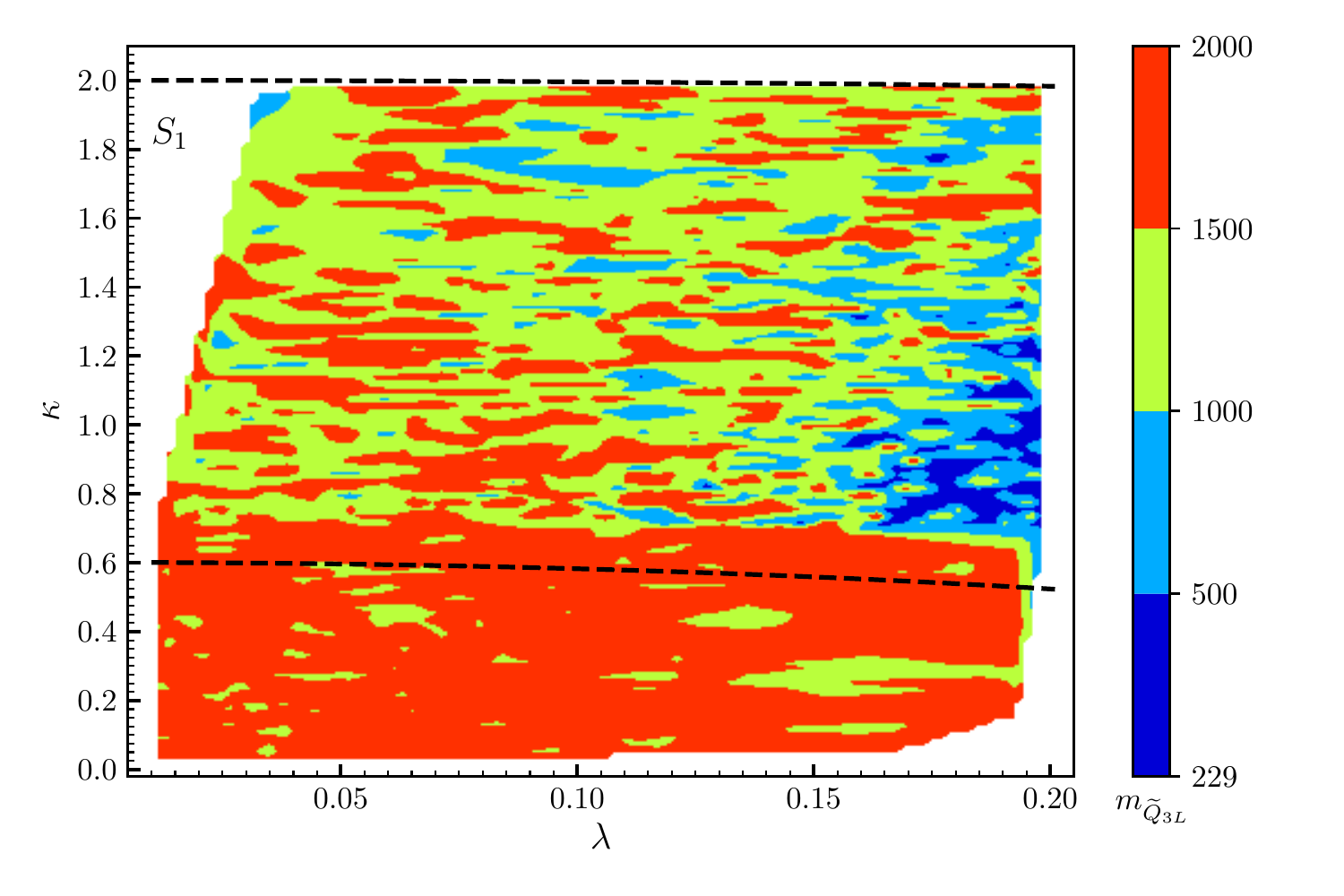}
  \caption{\footnotesize{The same as in Fig.~\ref{S1-2D-Lambda-Kappa-vR}, but the colours indicate different low-energy values of the soft masses 
  $m_{\widetilde Q_{3L}}$.} }
\label{S1-2D-Lambda-Kappa-MQ3}
 \includegraphics[width=0.8\linewidth, height= 0.35\textheight]{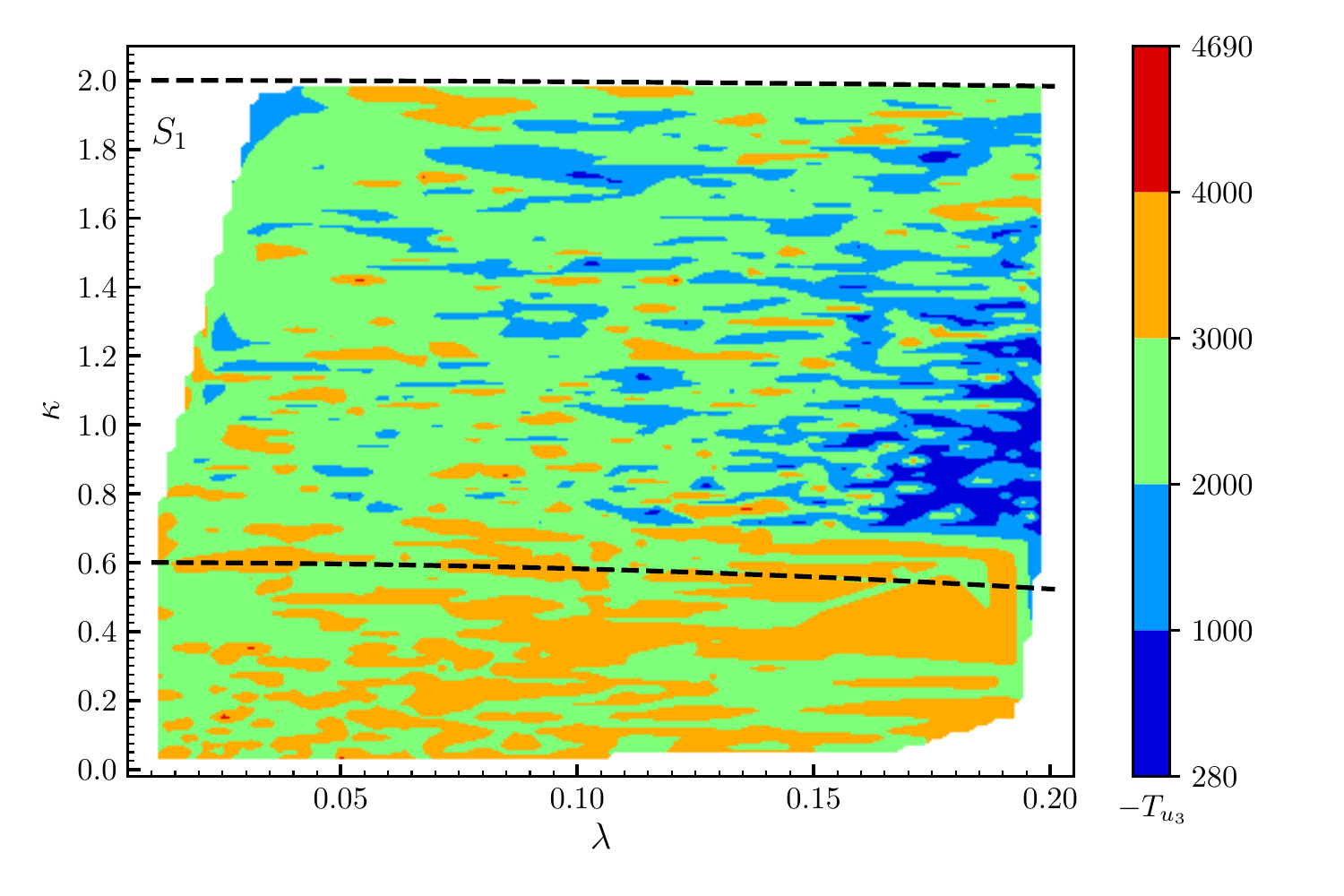}
  \caption{\footnotesize{The same as in Fig.~\ref{S1-2D-Lambda-Kappa-vR}, but the colours
  indicate different low-energy values of the 
 % third-generation squark 
  trilinear soft term $T_{u_3}$.}}
\label{S1-2D-Lambda-Kappa-Tu3}
\end{figure}

%\begin{figure}[t!]
% \centering
% \includegraphics[width=0.8\linewidth, height= 0.35\textheight]{Figures/small-lambda/3D/S1-2D-Lambda-Kappa-Tu3.pdf}
%  \caption{\footnotesize{The same as in Fig.~\ref{S1-2D-Lambda-Kappa-vR}, but the colours
%  indicate different low-energy values of the 
% % third-generation squark 
%  trilinear soft term $T_{u_3}$.}}
%\label{S1-2D-Lambda-Kappa-Tu3}
%\end{figure}

\begin{figure}[t!]
 \centering
 \includegraphics[width=0.8\linewidth, height= 0.35\textheight]{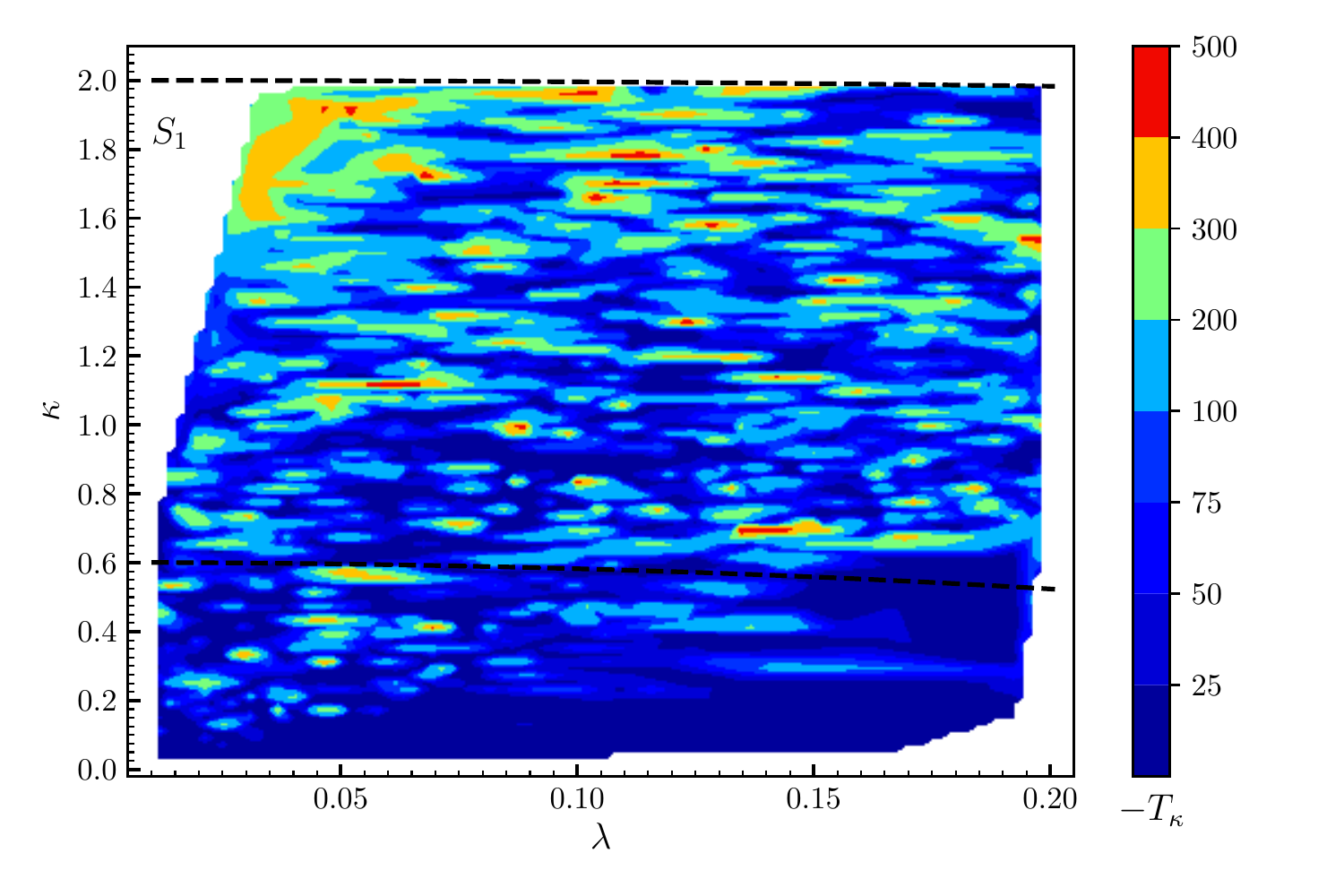}
  \caption{\footnotesize{The same as in Fig.~\ref{S1-2D-Lambda-Kappa-vR}, but the colours
  indicate different low-energy values of the trilinear soft terms $T_\kappa$.   }}
\label{S1-2D-Lambda-Kappa-Tkappa}
 \includegraphics[width=0.8\linewidth, height= 0.35\textheight]{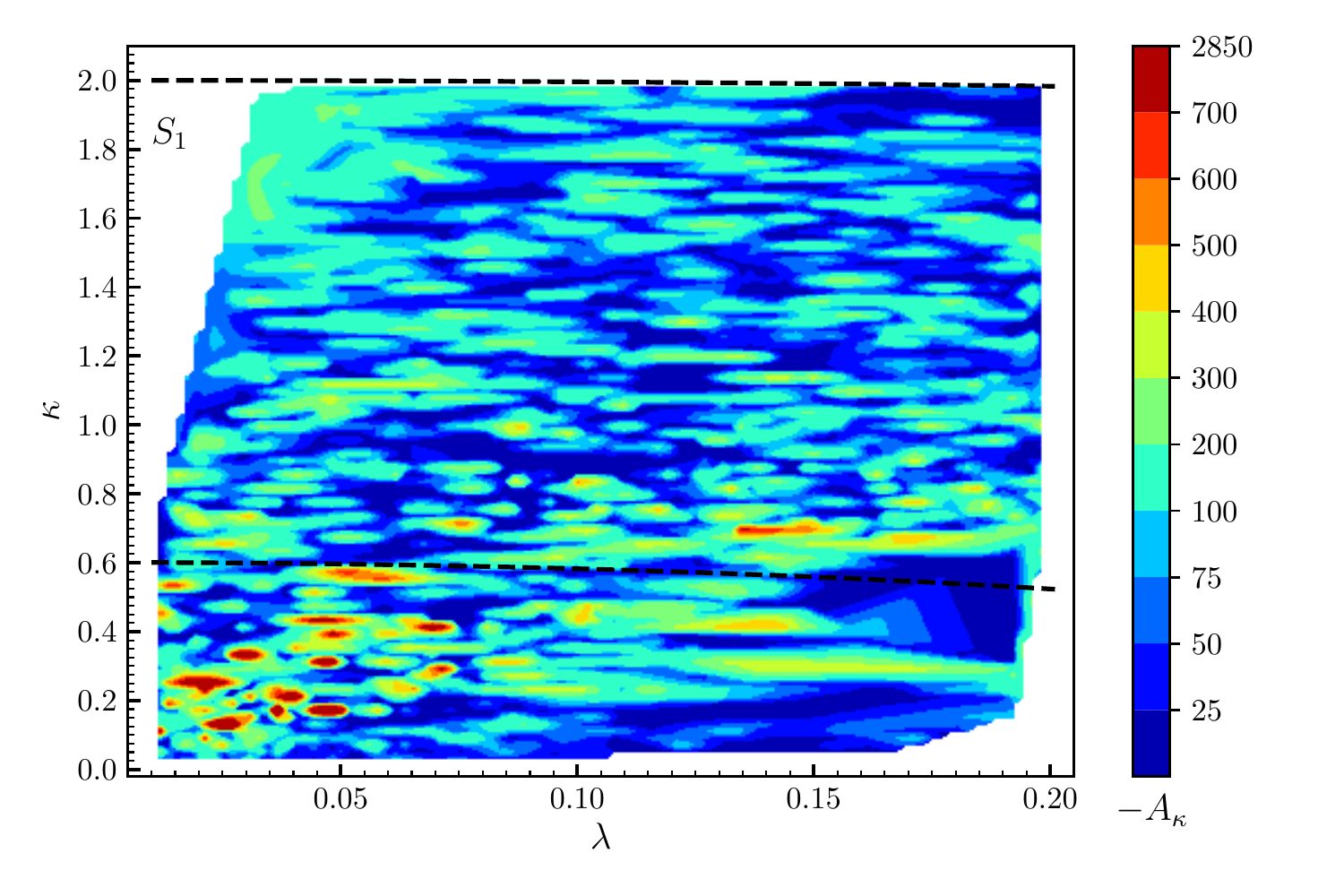}
  \caption{\footnotesize{The same as in Fig.~\ref{S1-2D-Lambda-Kappa-vR}, but the colours
  indicate different low-energy values of the trilinear soft terms $A_\kappa$, assuming the supergravity relation $A_\kappa = T_\kappa/\kappa$.  }}
\label{S1-2D-Lambda-Kappa-Akappa}
\end{figure}

%\begin{figure}[t!]
% \centering
% \includegraphics[width=0.8\linewidth, height= 0.35\textheight]{Figures/small-lambda/3D/S1-2D-Lambda-Kappa-Akappa.pdf}
%  \caption{\footnotesize{The same as in Fig.~\ref{S1-2D-Lambda-Kappa-vR}, but the colours indicate different low-energy values of the trilinear soft terms $A_\kappa$, assuming the supergravity relation $A_\kappa = T_\kappa/\kappa$.  }}
%\label{S1-2D-Lambda-Kappa-Akappa}
%\end{figure}

\clearpage

\subsection{
{\it Scan 2}
%Small to moderate
($0.2 \leq \lambda < 0.5$)}
%
%\subsection{Scan $S_2$}
\label{scan2}

\begin{figure}[H]
 \centering
 \includegraphics[width=0.8\linewidth, height= 0.35\textheight]{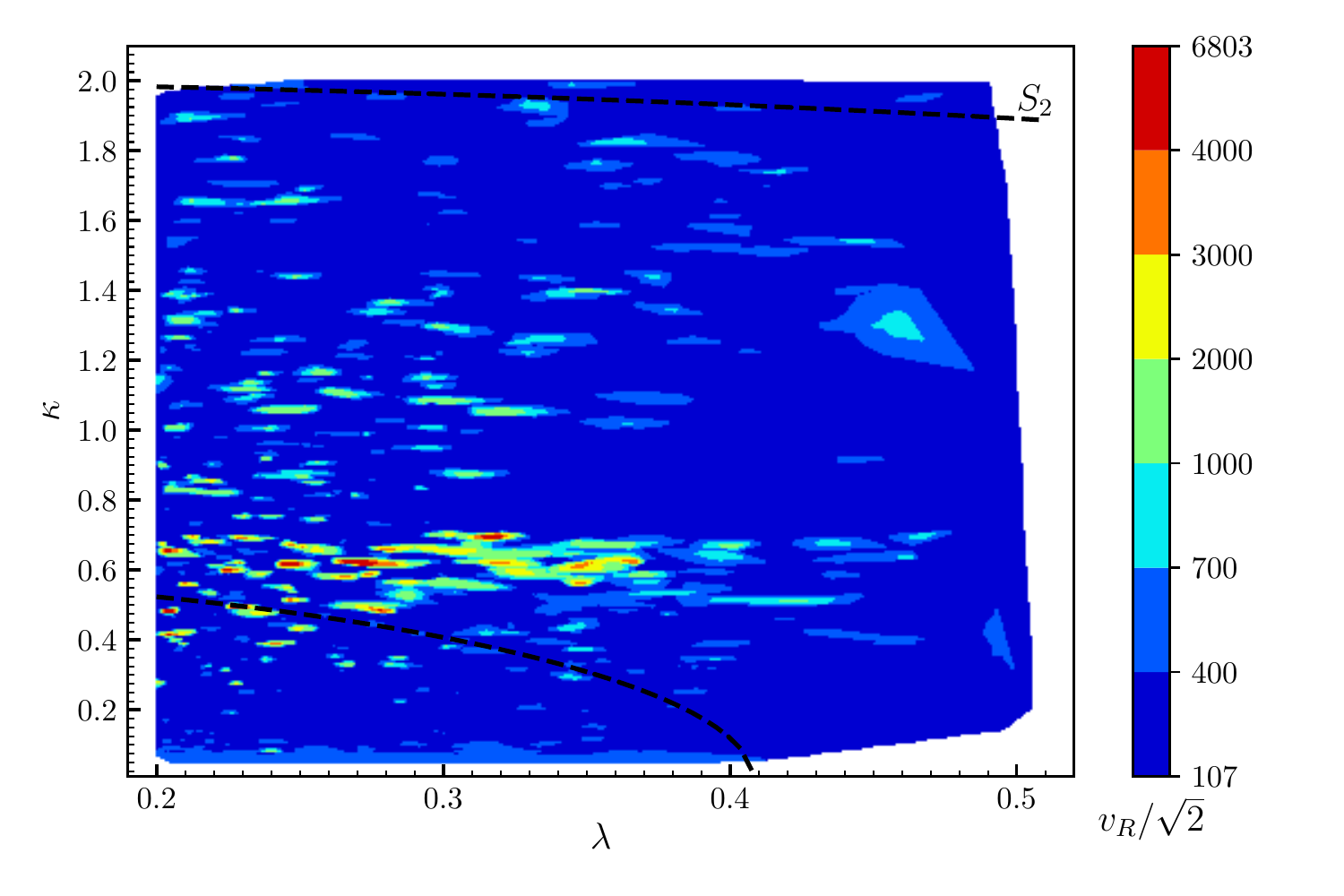}
  \caption{ \footnotesize{Viable points of the parameter space for $S_2$ in the $\kappa-\lambda$ plane.
 %   Universality of the parameters is assumed with 
%$\lambda_i=\lambda, \kappa_i=\kappa, 
%v_{iR}=v_R, T_{\kappa_i}=T_\kappa$, and 
%$T_{\lambda_i}=T_\lambda$. 
The points below 
the lower black dashed line fulfill the condition of 
Eq.~(\ref{perturbativity1}), where perturbativity is assumed up to the GUT scale. All points below the upper dashed line fulfill the condition of
Eq.~(\ref{perturbativity2}), where perturbativity is relaxed up to 10 TeV.
%The black dashed line fulfills the equality in 
%Eq.~(\ref{perturbativity1}), separating the
%perturbative region (up to the GUT scale) from the non-perturbative one.
The colours indicate different values of the right sneutrino VEVs $v_R/\sqrt 2$.  }}
\label{S2-2D-Lambda-Kappa-vR}
 \includegraphics[width=0.8\linewidth, height= 0.35\textheight]{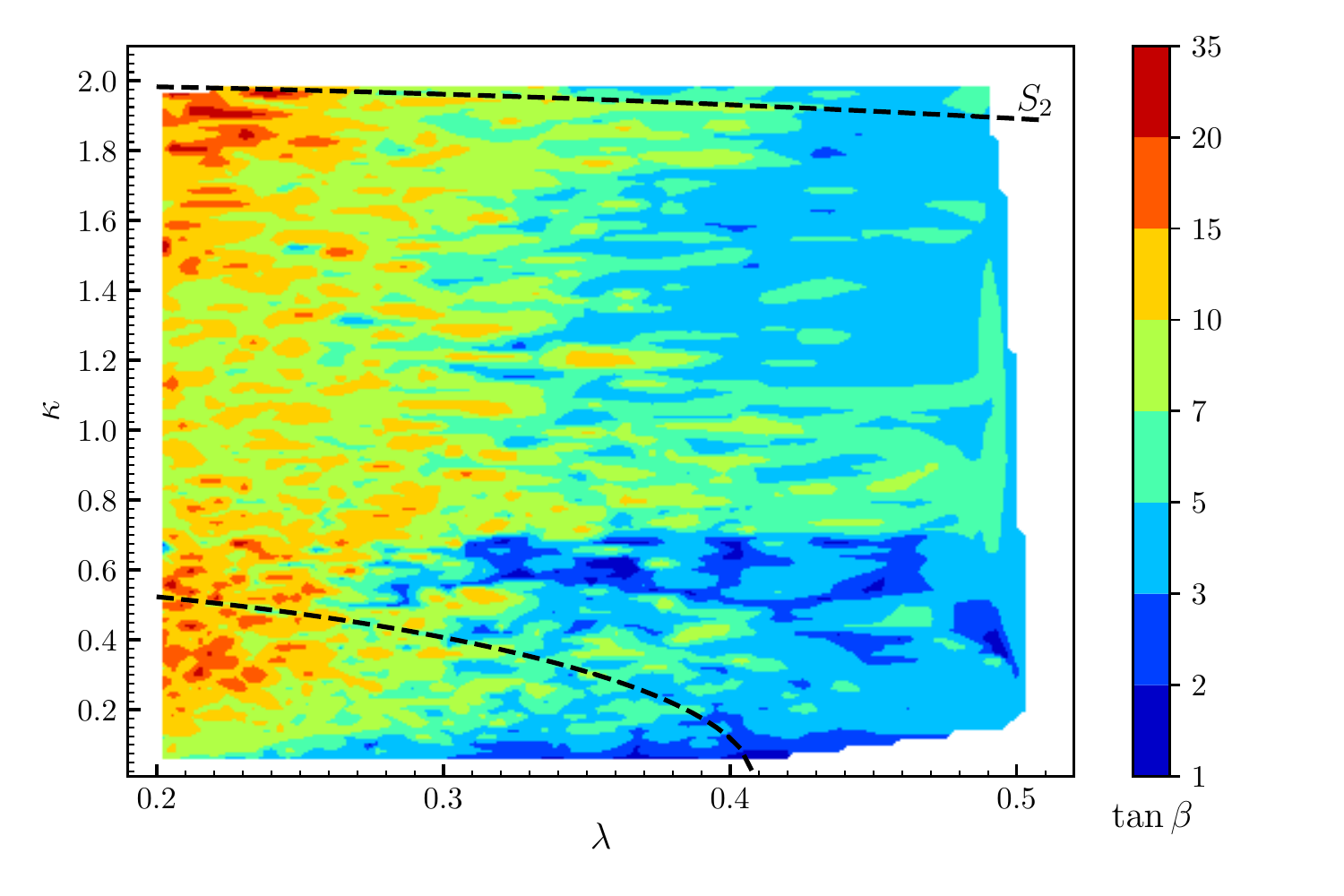}
  \caption{\footnotesize{The same as in Fig.~\ref{S2-2D-Lambda-Kappa-vR}, but the colours   indicate different values of $\tan\beta$. } }
\label{S2-2D-Lambda-Kappa-tanB}
\end{figure}

%\begin{figure}[t!]
% \centering
% \includegraphics[width=0.8\linewidth, height= 0.35\textheight]{Figures/moderate-lambda/3D/S2-2D-Lambda-Kappa-tanB.pdf}
%  \caption{The same as in Fig.~\ref{S2-2D-Lambda-Kappa-vR}, but the colours
%  indicate different values of $\tan\beta$.  }
%\label{S2-2D-Lambda-Kappa-tanB}
%\end{figure}

\begin{figure}[t!]
 \centering
 \includegraphics[width=0.8\linewidth, height= 0.35\textheight]{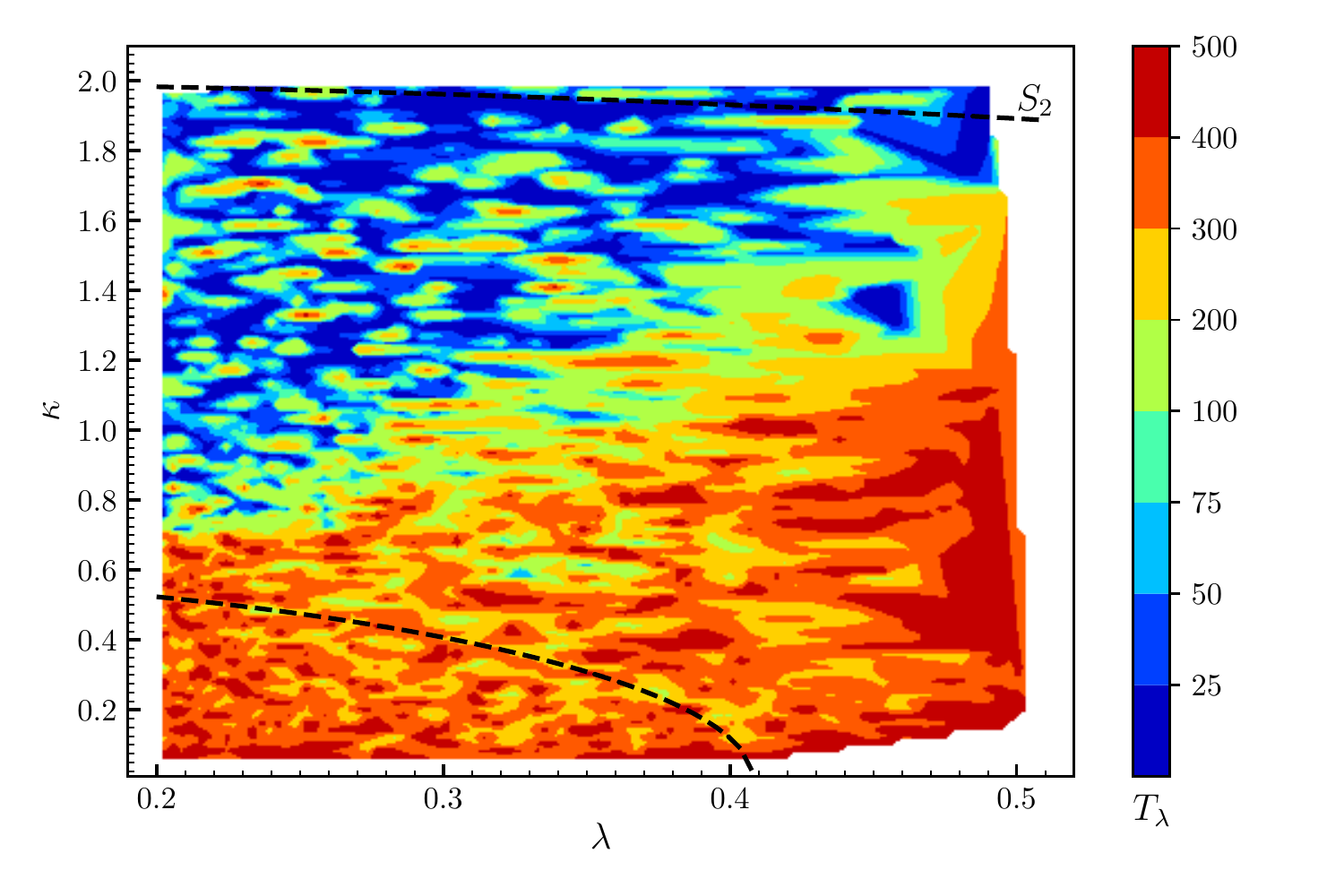}
  \caption{\footnotesize{The same as in Fig.~\ref{S2-2D-Lambda-Kappa-vR}, but the colours indicate different low-energy values of the trilinear soft terms $T_\lambda$.   }}
\label{S2-2D-Lambda-Kappa-Tlambda}
 \includegraphics[width=0.8\linewidth, height= 0.35\textheight]{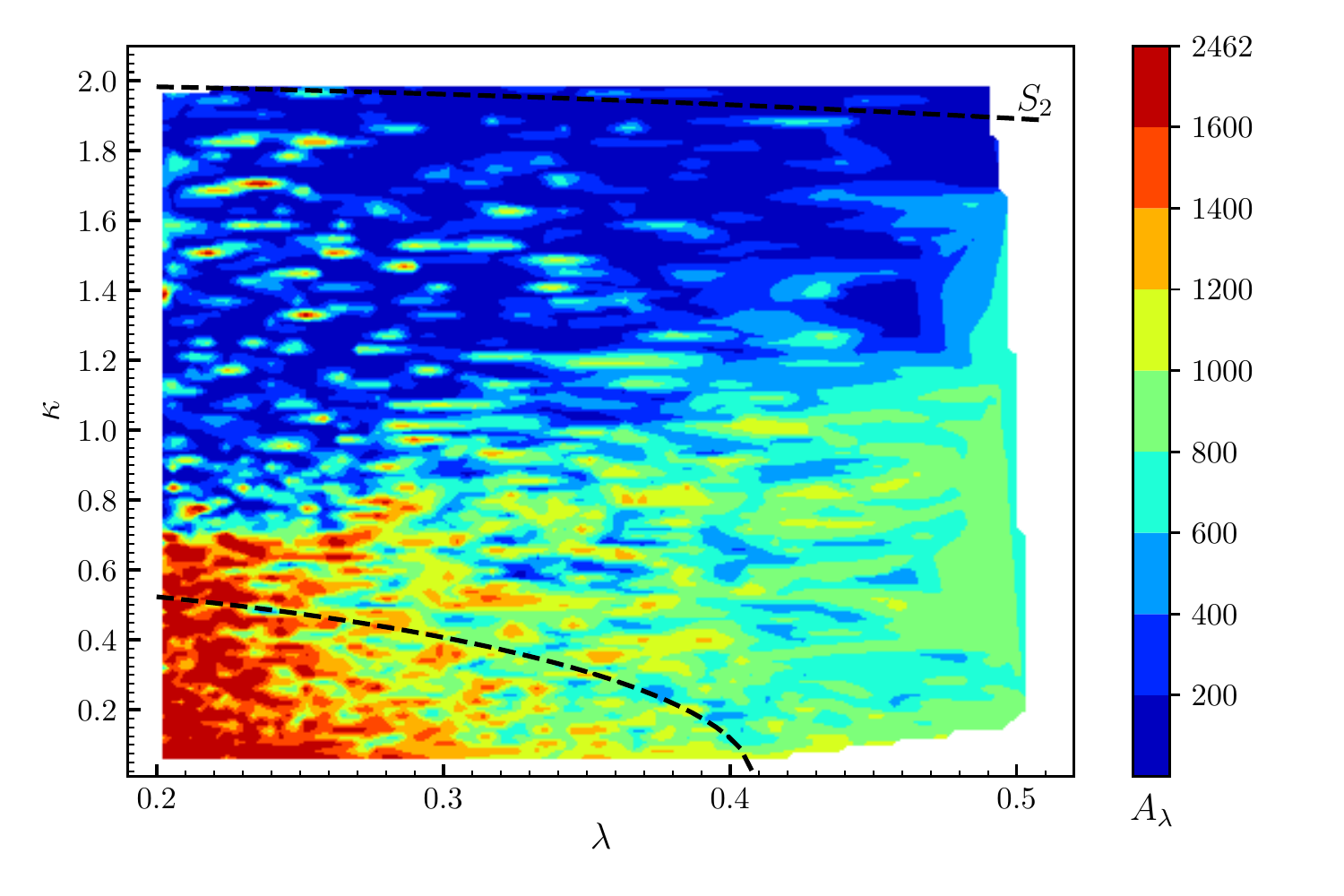}
  \caption{ \footnotesize{The same as in Fig.~\ref{S2-2D-Lambda-Kappa-vR}, but the colours indicate different low-energy values of the trilinear soft terms $A_\lambda$, assuming the supergravity relation $A_\lambda = T_\lambda/\lambda$. } }
\label{S2-2D-Lambda-Kappa-Alambda}
\end{figure}

%\begin{figure}[t!]
% \centering
% \includegraphics[width=0.8\linewidth, height= 0.35\textheight]{Figures/moderate-lambda/3D/S2-2D-Lambda-Kappa-Alambda.pdf}
% \caption{ The same as in Fig.~\ref{S2-2D-Lambda-Kappa-vR}, but the colours
%  indicate different low-energy values of the trilinear soft terms $A_\lambda$, assuming the supergravity relation $A_\lambda = T_\lambda/\lambda$.  }
%\label{S2-2D-Lambda-Kappa-Alambda}
%\end{figure}

%\clearpage

\begin{figure}[t!]
 \centering
 \includegraphics[width=0.8\linewidth, height= 0.35\textheight]{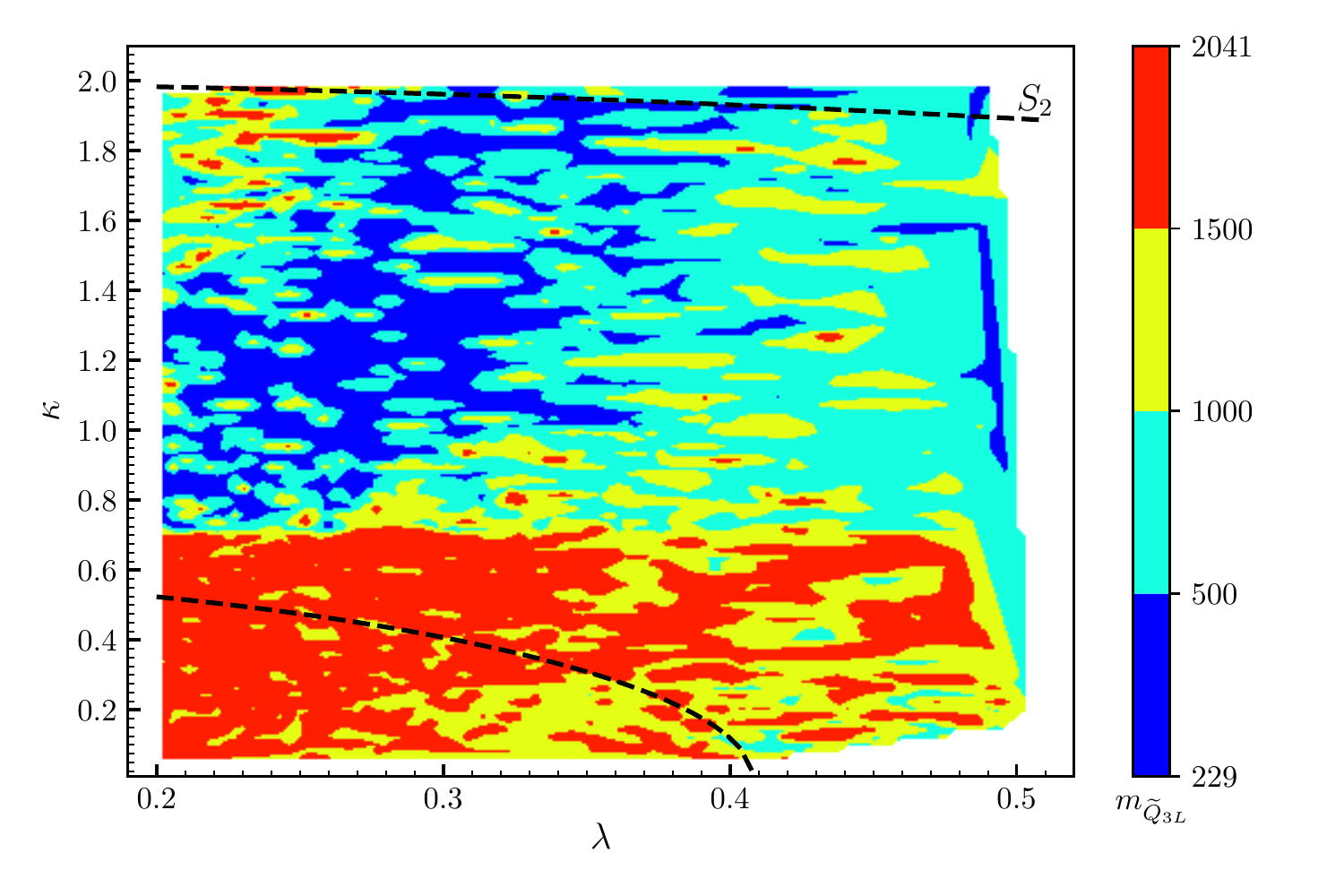}
  \caption{\footnotesize{The same as in Fig.~\ref{S2-2D-Lambda-Kappa-vR}, but the colours indicate different low-energy values of the soft masses 
  $m_{\widetilde Q_{3L}}$.  }}
\label{S2-2D-Lambda-Kappa-MQ3}
 \includegraphics[width=0.8\linewidth, height= 0.35\textheight]{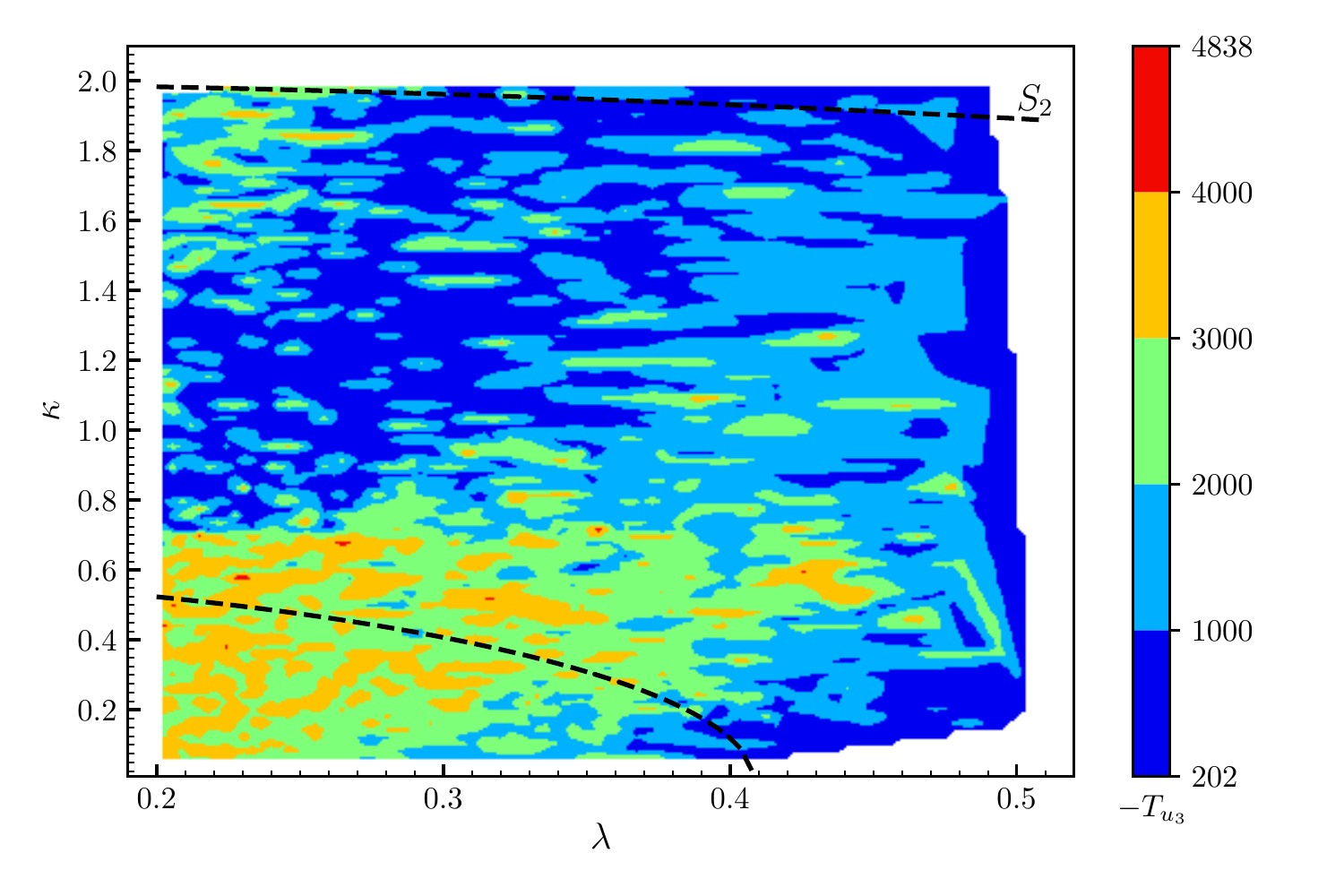}
  \caption{\footnotesize{The same as in Fig.~\ref{S2-2D-Lambda-Kappa-vR}, but the colours indicate different low-energy values of the 
 % third-generation squark 
  trilinear soft term $T_{u_3}$.  }}
\label{S2-2D-Lambda-Kappa-Tu3}
\end{figure}

%\begin{figure}[t!]
% \centering
% \includegraphics[width=0.8\linewidth, height= 0.35\textheight]{Figures/moderate-lambda/3D/S2-2D-Lambda-Kappa-Tu3.pdf}
%  \caption{The same as in Fig.~\ref{S2-2D-Lambda-Kappa-vR}, but the colours
%  indicate different low-energy values of the 
% % third-generation squark 
%  trilinear soft term $T_{u_3}$.  }
%\label{S2-2D-Lambda-Kappa-Tu3}
%\end{figure}

\begin{figure}[t!]
 \centering
 \includegraphics[width=0.8\linewidth, height= 0.35\textheight]{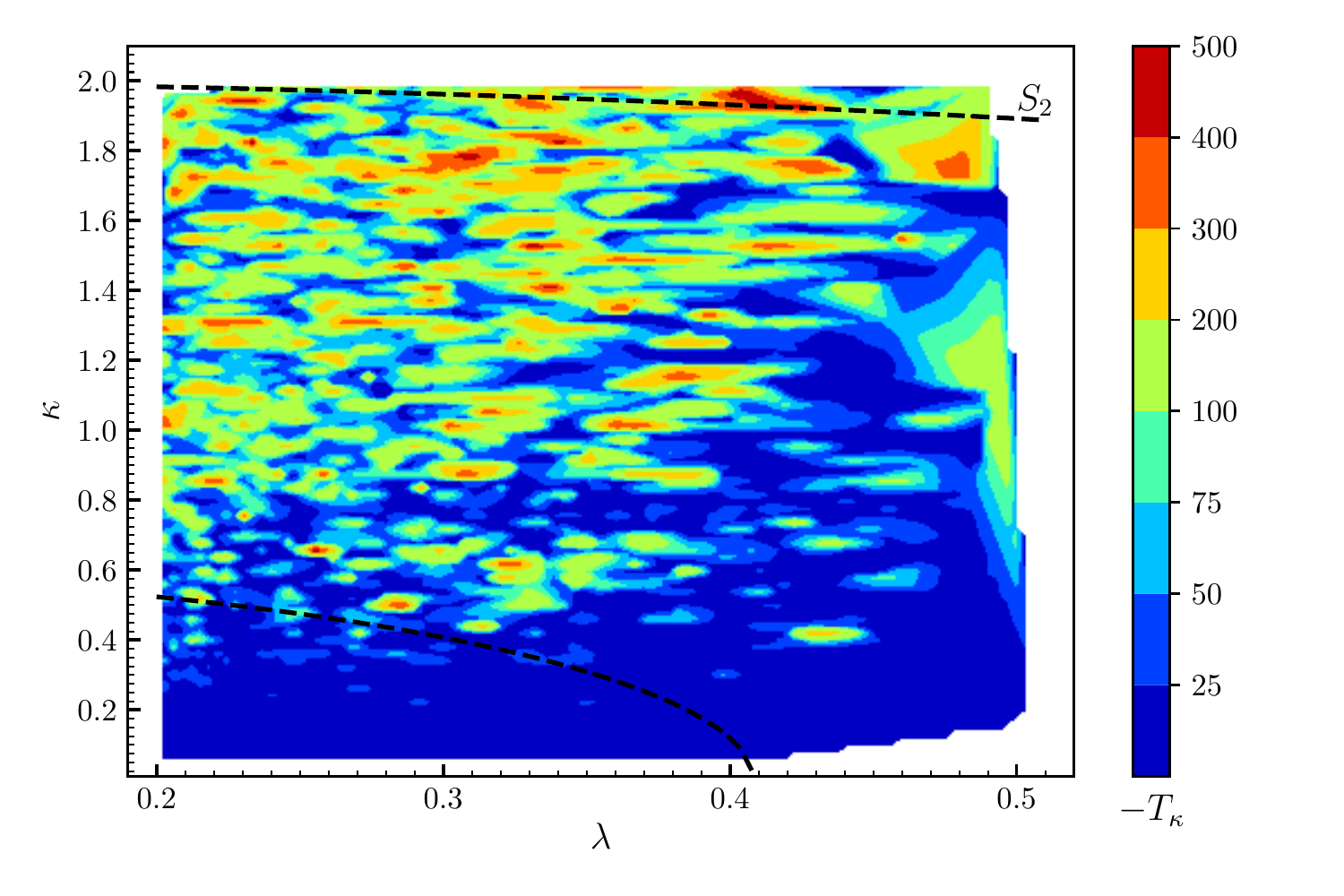}
  \caption{\footnotesize{ The same as in Fig.~\ref{S2-2D-Lambda-Kappa-vR}, but the colours indicate different low-energy values of the trilinear soft terms $T_\kappa$.   }}
\label{S2-2D-Lambda-Kappa-Tkappa}
 \includegraphics[width=0.8\linewidth, height= 0.35\textheight]{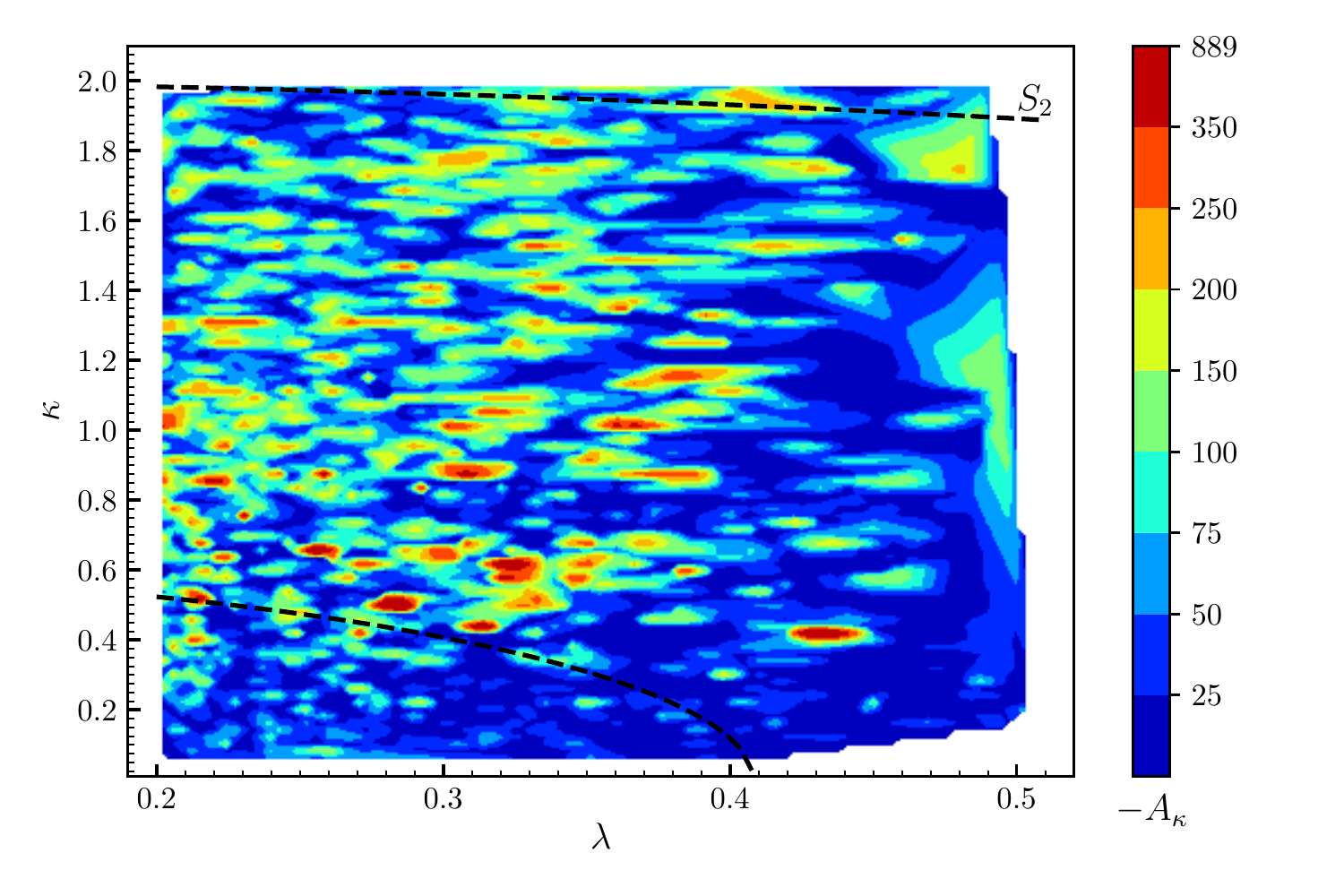}
  \caption{ \footnotesize{The same as in Fig.~\ref{S2-2D-Lambda-Kappa-vR}, but the colours indicate different low-energy values of the trilinear soft terms $A_\kappa$, assuming the supergravity relation $A_\kappa = T_\kappa/\kappa$. }}
\label{S2-2D-Lambda-Kappa-Akappa}
\end{figure}

%\begin{figure}[t!]
% \centering
% \includegraphics[width=0.8\linewidth, height= 0.35\textheight]{Figures/moderate-lambda/3D/S2-2D-Lambda-Kappa-Akappa.pdf}
%  \caption{ \footnotesize{The same as in Fig.~\ref{S2-2D-Lambda-Kappa-vR}, but the colours indicate different low-energy values of the trilinear soft terms $A_\kappa$, assuming the supergravity relation $A_\kappa = T_\kappa/\kappa$. }}
%\label{S2-2D-Lambda-Kappa-Akappa}
%\end{figure}

\clearpage

\subsection{
{\it Scan 3}
%Small to moderate
($0.5 \leq \lambda < 1.2$)}

%\subsection{Scan $S_3$}
\label{scan3}

\begin{figure}[H]
 \centering
 \includegraphics[width=0.8\linewidth, height= 0.35\textheight]{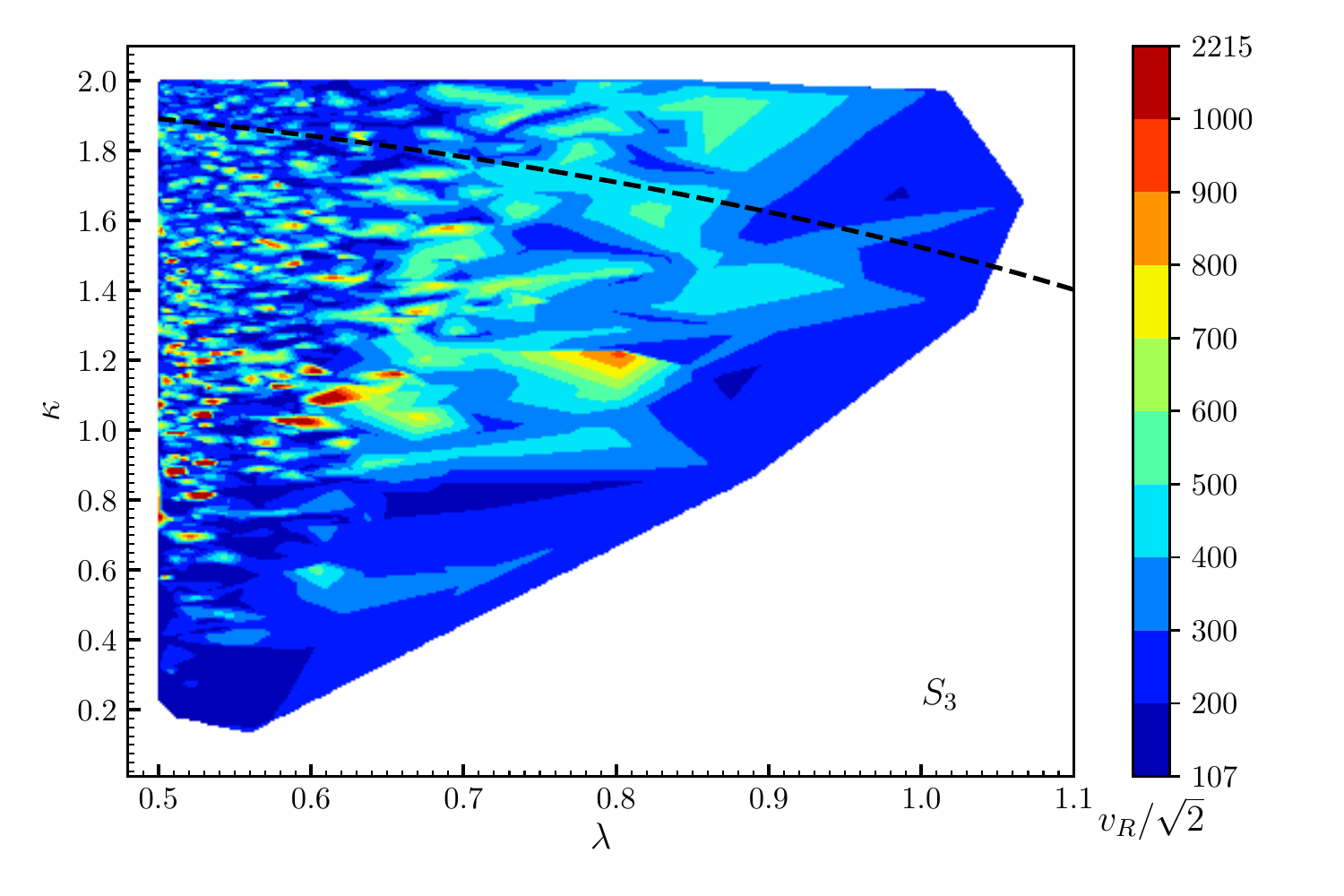}
  \caption{ \footnotesize{Viable points of the parameter space for $S_3$ in the $\kappa-\lambda$ plane. %Universality of the parameters is assumed with 
%$\lambda_i=\lambda, \kappa_i=\kappa, 
%v_{iR}=v_R, T_{\kappa_i}=T_\kappa$, and 
%$T_{\lambda_i}=T_\lambda$. 
%The points below 
%the lower black dashed line fulfill the condition of 
%Eq.~(\ref{perturbativity1}), where perturbativity is assumed up to the GUT scale. 
The points below the dashed line fulfill the condition of
Eq.~(\ref{perturbativity2}), where perturbativity is relaxed up to 10 TeV.
%The black dashed line fulfills the equality in 
%Eq.~(\ref{perturbativity1}), separating the
%perturbative region (up to the 10 TeV scale) from the non-perturbative one.
The colours indicate different values of the right sneutrino VEVs $v_R/\sqrt 2$.}}
\label{S3-2D-Lambda-Kappa-vR}
 \includegraphics[width=0.8\linewidth, height= 0.35\textheight]{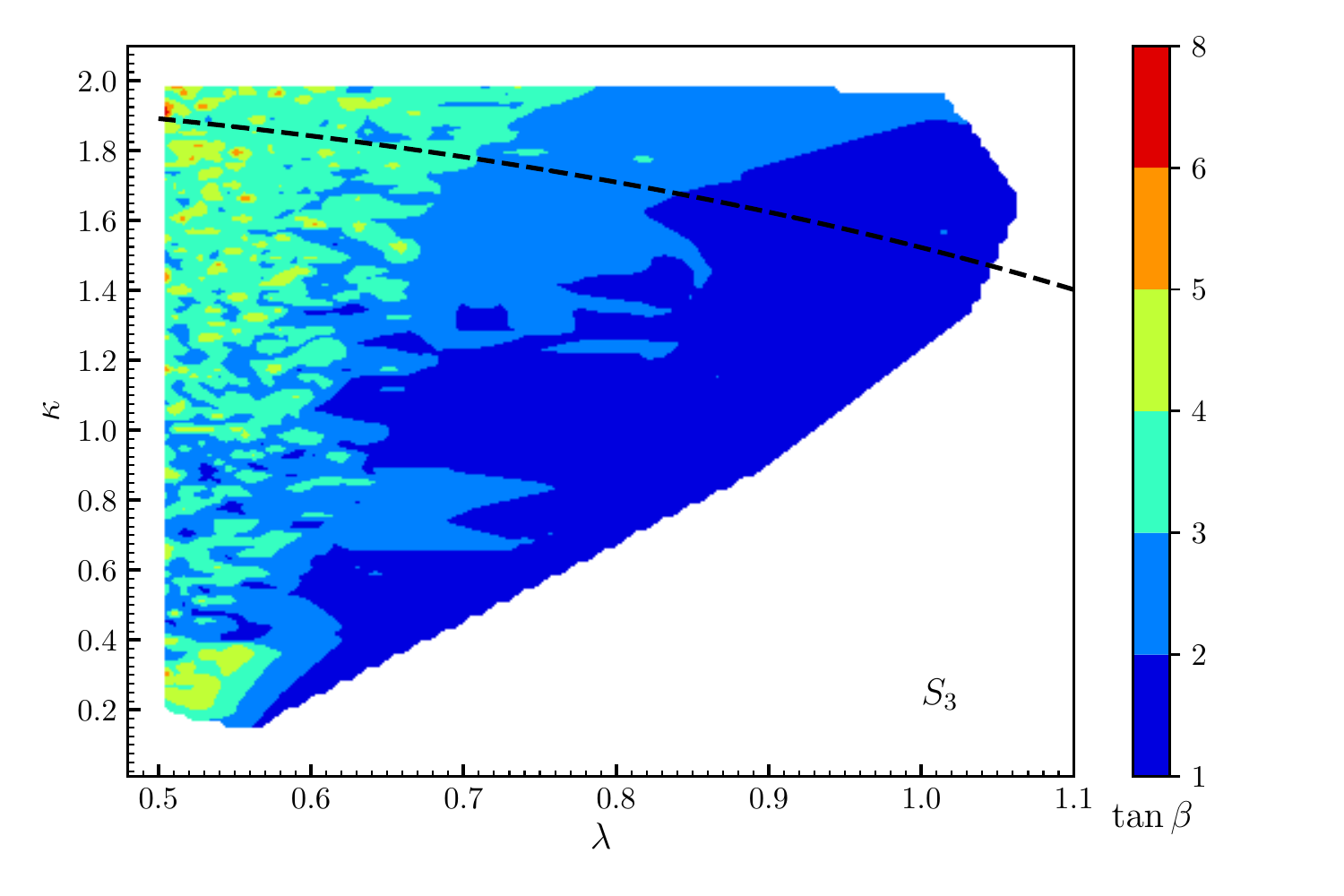}
  \caption{\footnotesize{The same as in Fig.~\ref{S3-2D-Lambda-Kappa-vR}, but the colours indicate different values of $\tan\beta$.  }}
\label{S3-2D-Lambda-Kappa-tanB}
\end{figure}

%\begin{figure}[t!]
% \centering
% \includegraphics[width=0.8\linewidth, height= 0.35\textheight]{Figures/large-lambda/3D/S3-2D-Lambda-Kappa-tanB.pdf}
%  \caption{\footnotesize{The same as in Fig.~\ref{S3-2D-Lambda-Kappa-vR}, but the colours indicate different values of $\tan\beta$.  }}
%\label{S3-2D-Lambda-Kappa-tanB}
%\end{figure}

\begin{figure}[t!]
 \centering
 \includegraphics[width=0.8\linewidth, height= 0.35\textheight]{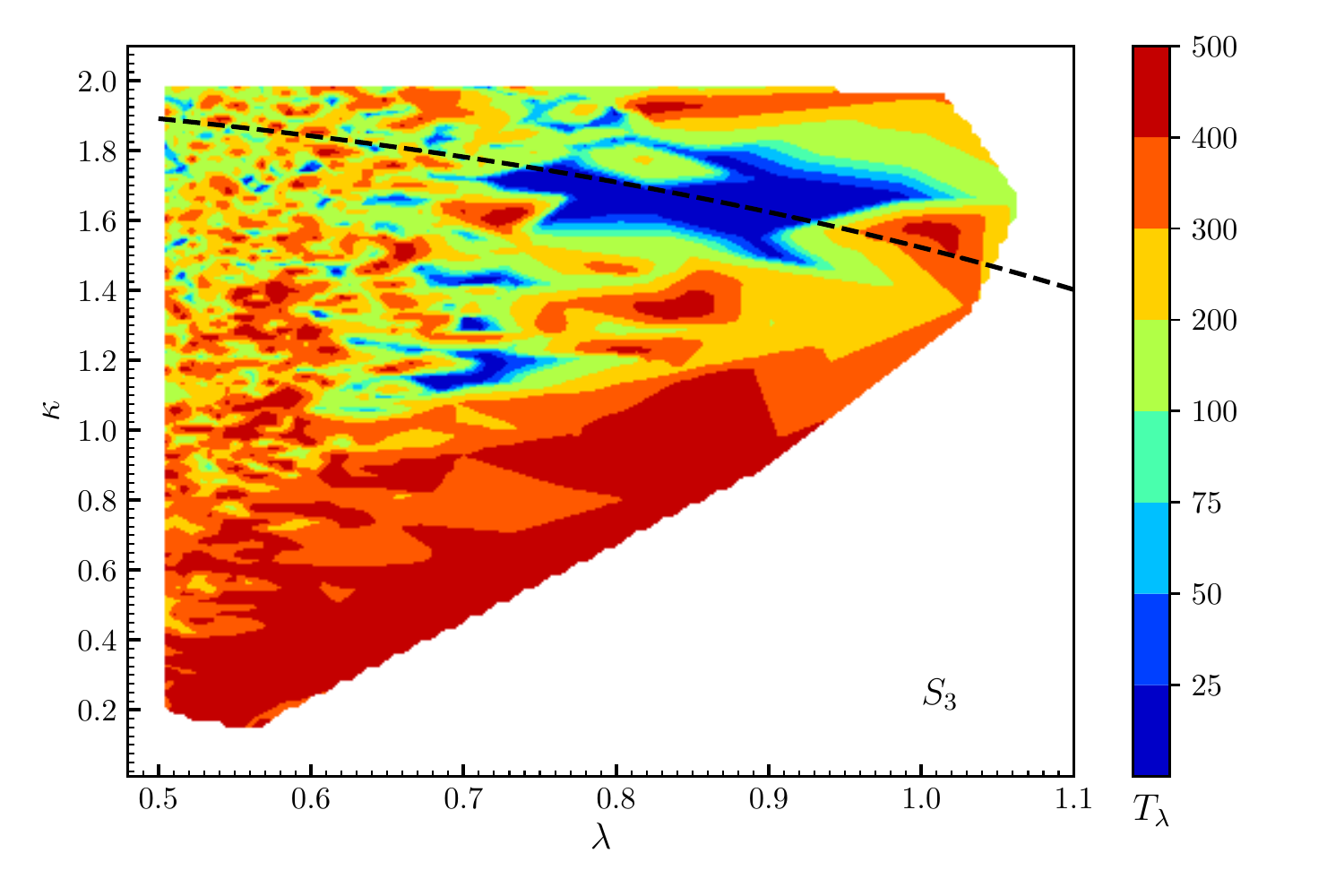}
  \caption{ \footnotesize{The same as in Fig.~\ref{S3-2D-Lambda-Kappa-vR}, but the colours
  indicate different low-energy values of the trilinear soft terms $T_\lambda$.  }}
\label{S3-2D-Lambda-Kappa-Tlambda}
 \includegraphics[width=0.8\linewidth, height= 0.35\textheight]{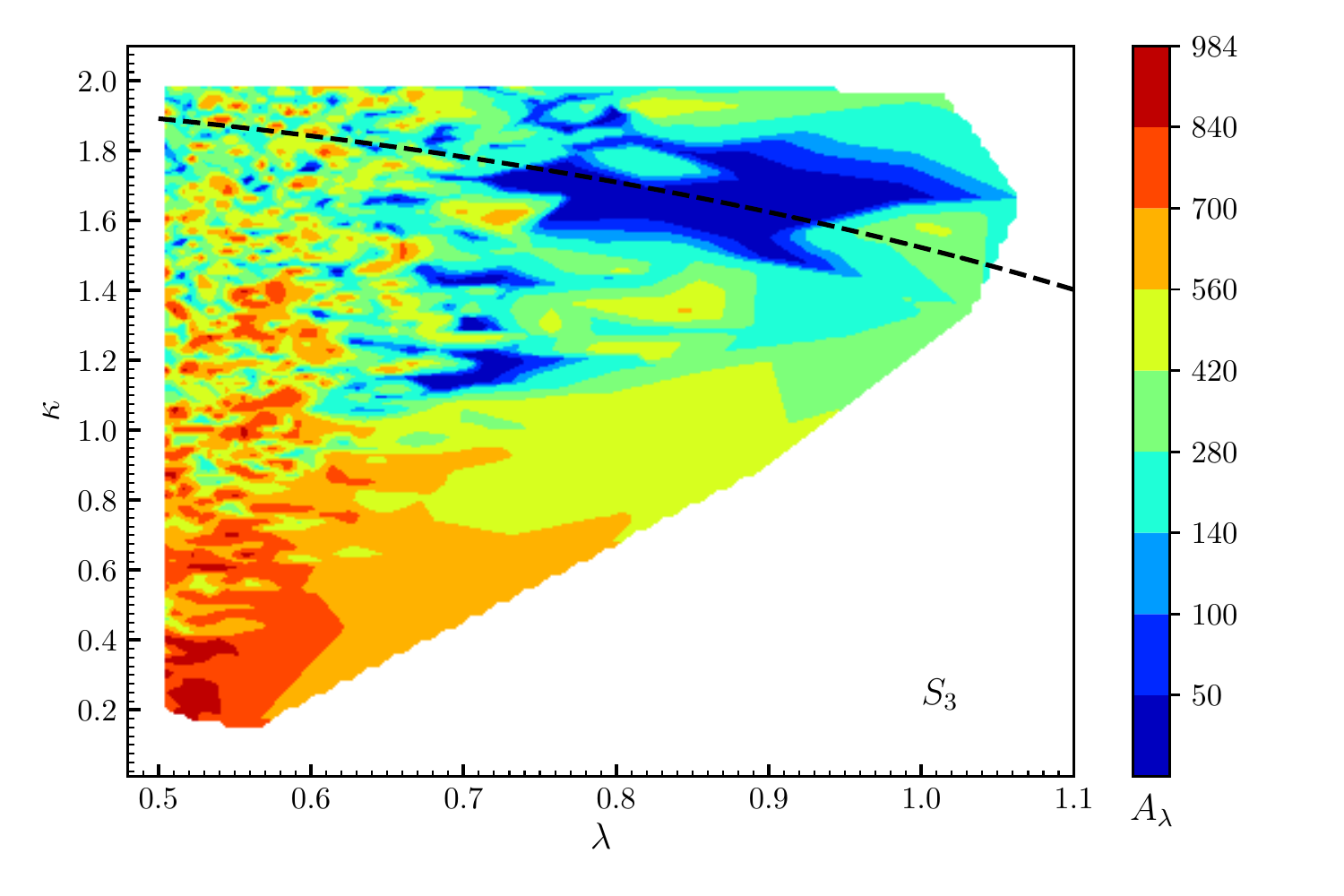}
  \caption{\footnotesize{The same as in Fig.~\ref{S3-2D-Lambda-Kappa-vR}, but the colours indicate different low-energy values of the trilinear soft terms $A_\lambda$, assuming the supergravity relation $A_\lambda = T_\lambda/\lambda$.  }  }
\label{S3-2D-Lambda-Kappa-Alambda}
\end{figure}

%\begin{figure}[t!]
% \centering
% \includegraphics[width=0.8\linewidth, height= 0.35\textheight]{Figures/large-lambda/3D/S3-2D-Lambda-Kappa-Alambda.pdf}
%  \caption{\footnotesize{The same as in Fig.~\ref{S3-2D-Lambda-Kappa-vR}, but the colours indicate different low-energy values of the trilinear soft terms $A_\lambda$, assuming the supergravity relation $A_\lambda = T_\lambda/\lambda$.  }  }
%\label{S3-2D-Lambda-Kappa-Alambda}
%\end{figure}

%\clearpage

\begin{figure}[t!]
 \centering
 \includegraphics[width=0.8\linewidth, height= 0.35\textheight]{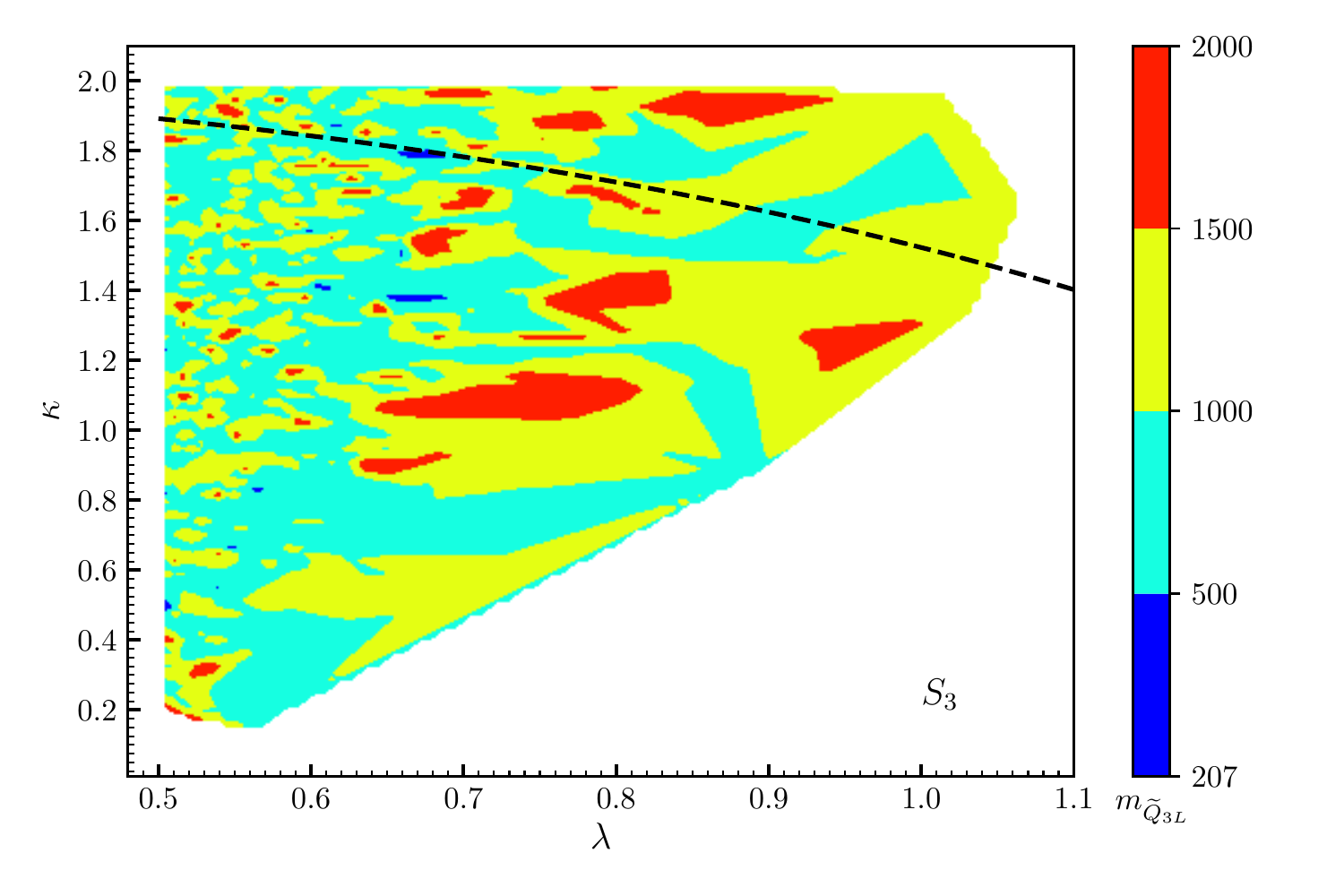}
  \caption{\footnotesize{The same as in Fig.~\ref{S3-2D-Lambda-Kappa-vR}, but the colours indicate different low-energy values of the soft masses 
  $m_{\widetilde Q_{3L}}$.   }}
\label{S3-2D-Lambda-Kappa-MQ3}
 \includegraphics[width=0.8\linewidth, height= 0.35\textheight]{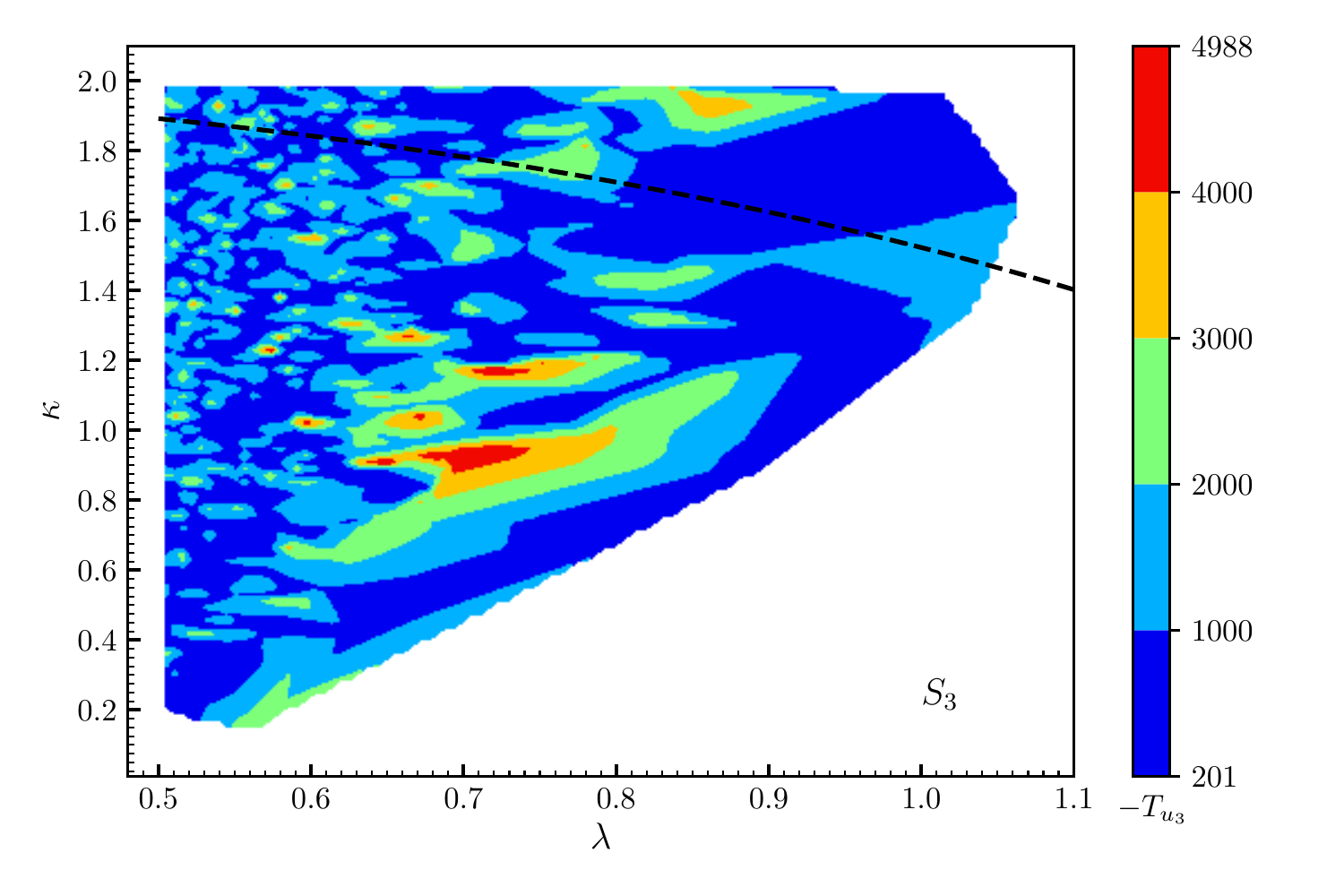}
  \caption{\footnotesize{The same as in Fig.~\ref{S3-2D-Lambda-Kappa-vR}, but the colours indicate different low-energy values of the 
 % third-generation squark 
  trilinear soft term $T_{u_3}$.   }}
\label{S3-2D-Lambda-Kappa-Tu3}
\end{figure}

%\begin{figure}[t!]
% \centering
% \includegraphics[width=0.8\linewidth, height= 0.35\textheight]{Figures/large-lambda/3D/S3-2D-Lambda-Kappa-Tu3.pdf}
%  \caption{\footnotesize{The same as in Fig.~\ref{S3-2D-Lambda-Kappa-vR}, but the colours indicate different low-energy values of the 
% % third-generation squark 
%  trilinear soft term $T_{u_3}$.   }}
%\label{S3-2D-Lambda-Kappa-Tu3}
%\end{figure}

\begin{figure}[t!]
 \centering
 \includegraphics[width=0.8\linewidth, height= 0.35\textheight]{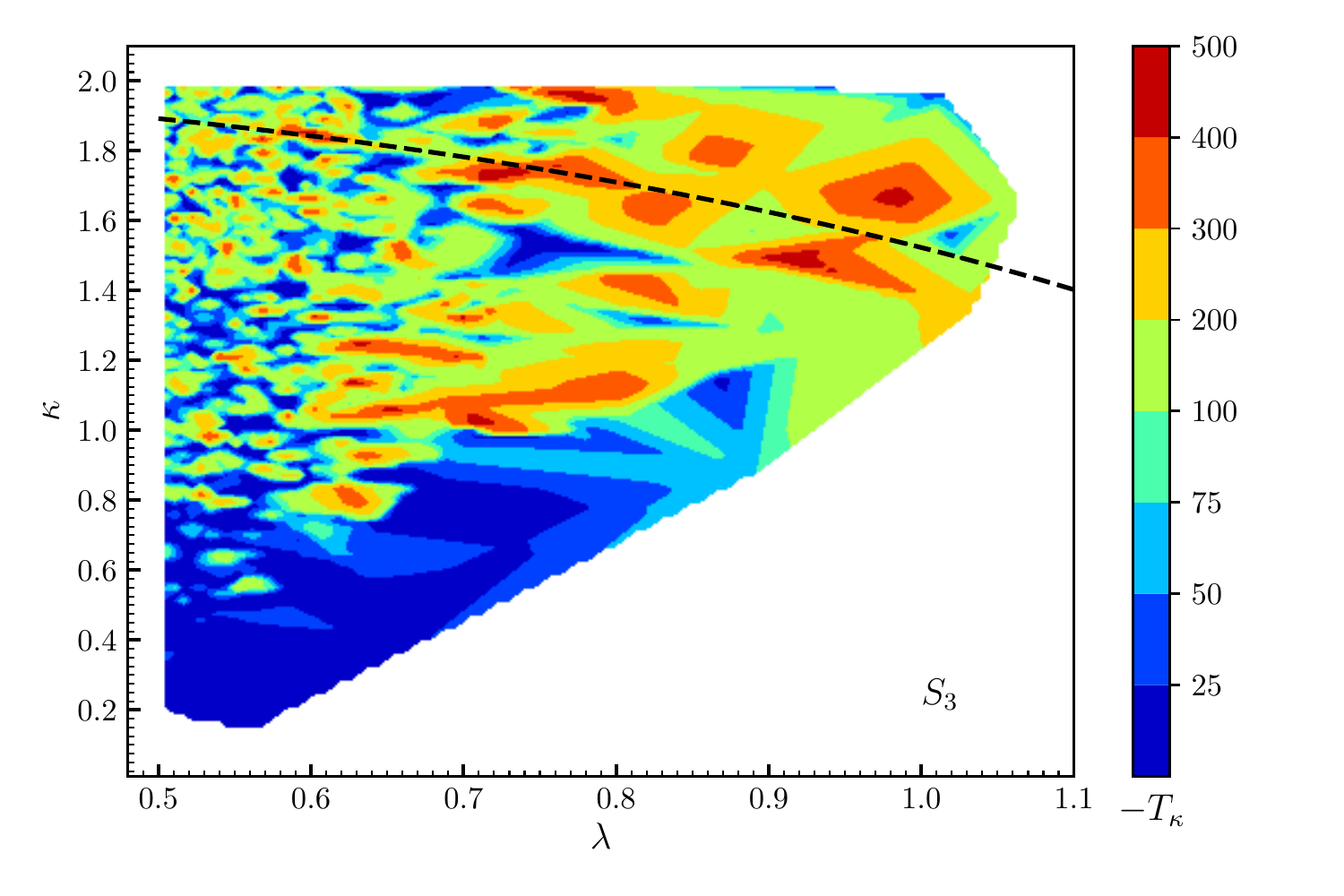}
  \caption{ \footnotesize{ The same as in Fig.~\ref{S3-2D-Lambda-Kappa-vR}, but the colours indicate different low-energy values of the trilinear soft terms $T_\kappa$.   }}
\label{S3-2D-Lambda-Kappa-Tkappa}
 \includegraphics[width=0.8\linewidth, height= 0.35\textheight]{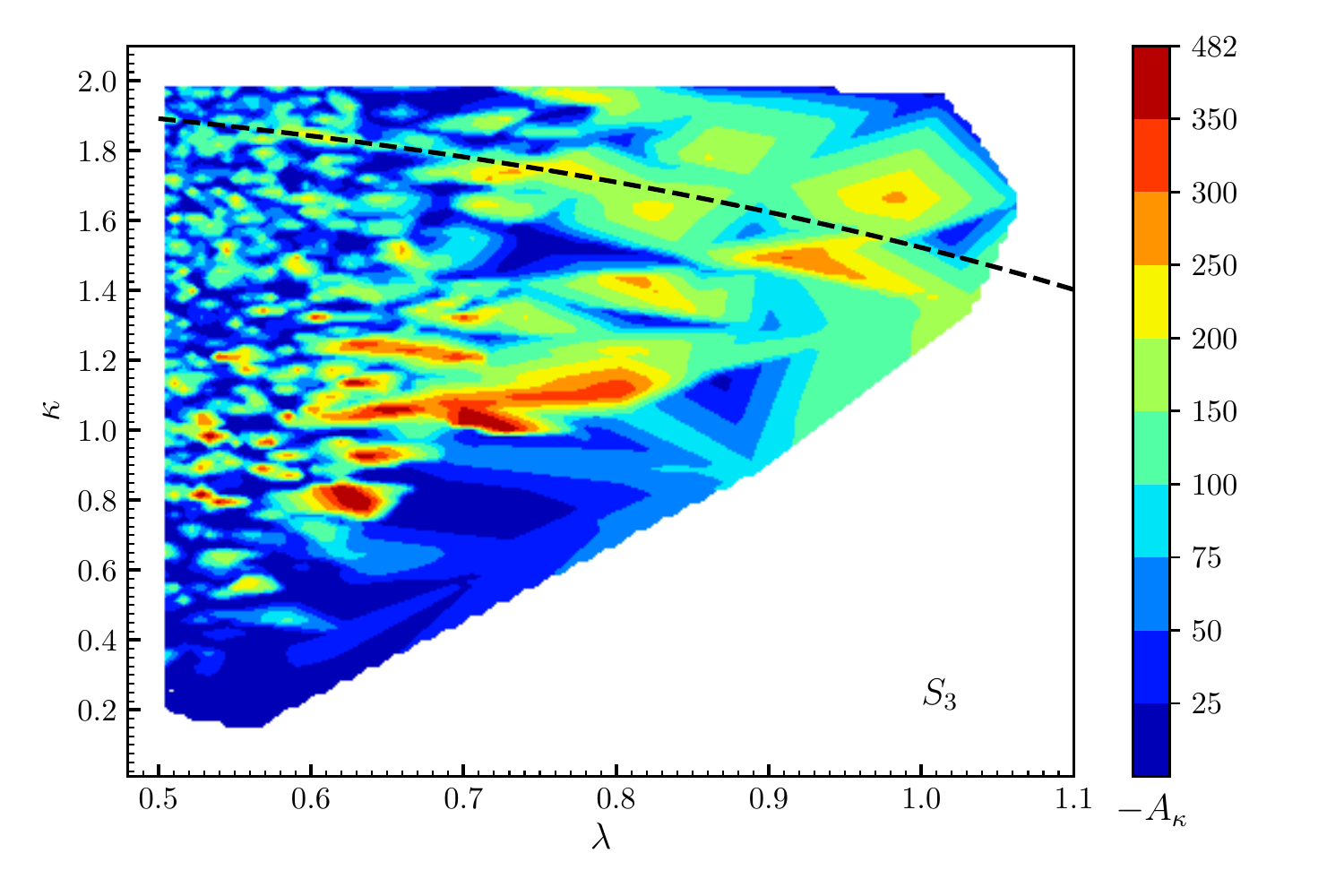}
  \caption{\footnotesize{ The same as in Fig.~\ref{S3-2D-Lambda-Kappa-vR}, but the colours indicate different low-energy values of the trilinear soft terms $A_\kappa$, assuming the supergravity relation $A_\kappa = T_\kappa/\kappa$.   }}
\label{S3-2D-Lambda-Kappa-Akappa}
\end{figure}

%\begin{figure}[t!]
% \centering
% \includegraphics[width=0.8\linewidth, height= 0.35\textheight]{Figures/large-lambda/3D/S3-2D-Lambda-Kappa-Akappa.pdf}
%  \caption{\footnotesize{ The same as in Fig.~\ref{S3-2D-Lambda-Kappa-vR}, but the colours indicate different low-energy values of the trilinear soft terms $A_\kappa$, assuming the supergravity relation $A_\kappa = T_\kappa/\kappa$.   }}
%\label{S3-2D-Lambda-Kappa-Akappa}
%\end{figure}

\clearpage
%\appendix

%\iffalse
%%%%%%%%%%%%%%%%%%%%%%%%%%%%%%%%%%%%%%%%%%%%%%%%%%%%%%%%%%%%%%%%%%%%%%%%%%%%%%%%
\section{ Benchmark points } 
\label{BPs}

Here we show for each scan several benchmark points discussed in the text.
{The SLHA-like output of {\tt SPheno} for all of them can be found in the website of the $\mu\nu$SSM Working Group, \url{http://dark.ft.uam.es/mununiverse/index.php/repository}}

%%%%%%%%%%%%%%%%%%%%%%%%%%%%%%%%%%%%%%%%%%%%%%%%%%%%
\subsection{ {\it Scan 1} ($0.01 \leq \lambda < 0.2$)}
%\subsection{Scan $S_1$}
\label{scan1bp}

%%%%%%%%%%%%%%%%%%%%%%%%%%%%%%%%%%%%%%%
\begin{table}[ht]
\begin{center}
\renewcommand{\arraystretch}{1.4}
\resizebox{15cm}{!}{
\begin{tabular}{|c|c|c|c|c|c|c|c|c|}
\clineB{1-9}{3}
\multicolumn{9}{|c|}{\bf S1-R1} \\ 
\clineB{1-9}{3}\hline
%%%%%%%%%
\multicolumn{9}{|c|}{$\lambda = 0.053$, \, $\kappa = 0.58$, \, $v_R/\sqrt 2 = 797.53\,$  ($\mu = 126.80$, \, $\mathcal{M} = 925.13$)}\\ 
\multicolumn{9}{|c|}{$\tan\beta = 13.39$, \, $T_\lambda = 4.53$,\, $-T_\kappa = 0.4$, \, $-T_{u_3} = 2597.94$, \, $M_{\widetilde Q_{3L}} = 1667.26$ }\\ \hline \hline
%%%%%%%%%
\multicolumn{9}{|c|}{$m_{h_1} (H^{\mathcal{R}}_u) = 123.86$, \, $m_{h_2} (\widetilde \nu^{\mathcal{R}}_{\tau L}) = 315.42$, \, $m_{h_3}(\widetilde \nu^{\mathcal{R}}_{\mu L}) = 632.79$, \, $m_{h_4}(\widetilde \nu^{\mathcal{R}}_R ) = 925.52 $}\\
\multicolumn{9}{|c|}{$m_{h_5}(\widetilde \nu^{\mathcal{R}}_R ) = 944.04$, \, $m_{h_6}(\widetilde \nu^{\mathcal{R}}_R ) = 961.74$, \, $m_{h_7}(H^{\mathcal{R}}_d) = 974.17$, \, $m_{h_8}(\widetilde \nu^{\mathcal{R}}_{e L}) = 1088.80$}\\ \hline \hline
%%%%%%%%%
\multicolumn{9}{|c|}{ $m_{A_2}(\widetilde \nu^{\mathcal{I}}_R ) =  36.75$, \, $m_{A_3} (\widetilde \nu^{\mathcal{I}}_R )= 40.72$, \, $m_{A_4}(\widetilde \nu^{\mathcal{I}}_R ) = 41.24$}\\
\multicolumn{9}{|c|}{$m_{A_5}(\widetilde \nu^{\mathcal{I}}_{\tau L}) = 315.42$, \, $m_{A_6}(\widetilde \nu^{\mathcal{I}}_{\mu L}) = 632.79$, \, $m_{A_7}(H^{\mathcal{I}}_u) = 970.86$, \, $m_{A_8}(\widetilde \nu^{\mathcal{I}}_{e L}) = 1088.80$}\\ \hline \hline
%%%%%%%%%
\multicolumn{9}{|c|}{ $m_{H^{-}_2}(\widetilde \tau_L) = 325.62$, \, $m_{H^{-}_3}(\widetilde \mu_L) = 638.64$, \, $m_{H^{-}_4} (H^-_d) = 973.49$}\\
\multicolumn{9}{|c|}{$m_{H^{-}_5}(\widetilde e_R) = 1004.43$, \, $m_{H^{-}_6}(\widetilde \tau_R) = 1005.10$, \, $m_{H^{-}_7}(\widetilde \mu_R) = 1005.10 $, \, $m_{H^{-}_8}(\widetilde e_L) = 1089.70 $}\\ \hline \hline
%%%%%%%%%
\multicolumn{9}{|c|}{ $|Z^H_{h_1 \widetilde{{\nu}}^{\mathcal{R}}_{eR}}|^2 = 0.00011\, \% $, \,$|Z^H_{h_1 \widetilde{{\nu}}^{\mathcal{R}}_{\mu R}}|^2 = 0.000097\, \% $, \,$|Z^H_{h_1 \widetilde{{\nu}}^{\mathcal{R}}_{\tau R}}|^2 = 0.000085\, \% $, \, 
$|Z^H_{h_1H_d^{\mathcal{R}}} |^2 = 0.58 \, \% $, \, $|Z^H_{h_1H_u^{\mathcal{R}}} |^2 = 99.41\, \% $}\\  %\hline \hline
%%%%%%%%%
\multicolumn{9}{|c|}{ $|Z^H_{h_4 \widetilde{{\nu}}^{\mathcal{R}}_{eR}}|^2 = 98.50\, \% $, \,$|Z^H_{h_4 \widetilde{{\nu}}^{\mathcal{R}}_{\mu R}}|^2 = 0.096\, \% $, \,$|Z^H_{h_4 \widetilde{{\nu}}^{\mathcal{R}}_{\tau R}}|^2 = 0.025\, \% $, \, $|Z^H_{h_4H_d^{\mathcal{R}}} |^2 = 0.013 \, \% $, \, $|Z^H_{h_4H_u^{\mathcal{R}}} |^2 = 0.0061\, \% $}\\  %\hline \hline
%%%%%%%%%
\multicolumn{9}{|c|}{ $|Z^H_{h_5 \widetilde{{\nu}}^{\mathcal{R}}_{eR}}|^2 = 0.29\, \% $, \,$|Z^H_{h_5 \widetilde{{\nu}}^{\mathcal{R}}_{\mu R}}|^2 = 95.61\, \% $, \,$|Z^H_{h_5 \widetilde{{\nu}}^{\mathcal{R}}_{\tau R}}|^2 = 0.25\, \% $, \, $|Z^H_{h_5H_d^{\mathcal{R}}} |^2 = 3.81 \, \% $, \, $|Z^H_{h_5H_u^{\mathcal{R}}} |^2 = 0.019\, \% $}\\  %\hline \hline
%%%%%%%%%
\multicolumn{9}{|c|}{ $|Z^H_{h_6 \widetilde{{\nu}}^{\mathcal{R}}_{eR}}|^2 = 0.43\, \% $, \,$|Z^H_{h_6 \widetilde{{\nu}}^{\mathcal{R}}_{\mu R}}|^2 = 2.04 \, \% $, \,$|Z^H_{h_6 \widetilde{{\nu}}^{\mathcal{R}}_{\tau R}}|^2 = 74.90\, \% $, \, $|Z^H_{h_6H_d^{\mathcal{R}}} |^2 = 22.50 \, \% $, \, $|Z^H_{h_6H_u^{\mathcal{R}}} |^2 = 0.12\, \% $}\\  \hline \hline
%%%%%%%%%
\multicolumn{9}{|c|}{ BR($h_1 \to bb$) = 0.51, \, BR($h_1 \to \tau\tau$) = 0.0854, \, BR($h_1 \to WW$) = 0.226 } \\
\multicolumn{9}{|c|}{ BR($h_1 \to ZZ$) = 0.0237, \,  BR($h_1 \to \gamma\gamma$) = 0.00282, \, 
%BR($h_1 \to \widetilde \chi^0 \widetilde \chi^0 $) = 0.0, 
BR($h_1 \to gg $) = 0.109}\\ \hline \hline
\multicolumn{9}{|c|}{ $\Gamma^{\text{tot}}_{h_1} = 3.19 \times 10^{-3}$ }\\ 
%%%%%%%%%
\clineB{1-9}{3}
\end{tabular} }
\renewcommand{\arraystretch}{1}
\end{center}
\caption{
Benchmark point {\bf S1-R1} from scan $S_1$, with the SM-like Higgs $h_1$ being the lightest scalar.
Input parameters at the low scale $M_{\text{EWSB}}$ are given in the first box,
where we also show for completeness $\mu=3\lambda v_R/\sqrt 2$ and ${\mathcal{M}=2\kappa v_R/\sqrt 2}$ since their values determine Higgsino and right-handed neutrino masses.
Scalar, pseudoscalar and charged Higgs masses are shown in the second, third and fourth boxes, respectively.
Scalar mass eigenstates are denoted by $h_{1,...,8}$, pseudoscalars by 
$A_{2,...,8}$ and charged Higgses
by $H^-_{2,...,8}$ associating the first states to the Goldstone bosons
eaten by the $Z$ and $W^{\pm}$.
Their dominant composition is written in brackets.
For the case of the SM-like Higgs 
%$h_1$ 
and singlet-like scalars 
%$h_{4,5,6}$
their main compositions
are broken down in the fifth box.
Relevant branching ratios for the SM-like Higgs $h_1$
%$h_{1}$
are shown in the sixth box.
Its decay width is shown in the seventh box.
VEVs, soft parameters, sparticle masses and decay widths 
are given in GeV. 
%The singlet component of the singlet-like scalar $h_6$ is $74.90\,\%$.
}
\label{S1-R1}
\end{table}

%%%%%%%%%%%%%%%%%%%%%%%%%%%%%%%%%%%%%%%
\begin{table}[ht]
\begin{center}
\renewcommand{\arraystretch}{1.4}
\resizebox{15cm}{!}{
\begin{tabular}{|c|c|c|c|c|c|c|c|c|}
\clineB{1-9}{3}
\multicolumn{9}{|c|}{\bf S1-R2} \\ 
\clineB{1-9}{3}\hline
%%%%%%%%%%%%%%%%%%%%%%%%%%%%
\multicolumn{9}{|c|}{$\lambda = 0.14$, \, $\kappa = 0.41$, \, $v_R/\sqrt 2 = 672.02\,$  ($\mu = 282.25$, \, $\mathcal{M} = 557.32$)}\\ 
\multicolumn{9}{|c|}{$\tan\beta = 10.07$, \, $T_\lambda = 234.70$,\, $-T_\kappa = 239.18$, \, $-T_{u_3} = 2504.00$,
\, $M_{\widetilde Q_{3L}} = 1205.73$ }\\ \hline \hline
%%%%%%%%%%%%%%%%%%%%%%%%%%%%
\multicolumn{9}{|c|}{$m_{h_1} (H^{\mathcal{R}}_u) = 122.46$, \, $m_{h_2} (\widetilde \nu^{\mathcal{R}}_{\tau L}) = 292.78$,
\, $m_{h_3}(\widetilde \nu^{\mathcal{R}}_R ) =  383.75$, \, $m_{h_4} (\widetilde \nu^{\mathcal{R}}_R )  =  399.78 $}\\
\multicolumn{9}{|c|}{$m_{h_5}(\widetilde \nu^{\mathcal{R}}_R ) = 415.60 $, \, $m_{h_6}(\widetilde \nu^{\mathcal{R}}_{\mu L} ) = 604.77$,
\, $m_{h_7} (\widetilde \nu^{\mathcal{R}}_{e L}) = 1013.23$, \, $m_{h_8} (H^{\mathcal{R}}_d) = 2377.17$}\\ \hline \hline
%%%%%%%%%%%%%%%%%%%%%%%%%%%%
\multicolumn{9}{|c|}{ $m_{A_2} (\widetilde \nu^{\mathcal{I}}_{\tau L}) = 292.78$, \, $m_{A_3}(\widetilde \nu^{\mathcal{I}}_{\mu L}) = 604.77$,
\, $m_{A_4} (\widetilde \nu^{\mathcal{I}}_R  )= 686.14$}\\
\multicolumn{9}{|c|}{$m_{A_5} (\widetilde \nu^{\mathcal{I}}_R ) = 686.25 $, \, $m_{A_6} (\widetilde \nu^{\mathcal{I}}_R ) = 687.67$,
\, $m_{A_7} (\widetilde \nu^{\mathcal{I}}_{e L}) = 1013.23$, \, $m_{A_8} (H^{\mathcal{I}}_u) = 2376.87$}\\ \hline \hline
%%%%%%%%%%%%%%%%%%%%%%%%%%%%
\multicolumn{9}{|c|}{ $m_{H^{-}_2}(\widetilde \tau_L) = 303.90$, \, $m_{H^{-}_3}(\widetilde \mu_L) = 610.30$,
\, $m_{H^{-}_4}(\widetilde e_R) = 1002.79$}\\
\multicolumn{9}{|c|}{$m_{H^{-}_5}(\widetilde \mu_R) = 1002.84$, \, $m_{H^{-}_6}(\widetilde \tau_R) = 1002.84$,
\, $m_{H^{-}_7} (\widetilde e_L) = 1016.50 $, \, $m_{H^{-}_8}  (H^-_d) = 2374.86 $}\\ \hline \hline
%%%%%%%%%%%%%%%%%%%%%%%%%%%%
\multicolumn{9}{|c|}{ $|Z^H_{h_1 \widetilde{{\nu}}^{\mathcal{R}}_{eR}}|^2 = 0.046\, \% $,
\,$|Z^H_{h_1 \widetilde{{\nu}}^{\mathcal{R}}_{\mu R}}|^2 = 0.037\, \% $,
\,$|Z^H_{h_1 \widetilde{{\nu}}^{\mathcal{R}}_{\tau R}}|^2 = 0.03\, \% $, \, 
$|Z^H_{h_1H_d^{\mathcal{R}}} |^2 = 0.97\, \% $, \, $|Z^H_{h_1H_u^{\mathcal{R}}} |^2 = 99.91\, \% $}\\  %\hline \hline
%%%%%%%%%
\multicolumn{9}{|c|}{ $|Z^H_{h_3 \widetilde{{\nu}}^{\mathcal{R}}_{eR}}|^2 = 98.26\, \% $,
\,$|Z^H_{h_3 \widetilde{{\nu}}^{\mathcal{R}}_{\mu R}}|^2 = 1.39\, \% $,
\,$|Z^H_{h_3 \widetilde{{\nu}}^{\mathcal{R}}_{\tau R}}|^2 = 0.30\, \% $,
\, $|Z^H_{h_3 H_d^{\mathcal{R}}} |^2 = 0.0089 \, \% $, \, $|Z^H_{h_3 H_u^{\mathcal{R}}} |^2 = 0.029\, \% $}\\  %\hline \hline
%%%%%%%%%
\multicolumn{9}{|c|}{ $|Z^H_{h_4 \widetilde{{\nu}}^{\mathcal{R}}_{eR}}|^2 = 1.19\, \% $,
\,$|Z^H_{h_4 \widetilde{{\nu}}^{\mathcal{R}}_{\mu R}}|^2 = 96.88 \, \% $,
\,$|Z^H_{h_4 \widetilde{{\nu}}^{\mathcal{R}}_{\tau R}}|^2 = 1.88\, \% $, 
\, $|Z^H_{h_4 H_d^{\mathcal{R}}} |^2 = 0.011 \, \% $, 
\, $|Z^H_{h_4 H_u^{\mathcal{R}}} |^2 = 0.032\, \% $}\\  %\hline \hline
%%%%%%%%%
\multicolumn{9}{|c|}{ $|Z^H_{h_5 \widetilde{{\nu}}^{\mathcal{R}}_{eR}}|^2 = 0.49\, \% $,
\,$|Z^H_{h_5 \widetilde{{\nu}}^{\mathcal{R}}_{\mu R}}|^2 = 1.67\, \% $,
\,$|Z^H_{h_5 \widetilde{{\nu}}^{\mathcal{R}}_{\tau R}}|^2 = 97.77\, \% $,
\, $|Z^H_{h_5 H_d^{\mathcal{R}}} |^2 = 0.018 \, \% $,
\, $|Z^H_{h_5 H_u^{\mathcal{R}}} |^2 = 0.04\, \% $}\\  \hline \hline
%%%%%%%%%%%%%%%%%%%%%%%%%%%%
\multicolumn{9}{|c|}{ BR($h_1 \to bb$) = 0.54, \, BR($h_1 \to \tau\tau$) = 0.0869, \, BR($h_1 \to WW$) = 0.209 } \\
\multicolumn{9}{|c|}{ BR($h_1 \to ZZ$) = 0.0209, \,  BR($h_1 \to \gamma\gamma$) = 0.00294, \, 
%BR($h_1 \to \widetilde \chi^0 \widetilde \chi^0 $) = 0.0, 
BR($h_1 \to gg $) = 0.112}\\ \hline \hline
\multicolumn{9}{|c|}{ $\Gamma^{\text{tot}}_{h_1} = 2.93 \times 10^{-3}$ }\\ 
%%%%%%%%%
\clineB{1-9}{3}
\end{tabular} }
\renewcommand{\arraystretch}{1}
\end{center}
\caption{The same as in Table~\ref{S1-R1}, but for another  
benchmark point {\bf S1-R2}.
%from scan $S_1$, 
%with the SM-like Higgs $h_4$  not being the lightest scalar.   
}
\label{S1-R2}
\end{table}

%%%%%%%%%%%%%%%%%%%%%%%%%%%%%%%%%%%%%%
\begin{table}[ht]
\begin{center}
\renewcommand{\arraystretch}{1.4}
\resizebox{15cm}{!}{
\begin{tabular}{|c|c|c|c|c|c|c|c|c|}
\clineB{1-9}{3}
\multicolumn{9}{|c|}{\bf S1-B1} \\ 
\clineB{1-9}{3}\hline
%%%%%%%%%
\multicolumn{9}{|c|}{$\lambda = 0.1$, \, $\kappa = 0.072$, \,$v_R/\sqrt 2 = 367.55$  ($\mu = 110.26$, \, $\mathcal{M} = 52.92$)}\\ 
\multicolumn{9}{|c|}{ $\tan\beta = 26.90$, \, $T_\lambda = 322.89$,\, $-T_\kappa = 1.22$, \, $-T_{u_3} = 3138.11$, \, $M_{\widetilde Q_{3L}} = 1903.57$ }\\ \hline \hline
%%%%%%%%%
\multicolumn{9}{|c|}{$m_{h_1} (\widetilde \nu^{\mathcal{R}}_R ) = 47.98$, \, $m_{h_2} (\widetilde \nu^{\mathcal{R}}_R ) = 49.27$, \, $m_{h_3}(\widetilde \nu^{\mathcal{R}}_R ) = 51.25$, \, $m_{h_4} (H^{\mathcal{R}}_u) = 125.03 $}\\
\multicolumn{9}{|c|}{$m_{h_5}(\widetilde \nu^{\mathcal{R}}_{\tau L}) = 225.07$, \, $m_{h_6}(\widetilde \nu^{\mathcal{R}}_{\mu L}) = 476.52$, \, $m_{h_7} (\widetilde \nu^{\mathcal{R}}_{e L}) = 776.81$, \, $m_{h_8}(H^{\mathcal{R}}_d)  = 3104.42$}\\ \hline \hline
%%%%%%%%%
\multicolumn{9}{|c|}{ $m_{A_2}(\widetilde \nu^{\mathcal{I}}_R ) =  36.26$, \, $m_{A_3} (\widetilde \nu^{\mathcal{I}}_R )= 36.27$, \, $m_{A_4}(\widetilde \nu^{\mathcal{I}}_R ) = 37.66$}\\
\multicolumn{9}{|c|}{$m_{A_5}(\widetilde \nu^{\mathcal{I}}_{\tau L}) = 225.07$, \, $m_{A_6}(\widetilde \nu^{\mathcal{I}}_{\mu L}) = 476.52$, \, $m_{A_7} (\widetilde \nu^{\mathcal{I}}_{e L}) = 776.81$, \, $m_{A_8} (H^{\mathcal{I}}_u) = 3104.36$}\\ \hline \hline
%%%%%%%%%
\multicolumn{9}{|c|}{ $m_{H^{-}_2}(\widetilde \tau_L) = 236.94$, \, $m_{H^{-}_3}(\widetilde \mu_L) = 482.31$, \, $m_{H^{-}_4}(\widetilde e_L) = 779.21$}\\
\multicolumn{9}{|c|}{$m_{H^{-}_5}(\widetilde \tau_R) = 1002.44$, \, $m_{H^{-}_6}(\widetilde \mu_R) = 1004.23$, \, $m_{H^{-}_7}(\widetilde e_R) = 1004.23 $, \, $m_{H^{-}_8} (H^-_d)= 3099.41 $}\\ \hline \hline
%%%%%%%%%
\multicolumn{9}{|c|}{ $|Z^H_{h_1 \widetilde{{\nu}}^{\mathcal{R}}_{eR}}|^2 = 85.15 \, \% $, \,$|Z^H_{h_1 \widetilde{{\nu}}^{\mathcal{R}}_{\mu R}}|^2 = 12.68\, \% $, \,$|Z^H_{h_1 \widetilde{{\nu}}^{\mathcal{R}}_{\tau R}}|^2 = 2.16\, \% $, \, 
$|Z^H_{h_1H_d^{\mathcal{R}}} |^2 = 0.0005 \, \% $, \, $|Z^H_{h_1H_u^{\mathcal{R}}} |^2 = 0.0012\, \% $}\\  %\hline \hline
%%%%%%%%%
\multicolumn{9}{|c|}{ $|Z^H_{h_2 \widetilde{{\nu}}^{\mathcal{R}}_{eR}}|^2 = 5.46 \, \% $, \,$|Z^H_{h_2 \widetilde{{\nu}}^{\mathcal{R}}_{\mu R}}|^2 = 67.36\, \% $, \,$|Z^H_{h_2 \widetilde{{\nu}}^{\mathcal{R}}_{\tau R}}|^2 = 27.17\, \% $, \, $|Z^H_{h_2H_d^{\mathcal{R}}} |^2 = 0.00083 \, \% $, \, $|Z^H_{h_2H_u^{\mathcal{R}}} |^2 = 0.002\, \% $}\\  %\hline \hline
%%%%%%%%%
\multicolumn{9}{|c|}{ $|Z^H_{h_3 \widetilde{{\nu}}^{\mathcal{R}}_{eR}}|^2 = 9.37 \, \% $, \,$|Z^H_{h_3 \widetilde{{\nu}}^{\mathcal{R}}_{\mu R}}|^2 = 19.94\, \% $, \,$|Z^H_{h_3 \widetilde{{\nu}}^{\mathcal{R}}_{\tau R}}|^2 = 70.65\, \% $, \, $|Z^H_{h_3H_d^{\mathcal{R}}} |^2 = 0.0074 \, \% $, \, $|Z^H_{h_3H_u^{\mathcal{R}}} |^2 = 0.017\, \% $}\\  %\hline \hline
%%%%%%%%%
\multicolumn{9}{|c|}{ $|Z^H_{h_4 \widetilde{{\nu}}^{\mathcal{R}}_{eR}}|^2 = 0.0066\, \% $, \,$|Z^H_{h_4 \widetilde{{\nu}}^{\mathcal{R}}_{\mu R}}|^2 = 0.0065 \, \% $, \,$|Z^H_{h_4 \widetilde{{\nu}}^{\mathcal{R}}_{\tau R}}|^2 = 0.0065\, \% $, \, $|Z^H_{h_4H_d^{\mathcal{R}}} |^2 = 0.14 \, \% $, \, $|Z^H_{h_4H_u^{\mathcal{R}}} |^2 = 99.84\, \% $}\\  \hline \hline
%%%%%%%%%
\multicolumn{9}{|c|}{ BR($h_4 \to bb$) = 0.41, \, BR($h_4 \to \tau\tau$) = 0.0854, \, BR($h_4 \to WW$) = 0.215 } \\
\multicolumn{9}{|c|}{ BR($h_4 \to ZZ$) = 0.0234, \,  BR($h_4 \to \gamma\gamma$) = 0.00243, \, 
BR($h_4 \to \widetilde \chi^0 \widetilde \chi^0 $) = 0.15, \,
BR($h_4 \to gg $) = 0.093}\\ \hline \hline
\multicolumn{9}{|c|}{ $\Gamma^{\text{tot}}_{h_4} = 3.82 \times 10^{-3}$ }\\ 
%%%%%%%%%
\clineB{1-9}{3}
\end{tabular} }
\renewcommand{\arraystretch}{1}
\end{center}
\caption{The same as in Table~\ref{S1-R1}, but for a  
benchmark point {\bf S1-B1} 
%from scan $S_1$, 
with the SM-like Higgs $h_4$  not being the lightest scalar.
%The same as in Table~\ref{S1-R1}, but for a different
%benchmark point. This scenario corresponds to the case with $m_{h_{1,2,3}} < %m_{\text{Higgs}}/2$.
}
\label{S1-B1}
\end{table}

\clearpage
%%%%%%%%%%%%%%%%%%%%%%%%%%
\begin{table}[H]
\begin{center}
\renewcommand{\arraystretch}{1.3}
\resizebox{15cm}{!}{
\begin{tabular}{|c|c|c|c|c|c|c|c|c|}
\clineB{1-9}{3}
\multicolumn{9}{|c|}{\bf S1-2h1} \\ 
\clineB{1-9}{3}\hline
\multicolumn{9}{|c|}{$\lambda = 0.02$, \, $\kappa = 0.036$, 
%\, $v_R = 2544.57$ }\\ 
\, $v_R/\sqrt 2 = 1799.28$ $(\mu = 107.95$,\, ${\mathcal{M}} = 129.54)$ }\\ 
\multicolumn{9}{|c|}{$\tan\beta = 32.54$,  \, $T_\lambda = 60.70$,\, $-T_\kappa = 1.46$, \, $-T_{u_3} =2632.61$, \, $m_{\widetilde Q_{3L}} = 1604.33$}
\\ \hline \hline
%%%
\multicolumn{9}{|c|}{$m_{h_1} (\widetilde \nu^{\mathcal{R}}_R
%{e R}
) = 119.69$, \,  $m_{h_2}(\widetilde \nu^{\mathcal{R}}_R
%{\mu R}
) = 122.54$, \, $m_{h_3} (H^{\mathcal{R}}_u) = 124.81$, \, $m_{h_4}(\widetilde \nu^{\mathcal{R}}_R
%{\tau R}
) = 125.56 $}\\
\multicolumn{9}{|c|}{$m_{h_5}(\widetilde \nu^{\mathcal{R}}_{\tau L}) = 488.88$, \, $m_{h_6}(\widetilde \nu^{\mathcal{R}}_{\mu L}) = 1034.75$, \, $m_{h_7}(\widetilde \nu^{\mathcal{R}}_{e L}) = 1693.70 $, \, $m_{h_8} (H^{\mathcal{R}}_d) = 3290.47$}\\ \hline  \hline
%%%
\multicolumn{9}{|c|}{ $m_{A_2}(\widetilde \nu^{\mathcal{I}}_{ R}) =  87.95$, \, $m_{A_3} (\widetilde \nu^{\mathcal{I}}_{ R} )= 87.952$, \, $m_{A_4}(\widetilde \nu^{\mathcal{I}}_{ R}) = 88.75$}\\
\multicolumn{9}{|c|}{$m_{A_5}(\widetilde \nu^{\mathcal{I}}_{\tau L}) = 488.88$, \, $m_{A_6}(\widetilde \nu^{\mathcal{I}}_{\mu L})  = 1034.75$, \, $m_{A_7}(\widetilde \nu^{\mathcal{I}}_{e L})= 1693.70$, \, $m_{A_8} (H^{\mathcal{I}}_d) = 3290.46 $}\\ \hline  \hline
%%%
\multicolumn{9}{|c|}{ $m_{H^{-}_2}(\widetilde \tau_L) = 495.75$, \, $m_{H^{-}_3}(\widetilde e_R) = 1003.31$, \, $m_{H^{-}_4} (\widetilde \mu_R) = 1003.31$}\\
\multicolumn{9}{|c|}{$m_{H^{-}_5}(\widetilde \tau_R)= 1003.53$, \, $m_{H^{-}_6} (\widetilde \mu_L) = 1038.37$, \, $m_{H^{-}_7}(\widetilde e_L) = 1695.31 $, \, $m_{H^{-}_8}   (H^-_d)= 3291.45 $}\\ \hline  \hline
%%%
\multicolumn{9}{|c|}{$|Z^H_{h_1 \widetilde{{\nu}}^{\mathcal{R}}_{eR}}|^2 = 99.24\, \% $, \,$|Z^H_{h_1 \widetilde{{\nu}}^{\mathcal{R}}_{\mu R}}|^2 = 0.44\, \% $, \, $|Z^H_{h_1 \widetilde{{\nu}}^{\mathcal{R}}_{\tau R}}|^2 = 0.1 \, \% $, \, $|Z^H_{h_1H_d^{\mathcal{R}}} |^2 = 0.000027\, \% $, \, $|Z^H_{h_1H_u^{\mathcal{R}}} |^2 = 0.2\, \% $}\\ 
\multicolumn{9}{|c|}{$|Z^H_{h_2 \widetilde{{\nu}}^{\mathcal{R}}_{eR}}|^2 = 0.35\, \% $, \,$|Z^H_{h_2 \widetilde{{\nu}}^{\mathcal{R}}_{\mu R}}|^2 = 98.15\, \% $, \, $|Z^H_{h_2 \widetilde{{\nu}}^{\mathcal{R}}_{\tau R}}|^2 = 0.43 \, \% $, \, $|Z^H_{h_2H_d^{\mathcal{R}}} |^2 = 0.0005\, \% $, \, $|Z^H_{h_2H_u^{\mathcal{R}}} |^2 = 1.04\, \% $}\\ 
\multicolumn{9}{|c|}{ $|Z^H_{h_3 \widetilde{{\nu}}^{\mathcal{R}}_{eR}}|^2 = 0.061\, \% $, \,$|Z^H_{h_3 \widetilde{{\nu}}^{\mathcal{R}}_{\mu R}}|^2 = 0.25\, \% $, \, $|Z^H_{h_3 \widetilde{{\nu}}^{\mathcal{R}}_{\tau R}}|^2 = 28.56 \, \% $, \, $|Z^H_{h_3H_d^{\mathcal{R}}} |^2 = 0.07\, \% $, \, $|Z^H_{h_3H_u^{\mathcal{R}}} |^2 = 71.00\, \% $}\\ 
\multicolumn{9}{|c|}{$|Z^H_{h_4 \widetilde{{\nu}}^{\mathcal{R}}_{eR}}|^2 = 0.34\, \% $, \,$|Z^H_{h_4 \widetilde{{\nu}}^{\mathcal{R}}_{\mu R}}|^2 = 1.15\, \% $, \, $|Z^H_{h_4 \widetilde{{\nu}}^{\mathcal{R}}_{\tau R}}|^2 = 70.89 \, \% $, \, $|Z^H_{h_4H_d^{\mathcal{R}}} |^2 = 0.03\, \% $, \, $|Z^H_{h_4H_u^{\mathcal{R}}} |^2 = 27.58\, \% $}\\ \hline  \hline
%%%
\multicolumn{9}{|c|}{ BR($h_1 \to bb$) = 0.186, \, BR($h_1 \to \tau\tau$) = 0.031, \, BR($h_1 \to WW$) = 0.39 }\\
\multicolumn{9}{|c|}{ BR($h_1 \to ZZ$) = 0.0354, \, BR($h_1 \to \gamma\gamma$) = 0.00899,\, 
%BR($h_1 \to \widetilde \chi^0 \widetilde \chi^0 $) < $10^{-8}$, 
BR($h_1 \to gg$) = 0.285}\\ \hline  
%%%
\multicolumn{9}{|c|}{ BR($h_2 \to bb$) = 0.38, \, BR($h_2 \to \tau\tau$) = 0.063, \, BR($h_2 \to WW$) = 0.312 }\\
\multicolumn{9}{|c|}{ BR($h_2 \to ZZ$) = 0.0315, \, BR($h_2 \to \gamma\gamma$) = 0.0049, \, 
%BR($h_2 \to \widetilde \chi^0 \widetilde \chi^0 $) < $10^{-8}$, 
BR($h_2 \to gg$) = 0.171}\\ \hline   \hline%BR($h_2 \to gg$) = 0.171
%%%
\multicolumn{9}{|c|}{ BR($h_3 \to bb$) = 0.49, \, BR($h_3 \to \tau\tau$) = 0.0813, \, BR($h_3 \to WW$) = 0.258 }\\
\multicolumn{9}{|c|}{ BR($h_3 \to ZZ$) = 0.0279, \, BR($h_3 \to \gamma\gamma$) = 0.003, \,  
%BR($h_3 \to \widetilde \chi^0 \widetilde \chi^0 $) < $10^{-8}$, 
BR($h_3 \to gg$) = 0.114}\\ \hline  \hline
%BR($h_3 \to gg$) = 0.114
%%%
\multicolumn{9}{|c|}{ BR($h_4 \to bb$) = 0.51, \, BR($h_4 \to \tau\tau$) = 0.0846, \, BR($h_4 \to WW$) = 0.250 }\\
\multicolumn{9}{|c|}{ BR($h_4 \to ZZ$) = 0.0275, \, BR($h_4 \to \gamma\gamma$) = 0.00265, \, 
%BR($h_4 \to \widetilde \chi^0 \widetilde \chi^0 $) < $10^{-8}$, 
BR($h_4 \to gg$) = 0.1} 
\\ \hline \hline 
%BR($h_4 \to gg$) = 0.1}\\ \hline \hline
\multicolumn{9}{|c|}{ $\Gamma^{\text{tot}}_{h_1} = 2.32 \times 10^{-6}$, \, $\Gamma^{\text{tot}}_{h_2} = 2.09 \times 10^{-5}$, \, $\Gamma^{\text{tot}}_{h_3} = 2.18 \times 10^{-3}$, \, $\Gamma^{\text{tot}}_{h_4} =  10^{-3}$ }\\  
%\hline  \hline
%%2.32
 \clineB{1-9}{3}
\end{tabular} }
\renewcommand{\arraystretch}{1}
\end{center}
\caption{
The same as in Table~\ref{S1-R1}, but for a  
benchmark point {\bf S1-2h1} 
%from scan $S_1$, 
with several singlet-like scalars $h_{1,2,4}$ of masses close to the mass of the SM-like Higgs $h_3$. 
Their branching ratios 
are shown in the sixth-nineth boxes,
and their decay widths in the tenth box.
%VEVs, soft parameters, sparticle masses and decay widths 
%are given in GeV.
}
  \label{S1-3-scalars}
\end{table}

%In this section give two examples of benchmark points that we
%found but were missed out in \cite{King:2014xwa}. 
%
%We also give examples benchmark points that are still compatible with the current Higgs data and that show  interesting features like having many scalars around 125 GeV. 
%Note when the mass of the these scalars are near that of SM Higgs within the mass resolution of the experiment their signals rates are superposed and thus could all be contributing to the resonance observed at 125 GeV.

%%%%%%%%%%%%%%%%%%%%%%%%%%%%%%%%%%%%%%%%%%%%%%%%%%%%%%%%%%%%%%%%%%%%%%%%%%%%%%%%
\clearpage

\subsection{ {\it Scan 2} ($0.2 \leq \lambda < 0.5$)}
%\subsection{Scan $S_1$}
\label{scan2bp}

\begin{table}[H]
\begin{center}
\renewcommand{\arraystretch}{1.4}
\resizebox{15cm}{!}{
\begin{tabular}{|c|c|c|c|c|c|c|c|c|}
\clineB{1-9}{3}
\multicolumn{9}{|c|}{\bf S2-R1} \\ 
\clineB{1-9}{3}\hline
\multicolumn{9}{|c|}{$\lambda = 0.25$, \, $\kappa = 0.56$, 
%\, $\tan\beta = 2.31$, 
%\, $v_R = 288.12$, 
\, $v_R/\sqrt 2 = 203.73$\ $(\mu = 152.79$,\, ${\mathcal{M}} = 228.17)$ 
%, \, $\mu = 152.79$ 
}\\ 
\multicolumn{9}{|c|}{$\tan\beta = 2.31$, \, $T_\lambda = 82.2$,\, $-T_\kappa = 0.56$, \, $-T_{u_3} = 1515$, \, 
$m_{\widetilde Q_{3L}} = 1700$} \\ \hline \hline
%%%
\multicolumn{9}{|c|}{$m_{h_1} (H^{\mathcal{R}}_u) = 125.17$, \, $m_{h_2}(\widetilde \nu^{\mathcal{R}}_{\tau L}) = 159.82$,\, $m_{h_3} (\widetilde \nu^{\mathcal{R}}_{ R}) = 227.54$, \, $m_{h_4}(\widetilde \nu^{\mathcal{R}}_{ R}) = 232.34$}\\
%%%
\multicolumn{9}{|c|}{$m_{h_5}(\widetilde \nu^{\mathcal{R}}_{ R}) =239.13 $, \, $m_{h_6}(\widetilde \nu^{\mathcal{R}}_{\mu L}) = 340.90$, \, $m_{h_7}(H^{\mathcal{R}}_d) = 433.89 $, \, $m_{h_8}(\widetilde \nu^{\mathcal{R}}_{e L}) = 557.08$}\\ \hline \hline
%%%
\multicolumn{9}{|c|}{ $m_{A_2}(\widetilde \nu^{\mathcal{I}}_{ R}) =  91.61$, \, $m_{A_3} (\widetilde \nu^{\mathcal{I}}_{ R})= 92.45$, \, $m_{A_4}(\widetilde \nu^{\mathcal{I}}_{ R}) = 110.37$}\\
%%%
\multicolumn{9}{|c|}{$m_{A_5}(\widetilde \nu^{\mathcal{I}}_{\tau L}) = 159.82 $, \, $m_{A_6}(\widetilde \nu^{\mathcal{I}}_{\mu L}) = 340.90$, \, $m_{A_7}(H^{\mathcal{I}}_d) = 424.37$, \, $m_{A_8}(\widetilde \nu^{\mathcal{I}}_{e L}) =557.08 $}\\ \hline \hline
%%%
\multicolumn{9}{|c|}{ $m_{H^{-}_2}(\widetilde \tau_L) = 173.21$, \, $m_{H^{-}_3}(\widetilde \mu_L) = 346.13$, \, $m_{H^{-}_4} (H^-_d) = 426.77$}\\
%%%
\multicolumn{9}{|c|}{$m_{H^{-}_5}(\widetilde e_L) = 558.46$, \, $m_{H^{-}_6}(\widetilde \tau_R) = 1003.45$, \, $m_{H^{-}_7}(\widetilde \mu_R) =1003.51 $, \, $m_{H^{-}_8}(\widetilde e_R) =1003.51 $}\\ \hline \hline
%%%
\multicolumn{9}{|c|}{ $|Z^H_{h_1 \widetilde{{\nu}}^{\mathcal{R}}_{eR}}|^2 = 2.85\, \% $, \,$|Z^H_{h_1 \widetilde{{\nu}}^{\mathcal{R}}_{\mu R}}|^2 = 2.76\, \% $, \,$|Z^H_{h_1 \widetilde{{\nu}}^{\mathcal{R}}_{\tau R}}|^2 = 2.68\, \% $, \, 
$|Z^H_{h_1H_d^{\mathcal{R}}} |^2 = 19.36 \, \% $, \, $|Z^H_{h_1H_u^{\mathcal{R}}} |^2 = 72.32\, \% $}\\  %\hline \hline
%%%
\multicolumn{9}{|c|}{ $|Z^H_{h_3 \widetilde{{\nu}}^{\mathcal{R}}_{eR}}|^2 = 84.34\, \% $, \,$|Z^H_{h_3 \widetilde{{\nu}}^{\mathcal{R}}_{\mu R}}|^2 = 12.37\, \% $, \,$|Z^H_{h_3 \widetilde{{\nu}}^{\mathcal{R}}_{\tau R}}|^2 = 2.44\, \% $, \, $|Z^H_{h_3H_d^{\mathcal{R}}} |^2 = 0.016 \, \% $, \, $|Z^H_{h_3H_u^{\mathcal{R}}} |^2 = 0.83\, \% $}\\  %\hline \hline
%%%
\multicolumn{9}{|c|}{ $|Z^H_{h_4 \widetilde{{\nu}}^{\mathcal{R}}_{eR}}|^2 = 4.50\, \% $, \,$|Z^H_{h_4 \widetilde{{\nu}}^{\mathcal{R}}_{\mu R}}|^2 = 66.66\, \% $, \,$|Z^H_{h_4 \widetilde{{\nu}}^{\mathcal{R}}_{\tau R}}|^2 = 27.59\, \% $, \, $|Z^H_{h_4H_d^{\mathcal{R}}} |^2 = 0.03 \, \% $, \, $|Z^H_{h_4H_u^{\mathcal{R}}} |^2 = 1.20\, \% $}\\  %\hline \hline
%%%
\multicolumn{9}{|c|}{ $|Z^H_{h_5 \widetilde{{\nu}}^{\mathcal{R}}_{eR}}|^2 = 6.88\, \% $, \,$|Z^H_{h_5 \widetilde{{\nu}}^{\mathcal{R}}_{\mu R}}|^2 =16.70\, \% $, \,$|Z^H_{h_5 \widetilde{{\nu}}^{\mathcal{R}}_{\tau R}}|^2 = 65.72\, \% $, \, $|Z^H_{h_5H_d^{\mathcal{R}}} |^2 = 0.4 \, \% $, \, $|Z^H_{h_5H_u^{\mathcal{R}}} |^2 = 10.29\, \% $}\\  \hline \hline
%%%
\multicolumn{9}{|c|}{ BR($h_1 \to bb$) = 0.5548, \, BR($h_1 \to \tau\tau$) = 0.0926, \, BR($h_1 \to WW$) = 0.219 } \\
\multicolumn{9}{|c|}{ BR($h_1 \to ZZ$) = 0.0239, \,  BR($h_1 \to \gamma\gamma$) = 0.00235, \, 
%BR($h_1 \to \widetilde \chi^0 \widetilde \chi^0 $) = 0.0, 
BR($h_1 \to gg $) = 0.088}\\ \hline \hline
\multicolumn{9}{|c|}{ $\Gamma^{\text{tot}}_{h_1} = 3.48 \times 10^{-3}$ }\\ 
%BR($h_1 \to gg$) = 0.0882
%%%%%
 \clineB{1-9}{3}
\end{tabular} }
\renewcommand{\arraystretch}{1}
\end{center}
\caption{
Benchmark point {\bf S2-R1} from scan $S_2$, with the SM-like Higgs $h_1$ being the lightest scalar.
%, with three scalars $h_{2,3,4}$ of masses close to the mass of the SM-like Higgs.
Input parameters at the low scale $M_{\text{EWSB}}$ are given in the first box,
where we also show for completeness $\mu=3\lambda v_R/\sqrt 2$ and ${\mathcal{M}=2\kappa v_R/\sqrt 2}$ since their values determine Higgsino and right-handed neutrino masses.
Scalar, pseudoscalar and charged Higgs masses are shown in the second, third and fourth boxes, respectively.
Scalar mass eigenstates are denoted by $h_{1,...,8}$, pseudoscalars by 
$A_{2,...,8}$ and charged Higgses
by $H^-_{2,...,8}$ associating the first states to the Goldstone bosons
eaten by the $Z$ and $W^{\pm}$.
%to $A_1$ and 
%$H^-_1$, and with their dominant composition written in brackets.
Their dominant composition is written in brackets.
For the case of the SM-like Higgs and singlet-like scalars their main compositions
are broken down in the fifth box.
Relevant branching ratios for the SM-like Higgs $h_{1}$
are shown in the sixth box.
%, sixth and seventh boxes.
%Their decay 
Its decay width is shown in the seventh box.
%In the eighth box,
%nineth box, 
%the 
%its components 
%of $h_1$ 
%and singlet-like scalars 
%are given.
VEVs, soft parameters, sparticle masses and decay widths 
are given in GeV.
%Benchmark points BP1 and BP2 are examples of the solutions we found in our analysis but that were
%missed out in \cite{King:2014xwa}. The $\mu_{eff}$, soft parameters and masses are in GeV.  The first and second boxes contain the benchmark point label and the input parameters respectively. The third, fourth, and fifth boxes show respectively the masses of the scalars, pseudoscalars and charged Higgs. In the last box gives the information on the branching fractions for the SM-like Higgs boson.
}
  \label{NonSK-benchmarks}
\end{table}

\begin{table}[H]
\begin{center}
\renewcommand{\arraystretch}{1.4}
\resizebox{15cm}{!}{
\begin{tabular}{|c|c|c|c|c|c|c|c|c|}
\clineB{1-9}{3}
%\multicolumn{9}{|c|}{ } \\ 
%\clineB{1-9}{3}\hline
%%%%%
%\clineB{1-9}{3}
\multicolumn{9}{|c|}{\bf S2-R2} \\ 
\clineB{1-9}{3}\hline
\multicolumn{9}{|c|}{$\lambda = 0.231$, \, $\kappa = 0.4$, \, 
%$\tan\beta = 1.08$, \,
%\, $v_R = 1096.77$, 
$v_R/\sqrt 2 = 775.53$\ $(\mu = 537.44$,\, ${\mathcal{M}} = 620.42)$
%, \,
%$\tan\beta = 1.08$, \,
}
\\ 
\multicolumn{9}{|c|}{$\tan\beta = 1.08$, \, $T_\lambda = 51.0$,\, $-T_\kappa = 31.93$, \,
$-T_{u_3} = 1961.5$, \,  
$m_{\widetilde Q_{3L}} = 1646.89$}
%\\
%\multicolumn{9}{|c|}{$\mu = 537.44$,\, ${\mathcal{M}} = 620.42$ }
\\ \hline \hline
\multicolumn{9}{|c|}{$m_{h_1} (H^{\mathcal{R}}_u) = 125.08$, \, $m_{h_2} (\widetilde \nu^{\mathcal{R}}_{\tau L}) = 280.48$, \, $m_{h_3}(\widetilde \nu^{\mathcal{R}}_{ R}) = 598.58$, \, $m_{h_4}(\widetilde \nu^{\mathcal{R}}_{\mu L}) = 609.21 $}\\
\multicolumn{9}{|c|}{$m_{h_5}(\widetilde \nu^{\mathcal{R}}_{ R}) = 613.25$, \, $m_{h_6}(\widetilde \nu^{\mathcal{R}}_{ R}) = 693.20 $, \, $m_{h_7}(H^{\mathcal{R}}_d) = 748.88 $, \, $m_{h_8}(\widetilde \nu^{\mathcal{R}}_{e L}) = 982.96 $}\\ \hline \hline
\multicolumn{9}{|c|}{ $m_{A_2}(\widetilde \nu^{\mathcal{I}}_{ R}) =  279.06$, \, $m_{A_3} (\widetilde \nu^{\mathcal{I}}_{ R})= 279.37$, \, $m_{A_4}(\widetilde \nu^{\mathcal{I}}_{\tau L}) = 280.48 $}\\
\multicolumn{9}{|c|}{ $m_{A_5}(\widetilde \nu^{\mathcal{I}}_{ R}) = 283.94$, \, $m_{A_6}(\widetilde \nu^{\mathcal{I}}_{\mu L}) = 609.21$, \, $m_{A_7}(H^{\mathcal{I}}_d) = 744.60$, \, $m_{A_8}(\widetilde \nu^{\mathcal{I}}_{e L}) = 982.96 $}\\ \hline \hline
\multicolumn{9}{|c|}{ $m_{H^{-}_2}(\widetilde \tau_L) = 282.95$, \, $m_{H^{-}_3}(\widetilde \mu_L) = 614.07$, \, $m_{H^{-}_4} (H^-_d) = 746.13 $}\\
\multicolumn{9}{|c|}{$m_{H^{-}_5}(\widetilde e_L) = 981.86 $, \, $m_{H^{-}_6}(\widetilde \tau_R) = 1003.57$, \, $m_{H^{-}_7}(\widetilde e_R) = 1003.60 $, \, $m_{H^{-}_8}(\widetilde \mu_R) = 1003.60$}\\ \hline \hline
%%%
\multicolumn{9}{|c|}{ $|Z^H_{h_1 \widetilde{{\nu}}^{\mathcal{R}}_{eR}}|^2 = 0.40 \, \% $, \,  $|Z^H_{h_1 \widetilde{{\nu}}^{\mathcal{ R}}_{\mu R}}|^2 = 0.33 \, \% $, \,  $|Z^H_{h_1 \widetilde{{\nu}}^{\mathcal{R}}_{\tau R}}|^2 = 0.27 \, \% $ , \,  %} \\  
%\multicolumn{9}{|c|}{ 
$|Z^H_{h_1H_d^{\mathcal{R}}} |^2 = 46.66 \, \% $, \, $|Z^H_{h_1H_u^{\mathcal{R}}} |^2 = 52.33\, \% $}\\  %\hline \hline
%%%
\multicolumn{9}{|c|}{ $|Z^H_{h_3 \widetilde{{\nu}}^{\mathcal{R}}_{eR}}|^2 = 66.18\, \% $, \,$|Z^H_{h_3 \widetilde{{\nu}}^{\mathcal{R}}_{\mu R}}|^2 = 30.21\, \% $, \,$|Z^H_{h_3 \widetilde{{\nu}}^{\mathcal{R}}_{\tau R}}|^2 = 3.6\, \% $, \, $|Z^H_{h_3H_d^{\mathcal{R}}} |^2 = 0.0048 \, \% $, \, $|Z^H_{h_3H_u^{\mathcal{R}}} |^2 = 0.0055\, \% $}\\  %\hline \hline
%%%
\multicolumn{9}{|c|}{ $|Z^H_{h_5 \widetilde{{\nu}}^{\mathcal{R}}_{eR}}|^2 = 5.53\, \% $, \,$|Z^H_{h_5 \widetilde{{\nu}}^{\mathcal{R}}_{\mu R}}|^2 = 37.14\, \% $, \,$|Z^H_{h_5 \widetilde{{\nu}}^{\mathcal{R}}_{\tau R}}|^2 = 57.3\, \% $, \, $|Z^H_{h_5H_d^{\mathcal{R}}} |^2 = 0.0053 \, \% $, \, $|Z^H_{h_5H_u^{\mathcal{R}}} |^2 = 0.0059\, \% $}\\  %\hline \hline
%%%
\multicolumn{9}{|c|}{ $|Z^H_{h_6 \widetilde{{\nu}}^{\mathcal{R}}_{eR}}|^2 = 27.87\, \% $, \,$|Z^H_{h_6 \widetilde{{\nu}}^{\mathcal{R}}_{\mu R}}|^2 = 32.29\, \% $, \,$|Z^H_{h_6 \widetilde{{\nu}}^{\mathcal{R}}_{\tau R}}|^2 = 38.80\, \% $, \, $|Z^H_{h_6H_d^{\mathcal{R}}} |^2 = 0.70 \, \% $, \, $|Z^H_{h_6H_u^{\mathcal{R}}} |^2 = 0.31\, \% $}\\  \hline \hline
%%%
%BR($h_1 \to gg$) = 0.109
\multicolumn{9}{|c|}{ BR($h_1 \to bb$) = 0.497, \, BR($h_1 \to \tau\tau$) = 0.0827, \, BR($h_1 \to WW$) = 0.256 }\\
\multicolumn{9}{|c|}{ BR($h_1 \to ZZ$) = 0.0279, \, BR($h_1 \to \gamma\gamma$) = 0.00293, \, 
%BR($h_1 \to \widetilde \chi^0 \widetilde \chi^0 $) = 0.0, \, 
BR($h_1 \to gg$) = 0.109}\\  \hline \hline
\multicolumn{9}{|c|}{ $\Gamma^{\text{tot}}_{h_1} = 3.20 \times 10^{-3}$ }\\ %\hline  \hline
 \clineB{1-9}{3}
\end{tabular} }
\renewcommand{\arraystretch}{1}
\end{center}
\caption{ The same as in Table~\ref{NonSK-benchmarks}, but for another 
benchmark point {\bf S2-R2}.
%from scan $S_2$.
%Benchmark points BP1 and BP2 are examples of the solutions we found in our analysis but that were
%missed out in \cite{King:2014xwa}. The $\mu_{eff}$, soft parameters and masses are in GeV.  The first and second boxes contain the benchmark point label and the input parameters respectively. The third, fourth, and fifth boxes show respectively the masses of the scalars, pseudoscalars and charged Higgs. In the last box gives the information on the branching fractions for the SM-like Higgs boson.
}
  \label{NonSK-benchmarks2}
\end{table}

%%%%%%%%%%%%%%%%%%%%%%%%%%%%%%%%%%
\begin{table}[ht]
\begin{center}
\renewcommand{\arraystretch}{1.4}
\resizebox{15cm}{!}{
\begin{tabular}{|c|c|c|c|c|c|c|c|c|}
\clineB{1-9}{3}
\multicolumn{9}{|c|}{\bf S2-B1} \\ 
\clineB{1-9}{3}\hline
%%%%%%%%%
\multicolumn{9}{|c|}{$\lambda = 0.33$, \, $\kappa = 0.09$, \, $v_R/\sqrt 2 = 286.3$  ($\mu = 283.44$, \, $\mathcal{M} = 51.53$) }\\ 
\multicolumn{9}{|c|}{$\tan\beta = 4.13$, \, $T_\lambda = 397.13$,\, $-T_\kappa = 3.06$, \, $-T_{u_3} = 1897.38$ , \, $M_{\widetilde Q_{3L}} = 1093.36$}\\ \hline \hline
%%%%%%%%%
\multicolumn{9}{|c|}{$m_{h_1} (\widetilde \nu^{\mathcal{R}}_R) = 87.99$, \, $m_{h_2} (\widetilde \nu^{\mathcal{R}}_R) = 90.62$, \, $m_{h_3}(\widetilde \nu^{\mathcal{R}}_R) = 91.29$, \, $m_{h_4} (H^{\mathcal{R}}_u) = 125.19 $}\\
\multicolumn{9}{|c|}{$m_{h_5}(\widetilde \nu^{\mathcal{R}}_{\tau L}) = 196.06$, \, $m_{h_6}(\widetilde \nu^{\mathcal{R}}_{\mu L}) = 417.45$, \, $m_{h_7}(\widetilde \nu^{\mathcal{R}}_{e L}) = 677.30$, \, $m_{h_8}(H^{\mathcal{R}}_d) = 1243.45$}\\ \hline \hline
%%%%%%%%%
\multicolumn{9}{|c|}{ $m_{A_2}(\widetilde \nu^{\mathcal{I}}_R) =  71.61$, \, $m_{A_3} (\widetilde \nu^{\mathcal{I}}_R)= 101.05$, \, $m_{A_4}(\widetilde \nu^{\mathcal{I}}_R) = 101.18$}\\
\multicolumn{9}{|c|}{$m_{A_5}(\widetilde \nu^{\mathcal{I}}_{\tau L}) = 196.06$, \, $m_{A_6}(\widetilde \nu^{\mathcal{I}}_{\mu L}) = 417.45$, \, $m_{A_7} (\widetilde \nu^{\mathcal{I}}_{e L})= 677.30$, \, $m_{A_8} (H^{\mathcal{I}}_d)= 1243.71$}\\ \hline \hline
%%%%%%%%%
\multicolumn{9}{|c|}{ $m_{H^{-}_2}(\widetilde \tau_L) = 208.79$, \, $m_{H^{-}_3}(\widetilde \mu_L) = 424.25$, \, $m_{H^{-}_4} (\widetilde e_L) = 680.79 $}\\
\multicolumn{9}{|c|}{$m_{H^{-}_5}(\widetilde \tau_R) = 1003.31$, \, $m_{H^{-}_6}(\widetilde \mu_R) = 1003.36$, \, $m_{H^{-}_7}(\widetilde e_R) = 1003.36 $, \, $m_{H^{-}_8}  (H^-_d)= 1234.29 $}\\ \hline \hline
%%%%%%%%%
\multicolumn{9}{|c|}{ $|Z^H_{h_1 \widetilde{{\nu}}^{\mathcal{R}}_{eR}}|^2 = 45.80\, \% $, \,$|Z^H_{h_1 \widetilde{{\nu}}^{\mathcal{R}}_{\mu R}}|^2 = 30.89\, \% $, \,$|Z^H_{h_1 \widetilde{{\nu}}^{\mathcal{R}}_{\tau R}}|^2 = 22.40\, \% $, \, $|Z^H_{h_1H_d^{\mathcal{R}}} |^2 = 0.68 \, \% $, \, $|Z^H_{h_1H_u^{\mathcal{R}}} |^2 = 0.21\, \% $}\\  %\hline \hline
%%%%%%%%%
\multicolumn{9}{|c|}{ $|Z^H_{h_2 \widetilde{{\nu}}^{\mathcal{R}}_{eR}}|^2 = 50.35\, \% $, \,$|Z^H_{h_2 \widetilde{{\nu}}^{\mathcal{R}}_{\mu R}}|^2 = 44.31\, \% $, \,$|Z^H_{h_2 \widetilde{{\nu}}^{\mathcal{R}}_{\tau R}}|^2 = 5.30\, \% $, \, $|Z^H_{h_2H_d^{\mathcal{R}}} |^2 = 0.0098 \, \% $, \, $|Z^H_{h_2H_u^{\mathcal{R}}} |^2 = 0.0071\, \% $}\\  %\hline \hline
%%%%%%%%%
\multicolumn{9}{|c|}{ $|Z^H_{h_3 \widetilde{{\nu}}^{\mathcal{R}}_{eR}}|^2 = 3.55\, \% $, \,$|Z^H_{h_3 \widetilde{{\nu}}^{\mathcal{R}}_{\mu R}}|^2 = 24.28\, \% $, \,$|Z^H_{h_3 \widetilde{{\nu}}^{\mathcal{R}}_{\tau R}}|^2 = 71.95\, \% $, \, $|Z^H_{h_3H_d^{\mathcal{R}}} |^2 = 0.008 \, \% $, \, $|Z^H_{h_3H_u^{\mathcal{R}}} |^2 = 0.0068\, \% $}\\  %\hline \hline
%%%%%%%%%
\multicolumn{9}{|c|}{ $|Z^H_{h_4 \widetilde{{\nu}}^{\mathcal{R}}_{eR}}|^2 = 0.11\, \% $, \,$|Z^H_{h_4 \widetilde{{\nu}}^{\mathcal{R}}_{\mu R}}|^2 = 0.14 \, \% $, \,$|Z^H_{h_4 \widetilde{{\nu}}^{\mathcal{R}}_{\tau R}}|^2 = 0.17\, \% $, \, $|Z^H_{h_4H_d^{\mathcal{R}}} |^2 = 5.34 \, \% $, \, $|Z^H_{h_4H_u^{\mathcal{R}}} |^2 = 94.22\, \% $}\\  \hline \hline
%%%%%%%%%
\multicolumn{9}{|c|}{ BR($h_4 \to bb$) = 0.47, \, BR($h_4 \to \tau\tau$) = 0.0748, \, BR($h_4 \to WW$) = 0.248 } \\
\multicolumn{9}{|c|}{ BR($h_4 \to ZZ$) = 0.0270, \,  BR($h_4 \to \gamma\gamma$) = 0.00280, \, 
BR($h_4 \to \widetilde \chi^0 \widetilde \chi^0 $) = $4.5 \times 10^{-2}$, \,
BR($h_4 \to gg $) = 0.106}\\ \hline \hline
\multicolumn{9}{|c|}{ $\Gamma^{\text{tot}}_{h_4} = 3.37 \times 10^{-3}$ }\\ 
%%%%%%%%%
\clineB{1-9}{3}
\end{tabular} }
\renewcommand{\arraystretch}{1}
\end{center}
\caption{The same as in Table~\ref{NonSK-benchmarks}, but for a benchmark point {\bf S2-B1} 
%from scan $S_2$, 
with the SM-like Higgs $h_4$ not being the lightest scalar.
}
\label{S2-B1}
\end{table}

%%%%%%%%%%%%%%%%%%%%%%%%%%%%%%%%%%%%%%%%%%%%%%%%%%%%%%%%%%%%%%%%%%%%%%%%%%%%%%%%
\begin{table}[!]
\begin{center}
\renewcommand{\arraystretch}{1.4}
\resizebox{15cm}{!}{
\begin{tabular}{|c|c|c|c|c|c|c|c|c|}
\clineB{1-9}{3}
\multicolumn{9}{|c|}{\bf S2-2h1} \\ 
\clineB{1-9}{3}\hline
\multicolumn{9}{|c|}{$\lambda = 0.39$, \, $\kappa = 0.08$, \,  
%\, $v_R = 493.2$ }\\
$v_R/\sqrt 2 = 348.75$ $(\mu = 408.03$,\, ${\mathcal{M}} = 55.8)$   }\\
\multicolumn{9}{|c|}{$\tan\beta = 2.25$, \, $T_\lambda = 418 $,\, $-T_\kappa = 0.52$, \, $-T_{u_3} = 294.42$, \,  $m_{\widetilde Q_{3L}} = 1433.3$}
%\\
%\multicolumn{9}{|c|}{$\mu = 408.03$,\, ${\mathcal{M}} = 55.8$ }
\\ \hline \hline
%%%
\multicolumn{9}{|c|}{$m_{h_1}(H^{\mathcal{R}}_u)  = 123.18$, \, $m_{h_2} (\widetilde \nu^{\mathcal{R}}_R
%{e R}
) = 125.98$, \, $m_{h_3}(\widetilde \nu^{\mathcal{R}}_R
%{\tau R}
) = 126.55$, \, $m_{h_4}(\widetilde \nu^{\mathcal{R}}_R
%{\mu R}
) = 139.93 $}\\
\multicolumn{9}{|c|}{$m_{h_5}(\widetilde \nu^{\mathcal{R}}_{\tau L}) = 212.71$, \, $m_{h_6} (\widetilde \nu^{\mathcal{R}}_{\mu L})= 460.55$, \, $m_{h_7}(\widetilde \nu^{\mathcal{R}}_{e L}) = 737.57 $, \, $m_{h_8} (H^{\mathcal{R}}_d) = 1088.04$}\\ \hline
%%%
\multicolumn{9}{|c|}{ $m_{A_2}(\widetilde \nu^{\mathcal{I}}_{ R}) =  77.24$, \, $m_{A_3} (\widetilde \nu^{\mathcal{I}}_{ R})= 126.19$, \, $m_{A_4}(\widetilde \nu^{\mathcal{I}}_{ R}) = 126.41$}\\
\multicolumn{9}{|c|}{$m_{A_5}(\widetilde \nu^{\mathcal{I}}_{\tau L}) = 212.71$, \, $m_{A_6}  (\widetilde \nu^{\mathcal{I}}_{\mu L}) = 460.55$, \, $m_{A_7}(\widetilde \nu^{\mathcal{I}}_{e L}) = 737.57$, \, $m_{A_8}(H^{\mathcal{I}}_d) = 1085.45 $}\\ \hline \hline
%%%
\multicolumn{9}{|c|}{ $m_{H^{-}_2}(\widetilde \tau_L) = 221.36$, \, $m_{H^{-}_3}(\widetilde \mu_L) = 463.71$, \, $m_{H^{-}_4} (\widetilde e_L) = 737.61$}\\
\multicolumn{9}{|c|}{$m_{H^{-}_5}(\widetilde \tau_R)= 1003.78$, \, $m_{H^{-}_6}  (\widetilde \mu_R) = 1003.81$, \, $m_{H^{-}_7}(\widetilde e_R) = 1003.81 $, \, $m_{H^{-}_8}(H^-_d) = 1083.20 $}\\ \hline \hline
%%%
\multicolumn{9}{|c|}{ $|Z^H_{h_1 \widetilde{{\nu}}^{\mathcal{R}}_{eR}}|^2 = 5.53\, \% $, \, $|Z^H_{h_1 \widetilde{{\nu}}^{\mathcal{R}}_{\mu R}}|^2 = 1.71\, \% $, \,  $|Z^H_{h_1 \widetilde{{\nu}}^{\mathcal{R}}_{\tau R}}|^2 = 0.3\, \% $, \,  $|Z^H_{h_1H_d^{\mathcal{R}}} |^2 = 14.03\, \% $, \, $|Z^H_{h_1H_u^{\mathcal{R}}} |^2 = 78.40\, \% $}\\ 
%%%
\multicolumn{9}{|c|}{ $|Z^H_{h_2 \widetilde{{\nu}}^{\mathcal{R}}_{eR}}|^2 = 59.00\, \% $, \, $|Z^H_{h_2 \widetilde{{\nu}}^{\mathcal{R}}_{\mu R}}|^2 = 36.82\, \% $, \,  $|Z^H_{h_2 \widetilde{{\nu}}^{\mathcal{R}}_{\tau R}}|^2 = 3.26\, \% $, \,  $|Z^H_{h_2H_d^{\mathcal{R}}} |^2 = 0.14\, \% $, \, $|Z^H_{h_2H_u^{\mathcal{R}}} |^2 = 0.76\, \% $}\\ 
%%%
\multicolumn{9}{|c|}{  $|Z^H_{h_3 \widetilde{{\nu}}^{\mathcal{R}}_{eR}}|^2 = 5.54\, \% $, \, $|Z^H_{h_3 \widetilde{{\nu}}^{\mathcal{R}}_{\mu R}}|^2 = 30.15\, \% $, \,  $|Z^H_{h_3 \widetilde{{\nu}}^{\mathcal{R}}_{\tau R}}|^2 = 63.54\, \% $, \,  $|Z^H_{h_3H_d^{\mathcal{R}}} |^2 = 0.12\, \% $, \, $|Z^H_{h_3H_u^{\mathcal{R}}} |^2 = 0.63\, \% $}\\
%%%
\multicolumn{9}{|c|}{  $|Z^H_{h_4 \widetilde{{\nu}}^{\mathcal{R}}_{eR}}|^2 = 29.73\, \% $, \, $|Z^H_{h_4 \widetilde{{\nu}}^{\mathcal{R}}_{\mu R}}|^2 = 31.13\, \% $, \,  $|Z^H_{h_4 \widetilde{{\nu}}^{\mathcal{R}}_{\tau R}}|^2 = 32.70\, \% $, \,  $|Z^H_{h_4H_d^{\mathcal{R}}} |^2 = 2.65\, \% $, \, $|Z^H_{h_4H_u^{\mathcal{R}}} |^2 = 3.77\, \% $}\\ \hline \hline
%%%
\multicolumn{9}{|c|}{ BR($h_1 \to bb$) = 0.487, \, BR($h_1 \to \tau\tau$) = 0.0798, \, BR($h_1 \to WW$) = 0.225 }\\
\multicolumn{9}{|c|}{ BR($h_1 \to ZZ$) = 0.0231, \, BR($h_1 \to \gamma\gamma$) = 0.0030, \, BR($h_1 \to \widetilde \chi^0 \widetilde \chi^0 $) = 0.0367, \, BR($h_1 \to gg$) = 0.120}\\ \hline \hline
%%%
\multicolumn{9}{|c|}{ BR($h_2 \to bb$) = 0.0133, \, BR($h_2 \to \tau\tau$) = 0.00218, \, BR($h_2 \to WW$) = 0.00791 }\\
\multicolumn{9}{|c|}{ BR($h_2 \to ZZ$) = 0.00087, \, BR($h_2 \to \gamma\gamma$) = 0.000084, \, BR($h_2 \to \widetilde \chi^0 \widetilde \chi^0 $) = 0.971, \, BR($h_2 \to gg$) = 0.00324}\\ \hline \hline
%%%
\multicolumn{9}{|c|}{ BR($h_3 \to bb$) = 0.0127, \, BR($h_3 \to \tau\tau$) = 0.00208, \, BR($h_3 \to WW$) = 0.00793 }\\
\multicolumn{9}{|c|}{ BR($h_3 \to ZZ$) = 0.00089, \, BR($h_3 \to \gamma\gamma$) = 0.00008, \, BR($h_3 \to \widetilde \chi^0 \widetilde \chi^0 $) = 0.972, \, BR($h_3 \to gg$) = 0.00309}\\ \hline  \hline
%%%
\multicolumn{9}{|c|}{ BR($h_4 \to bb$) = 0.054, \, BR($h_4 \to \tau\tau$) = 0.0089, \, BR($h_4 \to WW$) = 0.0447 }\\
\multicolumn{9}{|c|}{ BR($h_4 \to ZZ$) = 0.00588, \, BR($h_4 \to \gamma\gamma$) = 0.000115, \, BR($h_4 \to \widetilde \chi^0 \widetilde \chi^0 $) = 0.88, \, BR($h_4 \to gg$) = 0.00436}\\ \hline  \hline
\multicolumn{9}{|c|}{ $\Gamma^{\text{tot}}_{h_1} = 2.74 \times 10^{-3}$, \, $\Gamma^{\text{tot}}_{h_2} = 1.05 \times 10^{-3}$, \, $\Gamma^{\text{tot}}_{h_3} = 9.22\times 10^{-4}$ , \, $\Gamma^{\text{tot}}_{h_4} = 5.27\times 10^{-3}$}\\ 
%\hline  \hline
%%
 \clineB{1-9}{3}
\end{tabular} }
\renewcommand{\arraystretch}{1}
\end{center}
\caption{The same as in Table~\ref{NonSK-benchmarks}, but for a benchmark point 
{\bf S2-2h1}
%from scan $S_2$, 
with several singlet-like scalars $h_{2,3,4}$ of masses close to the mass of the SM-like Higgs $h_1$. 
Their branching ratios 
are shown in the sixth-nineth boxes,
and their decay widths in the tenth box.
%where the three scalars of masses close to the mass of the SM-like Higgs are $h_{1,2,3}$. 
}
  \label{S2-3-scalars}
\end{table}

%$|Z^H_{h \widetilde{{\nu}}^{\mathcal{R}}_{R}}|$ 

%%%%%%%%%%%%%%%%%%%%%%%%%%%%%%%%%%%%%%%%%%%%%%%%%%%%%%%%%%%%%%%%%%%%%%%%%%%%%%%%
\begin{table}[ht]
\begin{center}
\renewcommand{\arraystretch}{1.4}
\resizebox{15cm}{!}{
\begin{tabular}{|c|c|c|c|c|c|c|c|c|}
\clineB{1-9}{3}
\multicolumn{9}{|c|}{\bf S2-2h2} \\ 
\clineB{1-9}{3}\hline
%%%%%%%%%
\multicolumn{9}{|c|}{$\lambda = 0.37$, \, $\kappa = 0.103$, \, $v_R/\sqrt 2 = 299.91$  ($\mu = 332.90$, \, $\mathcal{M} = 61.78$)}\\ 
\multicolumn{9}{|c|}{$\tan\beta = 1.68$, \, $T_\lambda = 256.62$,\, $-T_\kappa = 1.75$, \, $-T_{u_3} = 458.51$, \, $M_{\widetilde Q_{3L}} = 1335.29$}\\ \hline \hline
%%%%%%%%%
\multicolumn{9}{|c|}{$m_{h_1} (\widetilde \nu^{\mathcal{R}}_R) = 97.36$, \, $m_{h_2} (\widetilde \nu^{\mathcal{R}}_R) = 98.28$, \, $m_{h_3}(\widetilde \nu^{\mathcal{R}}_R) = 124.54$, \, $m_{h_4} (H^{\mathcal{R}}_u) = 125.46 $}\\
\multicolumn{9}{|c|}{$m_{h_5}(\widetilde \nu^{\mathcal{R}}_{\tau L}) = 190.28$, \, $m_{h_6}(\widetilde \nu^{\mathcal{R}}_{\mu L}) = 415.06$, \, $m_{h_7} (\widetilde \nu^{\mathcal{R}}_{e L}) = 663.87$, \, $m_{h_8}(H^{\mathcal{R}}_d)  = 727.27$}\\ \hline \hline
%%%%%%%%%
\multicolumn{9}{|c|}{ $m_{A_2}(\widetilde \nu^{\mathcal{I}}_R) =  88.63$, \, $m_{A_3} (\widetilde \nu^{\mathcal{I}}_R)= 103.32$, \, $m_{A_4}(\widetilde \nu^{\mathcal{I}}_R) = 103.65$}\\
\multicolumn{9}{|c|}{$m_{A_5}(\widetilde \nu^{\mathcal{I}}_{\tau L}) = 190.28$, \, $m_{A_6}(\widetilde \nu^{\mathcal{I}}_{\mu L}) = 415.06$, \, $m_{A_7} (\widetilde \nu^{\mathcal{I}}_{e L}) = 663.87$, \, $m_{A_8} (H^{\mathcal{I}}_u) = 731.42$}\\ \hline \hline
%%%%%%%%%
\multicolumn{9}{|c|}{ $m_{H^{-}_2}(\widetilde \tau_L) = 198.51$, \, $m_{H^{-}_3}(\widetilde \mu_L) = 417.74$, \, $m_{H^{-}_4}(\widetilde e_L) = 664.57$}\\
\multicolumn{9}{|c|}{$m_{H^{-}_5} (H^-_d) = 725.66$, \, $m_{H^{-}_6} (\widetilde \tau_R) = 1003.00$, \, $m_{H^{-}_7}  (\widetilde \mu_R) = 1003.04 $, \, $m_{H^{-}_8} (\widetilde e_R) = 1003.04 $}\\ \hline \hline
%%%%%%%%%
\multicolumn{9}{|c|}{ $|Z^H_{h_1 \widetilde{{\nu}}^{\mathcal{R}}_{eR}}|^2 = 63.68 \, \% $, \,$|Z^H_{h_1 \widetilde{{\nu}}^{\mathcal{R}}_{\mu R}}|^2 = 31.91\, \% $, \,$|Z^H_{h_1 \widetilde{{\nu}}^{\mathcal{R}}_{\tau R}}|^2 = 4.38\, \% $, \, 
$|Z^H_{h_1H_d^{\mathcal{R}}} |^2 = 0.0025 \, \% $, \, $|Z^H_{h_1H_u^{\mathcal{R}}} |^2 = 0.001\, \% $}\\  %\hline \hline
%%%%%%%%%
\multicolumn{9}{|c|}{ $|Z^H_{h_2 \widetilde{{\nu}}^{\mathcal{R}}_{eR}}|^2 = 4.53 \, \% $, \,$|Z^H_{h_2 \widetilde{{\nu}}^{\mathcal{R}}_{\mu R}}|^2 = 34.80\, \% $, \,$|Z^H_{h_2 \widetilde{{\nu}}^{\mathcal{R}}_{\tau R}}|^2 = 60.64\, \% $, \, $|Z^H_{h_2H_d^{\mathcal{R}}} |^2 = 0.0025 \, \% $, \, $|Z^H_{h_2H_u^{\mathcal{R}}} |^2 = 0.01\, \% $}\\  %\hline \hline
%%%%%%%%%
\multicolumn{9}{|c|}{ $|Z^H_{h_3 \widetilde{{\nu}}^{\mathcal{R}}_{eR}}|^2 = 28.29 \, \% $, \,$|Z^H_{h_3 \widetilde{{\nu}}^{\mathcal{R}}_{\mu R}}|^2 = 30.03\, \% $, \,$|Z^H_{h_3 \widetilde{{\nu}}^{\mathcal{R}}_{\tau R}}|^2 = 31.55\, \% $, \, $|Z^H_{h_3H_d^{\mathcal{R}}} |^2 = 0.94 \, \% $, \, $|Z^H_{h_3H_u^{\mathcal{R}}} |^2 = 8.79\, \% $}\\  %\hline \hline
%%%%%%%%%
\multicolumn{9}{|c|}{ $|Z^H_{h_4 \widetilde{{\nu}}^{\mathcal{R}}_{eR}}|^2 = 3.30\, \% $, \,$|Z^H_{h_4 \widetilde{{\nu}}^{\mathcal{R}}_{\mu R}}|^2 = 3.07 \, \% $, \,$|Z^H_{h_4 \widetilde{{\nu}}^{\mathcal{R}}_{\tau R}}|^2 = 2.84\, \% $, \, $|Z^H_{h_4H_d^{\mathcal{R}}} |^2 = 25.91 \, \% $, \, $|Z^H_{h_4H_u^{\mathcal{R}}} |^2 = 64.85\, \% $}\\  \hline \hline
%%%%%%%%%
\multicolumn{9}{|c|}{ BR($h_3 \to bb$) = 0.286, \, BR($h_3 \to \tau\tau$) = 0.0828, \, BR($h_3 \to WW$) = 0.251 } \\
\multicolumn{9}{|c|}{ BR($h_3 \to ZZ$) = 0.0275, \,  BR($h_3 \to \gamma\gamma$) = 0.00263, \, 
BR($h_3 \to \widetilde \chi^0 \widetilde \chi^0 $) = $8.06  \times 10^{-4}$, \,
BR($h_3 \to gg $) = 0.1}\\ \hline \hline
%%%%%%%%%
\multicolumn{9}{|c|}{ BR($h_4 \to bb$) = 0.51, \, BR($h_4 \to \tau\tau$) = 0.0469 \, BR($h_4 \to WW$) = 0.359 } \\
\multicolumn{9}{|c|}{ BR($h_4 \to ZZ$) = 0.0384, \,  BR($h_4 \to \gamma\gamma$) = 0.00602, \, 
BR($h_4 \to \widetilde \chi^0 \widetilde \chi^0 $) = $2.9  \times 10^{-10}$, \,
BR($h_4 \to gg $) = 0.218}\\ \hline \hline
\multicolumn{9}{|c|}{ $\Gamma^{\text{tot}}_{h_3} = 2.01 \times 10^{-4}$, \, $\Gamma^{\text{tot}}_{h_4} = 3.13 \times 10^{-3}$ }\\ 
%%%%%%%%%
\clineB{1-9}{3}
\end{tabular} }
\renewcommand{\arraystretch}{1}
\end{center}
\caption{
The same as in Table~\ref{NonSK-benchmarks}, but for a benchmark point 
{\bf S2-2h2}
%from scan $S_2$, 
with several singlet-like scalars $h_{1,2,3}$ of masses close to the mass of the SM-like Higgs $h_4$. 
The branching ratios of $h_4$, and $h_3$ which contributes significantly to the superposition of signals,  
are shown in the sixth and seventh boxes,
and their decay widths in the eight box.
%$S_2$, Blue, Large singlet - doublet mixing, Superposition
}
\label{S2-ss1}
\end{table}

%%%%%%%%%%%%%%%%%%%%%%%%%%%%%%%%%%%%%%%%%%%%%%%%%%%%%%%%%%%%%%%%%%%%%%%%%%%%%%%%
\clearpage

\subsection{ {\it Scan 3} ($0.5 \leq \lambda < 1.2$)}
%\subsection{Scan $S_1$}
\label{scan3bp}

%%%%%%%%%%%%%%%%%%%%%%%%%%%%%%%%%%%%%%
\begin{table}[ht]
\begin{center}
\renewcommand{\arraystretch}{1.4}
\resizebox{15cm}{!}{
\begin{tabular}{|c|c|c|c|c|c|c|c|c|}
\clineB{1-9}{3}
\multicolumn{9}{|c|}{\bf S3-R1} \\ 
\clineB{1-9}{3}\hline
%%%%%%%%%
\multicolumn{9}{|c|}{$\lambda = 0.66$, \, $\kappa = 0.82$, \, $v_R/\sqrt 2 = 125.47$  ($\mu = 248.43$, \, $\mathcal{M} = 205.77$)   }\\ 
\multicolumn{9}{|c|}{$\tan\beta = 2.64$, \, $T_\lambda = 268.26$,\, $-T_\kappa = 1.00$, \, $-T_{u_3} = 234.54$ , \, $M_{\widetilde Q_{3L}} = 759.27$}\\ \hline \hline
%%%%%%%%%
\multicolumn{9}{|c|}{$m_{h_1}(\widetilde \nu^{\mathcal{R}}_{\tau L}) = 121.92$, \, $m_{h_2} (H^{\mathcal{R}}_u) = 125.44$, \, $m_{h_3}(\widetilde \nu^{\mathcal{R}}_R) = 165.23$, \, $m_{h_4} (\widetilde \nu^{\mathcal{R}}_R)  = 171.18$}\\
\multicolumn{9}{|c|}{$m_{h_5} (\widetilde \nu^{\mathcal{R}}_{\mu L}) =  271.50 $, \, $m_{h_6} (\widetilde \nu^{\mathcal{R}}_R) = 297.96$, \, $m_{h_7} (\widetilde \nu^{\mathcal{R}}_{e L}) = 440.96$, \, $m_{h_8}  (H^{\mathcal{R}}_d) = 1545.47$}\\ \hline \hline
%%%%%%%%%
\multicolumn{9}{|c|}{ $m_{A_2}(\widetilde \nu^{\mathcal{I}}_R) =  115.08$, \, $m_{A_3} (\widetilde \nu^{\mathcal{I}}_R)= 117.34$, \, $m_{A_4} (\widetilde \nu^{\mathcal{I}}_{\tau L}) = 121.91$}\\
\multicolumn{9}{|c|}{$m_{A_5} (\widetilde \nu^{\mathcal{I}}_R) = 198.37$, \, $m_{A_6}(\widetilde \nu^{\mathcal{I}}_{\mu L}) = 271.50$, \, $m_{A_7} (\widetilde \nu^{\mathcal{I}}_{e L}) = 440.96$, \, $m_{A_8}  (H^{\mathcal{I}}_d) = 622.33$}\\ \hline \hline
%%%%%%%%%
\multicolumn{9}{|c|}{ $m_{H^{-}_2}(\widetilde \tau_L) = 144.18$, \, $m_{H^{-}_3}(\widetilde \mu_L) = 278.88$, \, $m_{H^{-}_4} (\widetilde e_L) = 443.76 $}\\
\multicolumn{9}{|c|}{$m_{H^{-}_5} (H^-_d) = 586.40$, \, $m_{H^{-}_6}(\widetilde \mu_R) = 1002.61$, \, $m_{H^{-}_7}(\widetilde \tau_R) = 1002.64 $, \, $m_{H^{-}_8} (\widetilde e_R) = 1002.64 $}\\ \hline \hline
%%%%%%%%% ###----------
\multicolumn{9}{|c|}{ $|Z^H_{h_2 \widetilde{{\nu}}^{\mathcal{R}}_{eR}}|^2 = 2.44\, \% $, \,$|Z^H_{h_2 \widetilde{{\nu}}^{\mathcal{R}}_{\mu R}}|^2 = 3.01\, \% $, \,$|Z^H_{h_2 \widetilde{{\nu}}^{\mathcal{R}}_{\tau R}}|^2 = 3.22\, \% $, \, $|Z^H_{h_2H_d^{\mathcal{R}}} |^2 = 10.64 \, \% $, \, $|Z^H_{h_2H_u^{\mathcal{R}}} |^2 = 80.67\, \% $}\\  %\hline \hline
%%%%%%%%%  
\multicolumn{9}{|c|}{ $|Z^H_{h_3 \widetilde{{\nu}}^{\mathcal{R}}_{eR}}|^2 = 58.89\, \% $, \,$|Z^H_{h_3 \widetilde{{\nu}}^{\mathcal{R}}_{\mu R}}|^2 = 37.56\, \% $, \,$|Z^H_{h_3 \widetilde{{\nu}}^{\mathcal{R}}_{\tau R}}|^2 = 3.47\, \% $, \, $|Z^H_{h_3H_d^{\mathcal{R}}} |^2 =  0.05\, \% $, \, $|Z^H_{h_3H_u^{\mathcal{R}}} |^2 = 0.019\, \% $}\\  %\hline \hline
%%%%%%%%%
\multicolumn{9}{|c|}{ $|Z^H_{h_4 \widetilde{{\nu}}^{\mathcal{R}}_{eR}}|^2 = 5.48\, \% $, \,$|Z^H_{h_4 \widetilde{{\nu}}^{\mathcal{R}}_{\mu R}}|^2 = 29.15 \, \% $, \,$|Z^H_{h_4 \widetilde{{\nu}}^{\mathcal{R}}_{\tau R}}|^2 = 65.30\, \% $, \, $|Z^H_{h_4H_d^{\mathcal{R}}} |^2 = 0.042 \, \% $, \, $|Z^H_{h_4H_u^{\mathcal{R}}} |^2 = 0.0078\, \% $}\\ 
%%%%%%%%%
\multicolumn{9}{|c|}{ $|Z^H_{h_6 \widetilde{{\nu}}^{\mathcal{R}}_{eR}}|^2 = 32.73\, \% $, \,$|Z^H_{h_6 \widetilde{{\nu}}^{\mathcal{R}}_{\mu R}}|^2 = 30.04\, \% $, \,$|Z^H_{h_6 \widetilde{{\nu}}^{\mathcal{R}}_{\tau R}}|^2 = 27.96\, \% $, \, $|Z^H_{h_6H_d^{\mathcal{R}}} |^2 = 0.03 \, \% $, \, $|Z^H_{h_6H_u^{\mathcal{R}}} |^2 = 9.28\, \% $}\\  \hline \hline
%%%%%%%%%
\multicolumn{9}{|c|}{ BR($h_2 \to bb$) = 0.49, \, BR($h_2 \to \tau\tau$) = 0.0738, \, BR($h_2 \to WW$) = 0.263 } \\
\multicolumn{9}{|c|}{ BR($h_2 \to ZZ$) = 0.0288, \,  BR($h_2 \to \gamma\gamma$) = 0.00313, \, 
%BR($h_2 \to \widetilde \chi^0 \widetilde \chi^0 $) = 2.74 \times 10^{-2}, \,
BR($h_2 \to gg $) = 0.12}\\ \hline \hline
\multicolumn{9}{|c|}{ $\Gamma^{\text{tot}}_{h_2} = 3.02 \times 10^{-3}$ }\\ 
%%%%%%%%%
\clineB{1-9}{3}
\end{tabular} }
\renewcommand{\arraystretch}{1}
\end{center}
\caption{
Benchmark point {\bf S3-R1} from scan $S_3$, with the SM-like Higgs
$h_2$
being the lightest scalar.
Input parameters at the low scale $M_{\text{EWSB}}$ are given in the first box,
where we also show for completeness $\mu=3\lambda v_R/\sqrt 2$ and ${\mathcal{M}=2\kappa v_R/\sqrt 2}$ since their values determine Higgsino and right-handed neutrino masses.
Scalar, pseudoscalar and charged Higgs masses are shown in the second, third and fourth boxes, respectively.
Scalar mass eigenstates are denoted by $h_{1,...,8}$, pseudoscalars by 
$A_{2,...,8}$ and charged Higgses
by $H^-_{2,...,8}$ associating the first states to the Goldstone bosons
eaten by the $Z$ and $W^{\pm}$.
Their dominant composition is written in brackets.
For the case of the SM-like Higgs and singlet-like scalars their main compositions
are broken down in the fifth box.
Relevant branching ratios for the SM-like Higgs $h_{2}$
are shown in the sixth box.
Its decay width is shown in the seventh box.
VEVs, soft parameters, sparticle masses and decay widths 
are given in GeV. 
%\color{red}  $S_3$, Light Red 
}
\label{S3-R1}
\end{table}

%%%%%%%%%%%%%%%%%%%%%%%%%%%%%%%%%%%%%%%%%%
\begin{table}[ht]
\begin{center}
\renewcommand{\arraystretch}{1.4}
\resizebox{15cm}{!}{
\begin{tabular}{|c|c|c|c|c|c|c|c|c|}
\clineB{1-9}{3}
\multicolumn{9}{|c|}{\bf S3-B1} \\ 
\clineB{1-9}{3}\hline
%%%%%%%%%
\multicolumn{9}{|c|}{$\lambda = 0.52$, \, $\kappa = 0.21$, \, $v_R/\sqrt 2 = 140.25$  ($\mu = 218.79$, \, $\mathcal{M} = 58.90$)  }\\ 
\multicolumn{9}{|c|}{ $\tan\beta = 3.89$, \, $T_\lambda = 446.30$,\, $-T_\kappa = 0.15$, \, $M_{\widetilde Q_{3L}} = 806.74$, \, $-T_{u_3} = 358.58$ GeV }\\ \hline \hline
%%%%%%%%%
\multicolumn{9}{|c|}{$m_{h_1} (\widetilde \nu^{\mathcal{R}}_R) = 102.80$, \, $m_{h_2} (\widetilde \nu^{\mathcal{R}}_R) = 103.57$, \, $m_{h_3}(\widetilde \nu^{\mathcal{R}}_R) = 107.43$, \, $m_{h_4} (H^{\mathcal{R}}_u) = 125.30 $}\\
\multicolumn{9}{|c|}{$m_{h_5}(\widetilde \nu^{\mathcal{R}}_{\tau L}) = 138.25$, \, $m_{h_6}(\widetilde \nu^{\mathcal{R}}_{\mu L}) = 293.04$, \, $m_{h_7} (\widetilde \nu^{\mathcal{R}}_{e L})= 475.63$, \, $m_{h_8} (H^{\mathcal{R}}_d)= 901.46$}\\ \hline \hline
%%%%%%%%%
\multicolumn{9}{|c|}{ $m_{A_2}(\widetilde \nu^{\mathcal{I}}_R) =  89.92$, \, $m_{A_3} (\widetilde \nu^{\mathcal{I}}_R)= 104.62$, \, $m_{A_4}(\widetilde \nu^{\mathcal{I}}_R) = 105.02$}\\
\multicolumn{9}{|c|}{$m_{A_5}(\widetilde \nu^{\mathcal{I}}_{\tau L}) = 138.25$, \, $m_{A_6}(\widetilde \nu^{\mathcal{I}}_{\mu L}) = 293.04$, \, $m_{A_7}(\widetilde \nu^{\mathcal{I}}_{e L}) = 475.63$, \, $m_{A_8}(H^{\mathcal{I}}_d) = 903.92$}\\ \hline \hline
%%%%%%%%%
\multicolumn{9}{|c|}{ $m_{H^{-}_2}(\widetilde \tau_L) = 155.71$, \, $m_{H^{-}_3}(\widetilde \mu_L) = 301.67$, \, $m_{H^{-}_4} (\widetilde e_L) = 479.63 $}\\
\multicolumn{9}{|c|}{$m_{H^{-}_5}(\widetilde \tau_R) = 1002.8$, \, $m_{H^{-}_6}(\widetilde \mu_R) = 1002.84$, \, $m_{H^{-}_7}(\widetilde e_R) = 1002.84 $, \, $m_{H^{-}_8}  (H^-_d)= 1002.84 $}\\ \hline \hline
%%%%%%%%%
\multicolumn{9}{|c|}{ $|Z^H_{h_1 \widetilde{{\nu}}^{\mathcal{R}}_{eR}}|^2 = 45.80\, \% $, \,$|Z^H_{h_1 \widetilde{{\nu}}^{\mathcal{R}}_{\mu R}}|^2 = 30.89\, \% $, \,$|Z^H_{h_1 \widetilde{{\nu}}^{\mathcal{R}}_{\tau R}}|^2 = 22.40\, \% $, \, $|Z^H_{h_1H_d^{\mathcal{R}}} |^2 = 0.0032 \, \% $, \, $|Z^H_{h_1H_u^{\mathcal{R}}} |^2 = 0.21\, \% $}\\  %\hline \hline
%%%%%%%%%
\multicolumn{9}{|c|}{ $|Z^H_{h_2 \widetilde{{\nu}}^{\mathcal{R}}_{eR}}|^2 = 5.22\, \% $, \,$|Z^H_{h_2 \widetilde{{\nu}}^{\mathcal{R}}_{\mu R}}|^2 = 40.47\, \% $, \,$|Z^H_{h_2 \widetilde{{\nu}}^{\mathcal{R}}_{\tau R}}|^2 = 54.24\, \% $, \, $|Z^H_{h_2H_d^{\mathcal{R}}} |^2 = 0.0045 \, \% $, \, $|Z^H_{h_2H_u^{\mathcal{R}}} |^2 = 0.039\, \% $}\\  %\hline \hline
%%%%%%%%%
\multicolumn{9}{|c|}{ $|Z^H_{h_3 \widetilde{{\nu}}^{\mathcal{R}}_{eR}}|^2 = 24.39\, \% $, \,$|Z^H_{h_3 \widetilde{{\nu}}^{\mathcal{R}}_{\mu R}}|^2 = 30.71\, \% $, \,$|Z^H_{h_3 \widetilde{{\nu}}^{\mathcal{R}}_{\tau R}}|^2 = 40.66\, \% $, \, $|Z^H_{h_3H_d^{\mathcal{R}}} |^2 = 1.27 \, \% $, \, $|Z^H_{h_3H_u^{\mathcal{R}}} |^2 = 2.93\, \% $}\\  %\hline \hline
%%%%%%%%%
\multicolumn{9}{|c|}{ $|Z^H_{h_4 \widetilde{{\nu}}^{\mathcal{R}}_{eR}}|^2 = 0.75\, \% $, \,$|Z^H_{h_4 \widetilde{{\nu}}^{\mathcal{R}}_{\mu R}}|^2 = 0.64 \, \% $, \,$|Z^H_{h_4 \widetilde{{\nu}}^{\mathcal{R}}_{\tau R}}|^2 = 0.53\, \% $, \, $|Z^H_{h_4H_d^{\mathcal{R}}} |^2 = 6.84 \, \% $, \, $|Z^H_{h_4H_u^{\mathcal{R}}} |^2 = 91.20\, \% $}\\  \hline \hline
%%%%%%%%%
\multicolumn{9}{|c|}{ BR($h_4 \to bb$) = 0.51, \, BR($h_4 \to \tau\tau$) = 0.0787, \, BR($h_4 \to WW$) = 0.228 } \\
\multicolumn{9}{|c|}{ BR($h_4 \to ZZ$) = 0.0249, \,  BR($h_4 \to \gamma\gamma$) = 0.00239, \, 
BR($h_4 \to \widetilde \chi^0 \widetilde \chi^0 $) = $2.74 \times 10^{-2}$, \,
BR($h_4 \to gg $) = 0.1}\\ \hline \hline
\multicolumn{9}{|c|}{ $\Gamma^{\text{tot}}_{h_4} = 3.67 \times 10^{-3}$ }\\ 
%%%%%%%%%
\clineB{1-9}{3}
\end{tabular} }
\renewcommand{\arraystretch}{1}
\end{center}
\caption{
The same as in Table.~\ref{S3-R1}, but for a  
benchmark point {\bf S3-B1} 
%from scan $S_3$, 
with the SM-like Higgs $h_4$ not being the lightest scalar.
%$S_3$, Light Blue 
}
\label{S3-1sB1}
\end{table}

\begin{table}[ht]
\begin{center}
\renewcommand{\arraystretch}{1.4}
\resizebox{15cm}{!}{
\begin{tabular}{|c|c|c|c|c|c|c|c|c|}
\clineB{1-9}{3}
\multicolumn{9}{|c|}{\bf S3-2h1} \\ 
\clineB{1-9}{3}\hline
\multicolumn{9}{|c|}{$\lambda = 0.5$, \, $\kappa = 0.23$, 
%\, $v_R = 154.64$ }\\ 
\, $v_R/\sqrt 2 = 109.35$ $(\mu = 164.02$,\, ${\mathcal{M}} = 50.3)$ }\\ 
\multicolumn{9}{|c|}{$\tan\beta = 3.48$,  \, $T_\lambda = 309.70$, \, $-T_\kappa = 0.58$, \, $-T_{u_3} = 386.93$, \, $m_{\widetilde Q_{3L}} = 1340.97$}
\\ \hline \hline
%%%
\multicolumn{9}{|c|}{$m_{h_1} (\widetilde \nu^{\mathcal{R}}_R
%{\mu R}
) = 83.95 $, \, $m_{h_2}(\widetilde \nu^{\mathcal{R}}_{\tau L}) = 123.06$, \,  $m_{h_3}(\widetilde \nu^{\mathcal{R}}_R \,
%{e R}
) = 126.32$, \, $m_{h_4} (\widetilde \nu^{\mathcal{R}}_R
%{\tau R}
) = 126.76 $}\\
\multicolumn{9}{|c|}{ $m_{h_5}(H^{\mathcal{R}}_u) = 127.48 $, \, $m_{h_6}(\widetilde \nu^{\mathcal{R}}_{\mu L}) = 260.06$, \, $m_{h_7}(\widetilde \nu^{\mathcal{R}}_{e L}) = 423.60 $, \, $m_{h_8} (H^{\mathcal{R}}_d) = 619.46$}\\ \hline  \hline
%%%
\multicolumn{9}{|c|}{ $m_{A_2}(\widetilde \nu^{\mathcal{I}}_{ R}) =  101.28$, \, $m_{A_3}(\widetilde \nu^{\mathcal{I}}_{\tau L}) = 123.06$, \,  $m_{A_4} (\widetilde \nu^{\mathcal{I}}_{ R} )= 136.48$}\\
\multicolumn{9}{|c|}{$m_{A_5}(\widetilde \nu^{\mathcal{I}}_{ R}) = 136.87$, \, $m_{A_6}(\widetilde \nu^{\mathcal{I}}_{\mu L})  = 260.67$, \, $m_{A_7}(\widetilde \nu^{\mathcal{I}}_{e L})= 423.67$, \, $m_{A_8} (H^{\mathcal{I}}_d) = 622.49 $}\\ \hline  \hline
%%%
\multicolumn{9}{|c|}{ $m_{H^{-}_2}(\widetilde \tau_L) = 141.29$, \, $m_{H^{-}_3}(\widetilde \mu_L) = 269.37$, \, $m_{H^{-}_4} (\widetilde e_L) = 425.31$}\\
\multicolumn{9}{|c|}{$m_{H^{-}_5} (H^-_d) = 601.69$, \, $m_{H^{-}_6} (\widetilde \tau_R) = 1003.12$, \, $m_{H^{-}_7}(\widetilde \mu_L) = 1003.19 $, \, $m_{H^{-}_8} (\widetilde e_R)= 1003.19 $}\\ \hline  \hline
%%%
\multicolumn{9}{|c|}{ $|Z^H_{h_1 \widetilde{{\nu}}^{\mathcal{R}}_{eR}}|^2 = 32.12\, \% $, \, $|Z^H_{h_1 \widetilde{{\nu}}^{\mathcal{R}}_{\mu R}}|^2 = 31.73\, \% $, \,  $|Z^H_{h_1 \widetilde{{\nu}}^{\mathcal{R}}_{\tau R}}|^2 = 31.34\, \% $, \,  $|Z^H_{h_1H_d^{\mathcal{R}}} |^2 = 4.78\, \% $, \, $|Z^H_{h_1H_u^{\mathcal{R}}} |^2 = 1.42\, \% $}\\ 
%%%
\multicolumn{9}{|c|}{ $|Z^H_{h_3 \widetilde{{\nu}}^{\mathcal{R}}_{eR}}|^2 = 62.64\, \% $, \, $|Z^H_{h_3 \widetilde{{\nu}}^{\mathcal{R}}_{\mu R}}|^2 = 31.89\, \% $, \,  $|Z^H_{h_3 \widetilde{{\nu}}^{\mathcal{R}}_{\tau R}}|^2 = 5.07\, \% $, \,  $|Z^H_{h_3 H_d^{\mathcal{R}}} |^2 = 0.027\, \% $, \, $|Z^H_{h_3 H_u^{\mathcal{R}}} |^2 = 0.361\, \% $}\\ 
%%%
\multicolumn{9}{|c|}{ $|Z^H_{h_4 \widetilde{{\nu}}^{\mathcal{R}}_{eR}}|^2 = 3.73\, \% $, \, $|Z^H_{h_4 \widetilde{{\nu}}^{\mathcal{R}}_{\mu R}}|^2 = 34.82\, \% $, \,  $|Z^H_{h_4 \widetilde{{\nu}}^{\mathcal{R}}_{\tau R}}|^2 = 60.92\, \% $, \, $|Z^H_{h_4H_d^{\mathcal{R}}} |^2 = 0.035\, \% $, \, $|Z^H_{h_4H_u^{\mathcal{R}}} |^2 = 0.49\, \% $}\\ 
%%%
\multicolumn{9}{|c|}{  $|Z^H_{h_5 \widetilde{{\nu}}^{\mathcal{R}}_{eR}}|^2 = 0.058\, \% $, \, $|Z^H_{h_5 \widetilde{{\nu}}^{\mathcal{R}}_{\mu R}}|^2 = 0.10\, \% $, \,  $|Z^H_{h_5 \widetilde{{\nu}}^{\mathcal{R}}_{\tau R}}|^2 = 1.20\, \% $, \,  $|Z^H_{h_5H_d^{\mathcal{R}}} |^2 = 6.19\, \% $, \, $|Z^H_{h_5H_u^{\mathcal{R}}} |^2 = 92.45\, \% $}\\ \hline \hline
%%%
\multicolumn{9}{|c|}{ BR($h_1 \to bb$) = 0.858, \, BR($h_1 \to \tau\tau$) = 0.14, \, BR($h_1 \to WW$) = $3 \times 10^{-8}$ }\\
\multicolumn{9}{|c|}{ BR($h_1 \to ZZ$) = 0.0, \, BR($h_1 \to \gamma\gamma$) = 0.000038, \, BR($h_1 \to \widetilde \chi^0 \widetilde \chi^0 $) =$ 1.2\times 10^{-9}$, \, BR($h_1 \to gg$) = 0.000324}\\ \hline  \hline
%%%
\multicolumn{9}{|c|}{ BR($h_3 \to bb$) = 0.00023, \, BR($h_3 \to \tau\tau$) = 0.000037, \, BR($h_3 \to WW$) = 0.00014 }\\
\multicolumn{9}{|c|}{ BR($h_3 \to ZZ$) = 0.000016, \, BR($h_3 \to \gamma\gamma$) = 0.0000014, \, BR($h_3 \to \widetilde \chi^0 \widetilde \chi^0 $) = 0.99, \, BR($h_3 \to gg$) = 0.000058}\\ \hline  \hline
%BR($h_2 \to gg$) = 0.000058
%%%
\multicolumn{9}{|c|}{ BR($h_4 \to bb$) = 0.000314, \, BR($h_4 \to \tau\tau$) = 0.000051, \, BR($h_4 \to WW$) = 0.000213 }\\
\multicolumn{9}{|c|}{ BR($h_4 \to ZZ$) = 0.000024, \, BR($h_4 \to \gamma\gamma$) = 0.0000021, \, BR($h_4 \to \widetilde \chi^0 \widetilde \chi^0 $) = 0.99, \, BR($h_4 \to gg$) = 0.000082 }\\ \hline  \hline
%BR($h_3 \to gg$) = 0.000082
%%%
\multicolumn{9}{|c|}{ BR($h_5 \to bb$) = 0.37, \, BR($h_5 \to \tau\tau$) = 0.060, \, BR($h_5 \to WW$) = 0.293 }\\
\multicolumn{9}{|c|}{ BR($h_5 \to ZZ$) = 0.0337, \, BR($h_5 \to \gamma\gamma$) = 0.00280,\, BR($h_5 \to \widetilde \chi^0 \widetilde \chi^0 $) = 0.115, \, BR($h_5 \to gg$) = 0.1 }\\ \hline \hline
%BR($h_4 \to gg$) = 0.1
\multicolumn{9}{|c|}{ $\Gamma^{\text{tot}}_{h_1} = 7.85 \times 10^{-4}$, \, $\Gamma^{\text{tot}}_{h_3} = 2.58 \times 10^{-2}$, \, $\Gamma^{\text{tot}}_{h_4} = 2.42 \times 10^{-2}$, \, $\Gamma^{\text{tot}}_{h_5} =  3.64 \times 10^{-3}$ }\\  
%\hline  \hline
%%
 \clineB{1-9}{3}
\end{tabular} }
\renewcommand{\arraystretch}{1}
\end{center}
\caption{The same as in Table.~\ref{S3-R1}, but for the benchmark point
{\bf S3-2h1} 
%from scan $S_3$,
with several singlet-like scalars $h_{1,3,4}$ of masses close to the mass of the SM-like Higgs $h_5$. 
Their branching ratios 
are shown in the sixth-nineth boxes,
and their decay widths in the tenth box.
}
  \label{S3-3-scalars}
\end{table}

%%%%%%%%%%%%%%%%%%%%%%%%%%%%%%%%%%%%%%%
\begin{table}[ht]
\begin{center}
\renewcommand{\arraystretch}{1.4}
\resizebox{15cm}{!}{
\begin{tabular}{|c|c|c|c|c|c|c|c|c|}
\clineB{1-9}{3}
\multicolumn{9}{|c|}{\bf S3-2h2} \\ 
\clineB{1-9}{3}\hline
%%%%%%%%%
\multicolumn{9}{|c|}{$\lambda = 0.57$, \, $\kappa = 0.88$, \, $v_R/\sqrt 2 = 119.81$   ($\mu = 204.87$, \, $\mathcal{M} = 210.86$) }\\ 
\multicolumn{9}{|c|}{$\tan\beta =  4.30 $, \, $T_\lambda = 407.73$,\, $-T_\kappa = 237.76$, \, $-T_{u_3} = 1814.90$, \, $M_{\widetilde Q_{3L}} = 776.67$}\\ \hline \hline
%%%%%%%%%
\multicolumn{9}{|c|}{$m_{h_1}(\widetilde \nu^{\mathcal{R}}_R) = 126.32$, \, $m_{h_2} (H^{\mathcal{R}}_u) = 127.57$, \, $m_{h_3}(\widetilde \nu^{\mathcal{R}}_{\tau L}) = 130.17$, \, $m_{h_4} (\widetilde \nu^{\mathcal{R}}_R)  = 141.17$}\\
\multicolumn{9}{|c|}{$m_{h_5} (\widetilde \nu^{\mathcal{R}}_R) =  147.73 $, \, $m_{h_6} (\widetilde \nu^{\mathcal{R}}_{\mu L}) = 266.90$, \, $m_{h_7} (\widetilde \nu^{\mathcal{R}}_{e L}) = 435.57$, \, $m_{h_8}  (H^{\mathcal{R}}_d) = 896.77$}\\ \hline \hline
%%%%%%%%%
\multicolumn{9}{|c|}{ $m_{A_2}(\widetilde \nu^{\mathcal{I}}_{\tau L}) =  130.17$, \, $m_{A_3} (\widetilde \nu^{\mathcal{I}}_{\mu L})= 266.90$, \, $m_{A_4} (\widetilde \nu^{\mathcal{I}}_R) = 306.44$}\\
\multicolumn{9}{|c|}{$m_{A_5} (\widetilde \nu^{\mathcal{I}}_R) = 306.59$, \, $m_{A_6}(\widetilde \nu^{\mathcal{I}}_R) = 314.75$, \, $m_{A_7} (\widetilde \nu^{\mathcal{I}}_{e L}) = 435.57$, \, $m_{A_8}  (H^{\mathcal{I}}_d) = 892.60$}\\ \hline \hline
%%%%%%%%%
\multicolumn{9}{|c|}{ $m_{H^{-}_2}(\widetilde \tau_L) = 144.78$, \, $m_{H^{-}_3}(\widetilde \mu_L) = 274.71$, \, $m_{H^{-}_4} (\widetilde e_L) = 439.38 $}\\
\multicolumn{9}{|c|}{$m_{H^{-}_5} (H^-_d) = 863.35$, \, $m_{H^{-}_6}(\widetilde \mu_R) = 1002.71$, \, $m_{H^{-}_7}(\widetilde \tau_R) = 1002.76 $, \, $m_{H^{-}_8} (\widetilde e_R) = 1002.76 $}\\ \hline \hline
%%%%%%%%% ###---------- 
\multicolumn{9}{|c|}{ $|Z^H_{h_1 \widetilde{{\nu}}^{\mathcal{R}}_{eR}}|^2 = 44.66\, \% $, \,$|Z^H_{h_1 \widetilde{{\nu}}^{\mathcal{R}}_{\mu R}}|^2 = 18.82\, \% $, \,$|Z^H_{h_1 \widetilde{{\nu}}^{\mathcal{R}}_{\tau R}}|^2 = 10.25\, \% $, \, $|Z^H_{h_1 H_d^{\mathcal{R}}} |^2 = 0.17 \, \% $, \, $|Z^H_{h_1 H_u^{\mathcal{R}}} |^2 = 26.08\, \% $}\\ % \hline \hline
%%%%%%%%%
\multicolumn{9}{|c|}{ $|Z^H_{h_2 \widetilde{{\nu}}^{\mathcal{R}}_{eR}}|^2 = 12.27\, \% $, \,$|Z^H_{h_2 \widetilde{{\nu}}^{\mathcal{R}}_{\mu R}}|^2 = 6.54\, \% $, \,$|Z^H_{h_2 \widetilde{{\nu}}^{\mathcal{R}}_{\tau R}}|^2 = 4.49\, \% $, \, $|Z^H_{h_2H_d^{\mathcal{R}}} |^2 = 7.69 \, \% $, \, $|Z^H_{h_2H_u^{\mathcal{R}}} |^2 = 69.00\, \% $}\\  %\hline \hline
%%%%%%%%%  
\multicolumn{9}{|c|}{ $|Z^H_{h_4 \widetilde{{\nu}}^{\mathcal{R}}_{eR}}|^2 = 39.08\, \% $, \,$|Z^H_{h_4 \widetilde{{\nu}}^{\mathcal{R}}_{\mu R}}|^2 = 55.00\, \% $, \,$|Z^H_{h_4 \widetilde{{\nu}}^{\mathcal{R}}_{\tau R}}|^2 = 5.68\, \% $, \, $|Z^H_{h_4H_d^{\mathcal{R}}} |^2 =  0.099\, \% $, \, $|Z^H_{h_4H_u^{\mathcal{R}}} |^2 = 0.13\, \% $}\\  %\hline \hline
%%%%%%%%%
\multicolumn{9}{|c|}{ $|Z^H_{h_5 \widetilde{{\nu}}^{\mathcal{R}}_{eR}}|^2 = 2.88\, \% $, \,$|Z^H_{h_5 \widetilde{{\nu}}^{\mathcal{R}}_{\mu R}}|^2 = 18.52 \, \% $, \,$|Z^H_{h_5 \widetilde{{\nu}}^{\mathcal{R}}_{\tau R}}|^2 = 78.44\, \% $, \, $|Z^H_{h_5 H_d^{\mathcal{R}}} |^2 = 0.064 \, \% $, \, $|Z^H_{h_5 H_u^{\mathcal{R}}} |^2 = 0.092\, \% $}\\ 
\hline \hline
%%%%%%%%%
\multicolumn{9}{|c|}{ BR($h_1 \to bb$) = 0.127, \, BR($h_1 \to \tau\tau$) = 0.019, \, BR($h_1 \to WW$) = 0.51 } \\
\multicolumn{9}{|c|}{ BR($h_1 \to ZZ$) = 0.0572, \,  BR($h_1 \to \gamma\gamma$) = 0.00852, \, 
BR($h_1 \to gg $) = 0.226} \\ \hline \hline
%%%%%%%%%
\multicolumn{9}{|c|}{ BR($h_2 \to bb$) = 0.615, \, BR($h_2 \to \tau\tau$) = 0.0933, \, BR($h_2 \to WW$) = 0.195 } \\
\multicolumn{9}{|c|}{ BR($h_2 \to ZZ$) = 0.0225, \,  BR($h_2 \to \gamma\gamma$) = 0.00155, \, 
BR($h_2 \to gg $) = 0.0577} \\
\hline \hline
%%%%%%%%%%
%\multicolumn{9}{|c|}{ BR($h_4 \to bb$) = 0.777, \, BR($h_4 \to \tau\tau$) = 0.118, \, BR($h_4 \to WW$) = 0.073 } \\
%\multicolumn{9}{|c|}{ BR($h_4 \to ZZ$) = 0.0096, \,  BR($h_4 \to \gamma\gamma$) = 0.00114, \, 
%BR($h_4 \to gg $) = 0.0178} \\ 
%%%%%%%%%%
%\multicolumn{9}{|c|}{ BR($h_4 \to bb$) = 0.777, \, BR($h_4 \to \tau\tau$) = 0.118, \, BR($h_4 \to WW$) = 0.073 } \\
%\multicolumn{9}{|c|}{ BR($h_4 \to ZZ$) = 0.0096, \,  BR($h_4 \to \gamma\gamma$) = 0.00114, \, 
%BR($h_4 \to gg $) = 0.0178} \\ 
%%%%%%%%%%
%\multicolumn{9}{|c|}{ BR($h_5 \to bb$) = 0.7, \, BR($h_5 \to \tau\tau$) = 0.107, \, BR($h_5 \to WW$) = 0.1455 } \\
%\multicolumn{9}{|c|}{ BR($h_5 \to ZZ$) = 0.0178, \,  BR($h_5 \to \gamma\gamma$) = 0.00129, \, 
%BR($h_4 \to gg $) = 0.0194} \\ 
%\hline \hline
%%%%%%%%%
\multicolumn{9}{|c|}{ $\Gamma^{\text{tot}}_{h_1} = 4.47 \times 10^{-4}$, \,  $\Gamma^{\text{tot}}_{h_2} = 4.27 \times 10^{-3}$ }\\ 
%%%%%%%%%
\clineB{1-9}{3}
\end{tabular} }
\renewcommand{\arraystretch}{1}
\end{center}
\caption{
The same as in Table.~\ref{S3-R1}, but for the benchmark point
{\bf S3-2h2} 
%from scan $S_3$,
with several singlet-like scalars $h_{1,4,5}$ of masses close to the mass of the SM-like Higgs $h_2$. 
the branching ratios of $h_2$, and $h_1$ which contributes to the superposition of signals,  
are shown in the sixth and seventh boxes,
and their decay widths in the eight box.
%The same as in Table.~\ref{S3-R1}
%Benchmark point (S3-ss1) from scan $S_3$, with several scalars of masses close to the mass of the SM-like Higgs.
%Input parameters at the low scale $M_{\text{EWSB}}$ are given in the first box,
%where we also show for completeness $\mu=3\lambda v_R/\sqrt 2$ and ${\mathcal{M}=2\kappa v_R/\sqrt 2}$ since their values determine Higgsino and right-handed neutrino masses.
%Scalar, pseudoscalar and charged Higgs masses are shown in the second, third and fourth boxes, respectively.
%Scalar mass eigenstates are denoted by $h_{1,...,8}$, pseudoscalars by 
%$A_{2,...,8}$ and charged Higgses
%by $H^-_{2,...,8}$ associating the first states to the Goldstone bosons
%eaten by the $Z$ and $W^{\pm}$.
%Their dominant composition is written in brackets.
%For the case of the SM-like Higgs and singlet-like scalars their main compositions are broken down in the fifth box.
%\R{What about the pseudoscalars which are also very mixed.}
%Their branching ratios are shown in the sixth-nineth boxes.
%Their decay widths are shown in the tenth box.
%VEVs, soft parameters, sparticle masses and decay widths 
%are given in GeV.
}
\label{S3-ss1}
\end{table}
%%%%%%%%%%%%%%%%%%%%%%%%%%%%%%%%%%%%%%%%%%%%%%%%%%%%%%%%%%%%%%%%%%%%%%%%%%%%%%%%%%%%

\clearpage
%%%%%%%%%%%%%%%%%%%%%%%%%%%%%%%%%%%%%%%%%%%%%%%%%%%%%%%%%%%%%%%%%%%%%%%%%%
%%%%%%%%%%%%         Acknowledgments              %%%%%%%%%%%%%%%%%%%%%%%%
%%%%%%%%%%%%%%%%%%%%%%%%%%%%%%%%%%%%%%%%%%%%%%%%%%%%%%%%%%%%%%%%%%%%%%%%%%

\bibliographystyle{utphys}
\bibliography{higgsbib_v1}

\end{document}